\documentclass[twoside,12pt]{article}
\usepackage{epsfig}

\usepackage{cite}
\usepackage{amssymb}
\usepackage{graphicx}

\def\NPB{{\em Nucl. Phys.} B}

\def\PLB{{\em Phys. Lett.} B}

\def\PL{{\em Phys. Lett.}}
\def\PRL{{\em Phys. Rev. Lett.}}
\def\PREV{{\em Phys. Rev.}}
\def\PREP{{\em Phys. Rep.}}

\def\PRD{{\em Phys. Rev.} D}

\def\ZPC{{\em Z. Phys.} C}

\newcommand{\be}{\begin{equation}}
\newcommand{\ee}{\end{equation}}
\newcommand{\bea}{\begin{eqnarray}}
\newcommand{\eea}{\end{eqnarray}}
\newcommand{\nn}{\nonumber}
\def\myJournal#1#2#3#4{{#1} {#2} (#3) #4 }
\def\ARNPS{{\em Ann. Rev. Nucl. Part. Sci.}}
\def\MPL{{\em Mod. Phys. Lett.}}
\def\MP{{\em Int. J. Mod. Phys.}}
\def\RPP{{\em Rep. Prog. Phys.}}
\newcommand{\eqn}[1]{(\ref{#1})}
\newcommand{\bel}[1]{\be\label{#1}}
\newcommand{\ba}{\begin{array}{c}}
\newcommand{\bat}{\begin{array}{cc}}
\newcommand{\ea}{\end{array}}

\newcommand{\bi}{\begin{itemize}}
\newcommand{\ei}{\end{itemize}}
\def\etal{{\it et al}}
\newcommand{\Br}{\mathrm{Br}}
\newcommand{\lrder}{\stackrel{\leftrightarrow}{\partial}}

\newcommand{\cL}{{\cal L}}
\newcommand{\cJ}{{\cal J}}
\newcommand{\cM}{{\cal M}}
\newcommand{\cO}{{\cal O}}
\newcommand{\cP}{{\cal P}}

\newcommand{\cA}{{\cal A}}

\newcommand{\ie}{{\it i.e.},\ }

\begin{document}

\title{ \vspace{1cm}\huge\bf Precision Tau Physics}
\author{\\ \\ Antonio Pich 
\\
\\
Departament de F\'\i sica Te\`orica, IFIC, Universitat de Val\`encia -- CSIC\\
 Apartat Correus 22085, E-46071 Val\`encia, Spain}
\date{\mbox{$\phantom{M}$}}

\maketitle
\begin{abstract}
Precise measurements of the lepton properties provide stringent tests of the Standard Model and accurate determinations of its parameters. We overview the present status of $\tau$ physics, highlighting the most recent developments, and discuss the prospects for future improvements. The leptonic decays of the $\tau$ lepton probe the structure of the weak currents and the universality of their couplings to the $W$ boson. The universality of the leptonic $Z$ couplings has also been tested through $Z\to\ell^+\ell^-$ decays. The hadronic $\tau$ decay modes constitute an ideal tool for studying low-energy effects of the strong interaction in very clean conditions. Accurate determinations of the QCD coupling and the Cabibbo mixing $V_{us}$ have been obtained with $\tau$ data. The large mass of the $\tau$ opens the possibility to study many kinematically-allowed exclusive decay modes and extract relevant dynamical information. Violations of flavour and CP conservation laws can also be searched for with $\tau$ decays. Related subjects such as $\mu$ decays, the electron and muon anomalous magnetic moments, neutrino mixing and $B$-meson decays into $\tau$ leptons are briefly covered. Being one the fermions most strongly coupled to the scalar sector,  the $\tau$ lepton is playing now a very important role at the LHC as a tool to test the Higgs properties and search for new physics at higher scales.
\end{abstract}

\vfill\eject
\tableofcontents
\vfill\eject

\section{Introduction}
\label{sec:introduction}

Since its discovery \cite{PE:75} in 1975 at the SPEAR $e^+ e^-$ storage ring, the $\tau$ lepton has been a subject of extensive experimental study
\cite{PE:80,HP:88,BS:88,GP:88,KI:88,PI:90,PI:92,PE:92,RI:92,WS:93,GP:96,PI:98,Pich:1999uk,ST:00,PI:06,Davier:2005xq,Pich:2007cu,Pich:2009zza,Pich:2013kg}.
The very clean sample of boosted $\tau^+\tau^-$ events accumulated at the $Z$ peak, together with the large statistics collected in the $\Upsilon$ region, have considerably improved the statistical accuracy of the $\tau$ measurements and, more importantly, have brought a new level of systematic understanding, allowing us to make sensible tests of the $\tau$ properties.
On the theoretical side, a lot of effort has been invested to improve our understanding of the $\tau$ dynamics. The basic $\tau$ properties were already known, before its actual discovery \cite{PE:75}, thanks to the pioneering paper of Tsai \cite{TS:71}.
The detailed study of higher-order electroweak corrections and QCD contributions has promoted the physics of the $\tau$ lepton to the level of precision tests.

The $\tau$ lepton is a member of the third fermion generation which decays into particles belonging to the first and second ones. Thus, $\tau$ physics could provide some
clues to the puzzle of the recurring families of leptons and quarks.
In fact, one naively expects the heavier fermions to be more sensitive to
whatever dynamics is responsible for the fermion mass generation.
The pure leptonic or semileptonic character of $\tau$  decays
provides a clean laboratory to test the structure of the weak
currents  and the universality of their couplings to the gauge bosons.
Moreover, the  $\tau$ is the only known lepton massive enough to  decay  into  hadrons;
its  semileptonic decays are then an ideal tool for studying
strong interaction effects in  very clean conditions.

All experimental results obtained so far confirm the Standard Model (SM) scenario, in which the $\tau$ is a sequential lepton with its own quantum number and associated neutrino. The increased sensitivities of the most recent experiments result in interesting limits on possible new physics contributions to the $\tau$ decay amplitudes. In the following, the present knowledge on the $\tau$ lepton and the prospects for further improvements are analysed. Rather than giving a detailed review of experimental results, the emphasis is put on the physics which can be investigated with the $\tau$ data. Exhaustive information on more experimental aspects can be found in
Refs.~\cite{HFAG} and \cite{Beringer:1900zz}.

The leptonic $\tau$ decays can be accurately predicted in the SM. The relevant expressions are analysed in section~\ref{sec:LeptonDecays}, where they are compared with the most recent measurements of the $\mu$ and $\tau$ leptonic decay widths, and used to test the universality of the leptonic $W$ couplings in section~\ref{sec:universality},
which also includes the universality tests performed with $\pi$, $K$ and $W$ decays. The Lorentz structure of the leptonic charged-current interactions is further discussed in section~\ref{sec:current}. While the high-precision muon data shows nicely that the bulk of the $\mu$ decay amplitude is indeed of the predicted $V-A$ type, the Lorentz structure of the $\tau$ decay is not yet determined by data; nevertheless, useful constraints on hypothetical new-physics contributions have been established. Section~\ref{sec:nc} describes the leptonic electroweak precision tests performed at the $Z$ peak, confirming the family-universality of the leptonic $Z$ couplings and the existence of (only) three SM neutrino flavours.

The hadronic decays of the $\tau$ lepton allow us to investigate the hadronic weak currents and test low-energy aspects of the strong interaction. The exclusive decay modes are discussed in section~\ref{sec:Hadron}, which shows that at very low energies the chiral symmetry of QCD determines the coupling of any number of pseudoscalar mesons to the left-handed quark current. The measured hadronic distributions in $\tau$ decay provide crucial information on the resonance dynamics, which dominates at higher momentum transfer. Section~\ref{sec:inclusive} discusses the short-distance QCD analysis of the inclusive hadronic width of the $\tau$ lepton. The total hadronic width is currently known with four-loop accuracy, providing a very precise determination of the QCD coupling at the $\tau$ mass scale and, therefore, a very significant test of asymptotic freedom from its comparison with determinations performed at much higher energies.
The inclusive hadronic distribution gives, in addition, important information on non-perturbative QCD parameters. The semi-inclusive hadronic decay width into Cabibbo-suppressed modes is analysed in section~\ref{sec:Vus}, where a quite competitive determination of $|V_{us}|$ is obtained; the accuracy of this result could be considerably improved in the future with much higher statistics.

Together with hadronic $e^+e^-$ data, the hadronic $\tau$-decay distributions are needed to determine the SM prediction for the $\mu$ anomalous magnetic moment. Section~\ref{sec:dipole} presents an overview of the $e$, $\mu$ and $\tau$ magnetic, electric and weak dipole moments, which are expected to have a high sensitivity to physics beyond the SM. The $\tau$ lepton constitutes a superb probe to search for new-physics signals. The current status of CP-violating asymmetries in $\tau$ decays is described in section~\ref{sec:CPV}, while section~\ref{sec:B} discusses the production of $\tau$ leptons in $B$ decays, which is sensitive to new-physics contributions with couplings proportional to fermion masses. The large $\tau$ mass allows one to investigate lepton-flavour and lepton-number violation, through a broad range of kinematically-allowed decay modes,
complementing the high-precision searches performed in $\mu$ decay.
The current experimental limits are given in section~\ref{sec:LFV}; they provide stringent constraints on flavour models beyond the SM.

Processes with $\tau$ leptons in the final state are playing now an important role at the LHC, either to characterize the Higgs properties or to search for new particles at higher scales. The current status is briefly described in section~\ref{sec:LHC}, before concluding with a few summarizing comments in section~\ref{sec:outlook}.

\section{Lepton Decays}
\label{sec:LeptonDecays}

\begin{figure}[tb]\centering
\begin{minipage}[t]{.35\linewidth}\centering
\includegraphics[height=3cm]{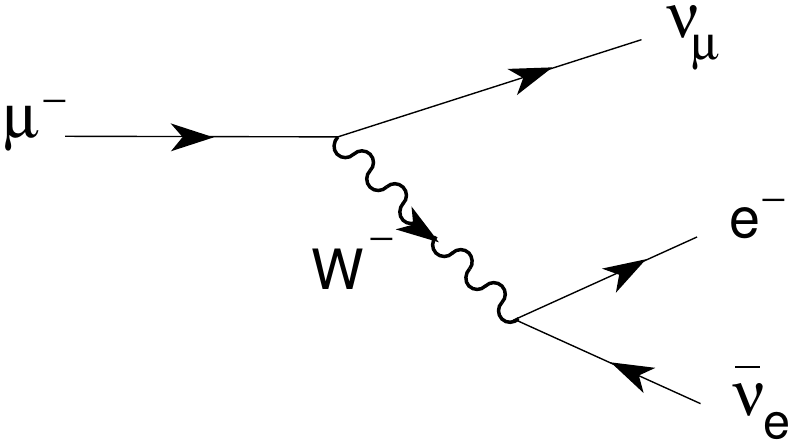}
\end{minipage}
\hskip 1.5cm
\begin{minipage}[t]{.45\linewidth}\centering
\includegraphics[height=3cm]{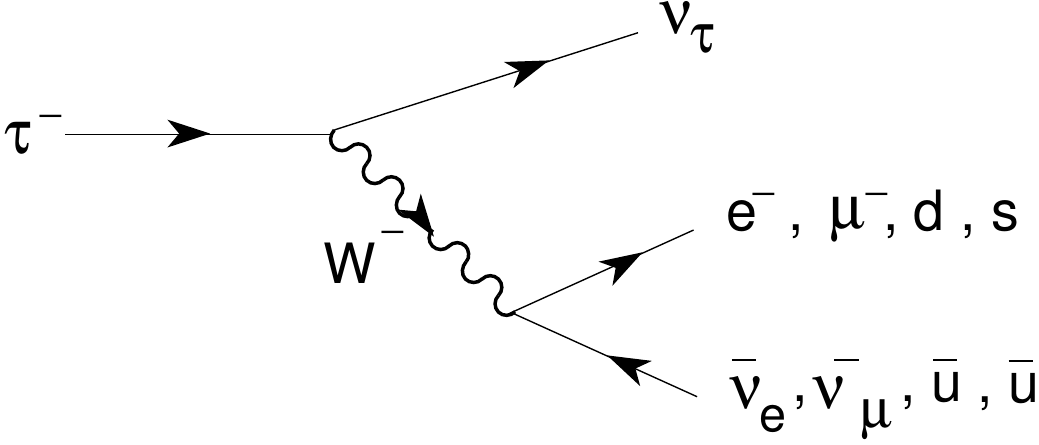}
\end{minipage}
\caption{Feynman diagrams for \ $\mu^-\to e^-\bar\nu_e\,\nu_\mu$ \
and \ $\tau^-\to\nu_\tau X^-$ ($X^-=e^-\bar\nu_e,\, \mu^-\bar\nu_\mu,\,
d\bar u,\, s\bar u$).}
\label{fig:mu_decay}
\end{figure}

The decays of the charged leptons, $\mu^-$ and $\tau^-$, proceed through the $W$-exchange diagrams shown in Fig.~\ref{fig:mu_decay}, with the universal SM strength associated with the charged-current interactions:
\bel{eq:L_cc}
\cL_{\mathrm{CC}} \, = \, - {g \over 2 \sqrt{2}} \, W_{\mu}^\dagger \,
   \left\{ \sum_\ell\, \bar{\nu}_\ell
      \gamma^{\mu} (1 - \gamma_5) \ell  \; + \; \bar{u} \gamma^{\mu}
      (1 - \gamma_5) \left( V_{ud}\, d + V_{us}\, s\right)
       \right\} \, + \, \mathrm{h.c.}\, .
\ee
The momentum transfer carried by the intermediate $W^-$ is very small compared to $M_W$. Therefore, the vector-boson propagator shrinks
to a point and can be well approximated through a local four-fermion interaction governed by the  Fermi coupling constant $G_F/\sqrt{2} = g^2/(8 M_W^2)$.
The leptonic decay widths are given by
\begin{equation}
\label{eq:leptonic}
\Gamma_{\ell\to \ell'} \, \equiv \,
\Gamma [\ell^-\to \ell'^-\bar\nu_{\ell'}\nu_\ell (\gamma)]  \, = \,
  {G_{\ell'\ell}^2 m_{\ell}^5 \over 192 \pi^3} \,
  f\!\left({m_{\ell'}^2 / m_{\ell}^2}\right) \, \left( 1 + \delta_{\mathrm{\scriptstyle RC}}^{\ell'\ell}\right) \,
\end{equation}
where
$\, f(x) = 1-8x+8x^3-x^4-12x^2\log{x}\,$, and  \cite{Behrends:1955mb,Berman:1958ti,Kinoshita:1958ru,Roos:1971mj,Marciano:1988vm,Ferroglia:1999tg,vRS:99,Steinhauser:1999bx,PC:08}
\bel{eq:qed_corr}
\delta_{\mathrm{\scriptstyle RC}}^{\ell'\ell} \, =\,
{\alpha\over 2\pi}\,\left[{25\over 4}-\pi^2 + \cO\left(\frac{m_{\ell'}^2}{ m_{\ell}^2}\right)\right]\, +\,\cdots
\ee
takes into account radiative QED corrections, which are known to $\cO(\alpha^2)$.
The tiny neutrino masses have been neglected and ($\gamma$)
represents additional photons or lepton pairs which have been included inclusively in
$\delta_{\mathrm{\scriptstyle RC}}^{\ell'\ell}$.
Higher-order electroweak corrections and the non-local structure of the $W$ propagator, are usually incorporated into the effective coupling \cite{Ferroglia:2013dga,Fael:2013pja}
\bel{eq:G_eff}
G_{\ell'\ell}^2\; =\; \left[\frac{g^2}{4\sqrt{2} M_W^2}\,\left( 1 + \Delta r\right)\right]^2\;
\left[ 1 +{3\over 5}{m_\ell^2\over M_W^2} +{9\over 5} {m_{\ell'}^2\over M_W^2}
+\cO\left(\frac{m_{\ell'}^4}{M_W^2m_\ell^2} \right)\right]\, ,
\ee
so that $G_{e\mu}$ coincides with the Fermi coupling defined in the $V-A$ theory.

Since $\tau_\mu^{-1} = \Gamma [\mu^-\to e^-\bar\nu_e\nu_\mu (\gamma)]$, the Fermi coupling is determined by the muon lifetime. The MuLan collaboration has recently achieved a very accurate measurement of $\tau_\mu$ with a precision of $1.0$ parts per million \cite{MuLan:10}, more than 15 times as precise as any previous experiment. It is the most accurate particle lifetime ever measured and, consequently, dominates the world average
$\tau_\mu=2.196\, 981\, 1\, (22)\times 10^{-6}$ s \cite{Beringer:1900zz}. Combined with the electron and muon masses, $m_e = 0.510\, 998\, 928\, (11)$ MeV and
$m_\mu = 105.658\, 371\, 5\, (35)$~MeV \cite{Beringer:1900zz}, it implies
\bel{eq:gf}
G_F \,\equiv\, G_{e\mu}
\, = \, (1.166\, 378\, 7\pm 0.000\, 000\, 6)\times 10^{-5} \:\mbox{\rm GeV}^{-2}\, .
\ee
%

%
\begin{table}[tb]
\caption{Average values of some basic $\tau$ parameters  \protect\cite{HFAG,Beringer:1900zz,Belous:2013dba}.}
\label{tab:parameters}
\renewcommand{\arraystretch}{1.1}
\centering
\vspace{0.2cm}
\begin{tabular}{|c|c|}
\hline & \\[-4mm]
$m_\tau$ & $(1776.82\pm 0.16)$~MeV \\
$\tau_\tau$ &\, $(290.29\pm 0.53)\times 10^{-15}$ s\,
\\
Br($\tau^-\to\nu_{\tau}e^-\bar{\nu}_e$)   
   & $(17.818\pm 0.041)\% $ \\
\, Br($\tau^-\to\nu_{\tau}\mu^-\bar{\nu}_\mu$)\,   
   & $(17.392\pm 0.040)\% $ \\
$B_\mu / B_e$ & $0.9761\pm 0.0028$ \\
Br($\tau^-\to\nu_\tau\pi^-$) & $(10.811\pm 0.053)\% $ \\
Br($\tau^-\to\nu_\tau K^-$) & $(0.6955\pm 0.0096)\% $ \\[1mm]
\hline
\end{tabular}
\end{table}
%

Owing to its much heavier mass,
the $\tau$ lepton has several final states which are kinematically allowed:
$\tau^-\to\nu_\tau e^-\bar\nu_e$,
$\tau^-\to\nu_\tau\mu^-\bar\nu_\mu$,
$\tau^-\to\nu_\tau d\bar u$ and $\tau^-\to\nu_\tau s\bar u$.
The universality of the $W$ couplings implies that all these
decay modes have equal amplitudes (if final fermion masses and
QCD interactions are neglected), except for an additional
$N_C |V_{ui}|^2$ factor ($i=d,s$) in the semileptonic
channels, where $N_C=3$ is the number of quark colours.
Taking into account the unitarity of the quark mixing matrix,
$|V_{ud}|^2 + |V_{us}|^2 = 1 - |V_{ub}|^2\approx 1$,
one easily gets a lowest-order estimate for the $\tau$ lifetime,
\bel{eq:tau_lifetime}
\tau_\tau\,\equiv\,\frac{1}{\Gamma(\tau)}\, \approx\,
\left\{\Gamma(\mu) \left({m_\tau\over m_\mu}\right)^5
\left[ 2 + N_C
\left( |V_{ud}|^2 + |V_{us}|^2\right)\right]\right\}^{-1}
\,\approx\, \frac{1}{5}\,\tau_\mu\,\left({m_\mu\over m_\tau}\right)^5
\, =\,3.3\times 10^{-13}\:\mathrm{s}\, ,
\ee
while the branching ratios for the different channels are expected to be approximately
($\ell=e, \mu$)
\bel{eq:naive}
 B_\ell \,\equiv\,
    \mathrm{Br}(\tau^- \rightarrow \nu_{\tau}\ell^- \bar{\nu}_\ell) \,
    \simeq \, {1 \over 5} \, = \, 20\% \, ,
\qquad\qquad\quad
 {\Gamma (\tau^- \rightarrow \nu_{\tau} +
  \mathrm{hadrons}) \over
        \Gamma (\tau^- \rightarrow \nu_{\tau} e^- \bar{\nu}_e ) }
     \, \simeq \, N_C \, = \, 3 \, .
\ee
The agreement with the experimental values given in table~\ref{tab:parameters} provides strong evidence for the quark colour degree of freedom.
The numerical differences are mainly due to the missing QCD corrections which
enhance the hadronic $\tau$ decay width by about 20\% (see section~\ref{sec:inclusive}).

Using the value of $G_F$ measured in $\mu$ decay and taking into account the final fermion masses and higher-order corrections, Eq.~\eqn{eq:leptonic} provides a precise relation between the $\tau$ lifetime and the leptonic branching ratios:
\bel{relation}
B_e \, = \, {B_\mu \over 0.972559\pm 0.000005} \, =\,
\frac{\tau_{\tau}}{(1632.9 \pm 0.6) \times 10^{-15}\, \mathrm{s}} \, .
\ee
The quoted errors reflect the present uncertainty of $0.16$ MeV
in the value of $m_\tau$.

\begin{figure}[tb]
\begin{center}
\includegraphics[height=5.5cm]{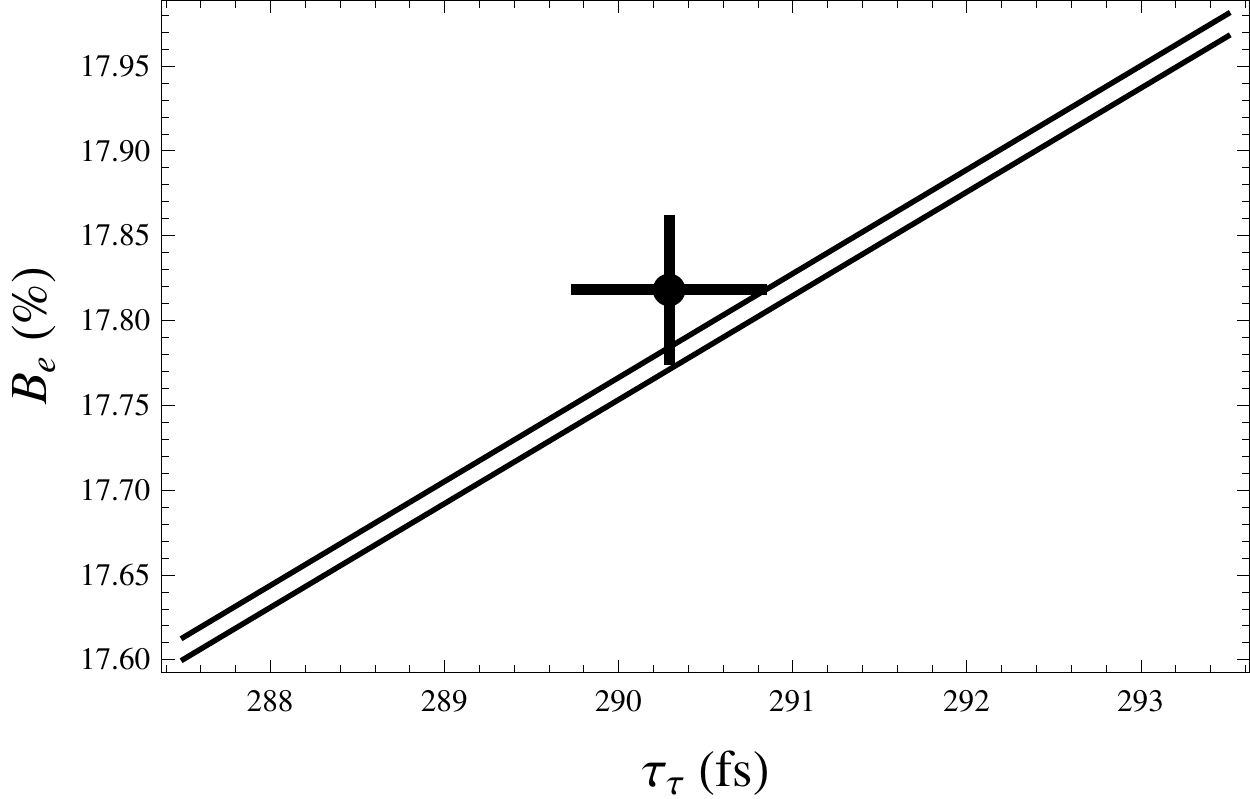}
\caption{Relation between $B_e$ and $\tau_\tau$. The diagonal
band corresponds to Eq.~(\protect\ref{relation}).}
\label{fig:BeLife}
\end{center}
\end{figure}

The predicted value of $B_\mu/B_e$ is in fair agreement with the measured ratio
given in table~\ref{tab:parameters}.  The small deviation ($1.3\,\sigma$) originates in the recent and precise BaBar measurement $B_\mu/B_e = 0.9796 \pm 0.0039$ \cite{Aubert:2009qj}, which is $1.8\,\sigma$ away from the theoretical value. The previous world-average without B-factory measurements, $B_\mu/B_e = 0.9725 \pm 0.0039$, was in perfect agreement with the SM prediction. As shown in Fig.~\ref{fig:BeLife}, the relation between $B_e$ and $\tau_\tau$ is also well satisfied by the present data. Notice, that this
relation is very sensitive to the value of the $\tau$ mass.

Using the relations in Eq.~\eqn{relation}, the measured values of $B_e$, $B_\mu$ and $\tau_\tau$ can be combined together to obtain an improved determination of the electronic branching fraction within the SM, \ie assuming the universality of the $W^\pm$ couplings:
\bel{eq:B_e_univ}
B_e^{\mathrm{uni}}\, =\, (17.818\pm 0.024)\%\, .
\ee

The $\tau$ leptonic branching fractions and the $\tau$ lifetime are
known with a precision of $0.2\%$, far away from the impressive $10^{-6}$ accuracy achieved for the muon lifetime. Belle has just released the first B-factory measurement of the $\tau$ lifetime, $\tau_\tau = (290.17\pm 0.53\pm 0.33)\times 10^{-15}$~s \cite{Belous:2013dba}, with a total uncertainty smaller than the PDG 2012 average $\tau_\tau = (290.6\pm 1.0)\times 10^{-15}$~s \cite{Beringer:1900zz}, leading to the new world average in table~\ref{tab:parameters}.
The Belle result is based on $e^+e^-\to\tau^+\tau^-$ events collected on the $\Upsilon(4S)$ resonance, where both $\tau$ leptons decay to $3\pi\nu_\tau$.

The $\tau$ mass is only known at the $10^{-4}$ level. Making an energy scan of $\sigma(e^+e^-\to\tau^+\tau^-)$ around the $\tau^+\tau^-$ production threshold \cite{Voloshin:2002mv,Smith:1993vp,RuizFemenia:2001fa},
the BES-III experiment aims to reach an accuracy better than 0.1~MeV \cite{Asner:2008nq}.
Uncertainties comparable to the $m_\tau$ world average have been already reached in
a preliminary analysis, using around $20\:\mathrm{pb}^{-1}$ of $\tau$ data \cite{Mo}.

\section{Lepton Universality}
\label{sec:universality}

In the SM all lepton doublets have identical couplings to the $W$ boson. Comparing the measured decay widths of leptonic or semileptonic decays which only differ in the lepton flavour, one can test experimentally that the $W$ interaction is indeed the same,
\ie that \ $g_e = g_\mu = g_\tau \equiv g\, $.
The $B_\mu/B_e$ ratio constrains $|g_\mu/g_e|$, while the
$B_e/\tau_\tau$ relation provides information on $|g_\tau/g_\mu|$.
The present results are shown in table~\ref{tab:ccuniv}, together with the constraints
obtained from leptonic $\pi$, $K$ and $W$ decays.

\begin{table}[th]\centering
\caption{Experimental determinations of the ratios \ $g_\ell/g_{\ell'}$.}
\label{tab:ccuniv}
\vspace{0.2cm}
\renewcommand{\arraystretch}{1.1}
\begin{tabular}{c@{\hspace{1cm}}c@{\hspace{0.7cm}}c@{\hspace{0.7cm}}c@{\hspace{0.7cm}}c@{\hspace{0.7cm}}c}
\hline &&&&&\\[-4mm] &
 $\Gamma_{\tau\to\mu}/\Gamma_{\tau\to e}$ &
 $\Gamma_{\pi\to\mu} /\Gamma_{\pi\to e}$ &
 $\Gamma_{K\to\mu} /\Gamma_{K\to e}$ &
 $\Gamma_{K\to\pi\mu} /\Gamma_{K\to\pi e}$ &
 $\Gamma_{W\to\mu} /\Gamma_{W\to e}$
\\ &&&&&\\[-4mm] \hline &&&&&\\[-4mm]
 $|g_\mu/g_e|$
 & $1.0018\; (14)$ & $1.0021\; (16)$ & $0.9978\; (20)$ & $1.0010\; (25)$ & $0.996\; (10)$
\\ &&&&&\\[-4mm] \hline\hline &&&&&\\[-4mm]  &
 $\Gamma_{\tau\to e}/\Gamma_{\mu\to e}$ &
 $\Gamma_{\tau\to\pi}/\Gamma_{\pi\to\mu}$ &
 $\Gamma_{\tau\to K}/\Gamma_{K\to\mu}$ &
 $\Gamma_{W\to\tau}/\Gamma_{W\to\mu}$
\\ &&&&&\\[-4mm] \hline &&&&&\\[-4mm]
 $|g_\tau/g_\mu|$
 & $1.0011\; (15)$ & $0.9962\; (27)$ & $0.9858\; (70)$ & $1.034\; (13)$
\\ &&&&&\\[-4mm] \hline\hline &&&&&\\[-4mm]  &
 $\Gamma_{\tau\to\mu}/\Gamma_{\mu\to e}$
 & $\Gamma_{W\to\tau}/\Gamma_{W\to e}$
\\ &&&&&\\[-4mm] \hline &&&&&\\[-4mm]
 $|g_\tau/g_e|$
 & $1.0030\; (15)$ & $1.031\; (13)$
\\ &&&&&\\[-4mm] \hline
\end{tabular}
\end{table}

The $\tau$ determination of $|g_\mu/g_e|$ is already as precise ($\sim 0.15\%$) as the one obtained from $\pi_{\ell 2}$ decays and comparable accuracies have been reached recently with $K_{\ell 2}$ and $K_{\ell 3}$ data. The ratios ($P=\pi,K$)
\bel{eq:P_l2}
R_{P\to e/\mu}\;\equiv\;\frac{\Gamma[P^-\to e^-\bar\nu_e (\gamma)]}{
\Gamma[P^-\to\mu^-\bar\nu_\mu (\gamma)]}\; =\; \left|\frac{g_e}{g_\mu}\right|^2\;
\frac{m_e^2}{m_\mu^2}\; \left( \frac{1 - m_e^2/m_P^2}{1 - m_\mu^2/m_P^2}\right)^2\;
\left( 1 +\delta R_{P\to e/\mu}\right)
\ee
have been calculated and measured with high accuracy. Within the SM, the leptonic decay rate of a pseudoscalar meson is helicity suppressed as a consequence of the $V-A$ structure of the charged currents, making these ratios sensitive probes of new-physics interactions.
The known radiative corrections $\delta R_{P\to e/\mu}$ include a summation of leading QED logarithms $\alpha^n\log^n{(m_\mu/m_e)}$ \cite{Marciano:1993sh,Finkemeier:1995gi} and a systematic two-loop calculation of $\cO(e^2 p^4)$ effects within Chiral Perturbation Theory \cite{Cirigliano:2007ga}. The comparison between the SM predictions \cite{Cirigliano:2007ga}
\bel{eq:R_Pl2_SM}
R_{\pi\to e/\mu}^{\mathrm{SM}}\; = \; (1.2352\pm 0.0001)\times 10^{-4}\, ,
\qquad\qquad
R_{K\to e/\mu}^{\mathrm{SM}}\; = \; (2.477\pm 0.001)\times 10^{-5}\, ,
\ee
and the experimental $R_{\pi\to e/\mu}$ \cite{Czapek:1993kc,Britton:1992pg,Bryman:1985bv}
and $R_{K\to e/\mu}$ ratios \cite{Lazzeroni:2012cx,Ambrosino:2009aa},
\bel{eq:pi_l2}
R_{\pi\to e/\mu}\; =\; (1.230\pm 0.004)\times 10^{-4}\, ,
\qquad\qquad
R_{K\to e/\mu}\; = \; (2.488\pm 0.010)\times 10^{-5}\, ,
\ee
gives the results quoted in table~\ref{tab:ccuniv}.
Ongoing experiments at PSI \cite{Pocanic:2009gd} and TRIUMF \cite{Malbrunot:2012zz}
are expected to improve the experimental precision of $R_{\pi\to e/\mu}$ from 0.3\% to 0.05\%.

The $K\to\pi\ell\bar\nu_\ell$ decays proceed without any helicity suppression; the differences between the muon and electron modes stem mainly from isospin and phase space
\cite{Cirigliano:2011ny}. Both sets of $K_{\ell 3}$ data are then used to determine the Cabibbo mixing between the quarks of the first and second generations. Comparing the $V_{us}$ values obtained from $K\to\pi e\bar\nu_e$ and $K\to\pi \mu\bar\nu_\mu$ decays, one obtains the corresponding $|g_\mu/g_e|$ value in table~\ref{tab:ccuniv} \cite{Antonelli:2010yf}.

The decay modes $\tau^-\to\nu_\tau\pi^-$ and $\tau^-\to\nu_\tau K^-$
can also be used to test universality through the ratios
\bel{eq:R_tP}
R_{\tau/P} \; \equiv\;
 {\Gamma(\tau^-\to\nu_\tau P^-) \over
 \Gamma(P^-\to \mu^-\bar\nu_\mu)}\; =\;
\Big\vert {g_\tau\over g_\mu}\Big\vert^2\; {m_\tau^3\over 2 m_P m_\mu^2}\;
{(1-m_P^2/ m_\tau^2)^2\over (1-m_\mu^2/ m_P^2)^2}\;
\left( 1 + \delta R_{\tau/P}\right)\, ,
\ee
where the dependence on the hadronic matrix elements (the so-called meson
decay constants $f_P$) factors out.
Owing to the different energy scales involved, the radiative
corrections to the $\tau^-\to\nu_\tau P^-$ amplitudes
are however not the same than the corresponding effects in
$P^-\to\mu^-\bar\nu_\mu$. The size of the relative
correction has been roughly estimated to be \cite{Marciano:1993sh,Decker:1994dd}:
\bel{eq:dR_tp_tk}
\delta R_{\tau/\pi} = (0.16\pm 0.14)\% \ ,
\qquad\qquad\qquad
\delta R_{\tau/K} = (0.90\pm 0.22)\%  \ .
\ee
Using these numbers, the measured $\tau^-\to\pi^-\nu_\tau$
and $\tau^-\to K^-\nu_\tau$ decay rates imply the
$|g_\tau/g_\mu|$ ratios given in table~\ref{tab:ccuniv}.

Due to the limited statistics available, the direct leptonic decays of the $W$ boson
only test universality at the 1\% level. The LEP data contains a slight excess of events in $W\to\nu_\tau\tau$, implying a $2.6\,\sigma$ ($2.4\,\sigma$) deviation from universality in $|g_\tau/g_\mu|$ ($|g_\tau/g_e|$) \cite{Beringer:1900zz,Schael:2013ita}.
This discrepancy cannot be easily understood,
given the stringent limits on $|g_\tau/g_{e,\mu}|$ from $W$-mediated decays \cite{Filipuzzi:2012mg}.
Future high-energy $e^+e^-$ colliders should clarify whether this is a real physical effect or just a statistical fluctuation.

The present data verify the universality of the leptonic
charged-current couplings to the 0.15\%
level. It is important to realize the complementarity of the
different universality tests. The pure leptonic decay modes probe
the couplings of a transverse $W$. In contrast,
the semileptonic decays $P^-\to \ell^-\bar\nu_\ell$ and $\tau^-\to\nu_\tau P^-$ are only
sensitive to the spin-0 piece of the charged current; thus,
they could unveil the presence of possible scalar-exchange
contributions with Yukawa-like couplings proportional to some
power of the charged-lepton mass.
One can easily imagine new-physics scenarios which would modify
differently the two types of leptonic couplings.
For instance, in the usual two-Higgs doublet models charged-scalar exchange
generates a correction to the ratio $B_\mu/B_e$, but
$R_{P\to e/\mu}$ remains unaffected \cite{Jung:2010ik}.

\section{Lorentz Structure of the Charged Current}
\label{sec:current}

Let us consider the leptonic decays $\ell^-\to\ell'^-\bar\nu_{\ell'}\nu_\ell $,
where the lepton pair ($\ell$, $\ell^\prime $)
may be ($\mu$, $e$), ($\tau$, $e$) or ($\tau$, $\mu$).
With high statistics, these leptonic decay modes allow us to investigate the Lorentz structure of the decay amplitudes through the analysis of the energy and angular distribution of the final charged lepton, complemented with polarization information whenever available.

The most general, local, derivative-free, lepton-number conserving,
four-lepton interaction Hamiltonian, consistent with locality and Lorentz invariance
\cite{MI:50,BM:57,KS:57,SCH:83,FGJ:86,FG:93,PS:95},
\be\label{eq:hamiltonian}
{\cal H} \; =\;  4\, \frac{G_{\ell'\ell}}{\sqrt{2}}\;
\sum_{n,\epsilon,\omega}\,
g^n_{\epsilon\omega}\,
\left[ \overline{\ell'_\epsilon}
\Gamma^n {(\nu_{\ell'})}_\sigma \right]\,
\left[ \overline{({\nu_\ell})_\lambda} \Gamma_n
	\ell_\omega \right]\, ,
\ee
contains ten complex coupling constants or, since a common phase is
arbitrary, nineteen independent real parameters
which could be different for each leptonic decay.
The subindices $\epsilon , \omega , \sigma, \lambda$ label
the chiralities (left-handed, right-handed)  of the  corresponding  fermions, and
$n=S,V,T$ the type of interaction: scalar ($\Gamma^S=I$), vector ($\Gamma^V=\gamma^\mu$), tensor
($\Gamma^T=\sigma^{\mu\nu}/\sqrt{2}$). For given $n, \epsilon ,
\omega $, the neutrino chiralities $\sigma $ and $\lambda$
are uniquely determined.
Taking out a common factor $G_{\ell'\ell}$, which is determined by the total
decay rate, the coupling constants $g^n_{\epsilon\omega}$
are normalized to \cite{FGJ:86}
\bea\label{eq:normalization}
1 & = & {1\over 4} \,\left( |g^S_{RR}|^2 + |g^S_{RL}|^2
    + |g^S_{LR}|^2 + |g^S_{LL}|^2 \right)
  \, + \, 3 \,\left( |g^T_{RL}|^2 + |g^T_{LR}|^2 \right)
\nn \\ & &
+\, \left( |g^V_{RR}|^2 + |g^V_{RL}|^2 + |g^V_{LR}|^2 + |g^V_{LL}|^2 \right)
\, .
\eea
Thus, $|g^S_{\epsilon\omega}|\leq 2$, $|g^V_{\epsilon\omega}|\leq 1$ and $|g^T_{\epsilon\omega}|\leq 1/\sqrt{3}$.
It is convenient to introduce the probabilities \cite{FGJ:86}
\bel{eq:Probabilities_Q}
Q_{\epsilon\omega}\; =\; {1\over 4} \, |g^S_{\epsilon\omega}|^2 + |g^V_{\epsilon\omega}|^2 +3 \,\left(1 - \delta_{\epsilon\omega}\right)\, |g^T_{\epsilon\omega}|^2
\ee
for the decay of an $\omega$-handed $\ell^-$ into an $\epsilon$-handed daughter lepton.
In the Standard Model, $g^V_{LL}  = 1$ and all other $g^n_{\epsilon\omega} = 0 $.

For an initial lepton polarization ${\cal P}_\ell$,
the final charged-lepton distribution in the decaying-lepton rest frame
is usually parametrized in the form  \cite{BM:57,KS:57}
\be\label{eq:spectrum}
{d^2\Gamma_{\ell\to \ell'} \over dx\, d\cos{\theta}}\; =\;
{m_\ell\,\omega^4 \over 2\pi^3}\; G_{\ell'\ell}^2\; \sqrt{x^2-x_0^2}\;
\left\{ F(x) - {\xi\over 3}\, {\cal P}_\ell\,\sqrt{x^2-x_0^2}
\,\cos{\theta}\, A(x)\right\}\, ,
\ee
where $\theta$ is the angle between the $\ell^-$ spin and the
final charged-lepton momentum,
$\, \omega \equiv (m_\ell^2 + m_{\ell'}^2)/2 m_\ell \, $
is the maximum $\ell'^-$ energy for massless neutrinos, $x \equiv E_{\ell'} /
\omega$ is the reduced energy, $x_0\equiv m_{\ell'}/\omega$
and
\bea\label{eq:Fx_Ax_def}
F(x)  & = &
  x (1 - x) + {2\over 9}\, \rho
 \left(4 x^2 - 3 x - x_0^2 \right)
+  \eta\, x_0 (1-x)
\, , \nn\\[3pt]
A(x) & = &
 1 - x   + {2\over 3}\,  \delta \left( 4 x - 4 + \sqrt{1-x_0^2}
\right)  \, .
\eea

For an unpolarized lepton $\ell$, the distribution is characterized by
the so-called Michel \cite{MI:50} parameter $\rho$
and the low-energy parameter $\eta$. Two more parameters, $\xi$
and $\delta$, can be determined when the initial lepton polarization is known.
If the polarization of the final charged lepton is also measured,
5 additional independent parameters
($\xi'$, $\xi''$, $\eta''$, $\alpha'$, $\beta'$) appear \cite{Beringer:1900zz}.
In the SM,
$\rho = \delta = 3/4$,
$\eta = \eta'' = \alpha' = \beta' = 0 $ and
$\xi = \xi' = \xi'' = 1 $.

For massless neutrinos, the total decay rate is given by \cite{PS:95}
\be\label{eq:gamma}
\Gamma_{\ell\to \ell'} \; = \; {\widehat G_{\ell'\ell}^2 m_\ell^5\over 192 \pi^3}\;
f\!\left({m_{\ell'}^2/ m_\ell^2}\right)\;
 \left( 1 + \delta_{\mathrm{\scriptstyle RC}}^{\ell'\ell}\right) \, ,
\ee
where
\be\label{eq:Ghat_def}
\widehat G_{\ell'\ell} \;\equiv\; G_{\ell'\ell} \;
\sqrt{1 + 4\,\eta\, {m_{\ell'}\over m_\ell}\,
{g\!\left( m_{\ell'}^2/ m_\ell^2 \right)\over
f\!\left( m_{\ell'}^2/ m_\ell^2 \right)}}
\, ,
\ee
$g(z) = 1 + 9 z - 9 z^2 - z^3 + 6 z (1+z) \log{z}$,
and the SM radiative correction $\delta_{\mathrm{\scriptstyle RC}}^{\ell'\ell}$ has been included.
Since we assume that the SM provides the dominant contribution
to the decay rate, any additional higher-order correction
beyond the effective Hamiltonian (\protect\ref{eq:hamiltonian})
would be a sub\-leading effect.

The normalization $G_{e\mu}$ corresponds to the Fermi coupling
$G_F$, measured in $\mu$ decay.
The $B_\mu/B_e$ and $B_e\tau_\mu/\tau_\tau$
universality tests, discussed in the previous section,
actually probe the ratios $|\widehat G_{\mu\tau}/\widehat G_{e\tau}|$
and $|\widehat G_{e\tau}/\widehat G_{e\mu}|$, respectively.
The experimental determination of $G_{e \mu}$ is then sensitive
to the uncertainty in $\eta_{\mu\to e}$.

In terms of the $g_{\epsilon\omega}^n$ couplings, the shape parameters in Eqs.~\eqn{eq:spectrum} and \eqn{eq:Fx_Ax_def} are:
\bea\label{eq:michel}
\rho  &  = &
{3\over 4}\, (\beta^+ + \beta^-) + (\gamma^+ + \gamma^-) \, ,
\nn\\
\xi &  = &  3\, (\alpha^- - \alpha^+) + (\beta^- - \beta^+)
  + {7\over 3}\, (\gamma^+ - \gamma^-) \, ,
\nn\\
\xi\delta & \!\!\! = & \!\!\!
 {3\over 4}\, (\beta^- - \beta^+) + (\gamma^+ - \gamma^-) \, ,
\nn\\
\eta &  = & \frac{1}{2}\, \mathrm{Re}\!\left[
g^V_{LL} g^{S\ast}_{RR} + g^V_{RR}  g^{S\ast}_{LL}
+ g^V_{LR} \left(g^{S\ast}_{RL} + 6\, g^{T\ast}_{RL}\right)
+ g^V_{RL} \left(g^{S\ast}_{LR} + 6\, g^{T\ast}_{LR}\right) \right] ,
\eea
where \cite{Rouge:2000um}
\bea\label{eq:abg_def}
\lefteqn{\alpha^+ \;\equiv\;
{|g^V_{RL}|}^2 + {1\over 16}\, {|g^S_{RL} + 6\, g^T_{RL}|}^2 \, , }
&\hskip 7cm &
\alpha^- \;\equiv\;
{|g^V_{LR}|}^2 + {1\over 16}\, {|g^S_{LR} + 6\, g^T_{LR}|}^2 \, ,
\nn\\
\lefteqn{\beta^+ \;\equiv\;
 {|g^V_{RR}|}^2 + {1\over 4}\, {|g^S_{RR}|}^2 \, , } &&
\beta^- \;\equiv\;  {|g^V_{LL}|}^2 + {1\over 4}\, {|g^S_{LL}|}^2 \, ,
\nn\\
\lefteqn{\gamma^+ \;\equiv\;
 {3\over 16}\, {|g^S_{RL} - 2\, g^T_{RL}|}^2 \, , } &&
\gamma^- \;\equiv\;
 {3\over 16}\, {|g^S_{LR} - 2\, g^T_{LR}|}^2 \, ,
\eea
are positive-definite combinations of decay constants, corresponding to
a final right-handed ($\alpha^+,\beta^+,\gamma^+$) or
left-handed ($\alpha^-,\beta^-,\gamma^-$) lepton.
The normalization constraint \eqn{eq:normalization} is equivalent to
$\alpha^+ + \alpha^- + \beta^+ + \beta^- + \gamma^+ + \gamma^- = 1$.
The probabilities $Q_{\epsilon\omega}$ can be extracted from the
measurable shape parameters:
\begin{eqnarray}\label{eq:Q_LL}
Q_{LL} & = &  \beta^- \; =\; {1 \over 4}\,\left(
-3 +{16\over 3}\rho -{1\over 3}\xi +{16\over 9}\xi\delta +\xi'+\xi''
\right)\, ,
\nn\\
Q_{RR} & = &  \beta^+ \; =\;  {1 \over 4}\,\left(
-3 +{16\over 3}\rho +{1\over 3}\xi -{16\over 9}\xi\delta -\xi'+\xi''
\right)\, ,
\nn\\
Q_{LR} & = & \alpha^- + \gamma^- \; =\; {1 \over 4}\,\left(
5 -{16\over 3}\rho +{1\over 3}\xi -{16\over 9}\xi\delta +\xi'-\xi''
\right)\, ,
\nn\\
Q_{RL} & = & \alpha^+ + \gamma^+ \; =\; {1 \over 4}\,\left(
5 -{16\over 3}\rho -{1\over 3}\xi +{16\over 9}\xi\delta -\xi'-\xi''
\right)\, . \quad
\end{eqnarray}
Upper bounds on any of these probabilities translate into corresponding limits for all couplings with the given chiralities. Particularly powerful are the relations:
\bea\label{eq:Q_R}
Q_{\ell_R} & \equiv & Q_{RR} + Q_{LR}
\; =\;  \frac{1}{2}\, \left[ 1 + \frac{\xi}{3} - \frac{16}{9} (\xi\delta)\right]
\, ,
\nn\\
Q_{\ell'_R} & \equiv & Q_{RR} +  Q_{RL} \; =\; {1\over 2} ( 1 -\xi') \, .
\eea
The total probability for the decay of an initial right-handed lepton is determined by
$\xi$ and $\xi\delta$, while a single polarization parameter fixes the probability to decay into a final right-handed lepton.
Other useful positive-definite quantities are:
\bea\label{eq:positiveQ}
{3\over 2} \beta^+ + 2 \gamma^-& =& \rho - \xi\delta\, ,
\nn\\
Q_{LR} +  Q_{RL} & =& \frac{1}{2}\, \left[ 5 - \frac{16}{3}\,\rho - \xi''\right] \, .
\eea
The first one provides direct bounds on $|g^V_{RR}|$, $|g^S_{RR}|$ and $|g^S_{LR} - 2\, g^T_{LR}|$, and shows that $\rho \geq \xi\delta$. A precise measurement of the polarization parameter $\xi''$ would imply, through the second identity, upper limits on all couplings  $g^n_{\epsilon\omega}$ with $\epsilon\not=\omega$.

For $\mu$ decay, where precise measurements of the polarizations of
both $\mu$ and $e$ have been performed, there exist \cite{FGJ:86}
upper bounds on $Q_{RR}$, $Q_{LR}$ and $Q_{RL}$, and a lower limit
on $Q_{LL}$. They imply corresponding upper bounds on the 8
couplings $|g^n_{RR}|$, $|g^n_{LR}|$ and $|g^n_{RL}|$.
The measurements of the $\mu^-$ and the $e^-$ do not allow to
determine $|g^S_{LL}|$ and $|g^V_{LL}|$ separately \cite{FGJ:86,JA:66}.
Nevertheless, a lower limit on $|g^V_{LL}|$ is obtained from the inverse muon decay because
$\sigma(\nu_\mu e^-\to\mu^-\nu_e)\propto |g^V_{LL}|^2$ \cite{FGJ:86}.
Since the helicity of the $\nu_\mu$ in pion decay is experimentally known \cite{RO:82,JO:88} to be $-1/2$ with high precision, $|2\, h_{\nu_\mu} +1| < 0.0041$ (90\% CL) \cite{FE:84},
the $|g^S_{LL}|$ contribution to $\sigma(\nu_\mu e^-\to\mu^-\nu_e)$ is negligible; contributions from other $g^n_{\epsilon\omega}$ couplings \cite{Mursula:1982em} are severely suppressed by the $\mu$-decay constraints. Once a lower bound on $|g^V_{LL}|$ has been set, the relation $Q_{LL}<1$ provides the upper limit $|g^S_{LL}|^2 < 4 \left( 1 - |g^V_{LL}|^2\right)$.
The present 90\% CL bounds on the $\mu$-decay couplings \cite{TWIST:2011aa,Gagliardi:2005fg,Mishra:1990yf,Vilain:1996yf}
are given in table~\ref{tab:mu_couplings}. These limits show nicely that the bulk of the $\mu$-decay transition amplitude is indeed of the predicted V$-$A type.

\begin{table}[tb]\centering
\caption{90\% CL experimental bounds for the $\mu^-\to e^-\bar\nu_e\nu_\mu$ couplings
\cite{FG:12}}
\label{tab:mu_couplings}
\vspace{0.2cm}
\renewcommand{\tabcolsep}{1.2pc} 
\renewcommand{\arraystretch}{1.2} 
\begin{tabular}{cccc}
\hline\\[-4mm]
 $|g_{RR}^S| < 0.035$ & $|g_{LR}^S| < 0.050$ & $|g_{RL}^S| < 0.412$ & $|g_{LL}^S| < 0.550$ \\
 $|g_{RR}^V| < 0.017$ & $|g_{LR}^V| < 0.023$ & $|g_{RL}^V| < 0.104$ & $|g_{LL}^V| > 0.960$ \\
 $|g_{RR}^T| \equiv 0$ & $|g_{LR}^T| < 0.015$ & $|g_{RL}^T| < 0.103$ & $|g_{LL}^T|\equiv 0$ \\[5pt]
 \multicolumn{4}{c}{$|g_{LR}^S + 6\, g_{LR}^T|< 0.143$ \hskip .75cm $|g_{LR}^S + 2\, g_{LR}^T|< 0.108$ \hskip .75cm  $|g_{LR}^S - 2\, g_{LR}^T|< 0.070$}\\
 \multicolumn{4}{c}{$|g_{RL}^S + 6\, g_{RL}^T|< 0.418$ \hskip .75cm
 $|g_{RL}^S + 2\, g_{RL}^T|< 0.417$ \hskip .75cm
 $|g_{RL}^S - 2\, g_{RL}^T|< 0.418$} \\[5pt]
 \multicolumn{4}{c}{$Q_{RR}+ Q_{LR}\, <\, 8.2\times 10^{-4}$}\\
\\[-4mm] \hline
\end{tabular}
\end{table}

The experimental analysis of the $\tau$ decay parameters is necessarily
different from the one applied to the muon, because of the much
shorter $\tau$ lifetime.
The measurement of the $\tau$ polarization and the parameters
$\xi$ and $\delta$ is still possible due to the fact that the spins
of the $\tau^+\tau^-$ pair produced in $e^+e^-$ annihilation
are strongly correlated
\cite{TS:71,KST:73,PS:77,KW:84,NE:91,GN:91,FE:90,Bernabeu:1990na,Alemany:1991ki,Davier:1992nw}.
Another possibility is to use the beam polarization, as done by SLD \cite{Abe:1997dy}.
However, the polarization of the charged lepton emitted in the $\tau$ decay
has never been measured. In principle, this could be done
for the decay $\tau^-\to\mu^-\bar\nu_\mu\nu_\tau$ by stopping the
muons and detecting their decay products \cite{FE:90}.
An alternative method would be \cite{SV:96} to use the radiative decays
$\tau\to \ell^-\bar\nu_\ell\nu_\tau\gamma$ ($\ell=e,\mu$),
since the distribution of the photons emitted by the daughter lepton is
sensitive to the lepton polarization.
The measurement of the inverse decay $\nu_\tau \ell^-\to\tau^-\nu_\ell$
looks far out of reach.

The experimental status on the $\tau$-decay Michel parameters \cite{Abe:1997dy,Heister:2001me,Abreu:2000sg,Ackerstaff:1998yk,Acciarri:1998as,Albrecht:1997gn,Alexander:1997bv,Janssen:1989wg,Ford:1987ha}
is shown in table~\ref{tab:tau_michel}. For comparison, the more accurate values measured in $\mu$ decay \cite{TWIST:2011aa,FG:12,Danneberg:2005xv,Balke:1988by,Beltrami:1987ne,Burkard:1985wn,Derenzo:1969za} are also given. Table~\ref{tab:Michel_tau} gives the resulting 95\% CL bounds on the $\tau$-decay couplings.

\begin{table}[bt]   
\caption{Michel parameters \protect\cite{Beringer:1900zz}.
The last column 
assumes identical couplings for $\ell=e,\mu$.
$\xi_{\mu\to e}$ refers to the product $\xi_{\mu\to e}\cP_\mu$,
where $\cP_\mu\approx 1$ is the longitudinal polarization
of the $\mu$ from $\pi$ decay.}
\label{tab:tau_michel}
\vspace{0.2cm}
\renewcommand{\tabcolsep}{1.2pc} 
\renewcommand{\arraystretch}{1.2} 
\centering
\begin{tabular}{|c|c|c|c|c|}
\hline &&&&\\[-5.5mm]
& $\mu^-\to e^-\bar\nu_e\nu_\mu$ & $\tau^-\to\mu^-\bar\nu_\mu\nu_\tau$ & $\tau^-\to e^-\bar\nu_e\nu_\tau$ & $\tau^-\to \ell^-\bar\nu_\ell\nu_\tau$
\\ &&&&\\[-5.5mm]\hline &&&&\\[-4mm]
$\rho$ & $0.74979\pm 0.00026$ & $0.763\pm 0.020$ & $0.747\pm 0.010$ &
$0.745\pm 0.008$
\\
$\eta$ & $0.057\pm 0.034$ & $0.094\pm 0.073$ & --- &
$0.013\pm 0.020$
\\
$\xi$ & $1.0009\; {}^{+\;\: 0.0016}_{-\;\: 0.0007}\;\;$ & $1.030\pm 0.059$ & $0.994\pm 0.040$ &
$0.985\pm 0.030$
\\
$\xi\delta$ & $0.7511\; {}^{+\;\: 0.0012}_{-\;\: 0.0006}\;\;$ & $0.778\pm 0.037$ & $ 0.734\pm 0.028$ & $ 0.746\pm 0.021$
\\
$\xi'$ & $1.00\pm 0.04$ & --- & --- & ---
\\
$\xi''$ & $0.65\pm 0.36$ & --- & --- & ---
\\[2mm] \hline
\end{tabular}
\end{table}

\begin{table}[tb]\centering
\caption{95\% CL experimental bounds for the leptonic $\tau$-decay couplings
\cite{Stahl:12}}
\label{tab:Michel_tau}
 \vspace{0.2cm}
 \renewcommand{\tabcolsep}{1.2pc} 
 \renewcommand{\arraystretch}{1.2} 
 \begin{tabular}{cccc}
 \hline \\[-5.5mm] \multicolumn{4}{c}{$\tau^-\to e^-\bar\nu_e\nu_\tau$}\\ \\[-5.5mm] \hline\\[-4mm]
 $|g_{RR}^S| < 0.70$ & $|g_{LR}^S| < 0.99$ & $|g_{RL}^S| \leq 2$
 & $|g_{LL}^S| \leq 2$ \\
 $|g_{RR}^V| < 0.17$ & $|g_{LR}^V| < 0.13$ & $|g_{RL}^V| < 0.52$
 & $|g_{LL}^V| \leq 1$ \\
 $|g_{RR}^T| \equiv 0$ & $|g_{LR}^T| < 0.082$ & $|g_{RL}^T| < 0.51$
 & $|g_{LL}^T|\equiv 0$ \\ \\[-4mm]
 \hline\hline \\[-5.5mm]
 \multicolumn{4}{c}{$\tau^-\to \mu^-\bar\nu_\mu\nu_\tau$}\\ \\[-5.5mm] \hline \\[-4mm]
 $|g_{RR}^S| < 0.72$ & $|g_{LR}^S| < 0.95$ & $|g_{RL}^S| \leq 2$
 & $|g_{LL}^S| \leq 2$ \\
 $|g_{RR}^V| < 0.18$ & $|g_{LR}^V| < 0.12$ & $|g_{RL}^V| < 0.52$
 & $|g_{LL}^V| \leq 1$ \\
 $|g_{RR}^T| \equiv 0$ & $|g_{LR}^T| < 0.079$ & $|g_{RL}^T| < 0.51$
 & $|g_{LL}^T|\equiv 0$ \\ \\[-4mm]
 \hline
 \end{tabular}
\end{table}

If lepton universality is assumed, the leptonic decay ratios $B_\mu/B_e$  and $B_e\tau_\mu/\tau_\tau$ provide limits on the low-energy parameter $\eta$.
The best sensitivity \cite{Stahl:1993yk} comes from $\widehat{G}_{\mu\tau}$,
where the term proportional to $\eta$ is not suppressed by
the small $m_e/m_\ell$ factor. The world-averaged value of the $B_\mu/B_e$ ratio implies
then:
\be\label{eq:eta_univ}
\eta_{\tau\to \ell} \; = \; 0.016\pm 0.013  \, ,
\ee
which only assumes $e/\mu$ universality.
This determination is more accurate that the $\mu\to e$ and $\tau\to\mu$
ones in table~\ref{tab:tau_michel}, obtained from the shape of the energy distribution.
It is also slightly more accurate than the value quoted in the table for $\tau\to\ell$.
A non-zero value of $\eta$ would show that there are at least two
different couplings with opposite chiralities for the charged leptons.
Assuming the V$-$A coupling $g_{LL}^V$ to be dominant, the
second one would be a scalar coupling $g^S_{RR}$.
To first order in new physics contributions,
$\eta\approx\mathrm{Re}(g^S_{RR})/2$;
Eq.~(\ref{eq:eta_univ}) puts then the 95\% CL bound:
$-0.019 \, <\mbox{\rm Re}(g^S_{RR}) < 0.083$.

\subsection{Model-Dependent Constraints}

The sensitivity of the present $\tau$ data is not good enough to get strong constraints from a completely general analysis of the four-fermion Hamiltonian. Nevertheless, better limits can be obtained within particular models. For instance, let us assume that there are no tensor couplings, \ie $g^T_{\epsilon\omega}=0$. This condition is satisfied in any model where the interactions are mediated by vector bosons and/or charged scalars.
In this case, the quantities  $(1-\frac{4}{3}\rho)$, $(1-\frac{4}{3}\xi\delta)$ and
$(1-\frac{4}{3}\rho) + \frac{1}{2} (1-\xi)$ reduce to sums of $|g^n_{\epsilon\omega}|^2$,
which are positive semidefinite; \ie in the absence of tensor couplings, $\rho\leq\frac{3}{4}$, $\xi\delta\leq\frac{3}{4}$ and $(1-\xi) > 2\, (\frac{4}{3}\rho - 1)$ \cite{PS:95}.
The corresponding limits on the couplings $g^n_{\epsilon\omega}$ are shown in table~\ref{tab:coup_notensor}. For $\mu$ decay one obtains a sizeable improvement of
$|g^V_{RL}|$ by a factor of 6, and a 4\% reduction of the $|g^S_{LR}|$ upper bound.
In the $\tau$ decay modes, stronger bounds are obtained for $|g^S_{RR}|$, $|g^S_{LR}|$ and $|g^V_{RL}|$.

\begin{table}[tb]
\caption{Limits on the couplings $g^n_{\epsilon\omega}$, assuming that there are no tensor couplings. The $\tau$-decay ($\mu$-decay) values are at 95\% CL (90\% CL).}
\label{tab:coup_notensor}
\centering
\vspace{0.2cm}
\renewcommand{\tabcolsep}{0.7pc} 
\renewcommand{\arraystretch}{1.2} 
\begin{tabular}{|c|cccc|cccc|}
\hline &&&&&&&&\\[-4mm]
& $|g^S_{RR}|$  & $|g^S_{LR}|$  & $|g^S_{RL}|$  & $|g^S_{LL}|$  &
$|g^V_{RR}|$  & $|g^V_{LR}|$  & $|g^V_{RL}|$  & $|g^V_{LL}|$
\\ &&&&&&&&\\[-4mm]\hline &&&&&&&&\\[-4mm] $\mu\to e$ &
$< 0.035$ & $< 0.048$ & $< 0.412$ & $< 0.550$  &
$< 0.017$ & $< 0.023$ & $< 0.017$ & $> 0.960$
\\ $\tau\to\mu$ &
$< 0.39$ & $< 0.39$ & $\leq 2$ & $\leq 2$ &
$< 0.18$ & $< 0.12$ & $< 0.14$ & $\leq 1$
\\ $\tau\to e$ &
$< 0.42$ & $< 0.42$ & $\leq 2$ & $\leq 2$ &
$< 0.17$ & $< 0.13$ & $< 0.13$ & $\leq 1$
\\  $\tau\to l$ &
$< 0.34$ & $< 0.34$ & $\leq 2$ & $\leq 2$ &
$< 0.17$ & $< 0.13$ & $< 0.12$ & $\leq 1$
\\[2mm] \hline
\end{tabular}
\end{table}

If one only considers $W$-mediated interactions, but admitting the possibility that the $W$ couples non-universally to leptons of any chirality, the effective couplings factorize into the product of two leptonic $W$ couplings, \ie $g^V_{\ell'_\epsilon \ell_\omega} = \kappa_{\ell'_\epsilon}^* \kappa_{\ell_\omega}^{\phantom{*}}$.
This implies additional relations among the couplings of the effective Hamiltonian, such as $g^V_{LR}\, g^V_{RL} = g^V_{LL}\, g^V_{RR}$ \cite{Mursula:1982em,Mursula:1984zb}, which hold within any of the three channels, $\mu\to e$, $\tau\to e$ and $\tau\to\mu$. Moreover, the effective couplings of the different decay modes get related through identities like
$g^V_{\mu_L \tau_L}\, g^V_{e_L \tau_R}  = g^V_{\mu_L \tau_R}\, g^V_{e_L \tau_L}$ \cite{PS:95}. The resulting limits are given in table~\ref{tab:W_couplings}. Notice the strong lower bounds on the $g^V_{LL}$ couplings, which are a consequence of the normalization condition~\eqn{eq:normalization}. If the lepton decays are mediated by a single vector-particle (the $W$), the effective interaction is constrained in all cases to be mostly left-handed with high accuracy.

\begin{table}[tb]
\caption{Bounds on the $g^V_{\epsilon \omega}$ couplings, assuming that (non-standard) $W$-exchange is the only relevant interaction. The $\tau$-decay ($\mu$-decay) limits are at 95\% CL (90\% CL). Numbers within parentheses use $\mu$-decay data through cross-channel identities.}
\label{tab:W_couplings}
\centering\vspace{0.2cm}
\renewcommand{\tabcolsep}{0.7pc} 
\renewcommand{\arraystretch}{1.2} 
\begin{tabular}{|c|cccc|}
\hline &&&&\\[-4mm] &
$|g^V_{RR}|$ & $|g^V_{LR}|$ & $|g^V_{RL}|$ & $|g^V_{LL}|$
\\ &&&&\\[-4mm]\hline &&&&\\[-4mm]
$\mu\to e$ & $<0.0004$ & $<0.023$ & $<0.017$ & $>0.999$
\\
$\tau\to\mu$ & $<0.017$ ($0.003$) & $<0.12$ & $<0.14$ ($0.023$) & $>0.983$
\\
$\tau\to e$ & $<0.017$ ($0.002$) & $<0.13$ & $<0.13$ ($0.017$) & $>0.983$
\\[2mm]  \hline
\end{tabular}
\end{table}

For $W$-mediated interactions, the hadronic decay modes $\tau^-\to\nu_\tau h^-$ can also be used to test the structure of the $\tau\nu_\tau W$ vertex, if one assumes that the $W$ coupling to the light quarks is the SM one.
The effective Hamiltonian contains only two vector couplings $g_\lambda$, with $\lambda$ being the $\tau$ (and $\nu_\tau$) chirality, with the normalization $|g_L|^2 + |g_R|^2 = 1$. The $\cP_\tau$ dependent part of the decay amplitude is then proportional to $\xi_h= |g_L|^2 - |g_R|^2$, which plays a role analogous to the leptonic decay constant $\xi$.
The parameter $\xi_h$ determines the mean $\nu_\tau$ helicity times a factor of $-2$; $\xi_h = 1$ in the SM.
The analysis of $\tau^+\tau^-$ decay correlations in leptonic-hadronic and hadronic-hadronic decay modes, using the $\pi$, $\rho$ and $a_1$ hadronic final states \cite{Abe:1997dy,Heister:2001me,Abreu:2000sg,Acciarri:1998as,Albrecht:1997gn,Alexander:1997bv,Thurn:1993qw,Asner:1999kj,Ackerstaff:1997dv,Albrecht:1992ka},
gives
$\xi_h = 0.995\pm 0.007$ \cite{{Stahl:12}}. This implies the 95\% CL bounds
$|g_L|> 0.995$ and $|g_R|< 0.10$.

\section{Neutral-Current Couplings}
\label{sec:nc}

\begin{figure}[tb]\centering
\includegraphics[width=12cm]{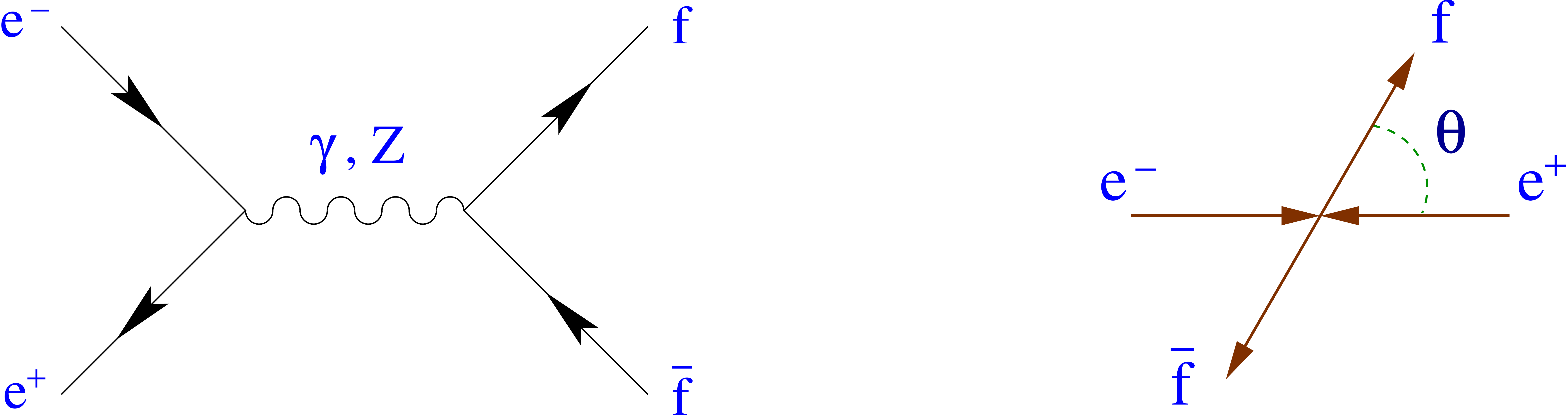}
\caption{Tree-level contributions to \ $e^+e^-\to\bar f f \,$ \
and kinematical configuration in the centre-of-mass system.}
\label{fig:eeZff}
\end{figure}

In the SM, tau pair production in $e^+e^-$ annihilation proceeds through the electromagnetic and weak neutral-current interactions, shown in Fig.~\ref{fig:eeZff}:  $e^+e^-\to\gamma^*,Z^*\to \tau^+\tau^-$. At high energies, where the $Z$ contribution is important, the study of the production cross section allows to extract information on the lepton electroweak parameters. The $Z$ coupling to the fermionic neutral current is given by \cite{Pich:2012sx}
\bel{eq:L_nc}
\cL_{\mathrm{NC}}^Z \; = \; - { g \over 2 \cos{\theta_W}} \;
     Z_\mu \;\sum_f\, \bar f \gamma^\mu (v_f - a_f \gamma_5) f \, ,
\ee
where
$v_f = T_3^f (1-4\, |Q_f|\sin^2{\theta_W})$ and $a_f=T_3^f$,
with $T_3^f=\pm\frac{1}{2}$ the corresponding weak isospin. Thus, the weak neutral couplings are predicted to be the same for all fermions with equal electric charge $Q_f$.

For unpolarized $e^+$ and $e^-$ beams, the differential $e^+e^-\to \ell^+\ell^-$
cross section can be written as
\bel{eq:dif_cross}
{d\sigma\over d\Omega}\; = \; {\alpha^2\over 8 s} \;
         \left\{ A \, (1 + \cos^2{\theta}) \, + B\,  \cos{\theta}\,
     - \, h_\ell\, \left[ C \, (1 + \cos^2{\theta}) \, +\, D \cos{\theta}
         \right] \right\} ,
\ee
where $h_\ell=\pm 1$ denotes the sign of the $\ell^-$ helicity and $\theta$ is
the scattering angle between $e^-$ and $\ell^-$ in the centre-of-mass reference system.
At lowest order,
\bea\label{eq:ABCD}
A & = & 1 + 2\, v_e v_\ell \,\mathrm{Re}(\chi)
 + \left(v_e^2 + a_e^2\right) \left(v_\ell^2 + a_\ell^2\right) |\chi|^2 \, ,
\\
B & = & 4\, a_e a_\ell \,\mathrm{Re}(\chi) + 8\, v_e a_e v_\ell a_\ell  |\chi|^2 \, ,
\\
C & = & 2\, v_e a_l \,\mathrm{Re}(\chi) + 2 \left(v_e^2 + a_e^2\right)
  v_\ell a_\ell |\chi|^2 \, ,
\\
D & = & 4\, a_e v_\ell \,\mathrm{Re}(\chi) + 4\, v_e a_e \left(v_\ell^2 +
      a_\ell^2\right) |\chi|^2 \, ,
\eea
and  $\chi$  contains the $Z$  propagator
\bel{eq:Z_propagator}
\chi \; = \; \frac{G_F M_Z^2}{2 \sqrt{2} \pi \alpha}
     \; \frac{s}{s - M_Z^2 + i s \Gamma_Z  / M_Z } \, .
\ee

The coefficients $A$, $B$, $C$ and $D$ can be experimentally determined, by measuring the total cross section, the forward-backward asymmetry, the polarization asymmetry and the forward-backward polarization asymmetry, respectively:
$$\sigma(s) \; = \; {4 \pi \alpha^2 \over 3 s } \, A \, ,
\hskip 2cm
\cA_{\mathrm{FB}}(s)\;\equiv\;\frac{N_F - N_B}{N_F + N_B}
\; =\;\frac{3}{8}\,\frac{B}{A}\, , $$
$$\cA_{\mathrm{Pol}}(s)  \;\equiv\;\frac{\sigma^{(h_\ell =+1)}
- \sigma^{(h_\ell =-1)}}{\sigma^{(h_\ell =+1)} + \sigma^{(h_\ell = -1)}}
\; = \;  - \frac{C}{A} \, , $$
\be\label{eq:A_FB_pol}
\cA_{\mathrm{FB,Pol}}(s) \;\equiv\;\frac{N_F^{(h_\ell =+1)} -
N_F^{(h_\ell = -1)} - N_B^{(h_\ell =+1)} + N_B^{(h_\ell = -1)}}{
N_F^{(h_\ell =+1)} + N_F^{(h_\ell = -1)} + N_B^{(h_\ell =+1)} + N_B^{(h_\ell = -1)}}
\; = \; -\frac{3}{8}\,\frac{D}{A}\, .
\ee
Here, $N_F$ and $N_B$ denote the number of negative leptons emerging in the forward and backward hemispheres, respectively, with respect to the electron direction.

For $s = M_Z^2$, the real part of the $Z$ propagator vanishes and the photon-exchange terms can be neglected in comparison with the $Z$-exchange contributions
($\Gamma_Z^2 / M_Z^2 \ll 1$). Eqs.~\eqn{eq:A_FB_pol} become then,
\bea\label{eq:A_pol_Z}
\sigma^{0,\ell}  \;\equiv\;  \sigma(M_Z^2)\;  =\;
 \frac{12 \pi}{M_Z^2 } \, \frac{\Gamma_e \Gamma_\ell}{\Gamma_Z^2}\, ,
&& \qquad\quad\;
\cA_{\mathrm{FB}}^{0,\ell}\;\equiv\;\cA_{FB}(M_Z^2)\; =\; \frac{3}{4}\,
\cP_e \cP_\ell \, ,
\nn\\
\cA_{\mathrm{Pol}}^{0,\ell}\; \equiv
  \cA_{\mathrm{Pol}}(M_Z^2)  = \cP_\ell \, ,
\quad && \qquad\quad
\cA_{\mathrm{FB,Pol}}^{0,\ell} \equiv\;
\cA_{\mathrm{FB,Pol}}(M_Z^2)\; =\; \frac{3}{4}\, \cP_e  \, ,
\eea
where $\Gamma_\ell$
is the $Z$ partial decay width to the $\ell^+\ell^-$ final state, and
\bel{eq:P_l}
\cP_\ell \; \equiv \; \frac{- 2 v_\ell a_\ell}{v_\ell^2 + a_\ell^2}
\ee
is the average longitudinal polarization of the lepton $\ell^-$,
which only depends on the ratio of the vector and axial-vector couplings.

Small higher-order corrections can produce large variations on the predicted lepton polarization because $|v_\ell| =\frac{1}{2}\, |1-4\,\sin^2{\theta_W}|\ll 1$. Therefore, $\cP_\ell$ is a very sensitive function of $\sin^2{\theta_W}$, providing an
interesting window to probe electroweak quantum effects.
Subtracting from the data initial-state QED corrections, $\gamma$-exchange and $\gamma$-$Z$ interference contributions, the tiny electroweak boxes and corrections for $s\not= M_Z^2$, one can use the equations given above as effective expressions in terms of loop-corrected
couplings or, equivalently, an effective electroweak mixing angle:
$\sin^2{\theta_{\mathrm{eff}}^{\mathrm{lept}}}\equiv \frac{1}{4}\, (1 - v_\ell/a_\ell)$.

The $Z$ partial decay width to the $\ell^+\ell^-$ final state,
\bel{eq:Z_l_QED}
\Gamma_\ell\;  \equiv\;\Gamma(Z\to \ell^+\ell^-)\; =\;
{G_F M_Z^3\over 6\pi\sqrt{2}} \; \left(v_\ell^2 + a_\ell^2\right)\;
\left(1 + \delta_{\mathrm{RC}}^Z\right)\, ,
\ee
determines the sum $(v_\ell^2 + a_\ell^2)$, while the ratio $v_\ell/a_\ell$
is derived from the asymmetries.\footnote{
The asymmetries determine two possible solutions: $v_\ell/a_\ell = -(1\pm\sqrt{1-\cP_\ell^2})/\cP_\ell$.
This ambiguity can be solved with the energy dependence of $\cA_{\mathrm{FB}}(s)$
\cite{ALEPH:2005ab} or through the measurement of the transverse spin-spin correlation
of the two taus in $Z\to\tau^+\tau^-$ which determines $C_{TT}= (a_\ell^2 - v_\ell^2)/(a_\ell^2 + v_\ell^2)$ \cite{Bernabeu:1990na}. The experimental value
$C_{TT}= 1.01\pm 0.12$ \cite{Barate:1997mz,Abreu:1997vp}
requires  $|v_\tau/a_\tau|<< 1$.}
The absolute signs of $v_\ell$ and $a_\ell$ are established by the convention $a_e<0$.

The measurement of the final polarization asymmetries can (only) be done for
$\ell=\tau$, because the spin polarizations of the $\tau^+$ and $\tau^-$
are reflected in the distorted distribution of their decay products.
Therefore, $\cP_\tau$ and $\cP_e$ can be determined from a
measurement of the spectrum of the final charged particles in the
decay of one $\tau$, or by studying the correlated distributions
between the final products of both taus \cite{KW:84,Alemany:1991ki,Davier:1992nw}.

With polarized $e^+e^-$ beams, which were available at SLC, one
can also study the left-right asymmetry between the cross sections
for initial left- and right-handed electrons, and the corresponding
forward-backward left-right asymmetry:
\bel{eq:A_LR}
\cA_{\mathrm{LR}}^0\;\equiv\;\cA_{\mathrm{LR}}(M_Z^2)\; = \;
\frac{\sigma_L(M_Z^2)- \sigma_R(M_Z^2)}{\sigma_L(M_Z^2) + \sigma_R(M_Z^2)}
\; = \; - \cP_e \,  ,
\qquad\quad
\cA_{\mathrm{FB,LR}}^{0,\ell} \;\equiv\;
\cA_{\mathrm{FB,LR}}(M_Z^2)\; =\; - {3 \over 4}\, \cP_\ell \, .
\ee
At the $Z$ peak, $\cA_{\mathrm{LR}}^0$ measures
the average initial lepton polarization, $\cP_e$,
without any need for final particle identification,
while $\cA_{\mathrm{FB,LR}}^{0,\ell}$ provides a direct determination
of the final fermion polarization.

\begin{table}[tb]
\centering
\caption{Measured values of the leptonic $Z$ decay widths, forward-backward asymmetries and longitudinal polarization of the final $\ell^-$ from polarization asymmetries \cite{Beringer:1900zz,ALEPH:2005ab,ALEPH:2010aa}. The last column shows the combined result (for a massless lepton) assuming lepton universality. The resulting effective vector and axial-vector $Z$ couplings are also given.
\label{tab:LEP_asym}}
\vspace{0.2cm}
\renewcommand{\tabcolsep}{0.7pc} 
\renewcommand{\arraystretch}{1.2} 
\begin{tabular}{|c|cccc|}
\hline
& $e$ & $\mu$ & $\tau$ & $\ell$
\\ \hline &&&&\\[-4mm]
$\Gamma_\ell$ \, (MeV) & $83.91\; (12)$
& $83.99\; (18)$ & $84.08\; (22)$ & $83.984\; (86)$
\\
$\cA_{\mathrm{FB}}^{0,\ell}$ \, (\%) & $1.45\; (25)$
& $1.69\; (13)$ & $1.88\; (17)$ & $1.71\; (10)$
\\
$\cP_\ell$ & $-0.1515\; (19)$
& $-0.142\; (15)$ & $-0.143\; (4)$ &   $-0.1499\; (18)$
\\ &&&&\\[-4mm] \hline &&&&\\[-4mm]
$v_\ell$ & $-0.03817\; (47)$ & $-0.0367\; (23)$ & $-0.0366\; (10)$ & $-0.03783\; (41)$
\\
$a_\ell$ & $-0.50111\; (35)$ & $-0.50120\; (54)$ & $-0.50204\; (64)$ & $-0.50123\; (26)$
\\[2mm] \hline\hline \multicolumn{5}{|c|}{}\\[-4mm]
\multicolumn{5}{|c|}{$v_\mu/v_e = 0.961\; (61) \qquad\qquad\qquad a_\mu/a_e = 1.0002\; (13)$}
\\
\multicolumn{5}{|c|}{$v_\tau/v_e = 0.959\; (29) \qquad\qquad\qquad a_\tau/a_e = 1.0019\; (15)$}
\\[2mm] \hline
\end{tabular}
\end{table}
%

\begin{figure}[tb]\centering
\begin{minipage}[c]{.45\linewidth}\centering
\includegraphics[height=7.5cm]{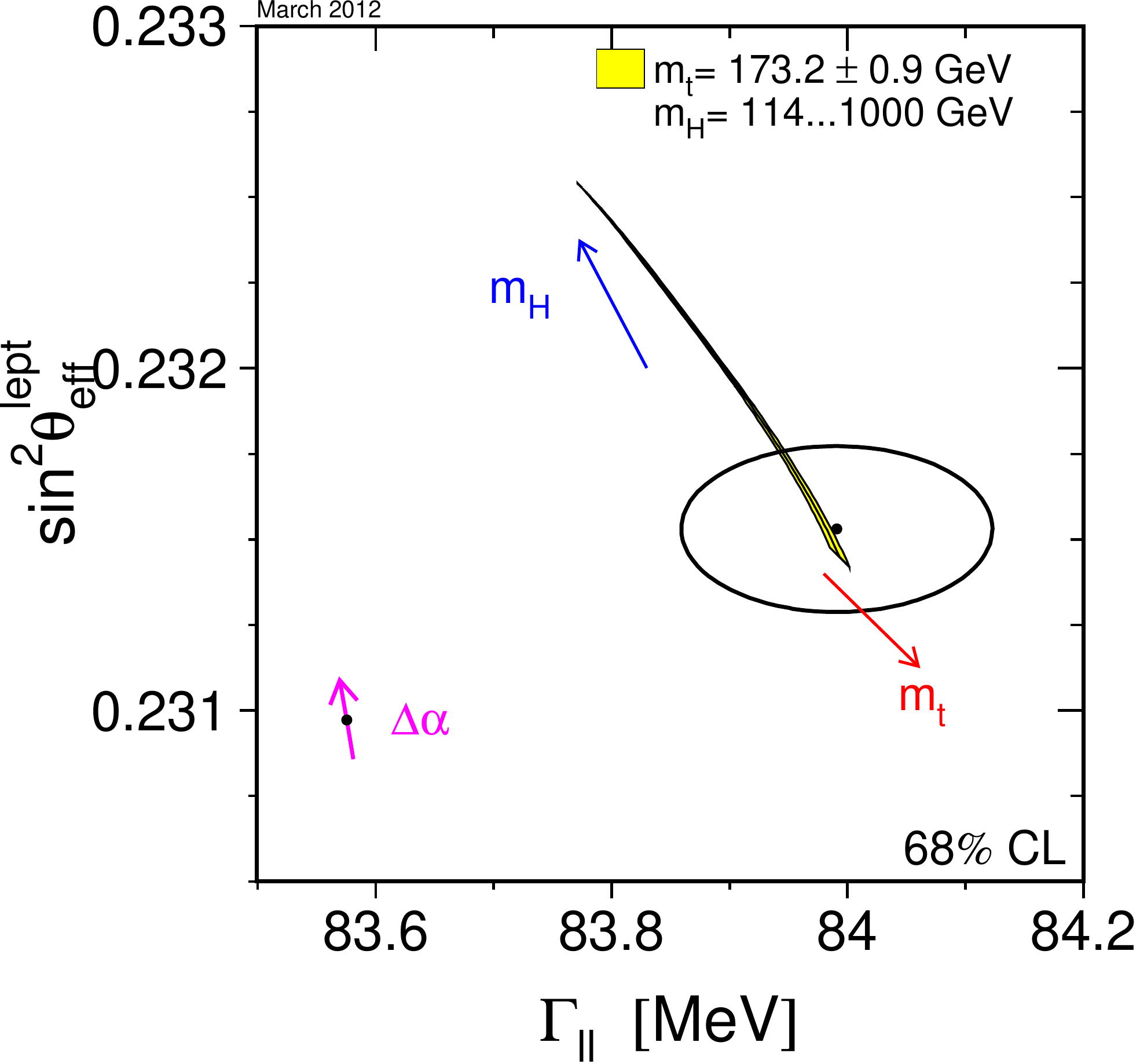}
\end{minipage}
\hskip 1.5cm
\begin{minipage}[c]{.45\linewidth}\centering
\includegraphics[height=7.5cm]{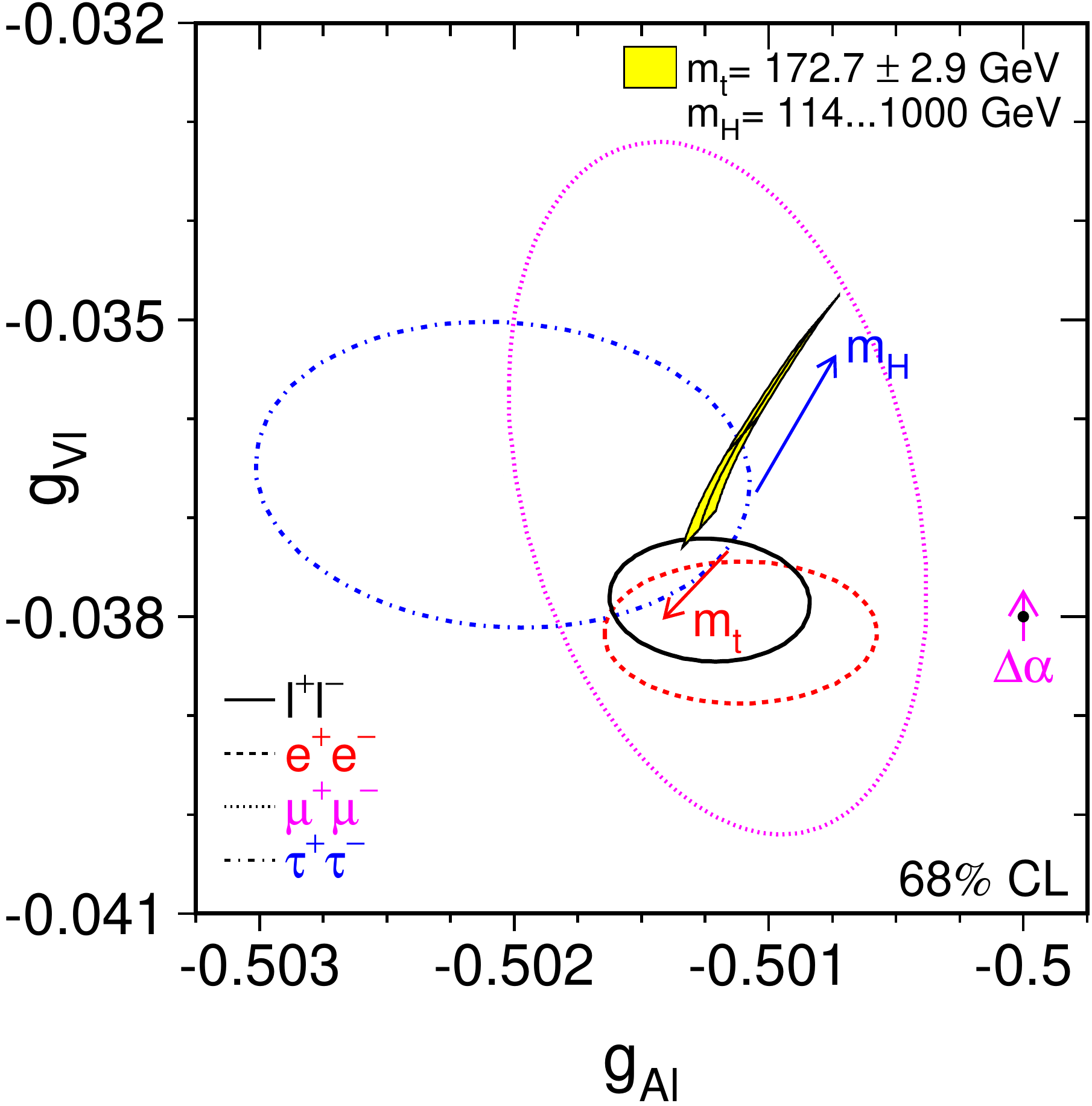}
\end{minipage}
\caption{Combined LEP and SLD measurements of $\sin^2{\theta\hskip .2pt^{\mathrm{lept}}_{\mathrm{eff}}}$ and
$\Gamma_\ell$ (left) and the corresponding effective vector and
axial-vector couplings $v_\ell$ and $a_\ell$ (right). The shaded region
shows the SM prediction. The arrows point in the direction of
increasing values of $m_t$ and $M_H$. The point shows the predicted
values if, among the electroweak radiative corrections, only the
photon vacuum polarization is included. Its arrow indicates the
variation induced by the uncertainty in $\alpha(M_Z^2)$
\cite{ALEPH:2005ab,ALEPH:2010aa}.} \label{fig:Zcouplings}
\end{figure}

Table~\ref{tab:LEP_asym} shows the present experimental results for the leptonic $Z$ decay widths and asymmetries, and the corresponding effective vector and axial-vector couplings of the $Z$ boson. Lepton universality appears to be well satisfied, although there is a $1.9\,\sigma$ difference between the values of $\cP_e$ (dominated by the SLD measurement of $\cA_{LR}^0$) and $\cP_\tau$ (dominated by the measured $\cA_{\mathrm{Pol}}^{0,\tau}$ at LEP).
Assuming universality,
the independent determinations of $\cP_\ell$ from the forward-backward asymmetries,
$|\cP_\ell^{\mathrm{FB}}|= (4\,\cA_{\mathrm{FB}}^{0,\ell}/3)^{1/2}
= 0.151\pm 0.004$, the LEP $\tau$ polarization asymmetries, $\cP_\ell^{\mathrm{LEP}} = -0.1465\pm 0.0033$, and the SLD left-right asymmetries, $\cP_\ell^{\mathrm{SLD}} = -0.1513\pm 0.0021$, are in good agreement, giving the combined average
$\cP_\ell = -0.1501\pm 0.0016$ \cite{ALEPH:2005ab,ALEPH:2010aa}.

The resulting 68\% probability contours are shown in Figs.~\ref{fig:Zcouplings}.
The left figure compares $\Gamma_\ell$ and the effective electroweak mixing angle determined from all LEP-I and SLD asymmetries, $\sin^2{\theta
\hskip .2pt^{\mathrm{lept}}_{\mathrm{eff}}} = 0.23153 \pm 0.00016$ \cite{ALEPH:2005ab,ALEPH:2010aa}, with the SM prediction as a function of $m_t$ and $M_H$. The right figure gives the corresponding effective vector and axial-vector couplings $v_\ell$ and $a_\ell$ The precision electroweak measurements require a low value of the Higgs mass, in nice agreement with the recent LHC discovery of a Higgs-like boson with $M_H = 126$~GeV \cite{Aad:2012tfa,Chatrchyan:2012ufa}. These figures provide also strong evidence of the electroweak radiative corrections. The good agreement with the SM predictions is lost if only the QED vacuum polarization contribution is taken into account, as indicated by the point with an arrow. Notice that the uncertainty induced by the input value of $\alpha(M_Z^2)^{-1}=128.95\pm 0.05$ is sizeable. The measured couplings of the three charged leptons confirm lepton universality in the neutral-current sector. The solid contour combines the three determinations, assuming universality.

The measurement of $\cA_{\mathrm{Pol}}^{0,\tau}$ and
$\cA^{0,\tau}_{\mathrm{FB,Pol}}$ assumes that the $\tau$ decay
proceeds through the SM charged-current interaction.
A more general analysis should take into account the fact that the
$\tau$ decay width depends on the product $\xi\cP_\tau$, where $\xi$
is the corresponding Michel parameter in leptonic decays or
the equivalent quantity $\xi_h$ in the semileptonic modes.
A separate measurement of $\xi$ and $\cP_\tau$ can be performed, analysing
the correlated distribution of the $\tau^+\tau^-$ decays \cite{Alemany:1991ki,Davier:1992nw}.
The LEP collaborations found in this way
$\cP_\tau = -0.132\pm 0.019$  (ALEPH) \cite{Buskulic:1994hi},
$\cP_\tau = -0.131\pm 0.014$ (DELPHI) \cite{Abreu:2000sg}
and  $\cP_\tau = -0.164\pm 0.016$ (L3) \cite{Acciarri:1998as},
in agreement with the more precise value quoted in table~\ref{tab:LEP_asym}.

The neutrino couplings can also be determined from the invisible $Z$
decay width, by assuming three identical neutrino generations with
left-handed couplings, and fixing the sign from neutrino scattering
data: $v_{\nu_\ell} = a_{\nu_\ell} = 0.50076\pm 0.00076$. Alternatively, one can use the SM prediction for $\Gamma_{\mathrm{inv}}$ to get a determination of the number of light neutrino flavours \cite{ALEPH:2005ab,ALEPH:2010aa}:
\bel{eq:Nnu} N_\nu = 2.9840\pm 0.0082\, . \ee
The universality of the neutrino couplings has been tested with $\nu_\mu e$ and $\nu_e e$ scattering data, which fix the $\nu_\mu$ and $\nu_e$ couplings to the $Z$: \
$v_{\nu_\mu} =  a_{\nu_\mu} = 0.502\pm 0.017$ \ and \
$v_{\nu_e} =  a_{\nu_e} = 0.528\pm 0.085$ \cite{Vilain:1993xb}.

The first direct observation of the $\tau$ neutrino was accomplished at Fermilab, twenty-five years after the $\tau$ discovery, using the $\nu_\tau$ production chain
$p+N\to D_s+X$, followed by $D_s\to\tau^-\bar\nu_\tau$ and $\tau^-\to\nu_\tau+Y$.
The DONUT experiment observed a total of nine $\nu_\tau + N\to\tau+Z$ events \cite{Kodama:2000mp}, making manifest the existence of a third neutrino associated with the tau flavour.

\section{Hadronic Decays}
\label{sec:Hadron}

The $\tau$ is the only known lepton massive enough to decay into hadrons. Its semileptonic decays are ideally suited to investigate the hadronic weak currents and perform low-energy tests of the strong interaction. The decay $\tau^-\to\nu_\tau H^-$ probes the  matrix element of the left-handed charged current between the vacuum and the final hadronic state $H^-$,
\bel{eq:Had_matrix}
\langle H^-| \left( V_{ud}^*\,\bar d + V_{us}^*\,\bar s\right)\gamma^\mu \left( 1-\gamma_5\right) u\, | 0 \rangle\, .
\ee
Contrary to the well-known process\ $e^+e^-\to\gamma^*\to\mathrm{hadrons}$,
which only tests the electromagnetic vector current, the semileptonic
$\tau$ decays offer the possibility to study the properties of both
vector and axial-vector currents, with  Cabibbo-allowed and Cabibbo-suppressed final states \cite{PI:89}.

For the Cabibbo-allowed modes with $J^P = 1^-$, the matrix element of the vector charged current can also be obtained, through an isospin rotation, from the isovector part of the $e^+ e^-$ annihilation cross section into hadrons, which measures the hadronic matrix element of the  $I=1$ component of the electromagnetic current,
\bel{eq:em_matrix}
\langle V^0|(\bar u \gamma^\mu u - \bar d \gamma^\mu d)|0\rangle\, .
\ee
The $\tau\to \nu_\tau V^-$ decay width is then expressed as an integral over
the cross section of the corresponding $e^+ e^-\to V^0$ process  \cite{TS:71,ThS:71}:
\bel{eq:cvc}
R_{\tau\to V}  \;\equiv\;
\frac{\Gamma (\tau^-\to\nu_\tau V^-)}{\Gamma_{\tau\to e}}\;
=\;  {3 \cos^2{\theta_C} \over 2 \pi \alpha^2 m_\tau^8 }\; S_{\mathrm{EW}}\;
   \int_0^{m_\tau^2}  ds \; (m_\tau^2 - s)^2\, (m_\tau^2 + 2 s) \, s \;
    \sigma^{I=1}_{e^+ e^- \to V^0}(s) \, ,
\ee
where the factor $S_{\mathrm{EW}} =1.0201\pm 0.0003$ contains the renormalization-group improved electroweak correction, including a next-to-leading order resummation of large logarithms \cite{Marciano:1988vm,Braaten:1990ef,Erler:2002mv}.
The available data on\ $e^+ e^- \to V^0$, can then be compared with the corresponding $\tau\to\nu_\tau V^-$ decay width and invariant-mass distribution
\cite{Gilman:1977cc,Gilman:1984ry,Gilman:1987my,Kuhn:1990ad,Eidelman:1990pb,Narison:1993sx}.
The $e^+e^-$ data contain in addition an isoscalar ($I=0$) component, which cannot be accessed through $\tau$ decays, and needs to be subtracted in \eqn{eq:cvc}.

Hadronic $\tau$ decays and the $e^+e^-$ annihilation into hadrons offer very good laboratories to improve our present understanding of the non-perturbative QCD dynamics. The general form factors characterizing the hadronic amplitudes can be experimentally extracted from the Dalitz-plot distributions of the final hadrons \cite{Kuhn:1992nz}.
The exhaustive analysis of these processes provides a very valuable data basis to confront with theoretical models.

\subsection{Chiral Dynamics}

In the absence of quark masses the QCD Lagrangian splits into two independent chirality (left/right) sectors, with their own quark flavour symmetries. With three light quarks ($u$, $d$, $s$), the QCD Lagrangian is then approximately invariant under chiral $SU(3)_L\otimes SU(3)_R$ rotations in flavour space. However, the vacuum is not symmetric under the chiral group. Thus, chiral symmetry breaks down to the usual eightfold-way $SU(3)_V$, generating the appearance of eight Goldstone bosons in the hadronic spectrum, which can be identified with the lightest pseudoscalar octet; their small masses being generated by the quark mass matrix, which explicitly breaks chiral symmetry. The Goldstone nature of the pseudoscalar octet implies strong constraints on their interactions, which can be worked out through an expansion in powers of momenta over the chiral symmetry-breaking scale \cite{WE:79}. At low momentum transfer, the coupling of any number of Goldstones ($\pi $, $K$, $\eta$) to the V$-$A current can be rigorously calculated with the effective field theory techniques of Chiral Perturbation Theory  ($\chi$PT) \cite{Gasser:1984gg,Ecker:1994gg,Pich:1995bw,Bernard:2006gx}.

In the low-energy effective chiral realization of 3-flavour QCD, the vector
$V_\mu^{ij} = \bar q^j\gamma_\mu q^i$
and axial-vector $A_\mu^{ij} = \bar q^j\gamma_\mu\gamma_5 q^i$
($i,j = u,d,s$) quark currents take the form \cite{Pich:1995bw,PI:89}:
\bea\label{eq:V_chpt}
V_\mu & = & -i \left(\Phi\lrder_\mu\Phi\right)
\, +\, \cO(\Phi^4)\, +\, \cO(p^3)\,
\nn\\ && \mbox{}
-\, {i N_C\over 6\sqrt{2}\pi^2 f^3}\;\varepsilon_{\mu\nu\alpha\beta}
\, \left\{\partial^\nu\Phi\, \partial^\alpha\Phi\, \partial^\beta\Phi
\, +\, \cO(\Phi^5)\, +\, \cO(p^5)\right\} \, ,
\nn\\
A_\mu & = & -\sqrt{2} f\,\partial_\mu\Phi
\, +\, {\sqrt{2}\over 3 f}\,\left[\Phi,\left(\Phi\lrder_\mu\Phi\right)\right]\,
+\, \cO(\Phi^5)\, +\, \cO(p^3)
\nn\\ && \mbox{}
+\, {N_C\over 12\pi^2 f^4}\;\varepsilon_{\mu\nu\alpha\beta}\,\left\{
\partial^\nu\Phi\,\partial^\alpha\Phi\,\left(\Phi
\stackrel{\leftrightarrow}{\partial^\beta}\Phi\right)\,
+\, \cO(\Phi^6)\, +\, \cO(p^5)\right\} \, ,
\eea
where the odd-parity pieces proportional to the Levi-Civita pseudotensor are generated by the Wess--Zumino--Witten term of the chiral Lagrangian \cite{WZ:71,WI:83}, which incorporates the non-abelian chiral anomaly of QCD. The $3\times 3$ matrix
\bel{eq:Phi_matrix}
\Phi (x) \;\equiv\; {\vec{\lambda}\over\sqrt 2} \, \vec{\phi} \; = \;
\pmatrix{{1\over\sqrt 2}\pi^0 \, +
\, {1\over\sqrt 6}\eta_8
 & \pi^+ & K^+ \cr
\pi^- & - {1\over\sqrt 2}\pi^0 \, + \, {1\over\sqrt 6}\eta_8
 & K^0 \cr K^- & \bar K^0 & - {2 \over\sqrt 6}\eta_8 }
\ee
parametrizes the pseudoscalar octet fields. Thus, at lowest order in momenta, the couplings of the Goldstones to the electroweak currents can be calculated in a straightforward way in terms of the single parameter $f$. In particular, $f$ determines the pion decay constant in the massless quark limit: $f_\pi = f + \cO(m_q) = (92.2\pm 0.1)$~MeV.

At very low momentum transfer (low invariant mass of the final hadrons),
the $\tau^-\to\nu_\tau (P_1\ldots P_N)^-$ amplitudes ($P_i = \pi, K,\eta$)
can be easily obtained from Eqs.~\eqn{eq:V_chpt}, which provide the lowest-order contribution in the momentum expansion. The one-loop $\chi$PT corrections are known \cite{Gasser:1984gg,Ecker:1994gg,Pich:1995bw,Colangelo:1996hs,Ecker:2002cw,Unterdorfer:2002zg}
for the lowest-multiplicity states ($\pi$, $K$, $2\pi$, $K\bar K$, $K\pi$, $K\eta$, $3\pi$, $4\pi$).
Moreover, a two-loop calculation for the $2\pi$ decay mode is also available \cite{Colangelo:1996hs}. Therefore, exclusive hadronic $\tau$ decay data at low values of $q^2$ can be compared with rigorous QCD predictions.
There exist also well-grounded theoretical results (based on a $1/M_\rho$ expansion) for decays such as $\tau^-\to\nu_\tau(\rho\pi)^-$, $\tau^-\to\nu_\tau(K^*\pi)^-$ and $\tau^-\to\nu_\tau(\omega\pi)^-$, but only in the kinematical configuration where the pion is soft in the vector-meson rest frame \cite{Davoudiasl:1995ed}.
Tau decays involve, however, high values of momentum transfer where the $\chi$PT predictions no longer apply. The relevant hadronic dynamics is governed by the non-perturbative regime of QCD, in the resonance region, which makes very difficult
to perform first-principle calculations for exclusive decays. Nevertheless, one can still construct reasonable approximations, taking into account the low-energy chiral theorems.

In the first theoretical analyses of hadronic $\tau$ decays, the chiral predictions were often extrapolated to higher values of $q^2$ by suitable final-state-interaction enhancements, through resonance form factors normalized to one at zero invariant mass
\cite{PI:89,Kuhn:1990ad,Fischer:1979fh,Kramer:1984pm,Pich:1987qq,Braaten:1987jh,GomezCadenas:1990uj}.
This simple prescription takes into account phenomenologically the resonance structures present in each channel and provides a useful description of the $\tau$ amplitudes in terms of a few hadronic parameters which can be adjusted to the data. Therefore it has been incorporated into the TAUOLA Monte Carlo library \cite{Jadach:1990mz} and has been extensively used to study the main $\tau$ decay modes \cite{Decker:1992kj,Decker:1992rj,Decker:1993ay,Decker:1994af,Finkemeier:1995sr}. However, the resonance modelization is too naive to be considered as an actual implementation of the QCD dynamics. The addition of resonance form factors to the chiral low-energy amplitudes does not guarantee that the chiral symmetry constraints on the resonance couplings are correctly implemented \cite{Portoles:2007cx}. Quite often, the numerical predictions could be drastically changed by varying some free parameter or modifying the form-factor ansatz.

The proper way of including higher-mass states into the effective chiral theory was
elaborated in Refs.~\cite{Ecker:1988te,Ecker:1989yg} and has been further developed
in more recent works \cite{Pich:2002xy,RuizFemenia:2003hm,Cirigliano:2006hb,Pich:2010sm,Pich:2008jm,Rosell:2005ai,Kampf:2011ty}. Using the techniques of the resulting Resonance Chiral Theory (R$\chi$T), it is possible to perform a systematic analysis of the $\tau$ decay amplitudes within an effective field theory realization of the underlying QCD Lagrangian.

\subsection{Two-body Semileptonic Decays}

For the decay modes with lowest multiplicity, $\tau^-\to\nu_\tau\pi^-$  and $\tau^-\to\nu_\tau K^-$, the  relevant matrix  elements
\bel{eq:f_pi_K}
\langle \pi^-(p)|\bar d \gamma^\mu\gamma_5 u | 0 \rangle \; =\;
 -i \sqrt{2}\, f_\pi\, p^\mu \, ,
\qquad\qquad\qquad
\langle K^-(p)|\bar s \gamma^\mu\gamma_5 u | 0 \rangle \; =\;
-i \sqrt{2}\, f_K\, p^\mu \, ,
\ee
are  already  known  from  the  measured  decays $\pi^-\to\mu^-\bar\nu_\mu$  and  $K^-\to\mu^-\bar\nu_\mu$. The corresponding $\tau$ decay widths can then be predicted
rather accurately through Eq.~\eqn{eq:R_tP}. As shown in table~\ref{tab:ccuniv}, these predictions are in good agreement with the measured values, and provide a quite precise test of charged-current universality.

Alternatively, the ratio between the measured $\tau^-\to\nu_\tau K^-$ and $\tau^-\to\nu_\tau \pi^-$ decay widths can be used to determine the ratio of the corresponding hadronic matrix elements:
\bel{eq:theta_C}
\frac{|V_{us}|\, f_K}{|V_{ud}|\, f_\pi} \; =\;
\frac{m^2_\tau - m^2_\pi}{m^2_\tau - m^2_K} \;
\left\{\frac{\mathrm{Br}(\tau^-\to\nu_\tau K^- )}{\mathrm{Br}(\tau^-\to\nu_\tau\pi^- )}
\, \frac{1+ \delta R_{\tau/\pi}}{1+ \delta R_{\tau/K}}\,
\frac{1}{1 + \delta R_{K/\pi}}\right\}^{1/2}
\; = \; 0.2737\pm 0.0021 \, .
\ee
We have used the radiative corrections in Eq.~\eqn{eq:dR_tp_tk}, together with the
corresponding correction to the meson decay ratio
$R_{K/\pi}\equiv\Gamma (K^- \to \mu^-\bar\nu_\mu ) / \Gamma (\pi^-\to\mu^-\bar\nu_\mu )$:
\bel{eq:deltaR_Kpi}
\delta R_{K/\pi}\; =\; -(0.0069\pm 0.0017) - (0.0044\pm 0.0015)\; =\; -(0.0113\pm 0.0023)
\, .
\ee
The first number is of electromagnetic origin \cite{Cirigliano:2011tm,Marciano:2004uf}
, while the second accounts for the strong isospin-breaking correction \cite{Cirigliano:2011tm}, so that $f_K$ and $f_\pi$ denote the meson decay constants in the isospin limit. The result \eqn{eq:theta_C} is consistent with the value
$(|V_{us}|\, f_K)/(|V_{ud}|\, f_\pi) = 0.2763\pm 0.0005$, obtained from
$R_{K/\pi}$ \cite{Cirigliano:2011tm}, but it has a much larger uncertainty.
Taking the lattice average $f_K/f_\pi = 1.193\pm 0.005$ \cite{Colangelo:2010et} and
$|V_{ud}| = 0.97425\pm 0.00022$ from super-allowed nuclear beta decays \cite{Towner:2010zz}, one gets then a determination of the Cabibbo mixing:
\bel{eq:Cabibbo}
\frac{|V_{us}|}{|V_{ud}|}\; =\; \left\{ \ba 0.2294\pm 0.0020 \\[3pt] 0.2316\pm 0.0011
\ea\right.\, ,
\qquad\qquad
|V_{us}|\; =\; \left\{ \ba 0.2235\pm 0.0019 \\[3pt] 0.2256\pm 0.0010
\ea\right.
\qquad \ba (\tau\to K/\pi) \\[3pt] (K/\pi\to\mu) \ea\, .
\ee

\subsection{Decays into Two Hadrons}

The decay into two pseudoscalar mesons, $\tau^-\to\nu_\tau P^-P^{'0}$, is mediated by the vector current. The relevant hadronic matrix element can be parametrized in terms of two form factors:
\be
\langle P^- P^{'0}|\,\bar d_i\gamma^\mu u \, |0\rangle\; =\; C_{PP'}\,\left\{
\left( p_--p_0-\frac{\Delta_{PP'}}{s}\, q\right)^\mu\, F^{PP'}_V(s)\, +\,
\frac{\Delta_{PP'}}{s}\, q^\mu\; F^{PP'}_S(s)\right\}\, ,
\ee
where $p_-^\mu$ and $p_0^\mu$ are the momenta of the charged and neutral pseudoscalars, respectively, $q^\mu = (p_-+p_0)^\mu$ is the momentum transfer and $s=q^2$.
The two Lorentz structures correspond to $J^P=1^-$ and $0^+$ transitions. The
scalar contribution is suppressed by the mass-squared difference $\Delta_{PP'} = m_{P^-}^2-m^2_{P^{'0}}$ because the vector current is conserved in the limit of equal quark masses. The global normalization coefficient $C_{PP'}$ has been chosen so that the
vector form factor $F^{PP'}_V(s)$ is one at lowest order in $\chi$PT. Thus,
\be
C_{\pi\pi} = \sqrt{2}\, ,\qquad
C_{K\bar K} = -1\, ,\qquad
C_{K\pi} = \frac{1}{\sqrt{2}}\, ,\qquad
C_{\pi \bar K} = -1\, ,\qquad
C_{K\eta_8} = \sqrt{\frac{3}{2}}\, ,
\ee
while $d_i$ refers to the corresponding down-type quark $d$ or $s$.

The hadronic invariant-mass distribution is given by
\begin{equation}\label{PP_spectrum}
 \frac{d\Gamma}{d s}\; =\; \frac{G_F^2|V_{ui}|^2 m_\tau^3}{768\pi^3}\;
S_{EW}^{\mathrm{had}}\; C_{PP'}^2\,
\biggl(1-\frac{s}{m_\tau^2}\biggr)^{2}\;\Biggl\{
\biggl(1+2\,\frac{s}{m_\tau^2} \biggr)\, \lambda_{PP'}^{3/2}\: |F_V^{PP'}(s)|^2
+ 3\,\frac{\Delta_{PP'}^2}{s^2}\,\lambda_{PP'}^{1/2}\:
 |F_S^{PP'}(s)|^2 \Biggr\}\, ,
\end{equation}
where $\lambda_{PP'}\equiv\lambda(s,m_{P^-}^2,m^2_{P^{'0}})/s^2$
and $S_{EW}^{\mathrm{had}}=1.0157\pm 0.0003$ accounts for the short-distance
electroweak corrections \cite{Marciano:1988vm,Braaten:1990ef,Erler:2002mv}.
Long-distance electromagnetic corrections and isospin-breaking contributions are channel dependent and have been only studied in a model-dependent way for the $\pi\pi$ \cite{Cirigliano:2002pv,Davier:2009ag} and $K\pi$ \cite{Antonelli:2013usa,Flores-Baez:2013eba}
final states.

\subsubsection{$\tau^-\to\nu_\tau \pi^-\pi^0$}
\label{subsubsec:2pi}

In the limit of isospin symmetry the two-pion final state does not receive any scalar contribution. The decay $\tau^-\to\nu_\tau \pi^-\pi^0$ is then governed by the
so-called pion form factor $F_\pi(s)\equiv F_V^{\pi\pi}(s)$.
A dynamical understanding of $F_\pi(s)$ can be achieved
\cite{Guerrero:1997ku,GomezDumm:2000fz,Pich:2001pj,SanzCillero:2002bs,Pich:2010sm}, using analyticity, unitarity and some general properties of QCD, such as chiral symmetry
\cite{Ecker:1994gg,Pich:1995bw,Bernard:2006gx} and the short-distance asymptotic behaviour
\cite{Ecker:1988te,Ecker:1989yg,Cirigliano:2006hb,Pich:2002xy}. Putting all these fundamental ingredients together, one can express the pion form factor in the simple form \cite{Guerrero:1997ku}
\begin{equation}\label{eq:PFF_GP}
F_\pi(s)\; =\; {M_\rho^2\over M_\rho^2 - s - i M_\rho \Gamma_\rho(s)}\;
\exp{\left\{-\frac{s}{96\pi^2f_\pi^2} \,\mathrm{Re}\, [A(s)]\right\}}\, ,
\end{equation}
where
\begin{equation}
 A(s) \;\equiv\; \log{\left({m_\pi^2\over
 M_\rho^2}\right)} + 8\, {m_\pi^2 \over s} - {5\over 3} + \sigma_\pi^3
 \log{\left({\sigma_\pi+1\over \sigma_\pi-1}\right)}
\end{equation}
contains the one-loop chiral logarithms \cite{Gasser:1984gg}, which account for the final-state interactions of the two pions,
$\sigma_\pi\equiv\sqrt{1-4m_\pi^2/s}$ and the off-shell $\rho$ width is given by $\Gamma_\rho(s) = \theta(s-4m_\pi^2)\,\sigma_\pi^3\, M_\rho\, s/(96\pi f_\pi^2)$ \cite{Guerrero:1997ku,GomezDumm:2000fz}.

\begin{figure}[tb]\centering
\begin{minipage}{0.45\textwidth}\centering
\includegraphics[angle=-90,width=8.5cm,clip]{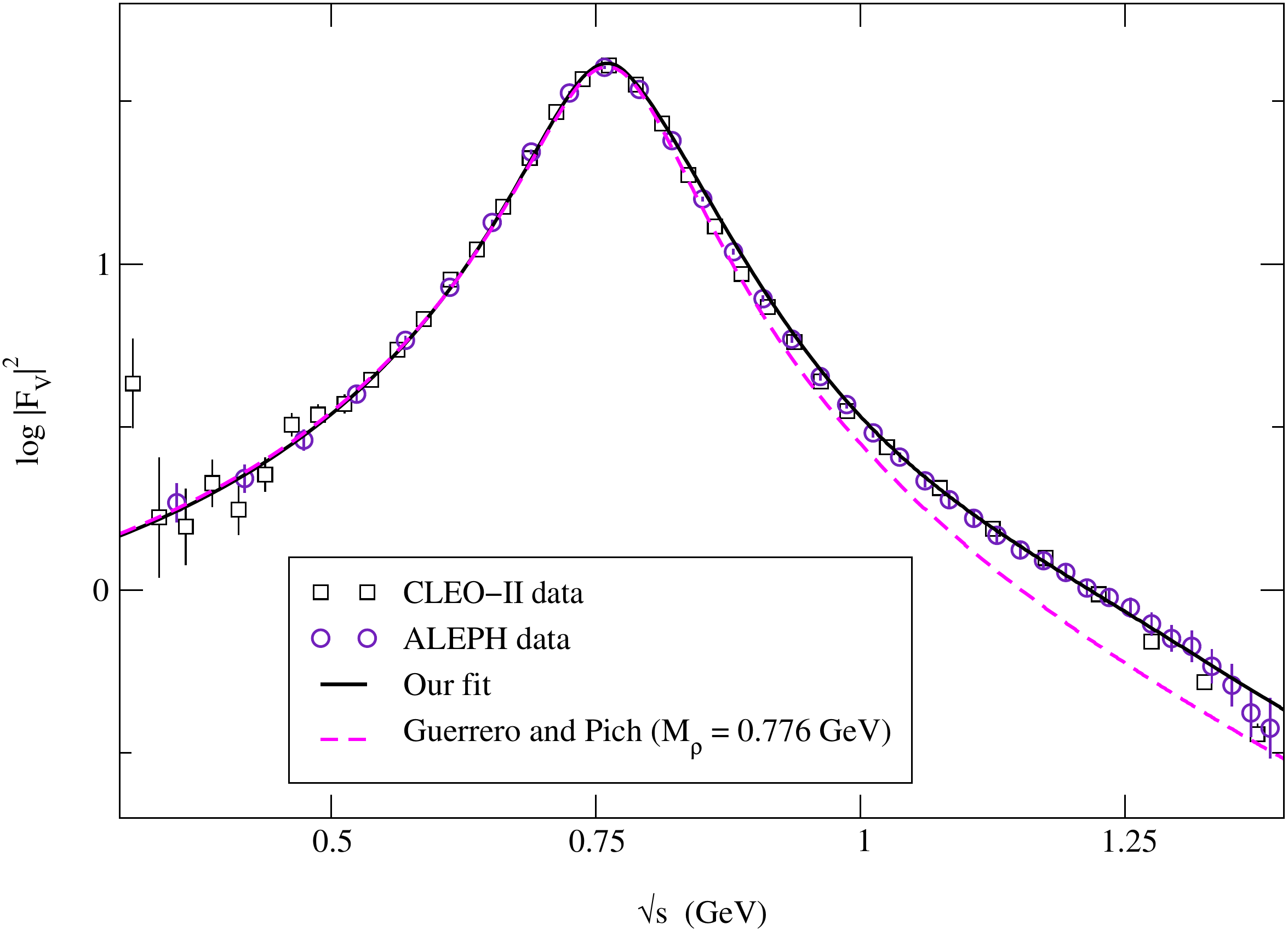}
\end{minipage}
\hskip 1.cm
\begin{minipage}{0.45\textwidth}\centering
\includegraphics[angle=-90,width=8cm,clip]{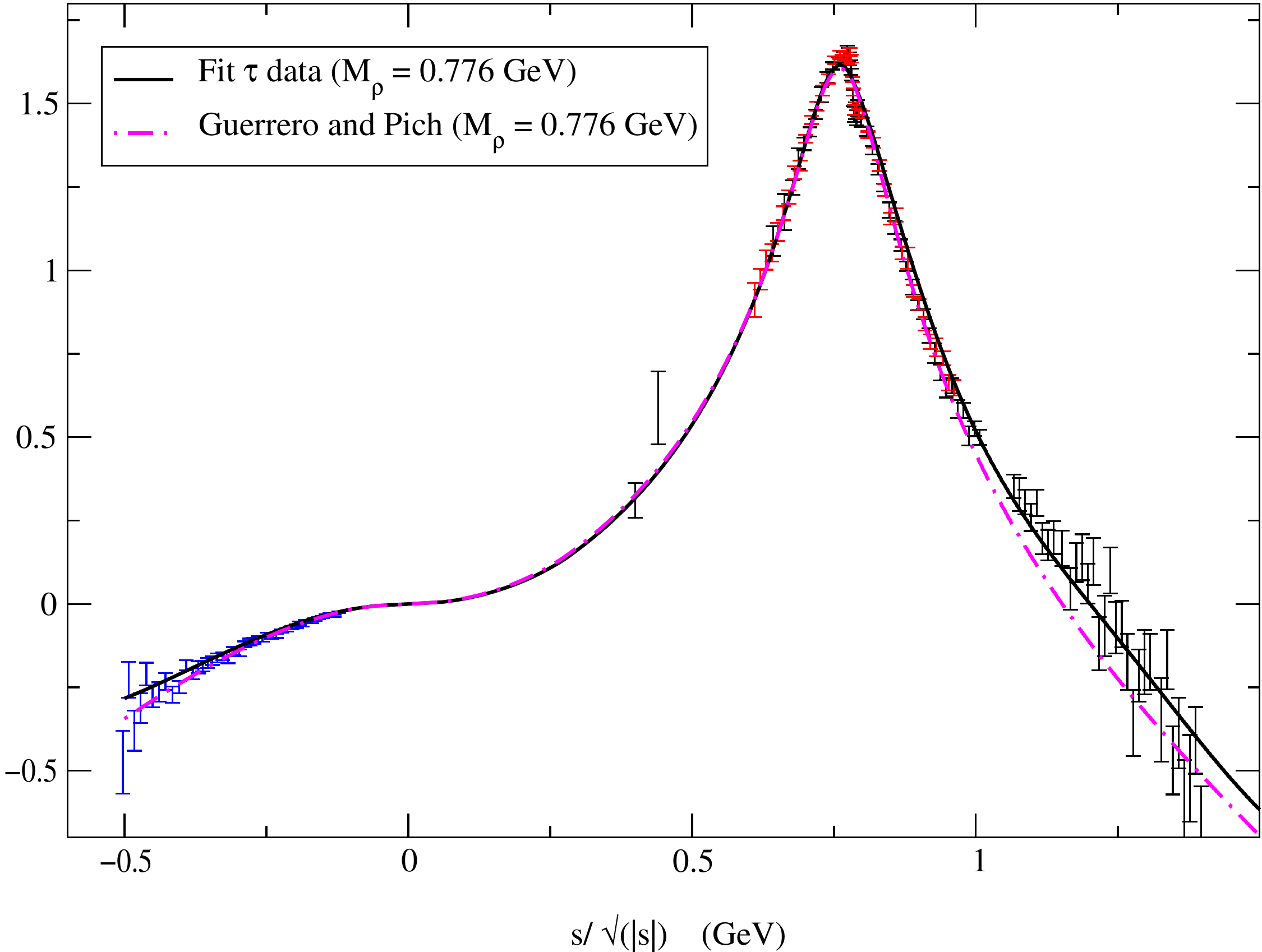}
\end{minipage}
\caption{Pion form factor from ALEPH \cite{Barate:1997hv} and CLEO \cite{Anderson:1999ui} $\tau$ data (left) and $e^+e^-\to\pi^+\pi^-$ \cite{Barkov:1985ac} and $e^-\pi^-\to e^-\pi^-$ \cite{Amendolia:1986wj} data (right), compared with theoretical
predictions \cite{Guerrero:1997ku,Pich:2001pj}. The dashed lines
correspond to the result in Eq.~(\ref{eq:PFF_GP}).}
\label{fig:pionth}
\end{figure}

In the limit of an infinite number of QCD colours \cite{Hooft:74,Witten:79}, $F_\pi(s)$ is described by an infinite sum of narrow-width vector resonance contributions \cite{Pich:2002xy,Dominguez:2001zu,Bruch:2004py}. Eq.~\eqn{eq:PFF_GP} only contains the contribution from the lightest $\rho$ state, which is the dominant one below 1 GeV. The normalization $F_\pi(0)=1$ is a consequence of the conservation of the electromagnetic current, while the short-distance properties of QCD require the form factor to vanish at infinite momentum. The large-$N_C$  propagator has been dressed with pion loop corrections, which are subleading in $1/N_C$, in such a way that a Taylor expansion in powers of $s/M_\rho^2$ correctly reproduces the rigorous one-loop $\chi$PT prediction at very low energies. The constraints from analyticity and unitarity allow us to perform a resummation of the one-loop chiral logarithms through an Omn\`es exponential \cite{Omnes:1958hv}; using a Dyson summation, the absorptive part of these corrections has been reabsorbed into the $\rho$ width in order to regulate the resonance pole.
Thus, Eq.~\eqn{eq:PFF_GP} extends the validity domain of the $\chi$PT prediction by combining together a series of basic theoretical requirements on $F_\pi(s)$. This analytical prediction, which only depends on $M_\rho$, $m_\pi$ and the pion decay constant $f_\pi$, is compared with the data in Fig.~\ref{fig:pionth}. The agreement is rather impressive and extends to negative $s$ values, where the $e^-\pi^-$ elastic data sits.

\begin{figure}[t]\centering
\begin{minipage}{8cm}\centering
\includegraphics[angle=-90,width=8cm,clip]{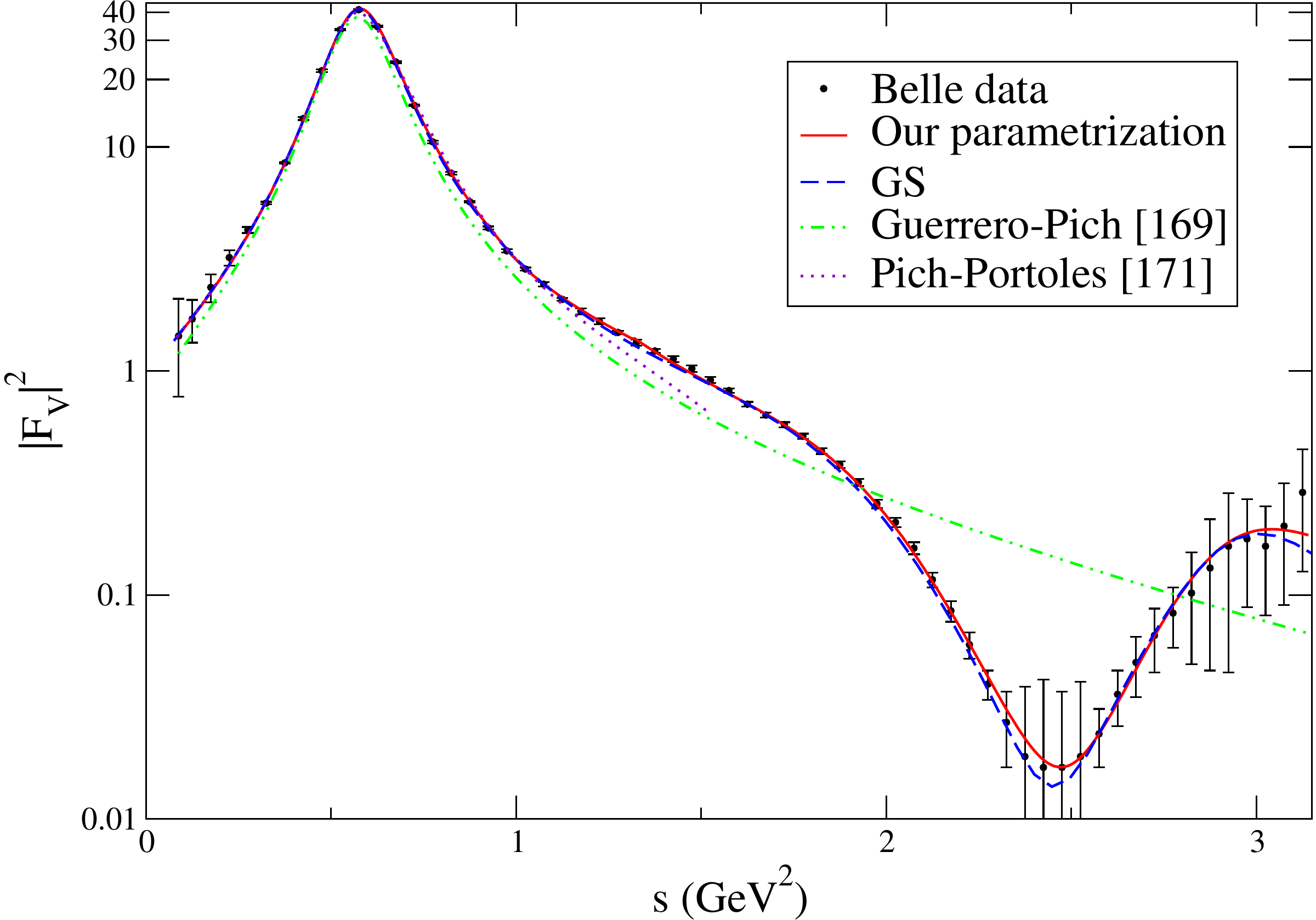}
\end{minipage}
\hskip .8cm
\begin{minipage}{9.cm}\centering
\includegraphics[width=9.cm]{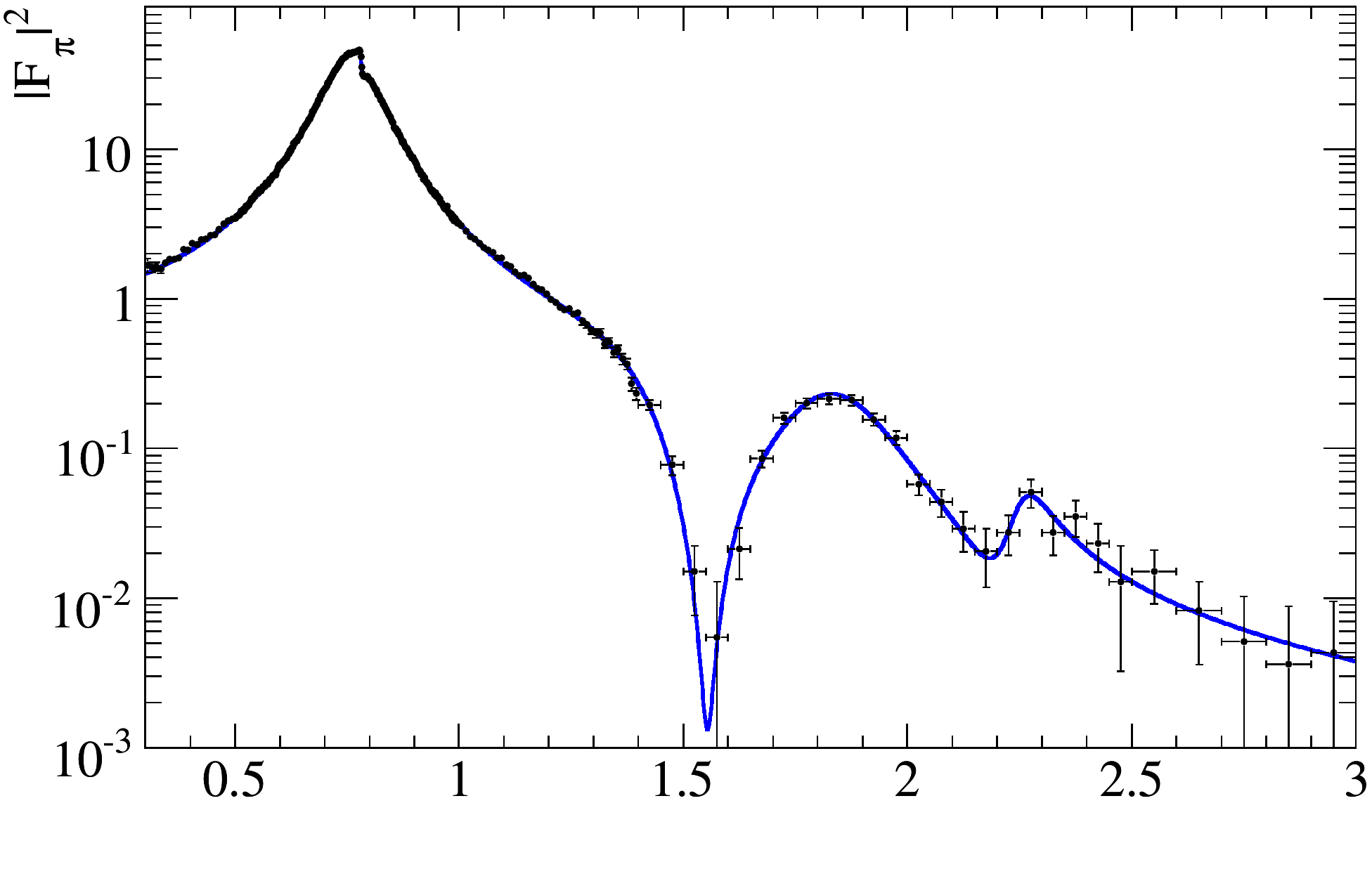}
  \put(-45,0){\fontsize{8}{9.6}\selectfont {$\sqrt{s'}$ (GeV)}}
\end{minipage}
\caption{Pion form factor extracted from Belle $\tau^-\to\nu_\tau \pi^-\pi^0$ data \cite{Fujikawa:2008ma} (left). The curves correspond to different phenomenological approximations,
including $\rho$, $\rho'$ and $\rho''$ contributions \cite{Dumm:2013zh}.
The BaBar measurement from $e^+e^-\to\pi^+\pi^-\gamma$ events \cite{Lees:2012cj} (right) shows also clear $\rho'''$ and $\omega$ ($I=0$) signals.}
\label{fig:BellePFF}
\end{figure}

The modifications induced by kaon loops, heavier $\rho$ resonance contributions and
additional next-to-leading in $1/N_C$ corrections can be easily
included; one gets then a more accurate approximation at the price of having more free parameters which decrease the predictive power \cite{Pich:2001pj,SanzCillero:2002bs,Pich:2010sm}. This gives a better
description of the $\rho'$ shoulder around 1.2 GeV (continuous lines
in Fig.~\ref{fig:pionth}).
A clear signal for the $\rho''(1700)$ resonance in
$\tau^-\to\nu_\tau \pi^-\pi^0$ events has been reported by Belle,
with a data sample 20 times larger than in previous experiments
\cite{Fujikawa:2008ma}. Fig.~\ref{fig:BellePFF} shows a recent fit to the Belle data, including the effects from the $\rho'$ and $\rho''$ states \cite{Dumm:2013zh}.

The $\tau^-\to\nu_\tau\pi^-\pi^0$ decay amplitude can be related
through an isospin rotation with the isovector piece of
$\sigma(e^+e^-\to\pi^+\pi^-)$. Thus, for $2m_\pi^2 <s< m_\tau^2$, the pion form factor can be obtained from the two sets of data. For many years there have been sizeable discrepancies between the different experimental measurements, suggesting that systematic uncertainties have been probably underestimated. The amplitudes measured in $e^+e^-$ annihilation experiments \cite{Barkov:1985ac,Lees:2012cj,Bisello:1988hq,Akhmetshin:2006bx,Achasov:2006vp,Babusci:2012rp}
are slightly lower than the corresponding $\tau$ decay measurements \cite{Barate:1997hv,Anderson:1999ui,Fujikawa:2008ma,Schael:2005am}, and the difference does not seem to be fully accounted for through the calculated isospin-violating effects \cite{Cirigliano:2002pv,Davier:2009ag}. In addition to the direct energy scan adopted in most $e^+e^-$ measurements, some recent experiments use
the so-called radiative return method to extract $\sigma(e^+e^-\to\pi^+\pi^-)$ from the
$e^+e^-\to\pi^+\pi^-\gamma$ data sample, probing different ranges of $\pi^+\pi^-$ invariant masses through the radiated photon (initial state radiation) \cite{Binner:1999bt,Rodrigo:2001kf,Actis:2010gg,Druzhinin:2011qd}. Using this method, the most recent BaBar data \cite{Lees:2012cj} has reduced considerably the tension with $\tau$-decay measurements, approaching the tau results, but discrepancies persist with the KLOE $e^+e^-\to\pi^+\pi^-\gamma$ data \cite{Babusci:2012rp}.

Taking into account isospin-breaking corrections \cite{Cirigliano:2002pv,Davier:2009ag}, one predicts from $e^+e^-$ data $\mathrm{Br}(\tau^-\to\nu_\tau \pi^-\pi^0) = (24.94\pm
0.25)\% $ \cite{Davier:2010nc}, which is $2.1\,\sigma$ smaller than the average of direct $\tau$ decay measurements $(25.504\pm 0.092)\% $ \cite{HFAG}. The largest (smallest) discrepancy of $2.7\,\sigma$ ($1.2\,\sigma$) between prediction and direct measurement is exhibited by KLOE (BaBar)  \cite{Davier:2010nc}. This discrepancy
translates into different estimates of the hadronic vacuum polarization contribution to the anomalous magnetic moment of the muon that will be discussed in section~\ref{sec:g-2}.
Accurate measurements of $F_\pi(s)$ are a critical ingredient of the SM prediction for the muon $g-2$, which reinforces the need for new precise $e^+e^-$ and $\tau$ data sets to clarify the present experimental situation.

\subsubsection{$\tau^-\to\nu_\tau (K\pi)^-$}

Owing to the different final pseudoscalar masses, the decays $\tau^-\to\nu_\tau K^-\pi^0$ and $\tau^-\to\nu_\tau \pi^-\bar K^0$ receive contributions from two different form factors with $J^P=1^-$ and $0^+$. The same vector and scalar form factors are probed in $K_{\ell 3}$ processes, but tau decays are sensitive to a much broader (and different) range of invariant hadronic masses.

\begin{figure}[tb]\centering
 \includegraphics[width=10cm,clip]{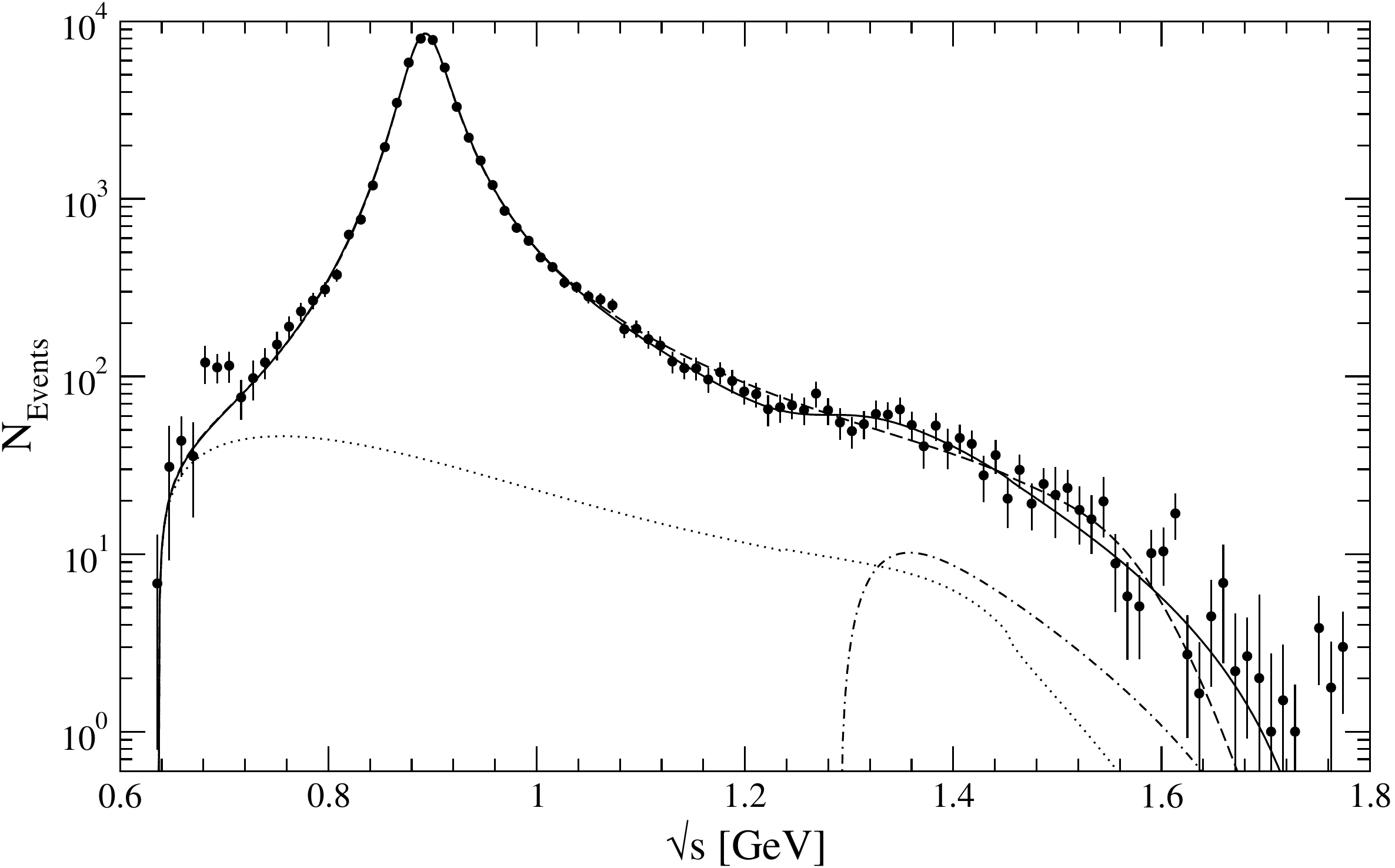}
 \caption{Distribution of $\tau^-\to\nu_\tau \pi^- K_S$ events measured by Belle \cite{Epifanov:2007rf}. The solid line shows the theoretical fit of Ref.~\cite{Jamin:2008qg}, including a R$\chi$T description of $F_V^{\pi K}(s)$ with two resonances and $F_S^{\pi K}(s)$ from \cite{Jamin:2006tj}. The scalar and $K^{*}(1410)$ contributions are indicated by the dotted and dash-dotted lines.}
 \label{fig:KpSpectrum}
\end{figure}

A detailed study of these decays has been made in Refs.~\cite{Finkemeier:1995sr,Antonelli:2013usa,Jamin:2008qg,Boito:2010me,Moussallam:2007qc}.
The vector form factor $F_V^{K\pi}(s)$ can be described in an analogous way to
$F_\pi(s)$, while the scalar component $F_S^{K\pi}(s)$ takes also
into account additional information from $K\pi$ scattering data
through dispersion relations \cite{Jamin:2001zq,Jamin:2006tj,Buettiker:2003pp}.
Fig.~\ref{fig:KpSpectrum} compares the distribution of $\tau^-\to\nu_\tau \pi^- K_S$ events measured by Belle \cite{Epifanov:2007rf}, with a theoretical fit using the R$\chi$T description of $F_V^{\pi K}(s)$ with two resonances \cite{Jamin:2008qg}.
The scalar component is predicted to give only a small contribution to the total decay width, Br$[\tau\to\nu_\tau(K\pi)_{\rm S-wave}]=(3.88\pm 0.19)\times 10^{-4}$, but it is sizeable at very low invariant mass. As expected, the $K^*(892)$ resonance generates
the dominant contribution with a small correction from the $K^*(1410)$ state at higher invariant mass.
The Belle data shows a bump at 0.682--0.705 GeV (points 5, 6 and 7) which is not supported by the theoretical description and, therefore, it is not included in the fit; this bump does not seem to be present in the preliminary BaBar data \cite{Adametz:2011zz}.

The fit to the $\tau^-\to\nu_\tau \pi^- K_S$ spectrum allows one to predict the slope and curvature of the vector form factor, with a precision comparable (slightly better) to the one attained in $K_{\ell 3}$ decays, and the absolute normalization, \ie the
$\tau^-\to\nu_\tau \pi^- \bar K^0$ and $\tau^-\to\nu_\tau K^-\pi^0$ branching ratios \cite{Antonelli:2013usa,Jamin:2008qg,Boito:2010me,Moussallam:2007qc}. The most recent analysis makes a combined fit to $\tau^-\to\nu_\tau \pi^- K_S$ and $K_{\ell 3}$ data,
using dispersive parametrizations of the two form factors and including electromagnetic and isospin-violating corrections which are needed for $K_{\ell 3}$, and finds the predictions \cite{Antonelli:2013usa}:
\bea\label{eq:BrKpi_pred}
\mathrm{Br}(\tau^-\to\nu_\tau \pi^- \bar K^0)_{\mathrm{th}}& =& (0.857\pm 0.030)\%\, ,
\nonumber\\
\mathrm{Br}(\tau^-\to\nu_\tau K^-\pi^0)_{\mathrm{th}}& =& (0.471\pm 0.018)\%\, .
\eea
These numbers are slightly higher than the present experimental world averages,
$\mathrm{Br}(\tau^-\to\nu_\tau \pi^- \bar K^0) = (0.8206\pm 0.0182)\% $
and
$\mathrm{Br}(\tau^-\to\nu_\tau K^-\pi^0) = (0.4322\pm 0.0149)\% $ \cite{HFAG},
confirming the earlier observation made in Ref.\cite{Jamin:2008qg}.
The implications of these results for the determination of $|V_{us}|$ will be discussed later in section~\ref{sec:Vus}.

The decays $\tau\to K^-\eta\nu_\tau$ and $\tau\to K^-\eta'\nu_\tau$ have been also studied recently in Ref.~\cite{Escribano:2013bca}. These decay modes are quite suppressed. The first one has been measured at the B factories \cite{delAmoSanchez:2010pc,Inami:2008ar} with an average branching fraction
$\mathrm{Br}(\tau\to K^-\eta\nu_\tau) = (1.53\pm 0.08)\times 10^{-4}$ \cite{HFAG}, while only an upper limit exists in the $K^-\eta'$ case, $\mathrm{Br}(\tau\to K^-\eta'\nu_\tau) < 2.4 \times 10^{-6}$ (90\% CL) \cite{Lees:2012ks}.

\subsubsection{$\tau^-\to\nu_\tau \pi^-\eta$}

The $\pi^-\eta$ final state has $I^G=1^-$ and $J^P=0^+$ or $1^-$ (for S or P wave, respectively). Its observation in $\tau$ decays would therefore indicate either a violation of G-parity (the Cabibbo-allowed vector current has even G-parity) or new physics incorporating second-class currents.
In the SM, this decay can proceed through the violation of isospin provided by the light quark mass difference $m_d-m_u$ or through electromagnetic corrections, and it is expected to be very suppressed \cite{Pich:1987qq,Tisserant:1982fc,Bramon:1987zb}.
At lowest order in $\chi$PT the coupling to the vector current is governed by the well-known $\pi^0$--$\eta$ mixing, giving rise to constant (and equal) vector and scalar form factors:
\bel{eq:ff_etapi_LO}
F_V^{\pi\eta}(s)_{\mathrm{LO}}\; =\; F_S^{\pi\eta}(s)_{\mathrm{LO}}\; =\;
\frac{\sqrt{3}\, (m_d-m_u)}{4\, (m_s-\hat{m})}\;\approx\; 0.99\times 10^{-2}\, ,
\ee
where $\hat{m}=(m_d+m_u)/2$. This result gets significantly enhanced by next-to-leading-order chiral corrections and electromagnetic contributions. The value of the two form factors at $s=0$ can be related in a simple and very elegant way with the ratio
$F_V^{K^-\pi^0}(0)/F_V^{\pi^-\bar K^0}(0)$ \cite{Neufeld:1994eg}. Exploiting the $K_{\ell 3}$ information \cite{Antonelli:2010yf}, one can then obtain a very precise prediction of the $\pi\eta$ form factors at $s=0$ \cite{Descotes-Genon:2013uya}:
\bel{eq:ff_etapi}
F_V^{\pi\eta}(0)\; =\; F_S^{\pi\eta}(0)\; =\; (1.49\pm 0.23)\times 10^{-2}\, .
\ee

Inserting this information into subtracted dispersion relations, together with the known $\chi$PT constraints, the two form factors have been recently estimated in the relevant kinematical domain, implying the following $\tau$ decay branching fraction \cite{Descotes-Genon:2013uya}:
\bel{eq:BrEtapi_pred}
\mathrm{Br}(\tau^-\to\nu_\tau \pi^- \eta)_{\mathrm{th}}\; =\; (0.48\, {}^{+\, 0.30}_{-\, 0.20})\times 10^{-5}\, .
\ee
This theoretical prediction, which is dominated by the scalar contribution, is in the lower range of previous evaluations \cite{Pich:1987qq,Tisserant:1982fc,Bramon:1987zb,Neufeld:1994eg,Nussinov:2008gx,Paver:2010mz}
and a factor 20 smaller than the present experimental upper bound:
$\mathrm{Br}(\tau^-\to\nu_\tau \pi^-\eta) < 9.9\times 10^{-5}$  (95\% CL) \cite{delAmoSanchez:2010pc}. A slightly better experimental limit exists for the $\pi^-\eta'$ final state:
$\mathrm{Br}(\tau^-\to\nu_\tau \pi^-\eta') < 4.0\times 10^{-6}$  (90\% CL)
\cite{Lees:2012ks}.

\subsection{Higher-Multiplicity Decays}

Higher-multiplicity modes involve a richer dynamical structure, providing a very valuable experimental window into the non-perturbative hadronization of the QCD currents.
However, accounting for the strong final-state interaction is not an easy game when three or more hadrons are present. There exist tree-level R$\chi$T calculations for most $\tau$ decays into three mesons, and even some final states with four pseudoscalars have been estimated, but the corrections induced by chiral loops are not yet implemented (except for $\tau\to\nu_\tau 3\pi$ \cite{Colangelo:1996hs} and $\tau\to\nu_\tau 4\pi$  \cite{Ecker:2002cw}, at very low $q^2$). The present predictions correspond to the limit of an infinity number of QCD colours; the only subleading correction in the $1/N_C$ expansion which is taken into account is the finite width of the hadronic resonances. Although the accuracy of these R$\chi$T approximations is still moderate, they provide a direct connection with the fundamental QCD theory and constitute a very valuable starting point to analyse the measured observables. Including suitable modifications to account phenomenologically for the contributions not yet included, one gets useful parametrizations to fit the experimental data in a sensible way, which nowadays are being incorporated into the TAUOLA library \cite{Shekhovtsova:2012ra}. From the high-statistics $\tau$ decay (and $e^+e^-$) data samples, it is then possible to extract important information on the hadronic structure, allowing one to improve the theoretical tools and get a better control of the strong interaction in the resonance region.

\begin{figure}[tb]\centering
 \includegraphics[width=13cm,clip]{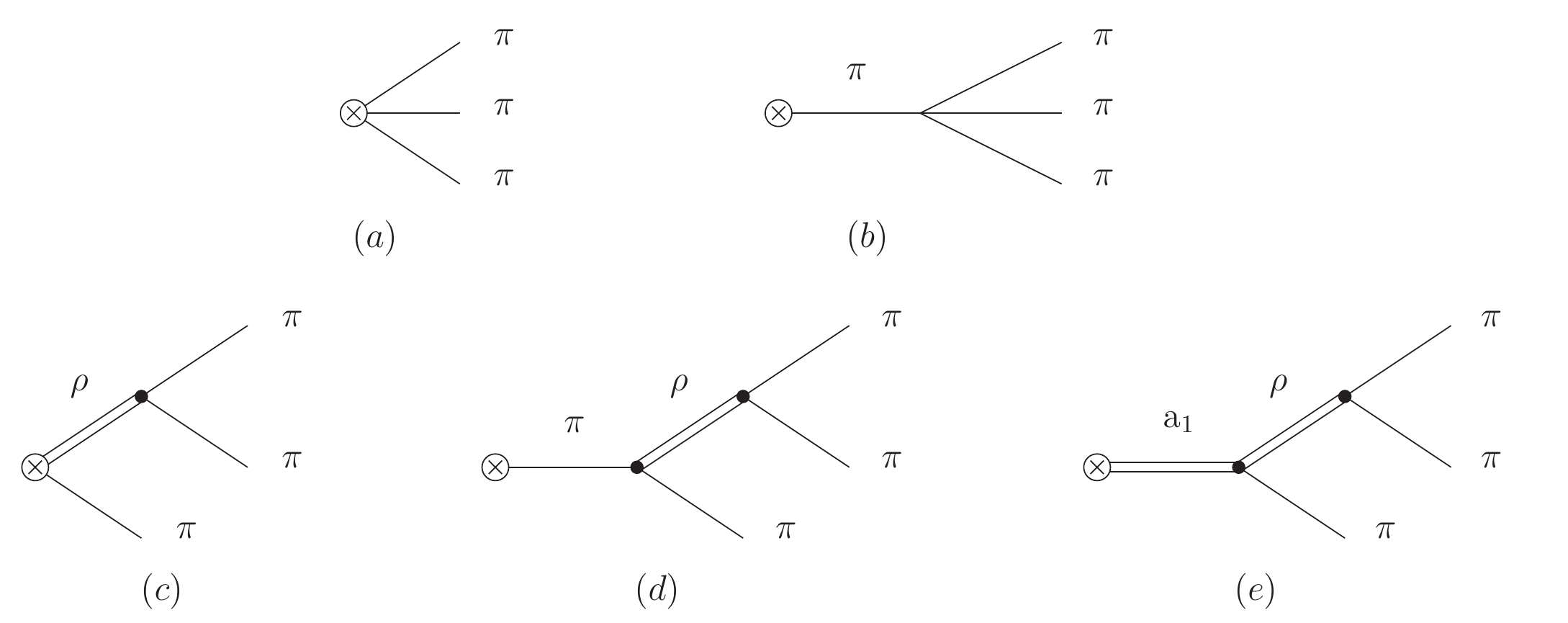}
 \caption{Tree-level Feynman diagrams contributing to $\tau^-\to\nu_\tau (3\pi)^-$.}
 \label{fig:TauTo3Pi}
\end{figure}

\begin{figure}[tb]\centering
 \includegraphics[angle=-90,width=10cm,clip]{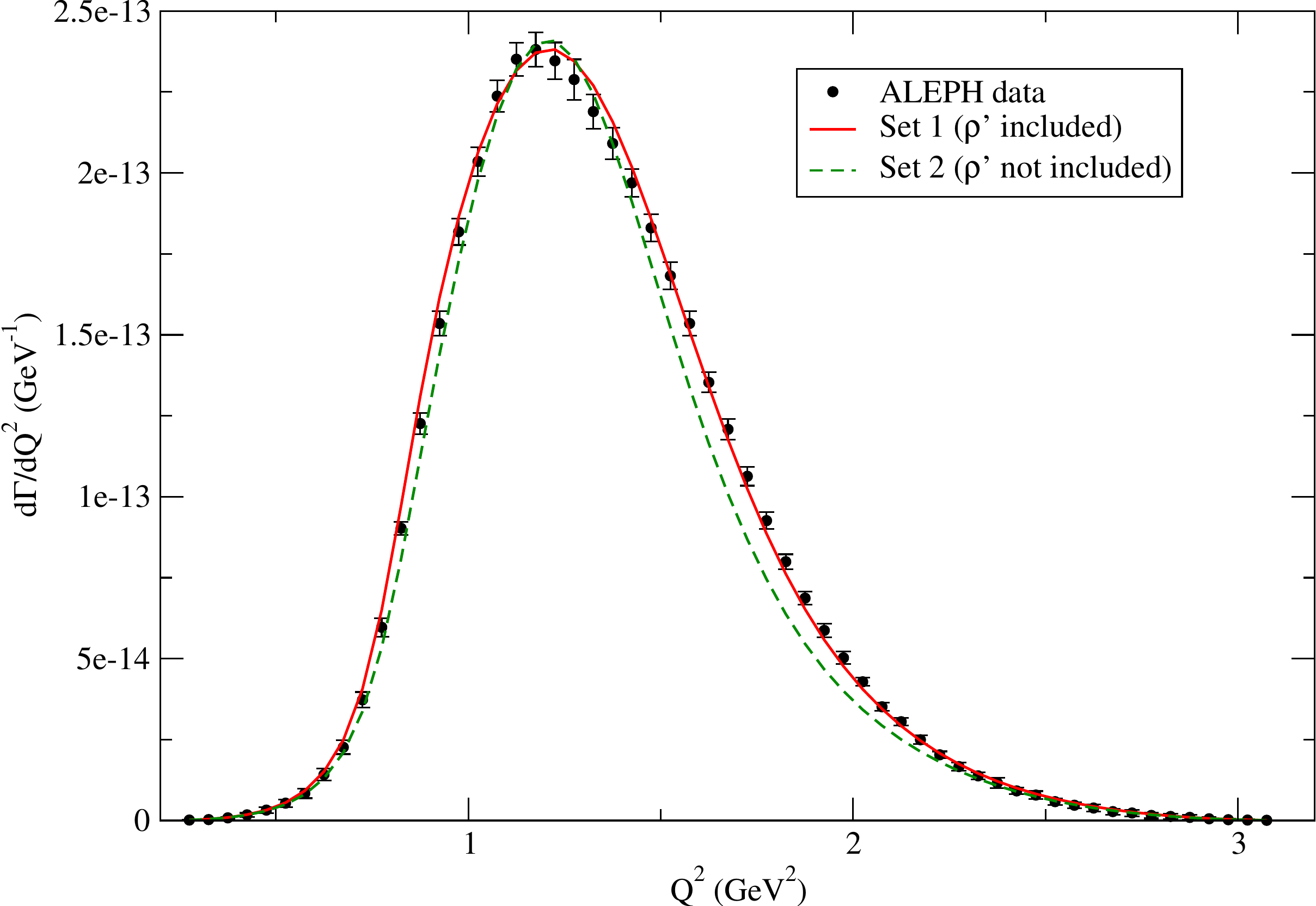}
 \caption{ALEPH $\tau^-\to\nu_\tau \pi^-\pi^+\pi^-$ data \cite{Barate:1998uf} compared
 with the fitted theoretical distribution \cite{Dumm:2009va}.}
\label{fig:TauTo3Pifit}
\end{figure}

The first detailed studies of
$\tau^-\to\nu_\tau \pi^+\pi^-\pi^-$ and $\tau^-\to\nu_\tau \pi^0\pi^0\pi^-$ were made in Refs.~\cite{Kuhn:1990ad,Fischer:1979fh}. A R$\chi$T
analysis was later performed  in Refs.~\cite{Dumm:2009va,GomezDumm:2003ku}, including a theoretical description of the measured \cite{Schael:2005am,Barate:1998uf,Buskulic:1993sv,Browder:1999fr,Ackerstaff:1997dv}
$J^P=1^+$ structure functions \cite{Kuhn:1992nz}. Fig.~\ref{fig:TauTo3Pi} shows the relevant tree-level Feynman diagrams with resonance exchanges. The theoretical analysis includes dynamical contributions from the $\rho(770)$, $\rho(1450)$ and $a_1(1260)$ resonances, taking into account their energy-dependent widths. This turns out to be an important physical effect for broad resonant structures, modifying in a sizeable way the values of their fitted mass and width parameters. The fit to the ALEPH data \cite{Barate:1998uf}, shown in Fig.~\ref{fig:TauTo3Pifit},  gives as central values $M_{a_1} = 1.120$~GeV and
$\Gamma_{a_1}(M_{a_1}^2) = 0.483$~GeV \cite{Dumm:2009va}, quite different from the numbers quoted by the PDG \cite{Beringer:1900zz} which do not take into account the energy dependence of the resonance width.

The $\tau^-\to\nu_\tau (3\pi)^-$ decay amplitudes contain two interfering contributions, corresponding to the exchange of the two identical pions in the final state, which generate
a parity-violating angular asymmetry \cite{Kuhn:1990ad,Kuhn:1982di,Feindt:1990ev}, making possible to determine the sign of the $\nu_\tau$ helicity to be $-1$ \cite{Asner:1999kj,Ackerstaff:1997dv,Albrecht:1992ka}.

Detailed phenomenological analyses exist for other decays such as
$\tau\to\nu_\tau K\bar K\pi$  \cite{Dumm:2009kj} and
$\tau\to\nu_\tau K 2\pi$ \cite{Roig:2013ts}, which involve both vector and axial-vector currents, and
$\tau \to \eta^{(\prime)} 2\pi \nu_\tau$ \cite{Dumm:2012vb}, which is driven by the vector current (up to tiny isospin-breaking effects).
The R$\chi$T analysis of $\tau\to\nu_\tau K\bar K\pi$  \cite{Dumm:2009kj} suggests that the vector contribution dominates in this mode, in agreement with the early study in Ref.~\cite{GomezCadenas:1990uj}, but at variance with Refs.~\cite{Finkemeier:1996hh,Davier:2008sk} which advocate a larger axial component.
A precise experimental determination of the hadronic invariant-mass distribution would help to disentangle the two contributions.
The vector-current amplitudes contributing to these decays can be also studied through the
corresponding $e^+e^-$ annihilation processes \cite{Dai:2013joa}.

The R$\chi$T techniques have been also applied to more involved transitions
such as $e^+e^-\to 3\pi$ \cite{Dai:2013joa},
$\tau\to\nu_\tau 4\pi$ and $e^+e^-\to 4\pi$  \cite{Ecker:2002cw,Unterdorfer:2002zg}
and radiative $\tau$ decays with one-meson in the final state\cite{Guo:2010dv,Guevara:2013wwa}.

\subsubsection{Experimental status}

%
\begin{table}[tbh]
\caption{Cabibbo-allowed hadronic $\tau$ branching ratios \protect\cite{HFAG}.
$h^\pm$ stands for $\pi^\pm$ or $K^\pm$.}
\label{tab:BR_CA}
\renewcommand{\arraystretch}{1.1}
\centering
\vspace{0.2cm}
\begin{tabular}{|cc|cc|}
\hline & &&\\[-4mm]
$X^-$ & $\mathrm{Br}(\tau^-\to\nu_\tau X^-)$ (\% ) & $X^-$ & $\mathrm{Br}(\tau^-\to\nu_\tau X^-)$ (\% )
\\[-4mm]& && \\ \hline & &&\\[-4mm]
$\pi^-$ & $(10.811\pm 0.053)$ &
$\pi^-\pi^+\pi^-$ {\small (ex. $K^0,\omega$)} & $(8.9719\pm 0.0511)$
\\
$\pi^-\pi^0$ & $(25.504\pm 0.092)$ &
$\pi^-\pi^+\pi^-\pi^0$ {\small (ex. $K^0,\omega$)} & $(2.7659\pm 0.0710)$
\\
$\pi^-2\pi^0$ {\small (ex. $K^0$)} & $(9.2414\pm 0.0997)$ &
$h^-h^+h^-2\pi^0$ {\small (ex. $K^0,\omega,\eta$)} & $(0.0973\pm 0.0354)$ 
\\
$\pi^-3\pi^0$ {\small (ex. $K^0$)} & $(1.0322\pm 0.0749)$ &
$h^-h^+h^-3\pi^0$  & $(0.0320\pm 0.0031)$  
\\
$h^-4\pi^0$ {\small (ex. $K^0,\eta$)} & $(0.1091\pm 0.0391)$ &
$\pi^-K^-K^+$ & $(0.1435\pm 0.0027)$
\\
$K^-K^0$ & $(0.1591\pm 0.0157)$ &
$\pi^-K^-K^+\pi^0$ & $(0.0061\pm 0.0018)$ 
\\
$K^-K^0\pi^0$ & $(0.1450\pm 0.0071)$ &
$3h^-2h^+$ {\small (ex. $K^0$)} & $(0.0823\pm 0.0031)$ 
\\
$\pi^-K_S^0 K_S^0$ & $(0.0240\pm 0.0050)$ & 
$3h^-2h^+\pi^0$ {\small (ex. $K^0$)} & $(0.0198\pm 0.0024)$ 
\\
$\pi^-K_S^0 K_L^0$ & $(0.1082\pm 0.0203)$ &
$\pi^-\pi^0\eta$ & $(0.1386\pm 0.0072)$
\\
$\pi^-K_L^0 K_L^0$ & $(0.0240\pm 0.0050)$ &  
$\pi^-\omega$ & $(1.9535\pm 0.0647)$
\\
$\pi^-K^0 \bar K^0\pi^0$ & $(0.0310\pm 0.0230)$ & 
$h^-\pi^0\omega$ & $(0.4049\pm 0.0418)$
\\
$a_1^-\;$ {\small $(\to\pi^-\gamma)$} & $(0.0400\pm 0.0200)$ & & 
\\[1.5mm] \hline \multicolumn{4}{|c|}{}\\[-4mm]
\multicolumn{4}{|c|}{$\mathrm{Br}(\tau^-\to\nu_\tau X^-)\, =\, (61.85\pm 0.11)\%$}
\\[1.5mm]\hline
\end{tabular}
\end{table}
%

%
\begin{table}[htb]
\caption{Cabibbo-suppressed hadronic $\tau$ branching ratios \protect\cite{HFAG}.
$h^\pm$ stands for $\pi^\pm$ or $K^\pm$.}
\label{tab:BR_CS}
\renewcommand{\arraystretch}{1.1}
\centering
\vspace{0.2cm}
\begin{tabular}{|cc|cc|}
\hline & &&\\[-4mm]
$X_s^-$ & $\mathrm{Br}(\tau^-\to\nu_\tau X_s^-)$ (\% )& $X_s^-$ & $\mathrm{Br}(\tau^-\to\nu_\tau X_s^-)$ (\% )
\\[-4mm]& && \\ \hline &&&\\[-4mm]
$K^-$ & $(0.6955\pm 0.0096)$ &
$K^-\eta$ & $(0.0153\pm 0.0008)$   
\\
$K^-\pi^0$ & $(0.4322\pm 0.0149)$ &
$K^-\pi^0\eta$ & $(0.0048\pm 0.0012)$  
\\
$K^- 2\pi^0$ {\small (ex. $K^0$)} & $(0.0630\pm 0.0222)$ &
$\pi^- \bar K^0\eta$ & $(0.0094\pm 0.0015)$  
\\
$K^- 3\pi^0$ {\small (ex. $K^0,\eta$)} & $(0.0419\pm 0.0218)$ & 
$K^-\omega$ & $(0.0410\pm 0.0092)$  
\\
$\pi^- \bar K^0$ & $(0.8206\pm 0.0182)$ &
$K^-\phi\;$ {\small $(\phi\to K\bar K)$} & $(0.0037\pm 0.0014)$ 
\\
$\pi^- \bar K^0\pi^0$ & $(0.3649\pm 0.0108)$ &
$K^- \pi^-\pi^+$ {\small (ex. $K^0,\omega$)} & $(0.2923\pm 0.0068)$
\\
$\pi^- \bar K^0 2\pi^0$ & $(0.0269\pm 0.0230)$ &  
$K^- \pi^-\pi^+\pi^0$ {\small (ex. $K^0,\omega,\eta$)} & $(0.0411\pm 0.0143)$ 
\\
$\bar K^0 h^-h^+h^-$ & $(0.0222\pm 0.0202)$ && 
\\[1.5mm] \hline \multicolumn{4}{|c|}{}\\[-4mm]
\multicolumn{4}{|c|}{$\mathrm{Br}(\tau^-\to\nu_\tau X_s^-)\, =\, (2.875\pm 0.050)\%$}
\\[1.5mm]\hline
\end{tabular}
\end{table}
%

A big effort is underway to fully understand the rich pattern of hadronic decay modes of the $\tau$ \cite{Beringer:1900zz,HFAG}. Tables~\ref{tab:BR_CA}~and \ref{tab:BR_CS} show the present world averages for the Cabibbo allowed and Cabibbo suppressed branching ratios, using the 40 ``base modes'' defined by the {\it Heavy Flavor Averaging Group} (HFAG)~\cite{HFAG}.The huge data samples accumulated at the B factories have allowed for a sizeable reduction of the statistical errors, so systematic uncertainties dominate in most cases.
The global HFAG fit to the world $\tau$ data has a $\chi^2/\mathrm{d.o.f.} = 143.5/118$, corresponding to a confidence level $\mathrm{CL}=5.5\%$ \cite{HFAG}.
Adding the two leptonic branching ratios, given in table~\ref{tab:parameters}, the sum of all fitted branching ratios is consistent with one \cite{HFAG},
\bel{eq:missingBr}
1 - \sum_i\mathrm{Br}(\tau^-\to\nu_\tau^- X_i)\; =\; (0.0704\pm 0.1060)\%\, ,
\ee
but the unitarity constraint is not used in the fit.

Several inconsistencies are known to exist in the $\tau$ branching fraction measurements, but the sources of these inconsistencies are unknown. In particular, the recent decrease of many experimental branching ratios is worrisome.
There are 20 branching fractions measured at the B factories for which older measurements exist. As pointed out by the PDG \cite{Beringer:1900zz},
18 of these 20 B-factory branching fractions are smaller than the previous non-B-factory values. The average normalized difference between the two sets of measurements is $-1.30\,\sigma$. Moreover, the BaBar and Belle results differ significantly for the 6 decay modes measured by both experiments. New measurements and refined analyses are clearly needed.

Recent progress includes the measurement of many high-multiplicity 3- and 5-prong decays
\cite{Lees:2012ks}, modes with $K_S$ [$\pi^-K_S (\pi^0),\pi^-K_S K_S (\pi^0), K^-K_S (\pi^0)$] \cite{Lees:2012de,Ryu:2013lca} and analyses of hadronic distributions in decays into two and three mesons \cite{Ryu:2013lca,Nugent:2013ij}.

\section{The Inclusive $\tau$ Hadronic Width}
\label{sec:inclusive}

The inclusive character of the total $\tau$ hadronic width renders possible \cite{Narison:1988ni,Braaten:1988hc,Braaten:1991qm} an accurate calculation of the ratio
[$(\gamma)$ represents additional photons or lepton pairs]
\be\label{eq:r_tau_def}
R_\tau \;\equiv\; \frac{\Gamma [\tau^-\rightarrow\nu_\tau\,\mathrm{hadrons}\, (\gamma)]}{ \Gamma [\tau^-\rightarrow\nu_\tau e^- {\bar \nu}_e (\gamma)]}
\, ,
\ee
using standard field theoretic methods.
The theoretical analysis involves the two-point correlation functions for
the vector $\, V^{\mu}_{ij} = \bar{\psi}_j \gamma^{\mu} \psi_i \, $
and axial-vector
$\, A^{\mu}_{ij} = \bar{\psi}_j \gamma^{\mu} \gamma_5 \psi_i \,$
colour-singlet quark currents ($i,j=u,d,s$; $\cJ = V,A$):
\be\label{eq:pi_v}
\Pi^{\mu \nu}_{ij,\cJ}(q)\; \equiv\;
 i \int d^4x \, e^{iqx}\;
\langle 0|T(\cJ^{\mu}_{ij}(x) \cJ^{\nu}_{ij}(0)^\dagger)|0\rangle \, ,
\ee
which have the Lorentz decompositions
\be\label{eq:lorentz}
\Pi^{\mu \nu}_{ij,\cJ}(q) \; = \;
  \left( -g^{\mu\nu} q^2 + q^{\mu} q^{\nu}\right) \; \Pi_{ij,\cJ}^{(1)}(q^2)
\; +\;   q^{\mu} q^{\nu} \, \Pi_{ij,\cJ}^{(0)}(q^2) \, ,
\ee
where the superscript $(J=0,1)$ denotes the angular momentum in the hadronic rest frame.

The imaginary parts of the correlators $\, \Pi^{(J)}_{ij,\cJ}(q^2) \, $
are proportional to the spectral functions for hadrons with the corresponding
quantum numbers.  The hadronic decay rate of the $\tau$
can be written as an integral of these spectral functions
over the invariant mass $s$ of the final-state hadrons:
\be\label{eq:spectral}
R_\tau  \; = \;
12 \pi \,\int^{m_\tau^2}_0 {ds \over m_\tau^2 } \;
 \left(1-{s \over m_\tau^2}\right)^2\,
\left[ \left(1 + 2\, {s \over m_\tau^2}\right)\,
 \mathrm{Im} \Pi^{(1)}(s) \, +\, \mathrm{Im} \Pi^{(0)}(s) \right]\,  .
\ee
The appropriate combinations of correlators are
\be\label{eq:pi}
\Pi^{(J)}(s)  \; \equiv  \;
  |V_{ud}|^2 \, \left( \Pi^{(J)}_{ud,V}(s) + \Pi^{(J)}_{ud,A}(s) \right)
\, + \,
|V_{us}|^2 \, \left( \Pi^{(J)}_{us,V}(s) + \Pi^{(J)}_{us,A}(s) \right)\, .
\ee

We can separate the inclusive contributions associated with
specific quark currents:
\be\label{eq:r_tau_v,a,s}
 R_\tau \; = \; R_{\tau,V} + R_{\tau,A} + R_{\tau,S}\, .
\ee
$R_{\tau,V}$ and $R_{\tau,A}$ correspond to the first two terms in \eqn{eq:pi}, while
$R_{\tau,S}$ contains the remaining Cabibbo-suppressed contributions.
Non-strange hadronic decays of the $\tau$ are resolved experimentally
into vector ($R_{\tau,V}$) and axial-vector ($R_{\tau,A}$)
contributions according to whether the hadronic final state includes an even or odd number of pions. Strange decays ($R_{\tau,S}$) are of course identified by the
presence of an odd number of kaons in the final state.

\begin{figure}[tb]\centering
\includegraphics[width=7cm,clip]{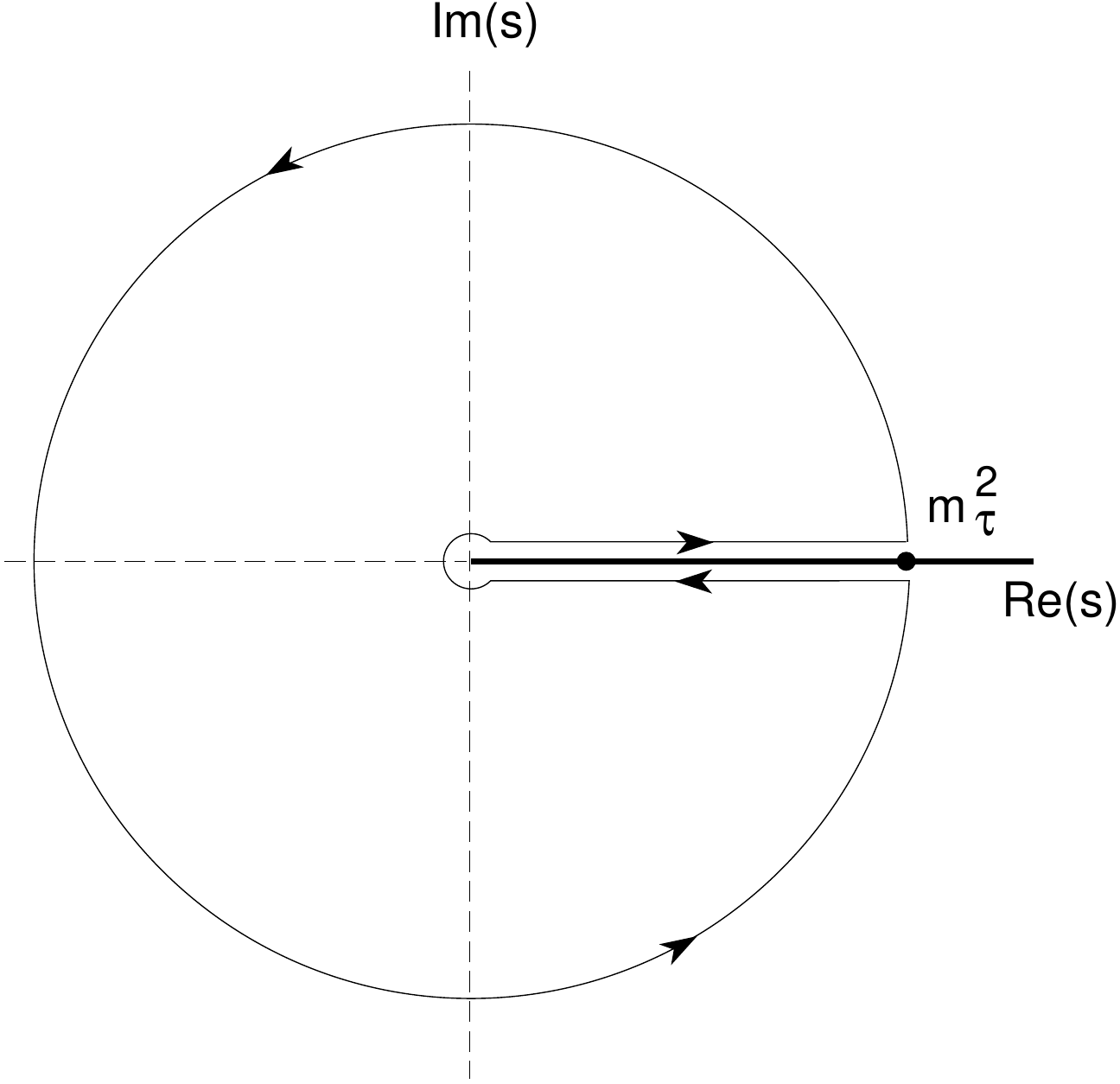}
\caption{Integration contour in the complex $s$ plane, used to obtain
Eq.~\protect\eqn{eq:circle}.}
\label{fig:contour}
\end{figure}

Since the hadronic spectral functions are sensitive to the non-perturbative
effects of QCD that bind quarks into hadrons, the integrand in
Eq.~\eqn{eq:spectral} cannot be calculated at present from QCD.
Nevertheless the integral itself can be calculated systematically by exploiting
the analytic properties of the correlators $\Pi^{(J)}(s)$. They are analytic
functions of $s$ except along the positive real $s$ axis, where their
imaginary parts have discontinuities. Using the closed contour in Fig.~\ref{fig:contour},
$R_\tau$ can then be expressed as a contour integral in the complex $s$ plane running
counter-clockwise around the circle $|s|=m_\tau^2$ \cite{Narison:1988ni,Braaten:1988hc,Braaten:1991qm}:
\bel{eq:circle}
 R_\tau\;  =\; 6 \pi i\, \oint_{|s|=m_\tau^2} {ds \over m_\tau^2} \;
 \left(1 - {s \over m_\tau^2}\right)^2\,
 \left[ \left(1 + 2\, {s \over m_\tau^2}\right)\, \Pi^{(0+1)}(s)
         - 2\, {s \over m_\tau^2}\, \Pi^{(0)}(s) \right] \, .
\ee

The advantage of expression \eqn{eq:circle} over \eqn{eq:spectral} for $R_\tau$
is that it requires the correlators only for complex $s$ of order $m_\tau^2$, which is significantly larger than the scale associated with non-perturbative effects in QCD.  The short-distance Operator Product Expansion (OPE) can therefore be used to organize
the perturbative and non-perturbative contributions to the correlators into a systematic expansion in powers of $1/s$ \cite{Shifman:1978bx},
\be\label{eq:ope}
 \Pi^{(J)}(s) \; =\; \sum_{D=2n}\,\sum_{\mathrm{dim}\, {\cal O} = D}\,
 \frac{{\cal C}^{(J)}(s,\mu) \,\langle {\cal O}(\mu)\rangle}{(-s)^{D/2}}\, ,
\ee
where the inner sum is over local gauge-invariant scalar operators of dimension $D=0,2,4\ldots$.
The parameter $\mu$ is an arbitrary factorization scale, which separates long-distance
non-perturbative effects, which are absorbed into the vacuum matrix elements
$\langle {\cal O}(\mu)\rangle $, from short-distance effects, which are included
in the Wilson coefficients ${\cal C}^{(J)}(s,\mu)$.
The $D=0$ term (unit operator) corresponds to the pure perturbative
contributions, neglecting quark masses. The leading quark-mass
corrections generate the $D=2$ term. The first dynamical operators
involving non-perturbative physics appear at $D=4$.

Inserting the functions \eqn{eq:ope} into \eqn{eq:circle} and evaluating the contour integral, $R_\tau$ can be expressed as an expansion in powers of $1/m_\tau^2$,
with coefficients that depend only logarithmically on $m_\tau$.
The uncertainties associated with the use of the OPE near the time-like axis are heavily suppressed because the integrand in Eq.~\eqn{eq:circle} includes a factor
$(1- s/m_\tau^2)^2$, which provides a double zero at $s=m_\tau^2$,
making negligible the contribution from the region near the branch cut.

It is convenient to express the corrections to $R_\tau$
from dimension-$D$ operators in terms of the
fractional corrections $\delta^{(D)}_{ij,\cJ}$ to the
naive contribution from the current with quantum numbers $ij,\cJ$ \cite{Braaten:1991qm}:
\be\label{eq:r_v}
R_{\tau,V/A}\;  = \; {3 \over 2}\, |V_{ud}|^2\,
   S_{\mathrm{EW}}\; \left( 1 + \delta_P +
      \sum_{D=2,4\ldots} \delta^{(D)}_{ud,V/A} \right)\,  ,
\ee
\be \label{eq:r_s}
R_{\tau,S}\;  =\;
 3\, |V_{us}|^2\, S_{\mathrm{EW}}\; \left( 1 + \delta_P +
  \sum_{D=2,4\ldots} \delta^{(D)}_{us} \right)\, ,
\ee
where
$\delta^{(D)}_{ij} = (\delta^{(D)}_{ij,V} + \delta^{(D)}_{ij,A})/2$
is the average of the vector and axial-vector corrections,
the factor $S_{\mathrm{EW}} =1.0201\pm 0.0003$\ contains the renormalization-group improved
electroweak correction \cite{Marciano:1988vm,Braaten:1990ef,Erler:2002mv},
and $\delta_P = \delta^{(0)}_{ij,\cJ}$ is the purely perturbative QCD correction, neglecting quark masses, which is the same for all the components of $R_\tau$.

In the chiral limit ($m_u=m_d=m_s=0$), the vector and axial-vector currents are conserved.
This implies  $s \,\Pi^{(0)}(s) = 0$. Therefore, only the correlator
$\Pi^{(0+1)}(s)$ contributes to Eq.~(\ref{eq:circle}).
Since $(1 - x)^2 (1 + 2 x) =
1 - 3 x^2 + 2x^3$ [$x\equiv s/m_\tau^2$],
Cauchy's theorem guarantees that, up to tiny logarithmic running corrections, the only non-perturbative contributions to the circle integration in (\ref{eq:circle}) originate from operators of dimensions $D=6$ and 8. The usually leading $D=4$ operators can only contribute to $R_\tau$ with an additional suppression factor of $\cO(\alpha_s^2)$, which makes their effect negligible \cite{Braaten:1991qm}.

The Cabibbo-allowed component of $R_\tau$,
\begin{equation}\label{eq:Rv+a}
 R_{\tau,V+A} \; =\; N_C\, |V_{ud}|^2\, S_{\mathrm{EW}}\; \left\{ 1 +
 \delta_{\mathrm{P}} + \delta_{\mathrm{NP}} \right\} \, ,
\end{equation}
is then a very clean observable to test perturbative QCD. Quark mass effects are tiny and the non-perturbative correction $\delta_{\mathrm{NP}} = \sum_{D\ge 2} \delta^{(D)}_{ud}$ is heavily suppressed by six powers of the $\tau$ mass. The dominant contribution is
the perturbative correction $\delta_{\mathrm{P}}\sim 20\%$, which
turns out to be very sensitive to $\alpha_s$, allowing for an accurate
determination of the fundamental QCD coupling \cite{Narison:1988ni,Braaten:1991qm}.
The recent calculation of the $\cO(\alpha_s^4)$ contribution to $\Pi^{(0+1)}(s)$ \cite{Baikov:2008jh} has triggered a renewed theoretical interest on the $\alpha_s(m_\tau^2)$ determination \cite{Davier:2005xq,Davier:2008sk,Beneke:2008ad,Beneke:2012vb,Caprini:2011ya,Abbas:2012fi,Abbas:2012py,Groote:2012jq,Maltman:2008nf,Boito:2012cr,Menke:2009vg,Narison:2009vy,Cvetic:2010ut,Pich:2011bb,Pich:2013sqa}, since it allows to push the accuracy to the four-loop level.

\subsection{Perturbative Contribution to $\boldmath R_\tau$}

In the chiral limit, the result is more conveniently expressed in terms of the
logarithmic derivative of the two-point correlation function of the vector (axial) current,
$\Pi(s)=\Pi^{(0+1)}_{i\not= j,V}(s)=\Pi^{(0+1)}_{i\not= j,A}(s)$, the so-called Adler function \cite{Adler:1974gd}, which satisfies an homogeneous renormalization-group equation:
\be\label{eq:d}
 D(s) \; \equiv\; - s {d \over d s } \Pi(s)\; =\;  \frac{N_C}{12 \pi^2}\; \left\{\, 1 + \sum_{n=1}  K_n
\left( {\alpha_s(-s)\over \pi}\right)^n \right\}\, .
\ee
With the choice of renormalization scale $\mu^2= - s$, all logarithmic corrections
proportional to powers of $\log{(-s/\mu^2)}$ have been summed into the running coupling.
For three active quark flavours and in the $\overline{\mathrm{MS}}$ scheme,
the known ($n\le 4$) $K_n$ coefficients take the values indicated in table~\ref{tab:Kcoeff}
\cite{Baikov:2008jh,Gorishnii:1990vf,Surguladze:1990tg}.

%
\begin{table}[tb]\centering
\caption{Perturbative coefficients of the Adler function expansion and FOPT approximation to $\delta_P$.}
\label{tab:Kcoeff}\vspace{0.2cm}
\renewcommand{\tabcolsep}{1.3pc} 
\renewcommand{\arraystretch}{1.2} 
\begin{tabular}{cccccc}
\hline
$n$ & 1 & 2 & 3 & 4 & 5
\\ \hline
$K_n$ & 1 & $1.6398$ & $6.3710$ & $49.0757$ &
\\
$g_n$ & 0 & $3.5625$ & $19.9949$ & $78.0029$ & $307.78$
\\
$r_n$ & 1 & 5.2023 & $26.3659$ & $127.079$ &
\\ \hline
\end{tabular}
\end{table}
%

The perturbative component of $R_\tau$ is given by
\be\label{eq:r_k_exp}
\delta_{\mathrm{P}} \; =\;
\sum_{n=1}  K_n \, A^{(n)}(\alpha_s)
\; =\; \sum_{n=1}\,  (K_n + g_n) \, a_\tau^n \;\equiv\;
\sum_{n=1}\,  r_n \, a_\tau^n \, ,
\ee
where the functions \cite{LeDiberder:1992te}
\be\label{eq:a_xi}
A^{(n)}(\alpha_s) \; = \; {1\over 2 \pi i}\,
\oint_{|s| = m_\tau^2} {ds \over s} \;
  \left({\alpha_s(-s)\over\pi}\right)^n\;
 \left( 1 - 2 {s \over m_\tau^2} + 2 {s^3 \over m_\tau^6}
         - {s^4 \over  m_\tau^8} \right)
\ee
are contour integrals in the complex plane, which only depend on
$a_\tau\equiv\alpha_s(m_\tau^2)/\pi$. Using the exact solution
(up to unknown $\beta_{n>4}$ contributions) for $\alpha_s(-s)$
given by the renormalization-group $\beta$-function equation,
they can be numerically computed with a very high accuracy \cite{LeDiberder:1992te}.
Table~\ref{tab:Afun} gives the numerical values
for $A^{(n)}(\alpha_s)$ ($n\le 4$) obtained at the one-, two-, three- and four-loop
approximations (i.e. $\beta_{n>1}=0$, $\beta_{n>2}=0$, $\beta_{n>3}=0$ and $\beta_{n>4}=0$,
respectively), together with the corresponding results for $\delta_{\mathrm{P}} = \sum_{n=1}^4\, K_n\, A^{(n)}(\alpha_s)$, taking $a_\tau=0.11$.
The perturbative convergence is very good and the results are stable under changes of the renormalization scale.
%

%
\begin{table}[tb]\centering
\caption{ Exact results
for $A^{(n)}(\alpha_s)$ ($n\le 4$) at different $\beta$-function approximations,
and corresponding values of \ $\delta_{\mathrm{P}} = \sum_{n=1}^4\, K_n\, A^{(n)}(\alpha_s)$,
for $a_\tau\equiv\alpha_s(m_\tau^2)/\pi=0.11$. The last row shows the FOPT estimates at $\cO(a_\tau^4)$.}
\label{tab:Afun}\vspace{0.2cm}
\newcommand{\cc}[1]{\multicolumn{1}{c}{#1}}
\renewcommand{\tabcolsep}{1.5pc} 
\renewcommand{\arraystretch}{1.2} 
\begin{tabular}{llllll}
\hline
& \cc{$A^{(1)}(\alpha_s)$} & \cc{$A^{(2)}(\alpha_s)$} & \cc{$A^{(3)}(\alpha_s)$}
 & \cc{$A^{(4)}(\alpha_s)$} & 
 \cc{$\delta_{\mathrm{P}}$}
\\ \hline
$\beta_{n>1}=0$ & $0.14828$ & $0.01925$ & $0.00225$ & $0.00024$ & 
$0.20578$ \\
$\beta_{n>2}=0$ & $0.15103$ & $0.01905$ & $0.00209$ & $0.00020$ & 
$0.20537$ \\
$\beta_{n>3}=0$ & $0.15093$ & $0.01882$ & $0.00202$ & $0.00019$ & 
$0.20389$ \\
$\beta_{n>4}=0$ & $0.15058$ & $0.01865$ & $0.00198$ & $0.00018$ & 
$0.20273$
\\ \hline
$\cO(a_\tau^4)$ &  $0.16115$ & $0.02431$ & $0.00290$ & $0.00015$ & 
$0.22665$
\\ \hline
\end{tabular}
\end{table}
%

The integrals $A^{(n)}(\alpha_s)$ can be expanded in powers of $a_\tau$,
$A^{(n)}(\alpha_s) = a_\tau^n + \cO(a_\tau^{n+1})$. One obtains in this way the naive perturbative expansion given in the r.h.s of Eq.~\eqn{eq:r_k_exp}.
This approximation is known as {\it fixed-order perturbation theory} (FOPT), while
the improved expression, keeping the non-expanded values of $A^{(n)}(\alpha_s)$, is usually called {\it contour-improved perturbation theory} (CIPT) \cite{LeDiberder:1992te,Pivovarov:1991rh}.

As shown in the last row of table~\ref{tab:Afun}, even at $\cO(a_\tau^4)$, FOPT gives a rather bad approximation to the integrals $A^{(n)}(\alpha_s)$, overestimating $\delta_{\mathrm{P}}$ by 11\% at $a_\tau = 0.11$. The long running of $\alpha_s(-s)$ along the circle $|s|=m_\tau^2$ generates the very large $g_n$ coefficients, given in table~\ref{tab:Kcoeff}, which depend on $K_{m<n}$ and $\beta_{m<n}$ \cite{LeDiberder:1992te}.
These corrections are much larger than the original $K_n$ contributions.
The origin of this bad behaviour can be understood analytically at one loop \cite{LeDiberder:1992te}. In FOPT one makes within
the contour integral the series expansion ($\log{(-s/m_\tau^2)} = i\phi$, $\phi\in [-\pi, \pi]$)
\bel{eq:fopt_ap}
\frac{\alpha_s(-s)}{\pi} \;\approx\; \frac{a_\tau}{1-i\beta_1 a_\tau\phi/2}\;\approx\; a_\tau \sum_n \left(\frac{i}{2}\beta_1 a_\tau\phi\right)^n \, ,
\ee
which is only convergent for $a_\tau < 0.14$. At the four-loop level the radius of convergence is slightly smaller than the physical value of $a_\tau$. Thus, FOPT gives rise to a non-convergent series, which suffers from a large renormalization-scale dependence. The long running along the circle makes compulsory to resum the large logarithms, $\log^n{(-s/m_\tau^2)}$, using the renormalization group. This is precisely what CIPT does, and results in a well-behaved perturbative series with a mild dependence on the renormalization scale.

\subsubsection{Higher-order perturbative contributions and renormalon hypothesis}

A lot of effort has been devoted to estimate the size of the unknown higher-order corrections to $\delta_{\mathrm{P}}$
\cite{Davier:2005xq,Davier:2008sk,Beneke:2008ad,Beneke:2012vb,Caprini:2011ya,Abbas:2012fi,Abbas:2012py,Groote:2012jq,Cvetic:2010ut,Pich:2011bb,Kataev:1995vh,Cvetic:2001ws,Ball:1995ni,Neubert:1995gd,Altarelli:1994vz,Jamin:2005ip,DescotesGenon:2010cr,Maxwell:2001he,Maxwell:1997yw,LovettTurner:1995ti,Maxwell:1989py}.
The perturbative expansion of the Adler function is expected to be an asymptotic series. If its Borel transform,
$B(t)\equiv\sum_{n=0} K_{n+1} t^n/n!$, were well-behaved, one could define $D(s)$ through the Borel integral
\bel{eq:borel}
D(s)\; =\; \frac{1}{4\pi^2}\,\left\{ 1 + \int_0^\infty dt\;
\mathrm{e}^{-\pi t/\alpha_s(s)}\, B(t)\right\}\, .
\ee
However, $B(t)$ has pole singularities at positive (infrared renormalons) and negative (ultraviolet renormalons) integer values of the variable
$u\equiv -\beta_1 t/2$, with the exception of $u=1$ \cite{Beneke:1998ui}.
The infrared renormalons at $u=+n$ are related to OPE corrections of dimension $D=2n$; an infrared singularity at $u=+1$ does not exist because there are no gauge-invariant operators with $D=2$.
The renormalon poles closer to the origin dominate the large-order behaviour of $D(s)$.

It has been argued that, once in the asymptotic regime (large $n$), the renormalonic behaviour of the $K_n$ coefficients could induce
cancellations with the running $g_n$ corrections, which would be missed by
CIPT. In that case, FOPT could approach faster the `true' result provided by the Borel summation of the full renormalon series \cite{Beneke:2008ad}.
This happens actually in the large-$\beta_1$ limit \cite{Ball:1995ni,Neubert:1995gd,Altarelli:1994vz},
which however does not approximate well the known $K_n$ coefficients.
A large $u=2$ renormalon effect could also induce such a cancellation, because the contour integral eliminates contributions from $D=4$ operators.
Models of higher-order corrections with this behaviour have been advocated, mixing different types of renormalons ($n=-1$, 2 and 3) plus a linear polynomial, and fitting the free parameters to the known $K_{n\le 4}$ coefficients with
the assumption $K_5=283$ \cite{Beneke:2008ad}. For a given value of $\alpha_s(m_\tau^2)$,
one gets in this way a larger  $\delta_{\mathrm{P}}$ than in CIPT. The result looks however model dependent \cite{Jamin:2005ip,DescotesGenon:2010cr}.

The implications of a renormalonic behaviour have also been studied in a more model-independent way, using an optimal conformal mapping in the Borel plane, which achieves the best asymptotic rate of convergence,
and properly implementing the CIPT procedure within the Borel transform \cite{Caprini:2011ya,Abbas:2012fi,Abbas:2012py}. Assuming that the known fourth-order series is already governed by the $u=-1$ and $u=2$ renormalons, the conformal mapping generates a full series expansion ($K_5=256$, $K_6=2929$ \ldots) which results, after Borel summation, in a larger value of $\delta_{\mathrm{P}}$; i.e., the $K_{n>4}$ terms give a positive contribution to $\delta_{\mathrm{P}}$. One obtains then numerical results close to the FOPT value.

Renormalons provide an interesting guide to possible higher-order corrections,
making apparent that the associated uncertainties have to be carefully estimated. However, one should keep in mind the adopted assumptions. In fact, there are no visible signs of renormalonic behaviour in the presently known series: the $n=-1$ ultraviolet renormalon is expected to dominate the asymptotic regime, implying an alternating series,
while all known $K_n$ coefficients have the same sign. One could either assume that renormalons only become relevant at higher
orders, for instance at $n=7$, and apply the conformal mapping with arbitrary input values for $K_5$ and $K_6$. Different assumptions about these two unknown coefficients would result in different predicted values for $\delta_{\mathrm{P}}$.

A different reshuffling of the perturbative series, not related to renormalons,
can be obtained \cite{Cvetic:2010ut} through an expansion in terms of the $\beta$ function and its derivatives, instead of the usual expansion in powers of the strong coupling.
This procedure results in a different estimate of higher-order corrections, leading to a weaker dependence on the renormalization scale and a value of $\delta_{\mathrm{P}}$ similar to the standard CIPT result.

\subsection{Non-Perturbative Corrections}

Compared with the perturbative uncertainties, the power-suppressed contributions to
$R_\tau$ are quite small. Quark mass effects  \cite{Braaten:1991qm,Chetyrkin:1993hi,Pich:1999hc,Baikov:2004tk}
are tiny for the Cabibbo-allowed current and amount to a negligible correction smaller than $10^{-4}$. Non-perturbative effects are suppressed by six powers of the
$\tau$ mass \cite{Braaten:1991qm}. Moreover, the $D=6$ contributions
to the vector and axial-vector correlators are expected to have opposite signs
(the vacuum-saturation approximation gives $\delta^{(6)}_{ud,V}/\delta^{(6)}_{ud,A}\sim -7/11$), leading to a partial cancellation in $\delta^{(6)}_{ud}$. Thus, the non-perturbative contribution to the total ($V+A$) Cabibbo-allowed decay width is smaller than the separate vector and axial-vector corrections.
The estimated theoretical value is $\delta_{\mathrm{NP}} = -0.007\pm 0.004$~\cite{Braaten:1991qm}.

The numerical size of non-perturbative effects can be determined from the measured invariant-mass distribution of the final hadrons in $\tau$ decay \cite{LeDiberder:1992fr}.
Although the distributions themselves cannot be predicted at present,
certain weighted integrals of the hadronic spectral functions can be
calculated in the same way as $R_\tau$.
The analyticity properties of $\Pi^{(J)}_{ij,\cJ}$ imply
\cite{PI:89,LeDiberder:1992fr}:
\bel{eq:weighted_integrals}
\int_0^{s_0} ds\; w(s)\; \mathrm{Im}\Pi^{(J)}_{ij,\cJ}\; =\;
\frac{i}{2}\; \oint_{|s|=s_0} ds\; w(s) \;\Pi^{(J)}_{ij,\cJ}\, ,
\ee
with $w(s)$ an arbitrary weight function without singularities in the
region $|s|\leq s_0$.
Generally speaking, the accuracy of the theoretical predictions can be
much worse than the one of $R_\tau$, because non-perturbative effects
are not necessarily suppressed; in fact, they can become very sizeable
with appropriately chosen weight functions. But this is precisely
what makes these integrals interesting: they can be used to measure the
parameters characterizing the non-perturbative dynamics and perform relevant tests of QCD in the strong-coupling regime. Notice that weights of the form $(s/m_\tau^2)^n$ project the OPE contribution of dimension $D=2n+2$.

\begin{figure}[t]\centering
\begin{minipage}{8.75cm}\centering
\includegraphics[width=8.75cm,clip]{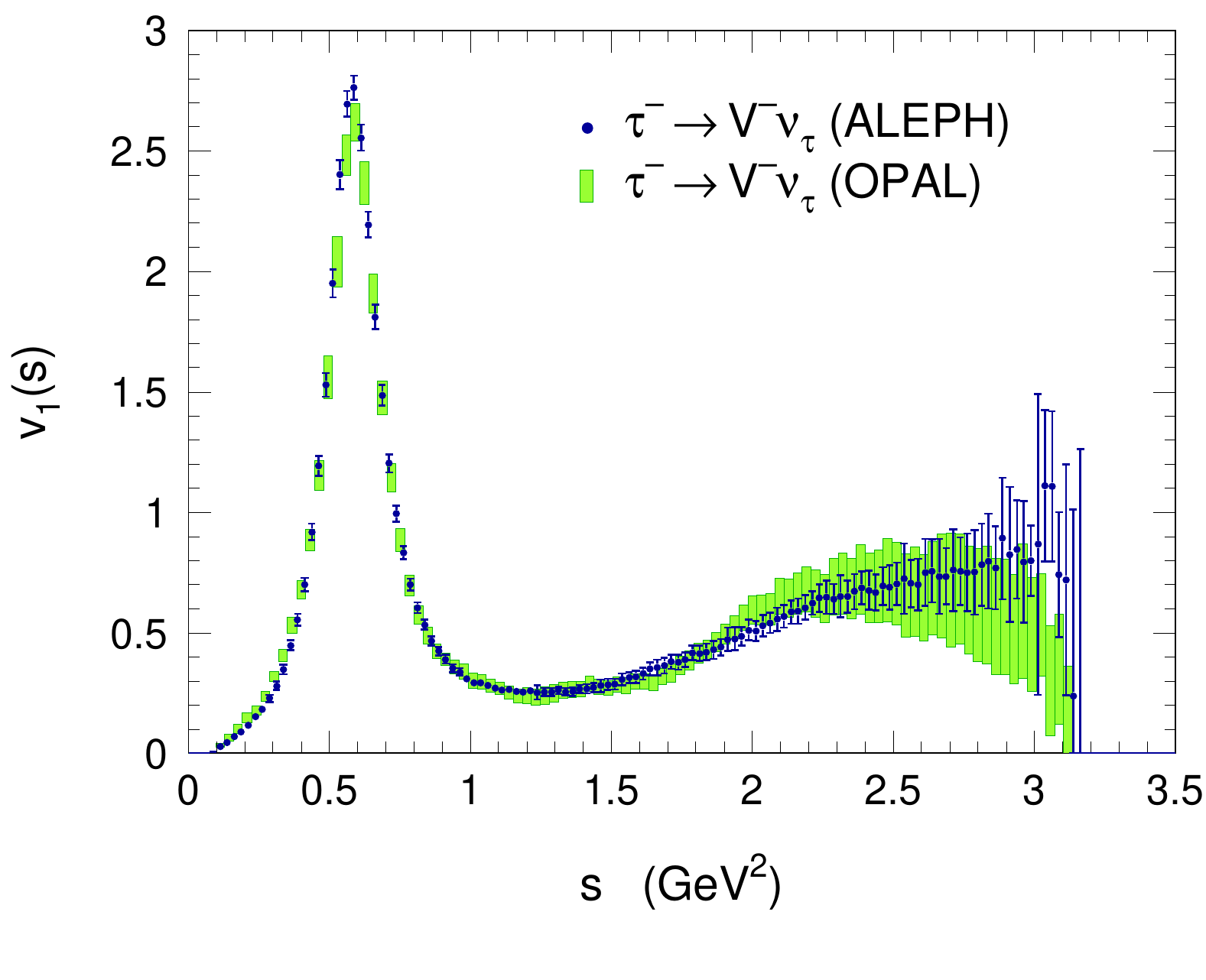}
\includegraphics[width=8.75cm,clip]{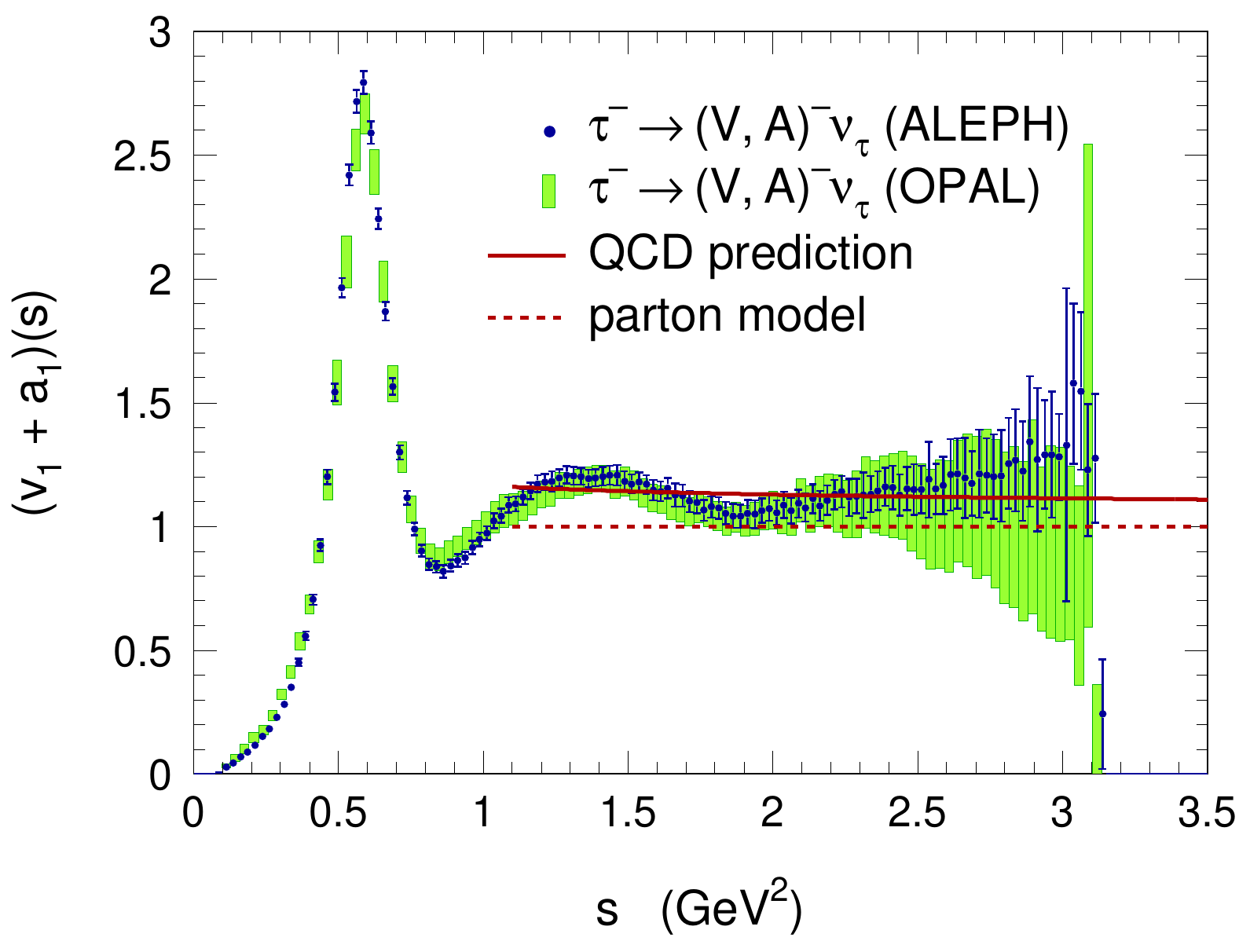}
\end{minipage}
\hskip .5cm
\begin{minipage}{8.75cm}\centering
\includegraphics[width=8.75cm]{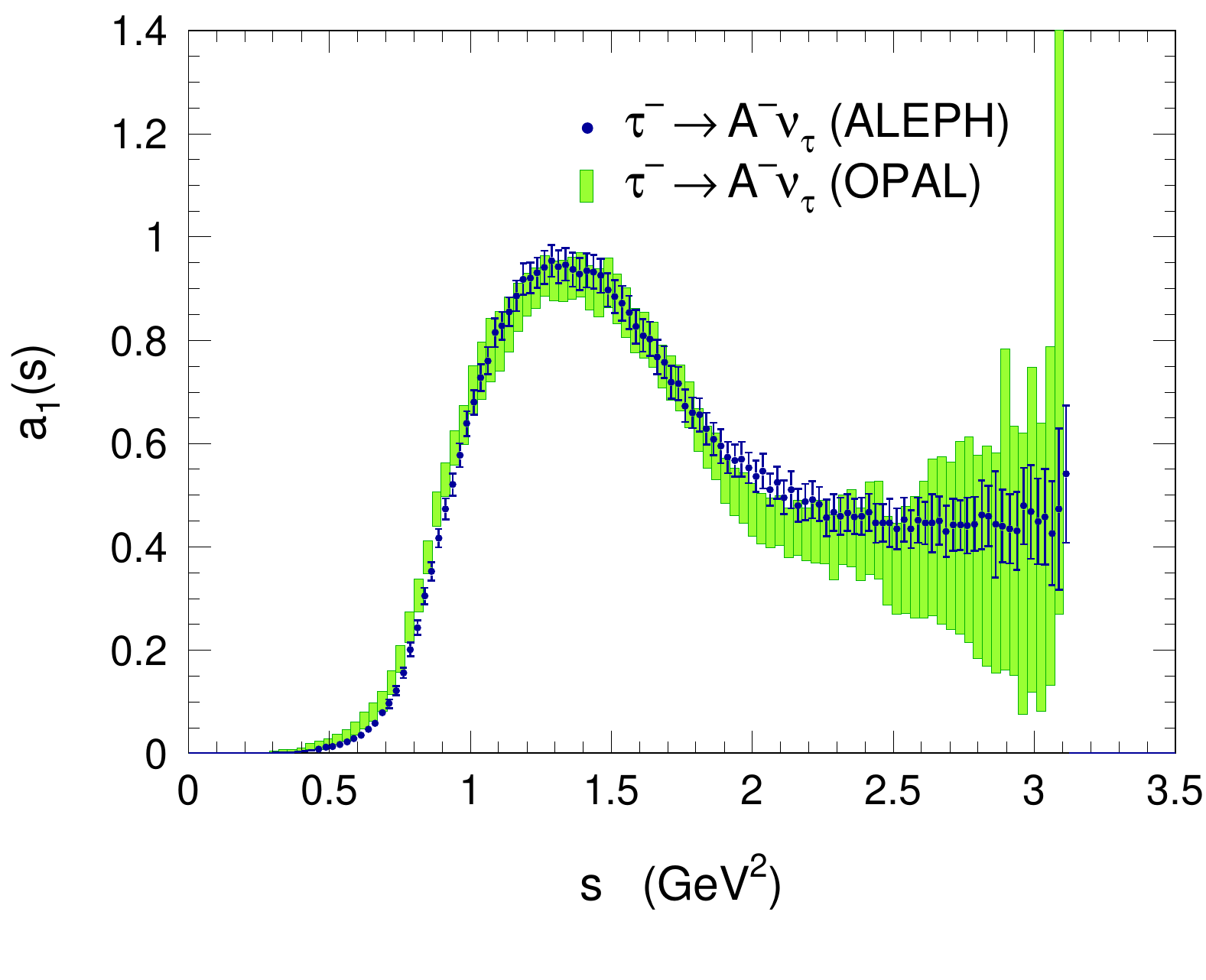}
\includegraphics[width=8.75cm,clip]{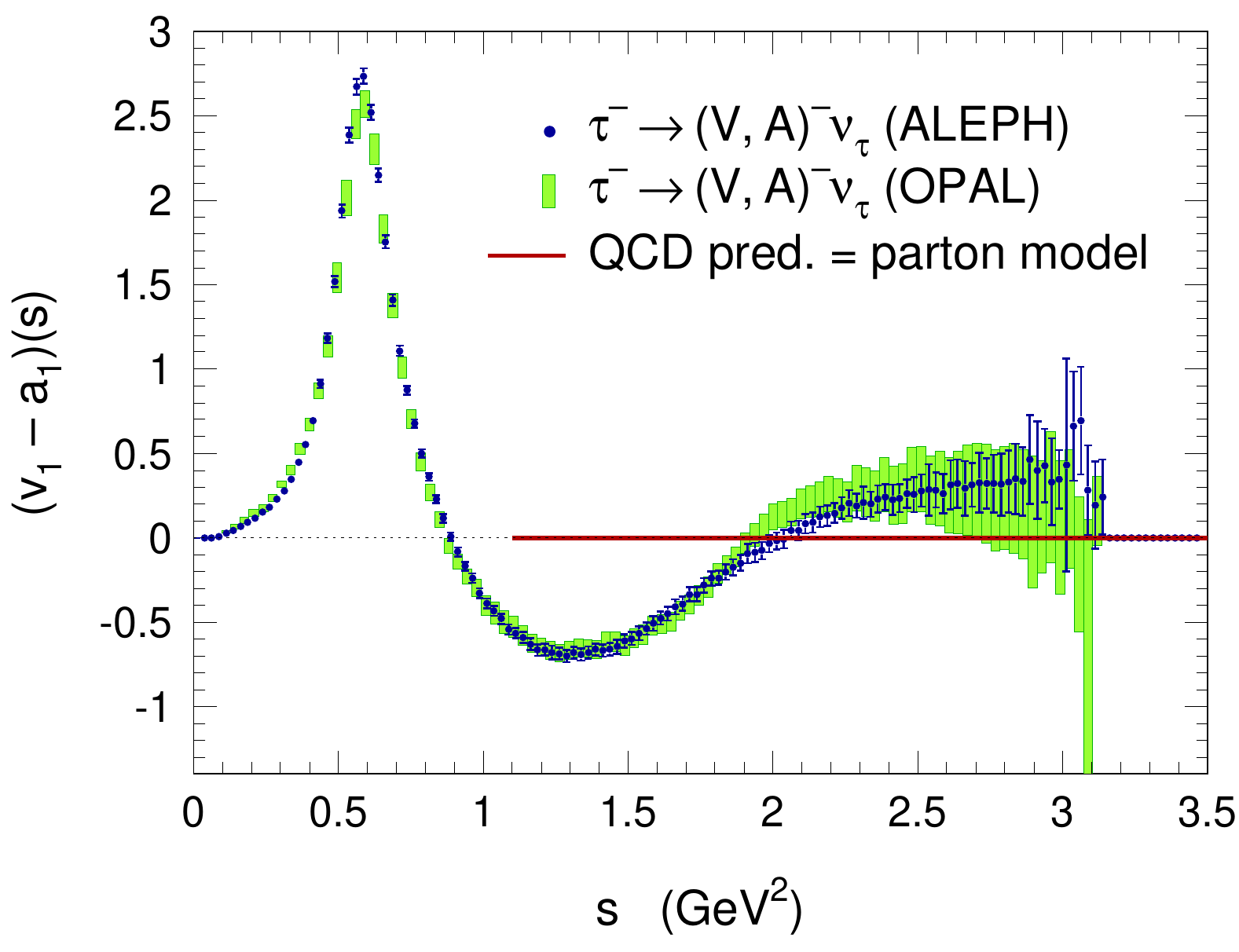}
\end{minipage}
\caption{Vector (upper-left) and axial (upper-right) spectral functions measured by ALEPH \cite{Schael:2005am} (blue data points), compared with the corresponding OPAL distributions \cite{Ackerstaff:1998yj} (green rectangular blocks). The lower plots show the sum (left) and difference (right) of the vector and axial spectral functions, together with the perturbative QCD predictions (continuous red lines). Figures taken from Ref.~\cite{Davier:2005xq}.}
\label{fig:spectralF}
\end{figure}

To perform an experimental analysis, it is convenient to use
moments of the directly measured invariant-mass distribution, for instance
\cite{LeDiberder:1992fr} ($k,l\ge 0$)
\be\label{eq:moments}
R^{kl}_{\tau,V+A}(s_0) \;\equiv\;\int_0^{s_0}\, ds
 \, \left(1 - {s\over s_0}\right)^k \left ( {s \over m_\tau^2} \right )^l\;
\frac{d  R_{\tau,V+A}}{d s}\, .
\ee
The factor $(1-s/s_0)^k$ supplements $(1-s/m_\tau^2)^2$ for
$s_0\not= m_\tau^2$,
in order to squeeze the integrand at the crossing of the positive real axis
and, therefore, improves the reliability of the OPE analysis; moreover, for
$s_0=m_\tau^2$ it reduces the contribution from the tail of the distribution,
which is badly defined experimentally.
A combined fit of different $R_{\tau,V+A}^{kl}(m_\tau^2)$ moments results in experimental
values for $\alpha_s(m_\tau^2)$ and for the coefficients of the inverse power corrections
in the OPE. $R_{\tau,V+A}^{00}(m_\tau^2) = R_{\tau,V+A}$ uses the overall normalization of the hadronic
distribution, while the ratios $D_{\tau,V+A}^{kl}(m_\tau^2) = R^{kl}_{\tau,V+A} (m_\tau^2)/R_{\tau,V+A}$ are based on the shape of the $s$ distribution and are more dependent on non-perturbative effects \cite{LeDiberder:1992fr}.

The spectral functions measured by ALEPH \cite{Schael:2005am} and OPAL \cite{Ackerstaff:1998yj} are shown in Figs.~\ref{fig:spectralF}, with the normalization
$v_1(s) = 2\pi\,\mathrm{Im}\Pi^{(1)}_{ud,V}(s)$ and
$a_1(s) = 2\pi\,\mathrm{Im}\Pi^{(1)}_{ud,A}(s)$. There is good agreement between both experiments, although the ALEPH results are more precise. The errors increase in the high $s$ region because of the kinematical phase-space suppression close to the end-point of the $\tau$ decay distribution. The $\rho$ and $a_1$ resonances are clearly visible. One also observes the expected smearing of resonance structures above 1~GeV in the $V+A$ distribution, where the data approaches the perturbative QCD prediction. The opposite behaviour is seen in the $V-A$ spectral function, which is a purely non-perturbative quantity. The total $V+A$ distribution was previously studied by CLEO \cite{Coan:1995nk}, but with much larger uncertainties.

The predicted suppression \cite{Braaten:1991qm} of the non-perturbative corrections to $R_\tau$ has been confirmed by ALEPH \cite{Schael:2005am,Barate:1998uf,Buskulic:1993sv}, CLEO \cite{Coan:1995nk} and OPAL \cite{Ackerstaff:1998yj}. The presently most complete and precise experimental analysis, performed with the ALEPH data, obtains
\cite{Davier:2005xq,Davier:2008sk}
\begin{equation}\label{eq:del_np}
 \delta_{\mathrm{NP}} \; =\; -0.0059\pm 0.0014 \, ,
\end{equation}
in good agreement with the theoretical expectations and previous experimental determinations. Larger non-perturbative
effects have been found for the separate currents,
$\delta_{\mathrm{NP}}^V  = 0.0189\pm 0.0025$ and
$\delta_{\mathrm{NP}}^A  = -0.0311\pm 0.0016$ \cite{Davier:2008sk},
with the expected size and signs.
Using the CIPT prescription, the fit to the whole ALEPH distribution gives
$\alpha_s(m_\tau^2) = 0.344\pm 0.009$, while the separate vector and axial data result in
$\alpha_s(m_\tau^2) = 0.347\pm 0.010$ and  $\alpha_s(m_\tau^2) = 0.335\pm 0.011$, respectively \cite{Davier:2008sk}.

\begin{figure}[t]\centering
\begin{minipage}{8.75cm}\centering
\includegraphics[width=8.75cm,clip]{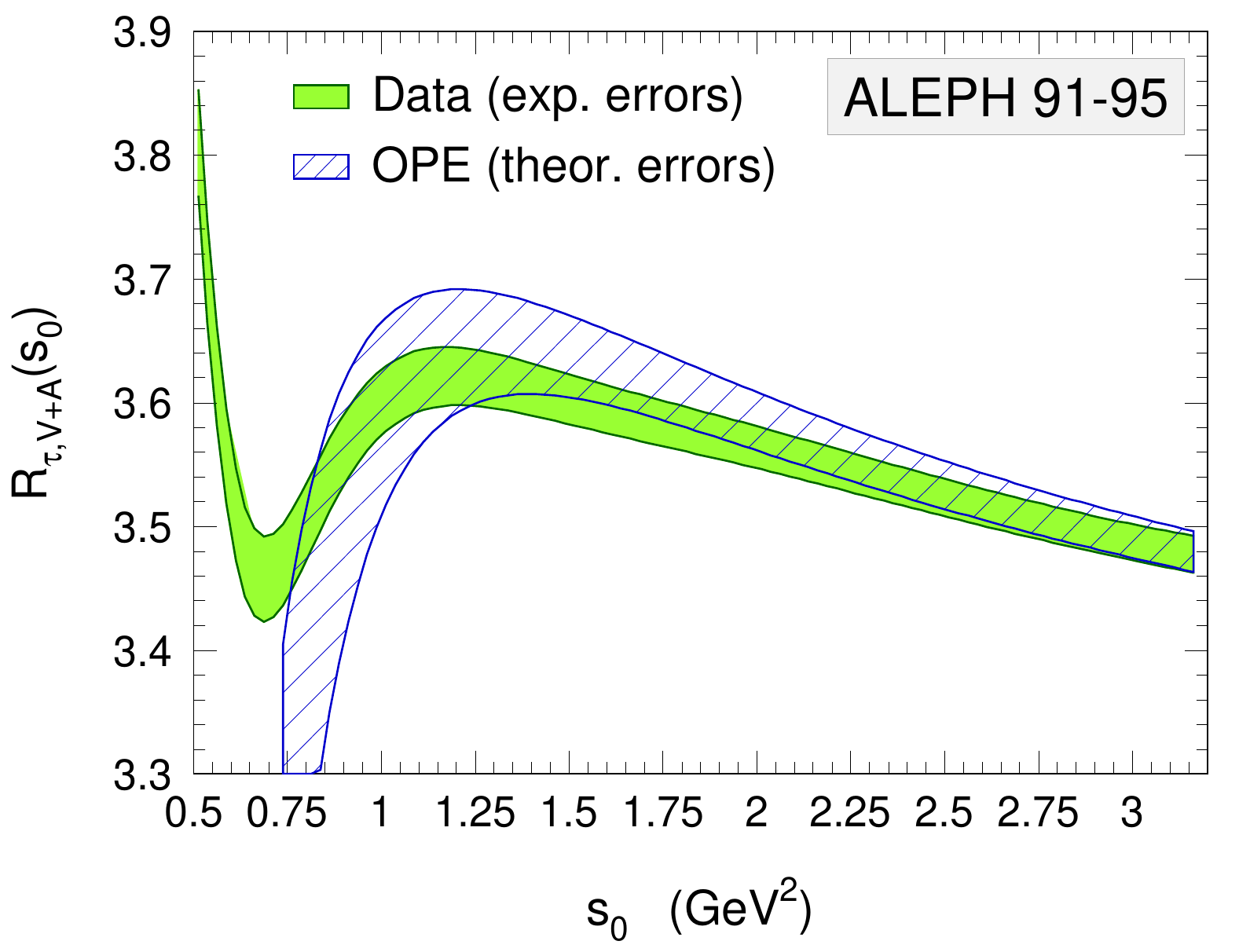}
\end{minipage}
\hskip .5cm
\begin{minipage}{8.75cm}\centering
\includegraphics[width=8.75cm]{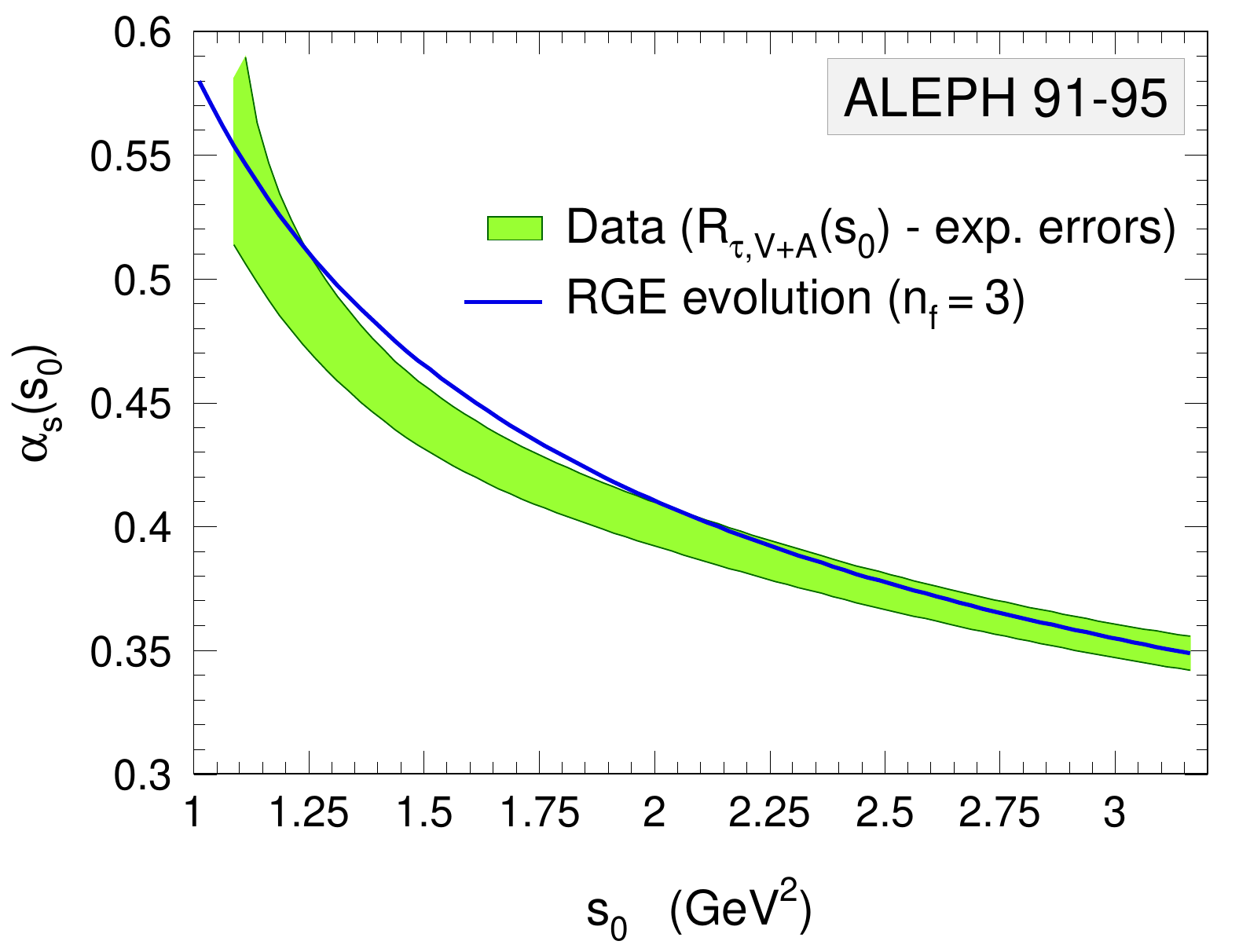}
\end{minipage}
\caption{Dependence on $s_0$ of $R_{\tau,V+A}(s_0)$ (left) and $\alpha_s(s_0)$ (right) \cite{Davier:2005xq}. The ALEPH fit (green bands) is compared with the theoretical CIPT prediction.}
\label{fig:alpha-s0}
\end{figure}

Evaluating the integral $R_{\tau,V+A}^{00}(s_0)$ with an arbitrary
upper limit of integration $s_0\leq m_\tau^2$,
one can test that the experimental $s_0$ dependence agrees well with the
theoretical predictions up to rather low values
of $s_0$ \cite{Davier:2005xq,Girone:1995xb}.
Equivalently,  from the measured $R_\tau^{00}(s_0)$ distribution one obtains
$\alpha_s(s_0)$ as a function of the scale $s_0$. As shown in Fig.~\ref{fig:alpha-s0},
the result exhibits an impressive agreement with the
running predicted at four-loop order by QCD.
A similar test was performed longtime ago for $R_{\tau,V}$,
using the vector spectral function measured in $e^+e^-\to$ hadrons, and
varying the value of the $\tau$ mass \cite{Narison:1993sx}.

All previous analyses were based on weight functions proportional to powers of $(1-s/s_0)^k$ (pinched weights), in order to suppress the OPE uncertainties near the real axis \cite{Davier:2005xq,Davier:2008sk,DescotesGenon:2010cr,Nason:1993ak,Balitsky:1993ki,Shifman:2000jv,Cata:2008ru,Jamin:2011vd}.
Using instead non-pinched weights, one can also analyse the role of duality violations in generic moments of the spectral distributions.
These effects are negligible in $R_{\tau,V+A}$, since they are smaller than the errors induced by $\delta_{\mathrm{NP}}$ which are in turn subdominant with respect to the leading perturbative uncertainties; however, they could be more relevant for other moments.
Although the present data are not good enough to perform accurate tests, interesting results start to emerge.
A recent fit to rescaled OPAL data (adjusted to reflect current values of exclusive hadronic $\tau$-decay branching ratios), with moments chosen to maximize duality violations, finds
$\delta_{\mathrm{NP}} = -0.003\pm 0.012$ and $\alpha_s(m_\tau^2) = 0.333\pm 0.018$ (CIPT+FOPT)
\cite{Boito:2012cr}, in agreement but less precise than the result obtained from the ALEPH
invariant-mass distribution.

\subsection{Updated Determination of the QCD Coupling}

The measured $\tau$ leptonic branching fractions imply a quite precise value for the normalized hadronic decay width: $R_\tau = (1-B_\mu-B_e)/B_e = 3.636\pm 0.011$. A slightly more accurate number can be obtained assuming lepton universality, \ie the relations in Eq.~\eqn{relation}. The information from the $\tau$ lifetime and leptonic branching fractions can then be combined to derive an improved value for the electron branching ratio, given in Eq.~\eqn{eq:B_e_univ}, and $R_\tau$:
%
\bel{eq:Rtau_univ1}
R_\tau^{\mathrm{uni}}\; =\; \frac{1}{B_e^{\mathrm{uni}}} - 1.972559 
\; =\; 3.6397\pm 0.0076\, .
\ee
The ratio $R_\tau$ can be also extracted from the sum of all hadronic decay modes. Using
$B_e^{\mathrm{uni}}$ and accounting for all statistical correlations \cite{HFAG}, one finds
\bel{eq:Rtau_univ}
R_\tau\; =\; \frac{\sum_{\mathrm{had}} \mathrm{Br}(\tau^-\to\nu_\tau X^-_{\mathrm{had}})}{B_e^{\mathrm{uni}}}
\; =\; 3.6326\pm 0.0084\, ,
\ee
in good agreement with the previous value and with a similar error.
The advantage of this last procedure is that, using the information given in tables~\ref{tab:BR_CA} and \ref{tab:BR_CS}, one can separate the Cabibbo-allowed and Cabibbo-suppressed components:
\bel{eq:R_tau_V+A_S}
R_{\tau,V+A}\; =\; 3.4712\pm 0.0079\, ,
\qquad\qquad\qquad
R_{\tau,S}\; =\; 0.1614\pm 0.0028\, .
\ee
Subtracting $R_{\tau,S}$ from $R_\tau^{\mathrm{uni}}$ in \eqn{eq:Rtau_univ1},
results in $R_{\tau,V+A} = 3.4783\pm 0.0081$,
consistent with \eqn{eq:R_tau_V+A_S} and of similar accuracy; however, this number would absorb the effect of any unobserved hadronic states (or unaccounted systematics) entirely in $R_{\tau,V+A}$, while they could also be strange final states. Therefore, we follow the procedure advocated by the HFAG \cite{HFAG} and use the direct determination of $R_{\tau,V+A}$ in Eq.~\eqn{eq:R_tau_V+A_S}. Taking $V_{ud} = 0.97425\pm 0.00022$ \cite{Towner:2010zz}, it implies
\bel{wq:delta_P_NP}
\delta_{\mathrm{P}} + \delta_{\mathrm{NP}}\; =\; 0.1950 \pm 0.0028\, .
\ee
Subtracting now the small non-perturbative contribution in Eq.~\eqn{eq:del_np}, one obtains %
\bel{wq:delta_P}
\delta_{\mathrm{P}} \; =\; 0.2009 \pm 0.0031\, .
\ee

The main uncertainty in the $\tau$ determination of the strong coupling originates in the treatment of higher-order perturbative corrections.
We estimate them \cite{Pich:2011bb} adding the fifth-order term $K_5 A^{(5)}(\alpha_s)$ with $K_5 = 275\pm 400$; the central value is in the range favoured by renormalon models, but our conservative uncertainty is large enough to allow for a correction of opposite sign. We also include the 5-loop variation with changes of the renormalization scale in the range $\mu^2/(-s)\in [0.4,\, 2.0]$. The error induced by the truncation of the $\beta$ function at fourth-order is estimated through the variation of the results at five loops, assuming $\beta_5=\pm\beta_4^2/\beta_3 = \mp 443$; in CIPT this slightly changes the values of $A^{(n)}(\alpha_s)$, while in FOPT it increases the scale dependence.
The FOPT procedure shows as expected a much more sizeable $\mu$ dependence, but it is less sensitive to the uncertainties on $\delta_{\mathrm{P}}$ and $K_5$. We add linearly the three theoretical uncertainties ($K_5$, $\mu$, $\beta_5$) and combine their sum in quadrature with the `experimental' error on $\delta_{\mathrm{P}}$.

\begin{figure}[t]\centering
\includegraphics[width=8.5cm,clip]{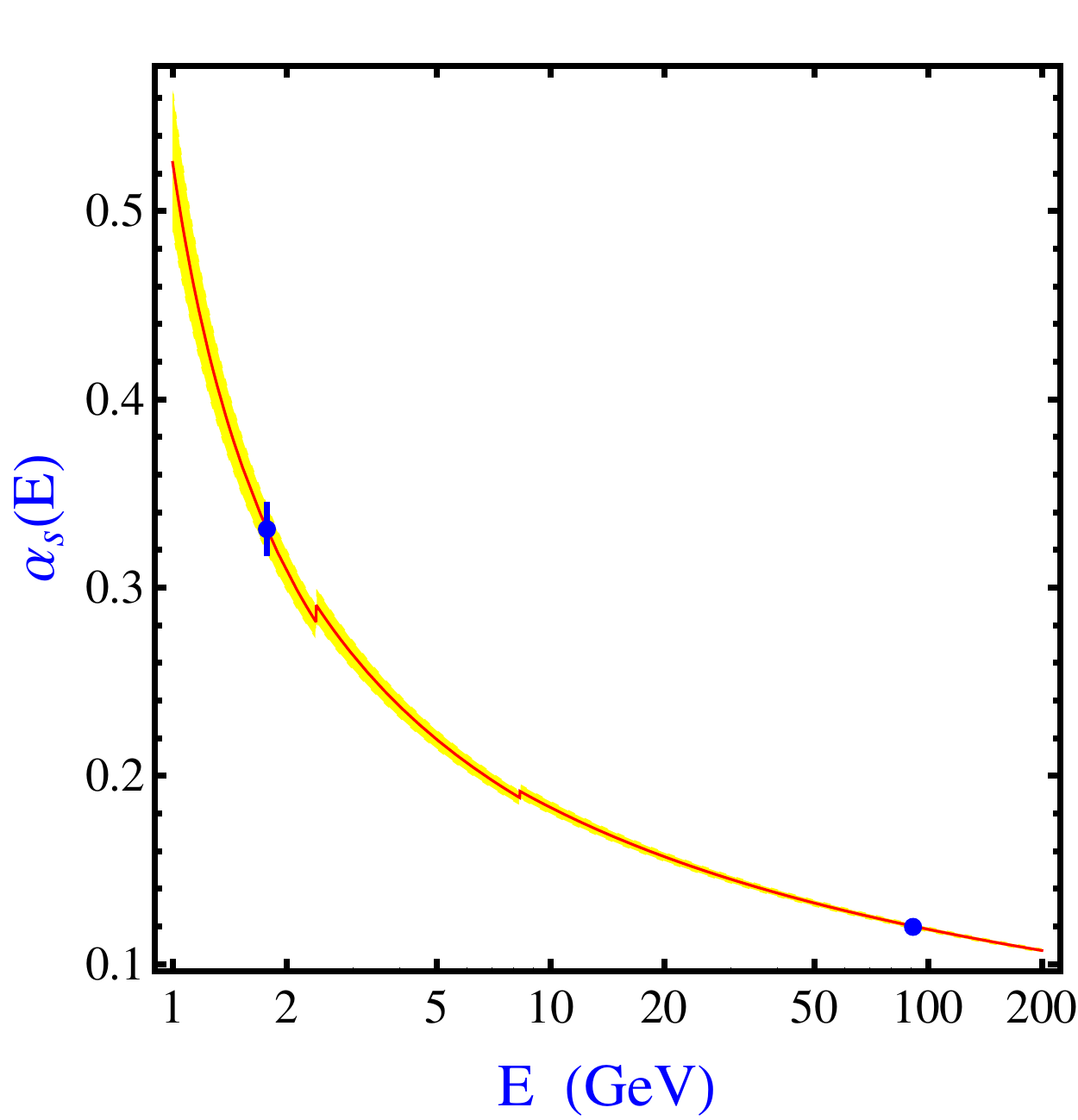}
\caption{Evolution of the strong coupling $\alpha_s(m_\tau^2)$ to higher energies and comparison with the direct measurement at $s=M_Z^2$.}
\label{fig:running-alpha}
\end{figure}

Using CIPT one gets $\alpha_s(m_\tau^2) = 0.341\pm 0.013$,
while FOPT would give
$\alpha_s(m_\tau^2) = 0.319\pm 0.014$  \cite{Pich:2013sqa}.
Combining the two results, but keeping conservatively the smallest error, we get
\begin{equation}\label{eq:alpha-result}
\alpha_s^{(n_f=3)}(m_\tau^2)\; =\; 0.331 \pm 0.013\, ,
\end{equation}
significantly larger ($16\,\sigma$) than the result obtained from the $Z$ hadronic width,
$\alpha_s^{(n_f=5)}(M_Z^2) = 0.1197\pm 0.0028$ \cite{Beringer:1900zz} ($n_f$ denotes the relevant number of quark flavours at the given energy scale).
After evolution up to the scale $M_Z$ \cite{Rodrigo:1997zd,vanRitbergen:1997va,Czakon:2004bu,Schroder:2005hy,Chetyrkin:2005ia}, the strong coupling in \eqn{eq:alpha-result} decreases to $\alpha_s^{(n_f=5)}(M_Z^2) = 0.1200\pm 0.0015$,
in excellent agreement with the direct measurement at the $Z$ peak.
The comparison of these two determinations, shown in Fig.~\ref{fig:running-alpha},
provides a beautiful test of the predicted QCD running;
\ie a very significant experimental verification of asymptotic freedom:
\be
\alpha_s^{(n_f=5)}(M_Z^2)\bigg|_\tau - \alpha_s^{(n_f=5)}(M_Z^2)\bigg|_Z
\; = \; 0.0003 \pm 0.0015_\tau \pm 0.0028_Z\, .
\ee

\subsection{Chiral Sum Rules}

For massless quarks, the chiral invariance of QCD guarantees that the two-point correlation function of a left-handed and a right-handed quark currents, $\Pi_{LR}(s) = \Pi^{(0+1)}_{ud,V}(s) - \Pi^{(0+1)}_{ud,A}(s)$,
vanishes identically to all orders in perturbation theory.
The non-zero value of $\Pi_{LR}(s)$ originates in the spontaneous breaking of chiral symmetry by the QCD vacuum. At large momenta, the corresponding OPE only receives contributions from operators with dimension $D\ge 6$,
which implies $\lim_{s\to\infty} s^2 \Pi_{LR}(s) = 0$.
The non-zero up and down quark masses induce tiny
corrections with dimensions two and four, which are negligible at high energies.
At very low momenta, $\chi$PT dictates the low-energy expansion of $\Pi_{LR}(s)$
in terms of the pion decay constant and the $\chi$PT couplings $L_{10}$ [$\cO(p^4)$] and $C_{87}$ [$\cO(p^6)$]; the needed expressions are known to two loops in the chiral expansion \cite{Amoros:1999dp}.

Analyticity relates the short- and long-distance regimes through the dispersion relation
\be\label{eq:dispersion}
\frac{1}{2\pi i}\, \oint_{|s|=s_0}  ds\; w(s)\, \Pi_{LR}(s)\; =\;
 -\int_{s_{th}}^{s_0} ds\; w(s)\, \rho(s)\,
 +\, 2 f_\pi^2\, w(m_\pi^2)\, +\, \mathrm{Res}\left[w(s)\,\Pi_{LR}(s),s=0\right]\, ,
\ee
where $\rho(s)\equiv\frac{1}{\pi}\,\mathrm{Im}\Pi_{LR}(s)$ and $w(s)$ is an arbitrary weight function that is analytic in the whole complex plane except at the origin, where it can have poles. The last term in (\ref{eq:dispersion}) accounts for the possible residue at the origin.

For $s_0\le m_\tau^2$, the integral along the real axis can be evaluated
with the measured tau spectral functions.
Taking $w(s) = s^n$ with $n\ge 0$, there is no residue at the origin and,
with $s_0$ large enough so that the OPE can be applied in the entire circle $|s|=s_0$,
the OPE coefficients are directly related to the spectral function integration. With $n=0$ and 1,
there is no OPE contribution in the chiral limit and one gets the celebrated first and second Weinberg sum rules \cite{Weinberg:1967kj}:
\bel{eq:WSR}
\lim_{s_0\to\infty}\, I_1(s_0) \; =\; f_\pi^2\, ,
\qquad\qquad\qquad\qquad
\lim_{s_0\to\infty}\, I_2(s_0) \; =\; f_\pi^2\, m_\pi^2\;\approx\; 0\, .
\ee
where
\bel{eq:In}
I_{n+1}(s_0) \; =\; \frac{1}{2}\,\int_0^{s_0} ds\; s^{n}\;\rho(s)\, .
\ee
For negative values of the integer $n$, the OPE does not contribute either while the residues at zero
are determined by the $\chi$PT low-energy couplings, which can be then experimentally determined \cite{Davier:1998dz}.

\begin{figure}[t]\centering
\begin{minipage}{8.75cm}\centering
\includegraphics[width=8.75cm,clip]{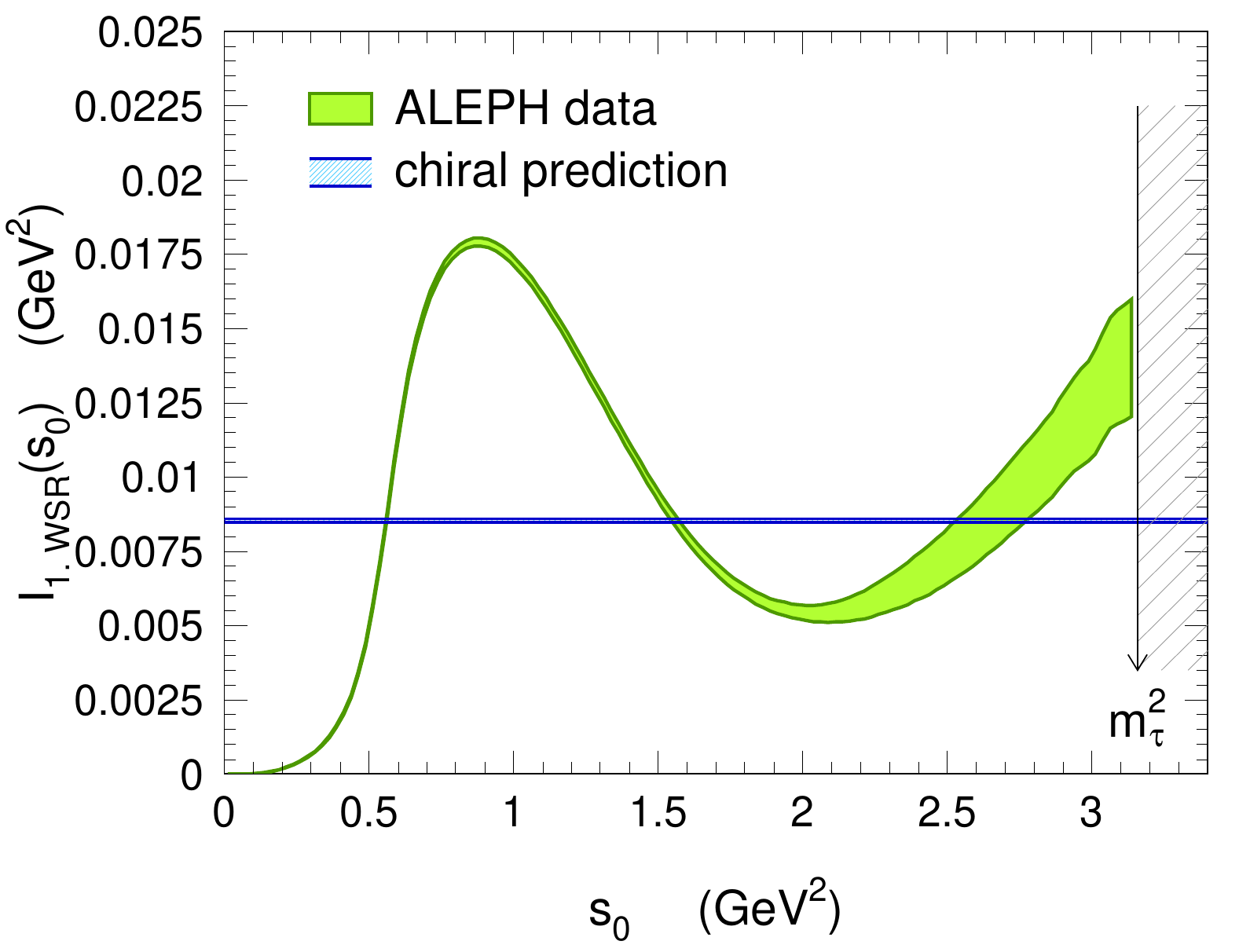} 
\includegraphics[width=8.5cm,clip]{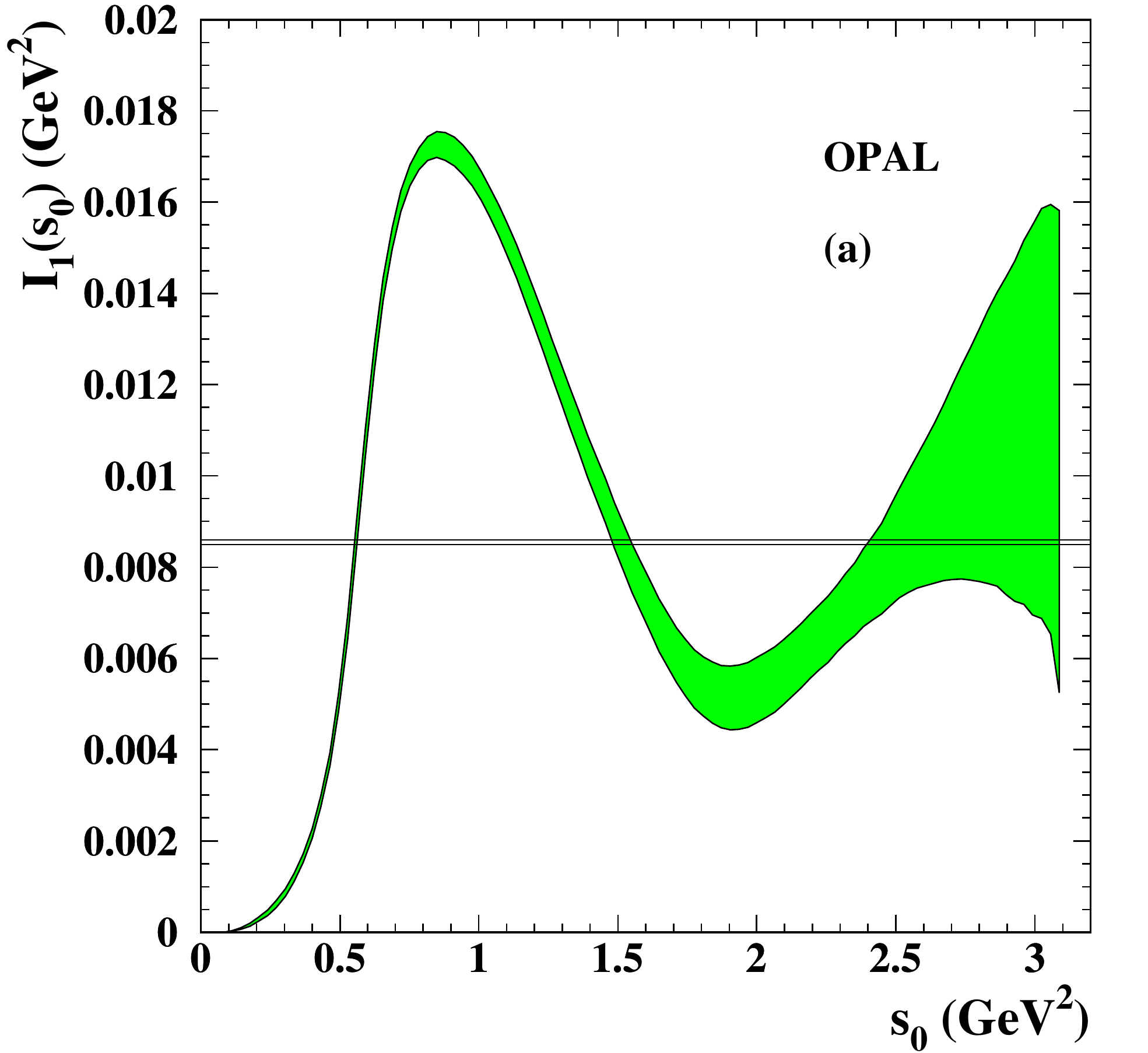}
\end{minipage}
\hskip .5cm
\begin{minipage}{8.75cm}\centering
\includegraphics[width=8.75cm]{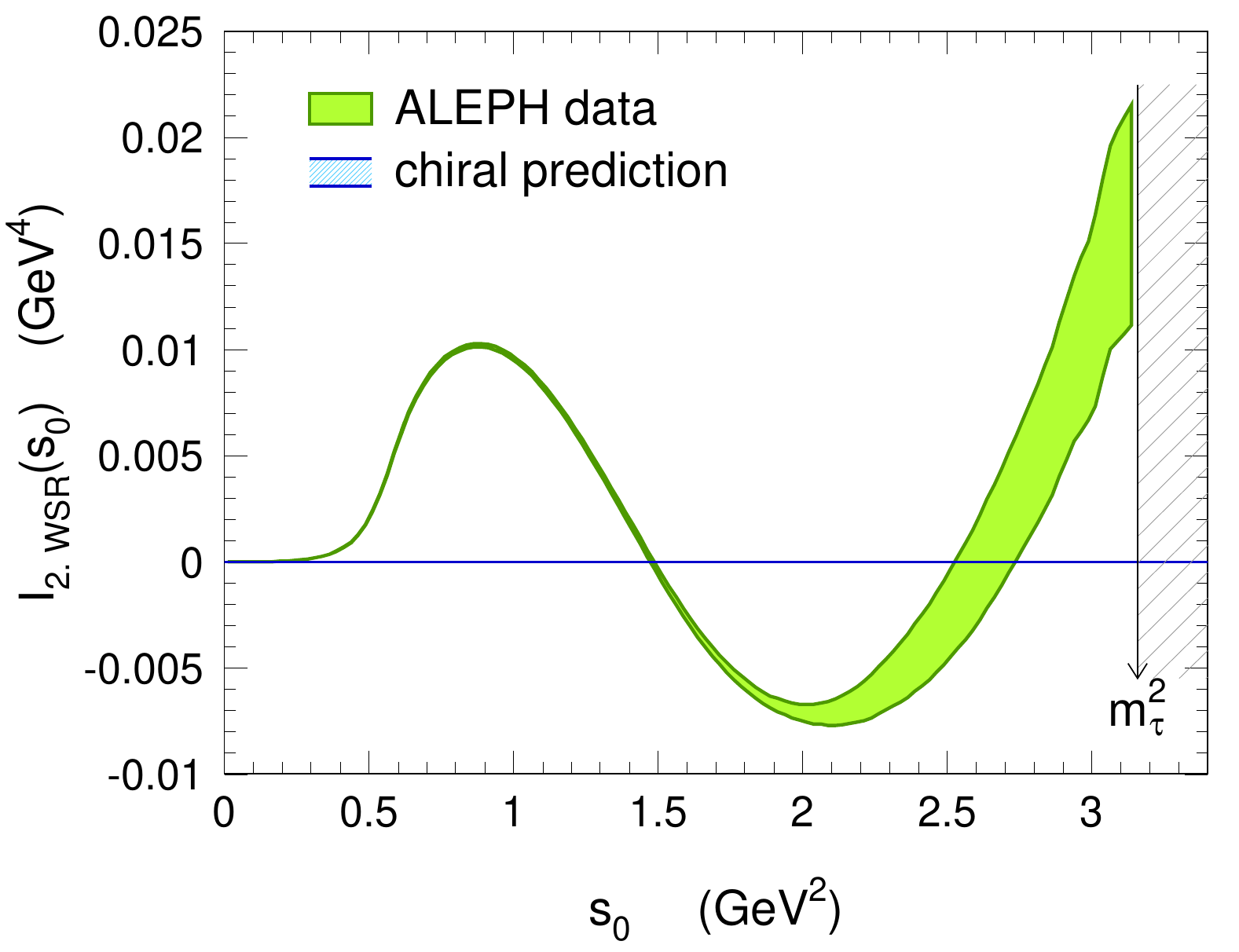} 
\includegraphics[width=8.5cm,clip]{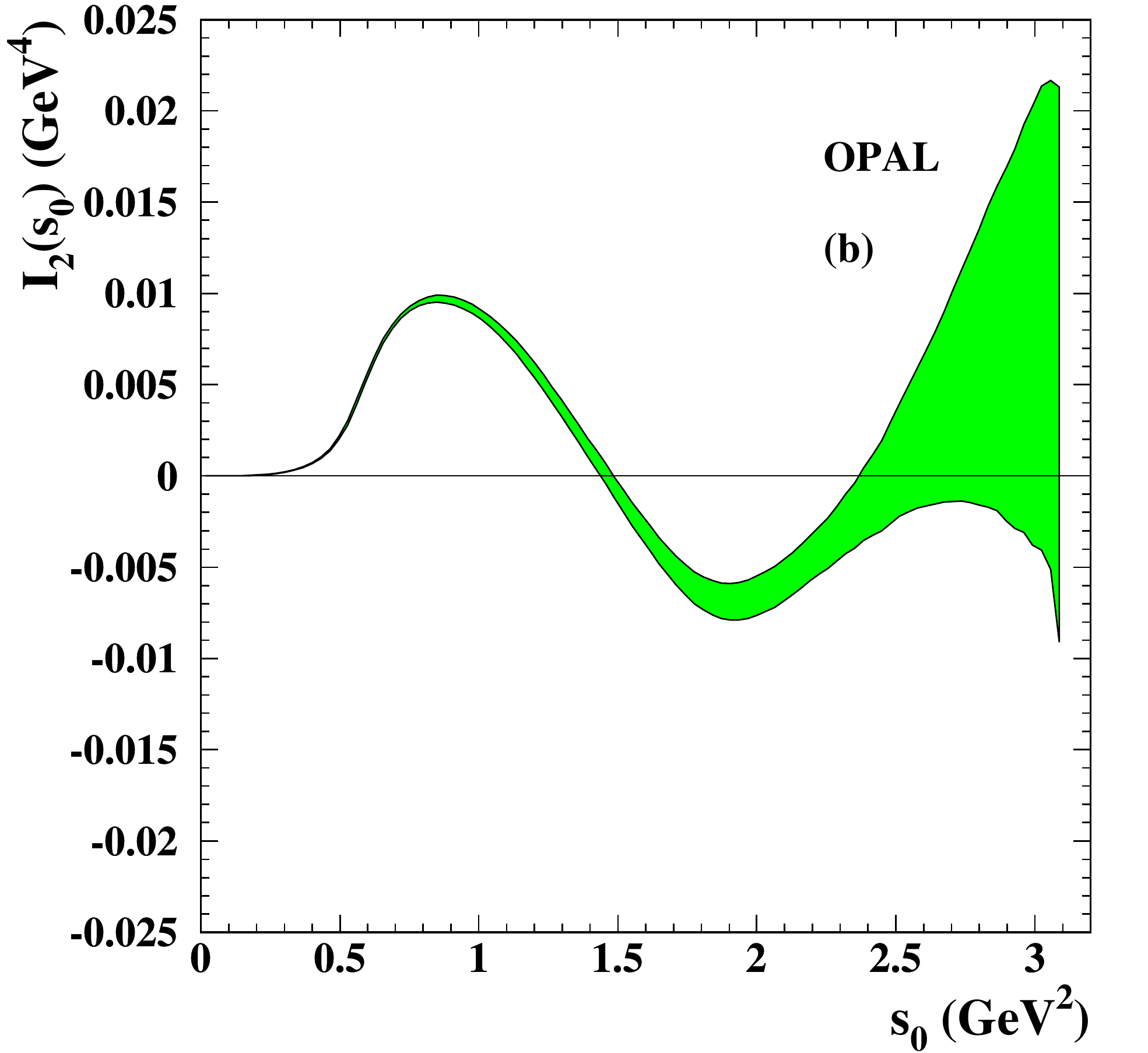}
\end{minipage}
\caption{First (left) and second (right) Weinberg Sum Rules as a function of $s_0$, using the ALEPH \cite{Davier:2005xq,Schael:2005am} (upper plots) and OPAL \cite{Ackerstaff:1998yj} (lower plots) spectral functions. The horizontal lines show the expected asymptotic result at $s_0\to\infty$.}
\label{fig:WSRs}
\end{figure}

The integrals $I_1(s_0)$ and $I_2(s_0)$, calculated with ALEPH \cite{Davier:2005xq,Schael:2005am} and OPAL \cite{Ackerstaff:1998yj} data, are shown in Figs.~\ref{fig:WSRs}. Although there is good agreement between the two sets of spectral functions, the ALEPH data are more precise, specially in the higher $s_0$ region close to the kinematical $\tau$ decay end-point. One clearly observes the $\rho$ and $a_1$ resonance structures, which generate oscillations around the expected asymptotic value. Within their large uncertainties, the OPAL data are already consistent with \eqn{eq:WSR} at $s_0=3~\mathrm{GeV}^2$; however, the more precise ALEPH data indicate that asymptotics is not yet reached. Therefore, additional information on the spectral function above $m_\tau^2$ is required.

The absence of perturbative contributions makes the relation (\ref{eq:dispersion}) and ideal tool to investigate possible quark-hadron duality effects, formally defined through
\cite{Shifman:2000jv,Cata:2008ru}
\be
\label{eq:DV}
\mathrm{DV}_w\;\equiv \;\frac{1}{2 \pi i} \, \oint_{|s|=s_0} ds\; w(s)\;
\left[ \Pi_{LR}(s) - \Pi^{\mathrm{OPE}}_{LR}(s)\right]
\; =\; \int^{\infty}_{s_0} ds \; w(s)\;\rho(s)\, .
\ee
This has been thoroughly studied in Refs.~\cite{GonzalezAlonso:2010xf},
using for the spectral function beyond $s_z \sim 2.1~\mathrm{GeV}$
the parametrization \cite{Shifman:2000jv,Cata:2008ru}
\be\label{eq:model}
\rho(s\ge s_z) \; =\; \kappa~ e^{-\gamma s} \sin(\beta (s-s_z))\, ,
\ee
and finding the region in the 4-dimensional $(\kappa,\gamma,\beta,s_z)$ parameter space
that is compatible with the most recent experimental ALEPH data \cite{Schael:2005am}, the first and second Weinberg sum rules \cite{Weinberg:1967kj} in \eqn{eq:WSR}
and the sum rule of Das et al. \cite{Das:1967it} that determines the pion electromagnetic mass difference:
\bel{eq:DGMLY}
\int_0^{\infty} ds\; s\;\log{\left(\frac{s}{\Lambda^2}\right)}\;\rho(s)|_{m_q=0}
\; =\; \left(m_{\pi^0}^2-m_{\pi^\pm}^2\right)_{\mathrm{EM}}\;\frac{8\pi}{3\alpha}\, f^2
\, .
\ee
The second Weinberg sum rule guarantees that this last integral does not depend on the arbitrary mass scale $\Lambda$ inside the logarithm.

Refs.~\cite{GonzalezAlonso:2010xf} perform a statistical analysis, scanning the parameter space $(\kappa,\gamma,\beta,s_z)$ and selecting those acceptable spectral functions which satisfy the experimental and theoretical constraints.
The differences among them determine how much freedom is left for the behaviour of the physical spectral function beyond the kinematical end of the $\tau$ data. For each acceptable spectral function one calculates the parameters $L_{10}$, $C_{87}$, $\cO_6$ and $\cO_8$, obtained through the dispersion relation (\ref{eq:dispersion}) with appropriate weight functions,
where $\cO_D$ is the coefficient of the $1/(-s)^{D/2}$ term in the OPE of $\Pi_{LR}(s)$.
The resulting statistical distributions determine their finally estimated values; the dispersion of the numerical results provides a good quantitative assessment of the actual uncertainties.
Pinched weight functions of the form $w(s) = s^n (s-s_z)^m$ ($m>0$), which vanish at $s=s_z$, are found to minimize the uncertainties from duality-violation effects, allowing for a more precise determination of the hadronic parameters;
one obtains \cite{GonzalezAlonso:2010xf}:
\be\label{eq:ChPT_Results}
L_{10}^r(M_\rho) \, =\, -(4.06\pm 0.39)\times 10^{-3}\, ,
\qquad\qquad
C_{87}^r(M_\rho) \, =\, (4.89\pm 0.19)\times 10^{-3}\;\mathrm{GeV}^{-2}\, ,
\ee
\be\label{eq:OPE_Results}
\cO_6 \, =\, (-4.3\,{}^{+\, 0.9}_{-\, 0.7})\times 10^{-3}\;\mathrm{GeV}^{6}\, ,
\qquad\qquad\qquad
\cO_8 \, =\, (-7.2\,{}^{+\, 4.2}_{-\, 5.3})\times 10^{-3}\;\mathrm{GeV}^{8}\, .
\ee

Duality-violation effects have very little impact on the determination of $L_{10}$ and $C_{87}$ because the corresponding sum rules are dominated by the low-energy region where the data sits. Thus, one obtains basically the same results with pinched and non-pinched weight functions. Similar values of $L_{10}$ and $C_{87}$ have been obtained in a recent analysis of the OPAL data \cite{Boito:2012nt}, using separate parametrizations of the type \eqn{eq:model} for the vector and axial spectral functions (\ie 4 additional parameters);
however, larger uncertainties are found because the OPAL distributions are less precise than the ALEPH ones.

The situation is different for $\cO_6$ and $\cO_8$, which are sensitive to the high-energy behaviour of the spectral function; pinched-weights provide then a much better accuracy.
This could explain the numerical differences among previous estimates
\cite{Schael:2005am,Ackerstaff:1998yj,Davier:1998dz,Bijnens:2001ps,Zyablyuk:2004iu,Rojo:2004iq,Narison:2004vz,Peris:2000tw,Friot:2004ba,Almasy:2008xu,Masjuan:2007ay,Cirigliano:2002jy,Cirigliano:2003kc,Dominguez:2003dr,Bordes:2005wv},
where duality violation uncertainties were not properly assessed.
Owing to the larger uncertainties in the higher $s$ region, the results obtained from OPAL data are less robust \cite{Boito:2012nt}; nevertheless, they are consistent with the ALEPH values in Eq.~\eqn{eq:OPE_Results}.

The determination \eqn{eq:ChPT_Results} of the two $\chi$PT couplings is in good agreement with (but more precise than) recent theoretical calculations, using R$\chi$T and large-$N_C$ techniques at the next-to-leading order, which predict \cite{Pich:2008jm}:
$L_{10}^r(M_\rho)  = -(4.4\pm 0.9)\times 10^{-3}$ and
$C_{87}^r(M_\rho)  = (3.6\pm 1.3)\times 10^{-3}\;\mathrm{GeV}^{-2}$. It also agrees with present lattice estimates of $L_{10}^r(M_\rho)$ \cite{Shintani:2008qe,Boyle:2009xi}.
The results (\ref{eq:OPE_Results}) fix with accuracy the value of $\cO_6$ and determine
the sign of $\cO_8$. This information is needed to calculate the electromagnetic penguin contribution to the CP-violating ratio $\varepsilon'_K/\varepsilon_K$ in neutral kaon decays.

\section{$V_{us}$ Determination}
\label{sec:Vus}

The separate measurement of the $|\Delta S|=1$ and $|\Delta S|=0$
tau decay widths provides a very clean determination of $V_{us}$ \cite{Gamiz:2004ar}.
If quark masses are neglected, the experimental ratio of the two decay widths directly measures $|V_{us}/V_{ud}|^2$.
Taking $V_{ud} = 0.97425\pm 0.00022$ \cite{Towner:2010zz} and the HFAG values in Eq.~\eqn{eq:R_tau_V+A_S}, one obtains
$|V_{us}|^{\mathrm{SU(3)}} = 0.210\pm 0.002$.

This rather remarkable determination is only slightly shifted by
the small SU(3)-breaking contributions induced by the strange quark mass.
These effects can be theoretically estimated through a careful QCD analysis of the difference
\cite{Pich:1999hc,Baikov:2004tk,Gamiz:2004ar,Chen:2001qf,Chetyrkin:1998ej,Korner:2000wd,Maltman:2006st,Kambor:2000dj,Maltman:1998qz}
\begin{equation}\label{eq:dRtau_def}
 \delta R_\tau  \;\equiv\;
 {R_{\tau,V+A}\over |V_{ud}|^2} - {R_{\tau,S}\over |V_{us}|^2}\, .
\end{equation}
Since the strong interactions are flavour blind, this quantity vanishes in the SU(3) limit.
The only non-zero contributions are proportional to powers of the quark
mass-squared difference $m_s^2-m_d^2$ or to vacuum expectation
values of SU(3)-breaking operators such as
$\delta O_4 \equiv \langle 0|m_s\bar s s - m_d\bar d d|0\rangle \approx (-1.4\pm 0.4)
\times 10^{-3}\; \mathrm{GeV}^4$ \cite{Pich:1999hc,Gamiz:2004ar}. The dimensions of these operators
are compensated by corresponding powers of $m_\tau^2$, which implies a strong
suppression of $\delta R_\tau$ \cite{Pich:1999hc}:
\be\label{eq:dRtau}
 \delta R_\tau \;\approx\;  24\, S_{\mathrm{EW}}\; \left\{ \frac{m_s^2(m_\tau^2)}{m_\tau^2} \; \left( 1-\epsilon_d^2\right)\,\Delta_{00}(a_\tau)
 - 2\pi^2\; {\delta O_4\over m_\tau^4} \; Q_{00}(a_\tau)\right\}\, ,
\ee
where $\epsilon_d\equiv m_d/m_s = 0.053\pm 0.002$ \cite{Leutwyler:1996qg}.

The perturbative QCD corrections $\Delta_{00}(a_\tau)$ and
$Q_{00}(a_\tau)$ are known to $O(\alpha_s^3)$ and $O(\alpha_s^2)$,
respectively \cite{Pich:1999hc,Baikov:2004tk}.
Separating their longitudinal ($J=0$) and transverse ($J=0+1$) contributions, as indicated in Eq.~\eqn{eq:circle},
\be
\Delta_{00}(a_\tau) \; = \; \frac{3}{4}\:\Delta_{00}^{(0+1)}(a_\tau) +
\frac{1}{4}\:\Delta_{00}^{(0)}(a_\tau)\, ,\ee
one finds a rather pathological behaviour for the longitudinal component $\Delta_{00}^{(0)}(a_\tau)$, with clear signs of being a non-convergent perturbative
series [$a_\tau = \alpha_s(m_\tau^2)/\pi\sim 0.1$]:
\bea
\Delta_{00}^{(0)}(a_\tau) & = & 1 + 9.333\: a_\tau + 109.989\: a_\tau^2 +\, 1322.52 \: a_\tau^3 +\cdots
\nonumber\\
\Delta_{00}^{(0+1)}(a_\tau) & = & 1 + 4\: a_\tau + 25.021\: a_\tau^2 +\, 148.25 \: a_\tau^3 +\cdots
\eea
Doing a CIPT resummation does not improve the bad behaviour of the longitudinal series.

Fortunately, the longitudinal contribution to
$\delta R_\tau$ can be estimated phenomenologically with good accuracy \cite{Gamiz:2004ar}, because it is dominated by far by the well-known $\tau\to\nu_\tau\pi$
and $\tau\to\nu_\tau K$ contributions. To estimate the remaining transverse
component, one needs an input value for the strange quark mass; we will adopt the lattice world average \cite{Colangelo:2010et,Laiho:2009eu}, but increasing conservatively its uncertainty to 6 MeV, \ie we take
$\overline{m}_s\equiv m_s^{\overline{\mathrm{MS}}}(2~\mathrm{GeV}) = (94\pm 6)~\mathrm{MeV}$. The numerical size of $\Delta_{00}^{(0+1)}(a_\tau)$ is estimated in a very conservative way, averaging the asymptotically summed CIPT and FOPT results and taking half of the difference as the uncertainty associated with the truncation of the series, giving the result \cite{Gamiz:2004ar,Gamiz:2013wn}
\bel{eq:deltaRtau}
\delta R_\tau\; =\; (0.1544\pm 0.0037)\, +\, (9.3\pm 3.4)\; \frac{\overline{m}_s^2}{1~\mathrm{GeV}^2}\,
+\, (0.0034\pm 0.0028)\; =\; 0.240\pm 0.032\, .
\ee
The first term contains the phenomenologically estimated longitudinal contribution
$\delta R_\tau^{L}$, the second gives the transverse $D=2$ contribution, while the third one accounts for the small higher-dimension corrections.

This SU(3)-breaking correction implies, through Eq.~\eqn{eq:dRtau_def},
\be\label{eq:Vus_det}
 |V_{us}| \; =\; \left(\frac{R_{\tau,S}}{\frac{R_{\tau,V+A}}{|V_{ud}|^2}-\delta
 R_{\tau,\mathrm{th}}}\right)^{1/2}
 \; =\;\,
  0.2173\pm 0.0020_{\mathrm{\, exp}}\pm 0.0010_{\mathrm{\, th}}
  \; =\; 0.2173\pm 0.0022\, .
\ee
This result is lower than the most recent
determination from $K_{\ell 3}$ decays, $|V_{us}|= 0.2238\pm 0.0011$ \cite{Cirigliano:2011ny,Bazavov:2012cd,Boyle:2013gsa}.

The branching ratios measured by BaBar and Belle are smaller than previous
world averages, which translates into smaller results for $R_{\tau,S}$ and $|V_{us}|$.
Using Eq.~\eqn{eq:R_tP} and the measured $K^-\to\mu^-\bar\nu_\mu$ decay width, one can determine the $\tau^-\to\nu_\tau K^-$ branching ratio with better accuracy than the present experimental world average; one finds
$\mathrm{Br}(\tau^-\to\nu_\tau K^-) = (0.713\pm 0.003)\%$ \cite{Antonelli:2013usa}, which is $1.7\,\sigma$ higher than the value quoted in table~\ref{tab:BR_CS}.
In a similar way, combining the measured spectra in $\tau^-\to\nu_\tau (K\pi)^-$ decays with $K_{\ell 3}$ data \cite{Antonelli:2013usa}, one predicts the $\tau^-\to\nu_\tau \bar K^0\pi^-$ and $\tau^-\to\nu_\tau K^-\pi^0$ branching ratios in Eq.~\eqn{eq:BrKpi_pred}, which are also $1.0\,\sigma$ and $1.6\,\sigma$ higher, respectively, than the direct $\tau$ decay measurements in table~\ref{tab:BR_CS}. Since these branching ratios are the three largest contributions to the Cabibbo-suppressed $\tau$ decay width, their slight underestimate has a very significant effect on the $V_{us}$ determination. Replacing the $\mathrm{Br}(\tau\to\nu_\tau K)$ and $\mathrm{Br}(\tau\to\nu_\tau K\pi)$ HFAG averages by these phenomenological estimates, and taking the remaining branching ratios from table~\ref{tab:BR_CS}, on gets the corrected inclusive results:
\bel{eq:R_S}
\mathrm{Br}(\tau^-\to\nu_\tau X^-_s)\; =\; (2.967\pm 0.060)\%\, ,
\qquad\qquad\qquad
R_{\tau,S}\; =\; 0.1665\pm 0.0034\, .
\ee
The slight increase in the total branching fraction into modes with non-zero strangeness,
$(0.092\pm 0.078)\%$, is well within the experimentally allowed range for unaccounted modes in Eq.~\eqn{eq:missingBr}. Moreover, the sum of this higher value for $R_{\tau,S}$ with
$R_{\tau,V+A}$ in \eqn{eq:R_tau_V+A_S}, $3.6377\pm 0.0086$, fits better with $R_\tau^{\mathrm{uni}}$ in \eqn{eq:Rtau_univ1}. Using as input \eqn{eq:R_S}, one gets a determination of $V_{us}$ in much better agreement with the kaon information:
\bel{eq:Vus_corr}
|V_{us}|\; =\; 0.2207\pm 0.0023_{\mathrm{\, exp}}\pm 0.0011_{\mathrm{\, th}}
\; =\; 0.2207\pm 0.0025\, .
\ee
Similar values are also obtained  combining the measured Cabibbo-suppressed $\tau$ distribution with electroproduction data \cite{Maltman:2008na,Maltman:2009bh}.
Contrary to $K_{\ell 3}$, the final error of the $V_{us}$ determination from
$\tau$ decay is dominated by the experimental uncertainties and, therefore, sizeable
improvements can be expected. Progress on the theoretical side requires a better understanding of the perturbative QCD corrections included in $\delta R_\tau$.

$|V_{us}|$ can also be obtained from exclusive modes, either from the ratio
$\Gamma(\tau\to\nu_\tau K)/\Gamma(\tau\to\nu_\tau \pi)$ or from $\Gamma(\tau\to\nu_\tau K)$, using the appropriate hadronic inputs from lattice calculations ($f_K/f_\pi$, $f_K$). The first method gives the result in Eq.~\eqn{eq:Cabibbo}, while
$|V_{us}| = 0.2214\pm 0.0022$ \cite{HFAG} is obtained from $\mathrm{Br}(\tau\to\nu_\tau K)$,
taking $\sqrt{2} f_K = 156.1\pm 1.1$~MeV \cite{Colangelo:2010et,Laiho:2009eu}.
These values are closer to the $K_{\ell 3}$ result.

\section{Electromagnetic and Weak Dipole Moments}
\label{sec:dipole}

A general description of the electromagnetic coupling of an on-shell spin-$\frac{1}{2}$ charged lepton to the virtual photon involves three different form factors:
\bel{eq:em_ff}
\cM_{\ell\bar \ell \gamma^*}
\; =\; e\, Q_\ell\, \varepsilon_\mu(q) \; \bar u_\ell(\vec{p}\, ')
\left[F_1(q^2)\,\gamma^\mu
+ i\, {F_2(q^2)\over 2 m_\ell}\, \sigma^{\mu\nu}q_\nu +
{F_3(q^2)\over 2 m_\ell}\, \sigma^{\mu\nu}\gamma_5\, q_\nu\right] u_\ell(\vec{p}) \, ,
\ee
where $q^\mu = (p'-p)^\mu$ is the incoming photon momentum and $Q_\ell = -1$. The only assumptions are Lorentz invariance and electromagnetic current conservation (required by gauge invariance). Owing to the conservation of the electric charge,
$F_1(0)=1$. At $q^2=0$, the other two form factors reduce to the lepton magnetic ($\mu_\ell$) and electric ($d_\ell$) dipole moments:
\bel{eq:DipoleMoments}
\mu_\ell\;\equiv\; \frac{e}{2 m_\ell} \; \frac{g^\gamma_\ell}{2}\; =\; \frac{e}{2 m_\ell}\, \left[ 1+F_2(0)\right]\, ,
\qquad\qquad\qquad\qquad
d_\ell^\gamma\; =\; \frac{e}{2 m_\ell} \; F_3(0)\, .
\ee
Similar expressions are defined in section~\ref{subsec:WDM} for the $\ell\bar\ell$ coupling of a virtual $Z$.

The $F_i(q^2)$ form factors are sensitive quantities to a possible lepton substructure.
Moreover, $F_3(q^2)$ violates $T$ and $P$ invariance; thus,
$d^{\gamma}_\ell$ constitutes a good probe of CP violation. Owing to their chiral-changing structure, the dipole moments may provide important insights on the mechanism responsible for mass generation. In general, one expects that a fermion of mass $m_f$ (generated by physics at some scale $M\gg m_f$) will have
induced dipole moments proportional to some power of $m_f/M$. Therefore, heavy fermions such as the $\tau$ should be a good testing ground for this kind of effects.

\subsection{Anomalous Magnetic Moments}
\label{sec:g-2}

The most stringent QED test \cite{Hughes:1999fp,Davier:2004gb,Passera:2004bj,Melnikov:2006sr,Miller:2007kk,Jegerlehner:2009ry,Roberts:2010zz,Miller:2012opa}
comes from the high-precision measurements of the electron \cite{Hanneke:2008tm,Odom:2006zz} and muon \cite{Bennett:2006fi} anomalous magnetic moments, $a_\ell\equiv (g^\gamma_\ell-2)/2\, $:
\bea\label{eq:a_e}
 a_e & =& (1\; 159\; 652\; 180.73\pm 0.28) \,\times\, 10^{-12}\, ,
\\ \label{eq:a_mu}
 a_\mu & =& (11\; 659\; 208.9\pm 6.3) \,\times\, 10^{-10}\, .
\eea
The $O(\alpha^5)$ calculation has been completed in both cases
\cite{Aoyama:2012wj}, with an impressive agreement with the measured $a_e$ value.
The dominant QED uncertainty is the input value of $\alpha$, therefore
$a_e$ provides the most accurate determination of the fine structure constant (0.25 parts per billion) \cite{Aoyama:2012wj},
\bel{eq:alpha}
\alpha^{-1} = 137.035\; 999\; 174 \,\pm\, 0.000\; 000\; 035\, ,
\ee
in agreement with the next most precise value (0.66 ppb)
$\alpha^{-1}_{\mathrm{Rb}} = 137.035\, 999\, 037 \pm 0.000\, 000\, 091$
\cite{Bouchendira:2010es}, deduced from a recent measurement of the ratio $h/m_{\mathrm{Rb}}$ between the Planck constant and the mass of the ${}^{87}$Rb atom.
The improved experimental accuracy on the electron anomalous magnetic moment is already sensitive to the hadronic contribution $\delta a_e^{\mathrm{QCD}}= (1.685\pm 0.033)\times 10^{-12}$, and approaching the level of the weak correction $\delta a_e^{\mathrm{ew}}= (0.0297\pm 0.0005)\times 10^{-12}$ \cite{Aoyama:2012wj}.

\begin{figure}[tbh]
\begin{center}
\includegraphics[width=10cm]{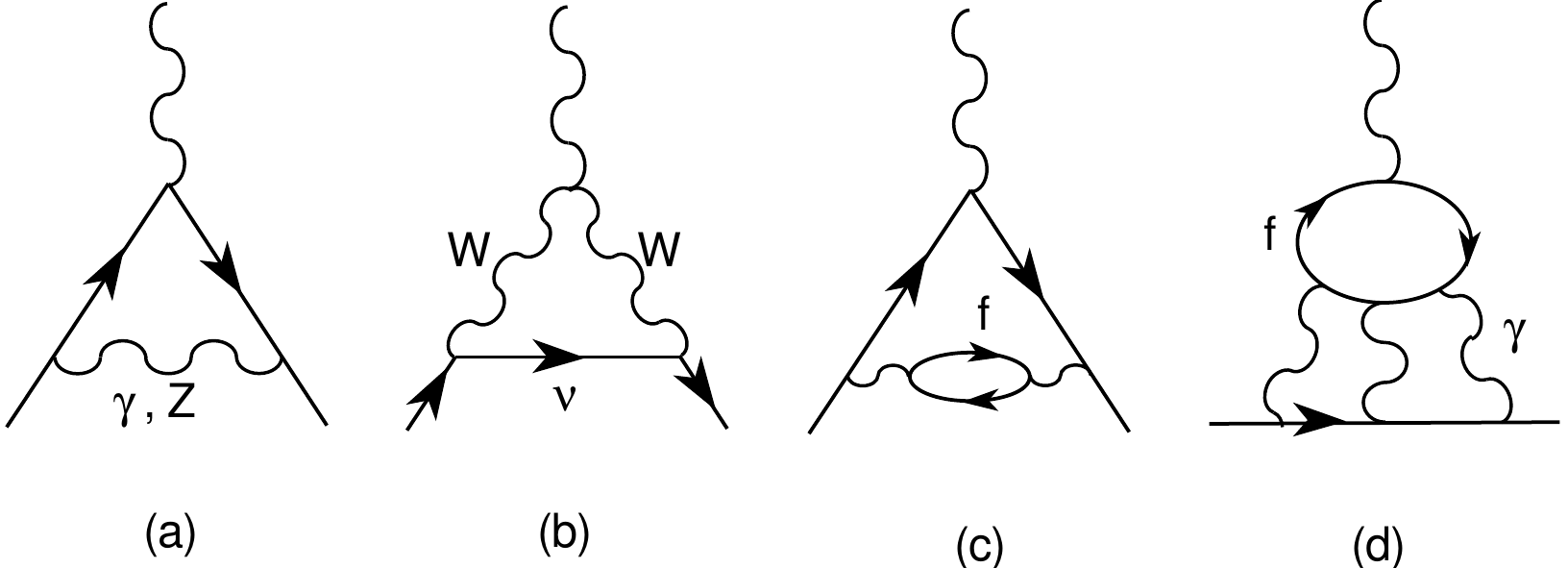}
\caption{Feynman diagrams contributing to the lepton anomalous magnetic moment.}
\label{fig:AnMagMom}
\end{center}
\end{figure}

\begin{figure}[t]\centering
\begin{minipage}{7cm}\centering
\includegraphics[height=6cm]{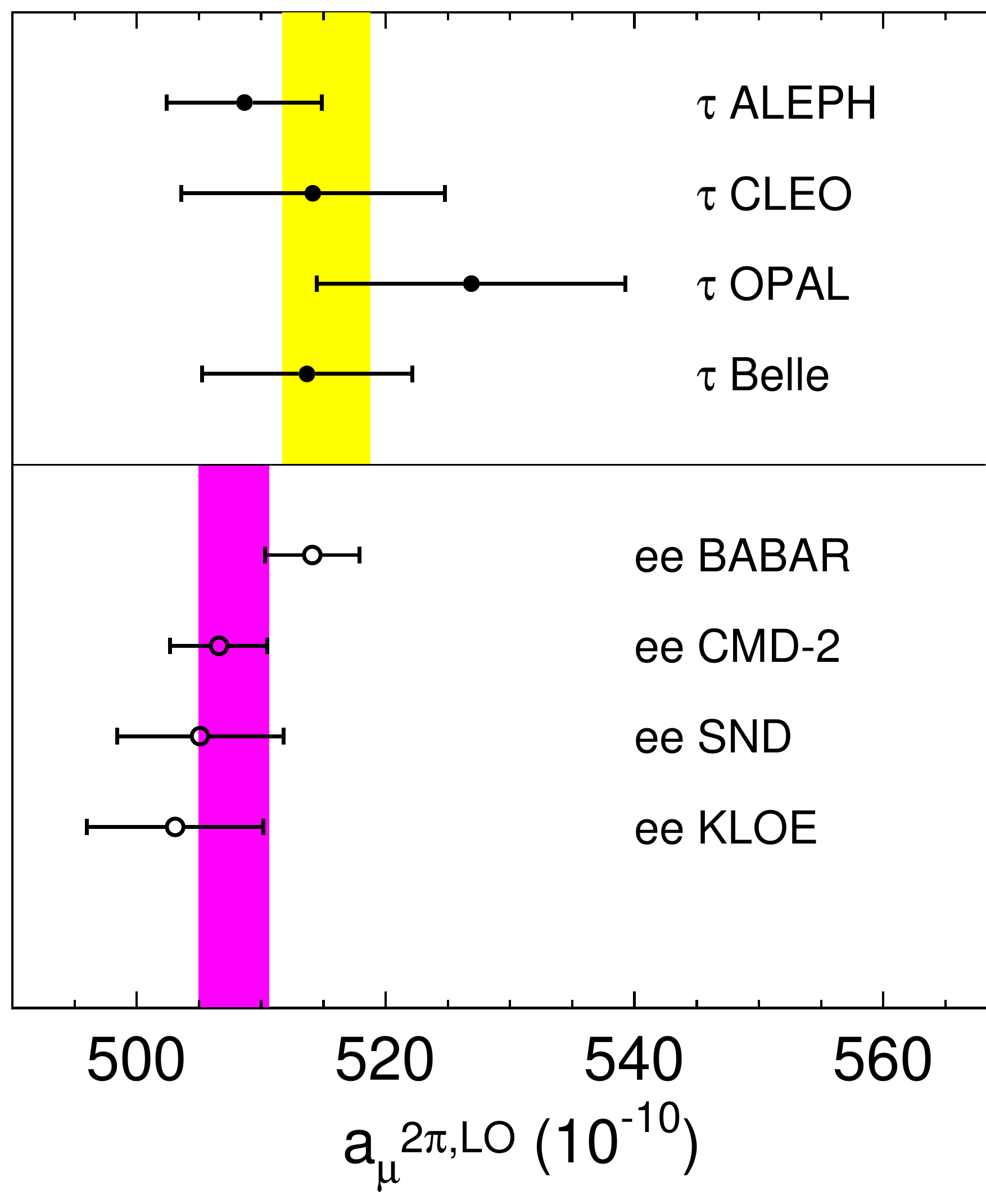}
\end{minipage}
\hskip 1cm
\begin{minipage}{9cm}\centering
\includegraphics[height=6cm,clip]{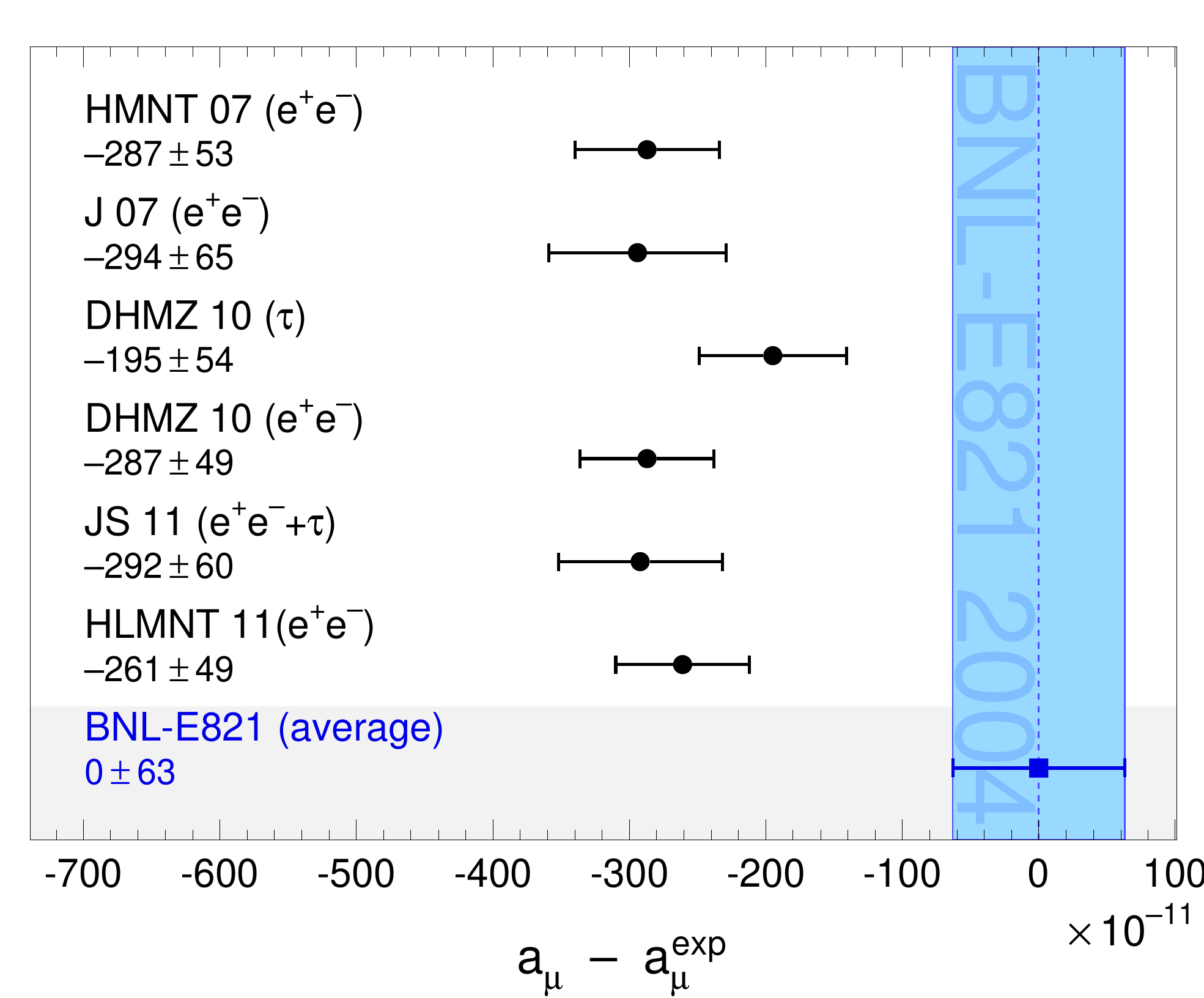}
\end{minipage}
\caption{$2\pi$ contribution to $\delta a_\mu^{\mathrm{hvp,LO}}$ (left), obtained from $\tau$-decay and $e^+e^-$ experiments \cite{Davier:2009ag,Davier:2009zi,Davier:2013wwg}. The right plot shows a recent compilation \cite{Zhang:2013oaw} of SM results for $a_\mu$
\cite{Davier:2010nc,Jegerlehner:2009ry,Hagiwara:2011af,Hagiwara:2006jt,Jegerlehner:2011ti},
compared with the experimental measurement \cite{Bennett:2006fi}. Figures taken from Refs.~\cite{Davier:2013wwg,Zhang:2013oaw}.}
\label{fig:g-2}
\end{figure}

The heavier muon mass makes $a_\mu$ much more sensitive to electroweak
corrections from virtual heavier states \cite{Czarnecki:1995wq,Czarnecki:2001pv,Czarnecki:2002nt,Gribouk:2005ee,Knecht:2002hr,Peris:1995bb,Heinemeyer:2004yq,Kukhto:1992qv,Gnendiger:2013pva}; compared to $a_e$, they scale as $m_\mu^2/m_e^2$.
The main theoretical uncertainty on the muon anomalous magnetic moment has a QCD origin. Since quarks have electric charge, virtual quark-antiquark pairs induce
{\it hadronic vacuum polarization} corrections to the photon
propagator (Fig.~\ref{fig:AnMagMom}.c). Owing to the
non-perturbative character of QCD at low energies, the light-quark
contribution cannot be calculated at present with the required precision. Fortunately,
this effect can be extracted from the measurement of the
cross section $\sigma(e^+e^-\to \mathrm{hadrons})$ and from the
invariant-mass distribution of the final hadrons in $\tau$ decays:
\bel{eq:hvp-formula}
\delta a_\mu^{\mathrm{hvp,LO}} \; =\;\frac{\alpha}{\pi}\,\int_0^\infty \frac{ds}{s}\; K(s/m_\mu^2)
\; e^2\,\frac{1}{\pi}\,\mathrm{Im}\,\Pi_{\mathrm{em}}(s)\, ,
\ee
where $\Pi_{\mathrm{em}}(s)$ is the QCD two-point correlation function of two electromagnetic currents. The kernel function \cite{BM:61,Brodsky:1967sr}
\be
K(s/m_\mu^2)\; =\; \int_0^1 dx\;\frac{x^2 (1-x)}{x^2 + \frac{s}{m_\mu^2}\, (1-x)}\, ,
\ee
is positive and monotonically decreasing as $K(s/m_\mu^2)\sim m_\mu^2/(3s)$ for increasing values $s/m_\mu^2$. Therefore, $\delta a_\mu^{\mathrm{hvp,LO}}$ is dominated by the low-energy spectral region; the largest contribution being the $2\pi$ final state.
As mentioned in section~\ref{subsubsec:2pi}, there is still a slight discrepancy between the $2\pi$ spectral functions extracted from $e^+e^-$ and $\tau$ data, which cannot be accounted for through isospin-breaking corrections
\cite{Cirigliano:2002pv,Davier:2009ag,Davier:2010nc,Davier:2009zi,FloresBaez:2006gf,Davier:2003pw,Alemany:1997tn}.
This translates into slightly different values for $\delta a_\mu^{\mathrm{hvp,LO}}$
\cite{Davier:2010nc}.
Additional disagreements among $e^+e^-$ experiments remain in several hadronic final states \cite{Davier:2013-ARNPS}, after the most recent BaBar \cite{Lees:2012cj,Lees:2011zi},
CMD-2 \cite{Akhmetshin:2006bx,Akhmetshin:2008gz},
CMD-3 \cite{Akhmetshin:2013xc},
KLOE \cite{Babusci:2012rp,Ambrosino:2008gb} and SND \cite{Achasov:2006vp,Achasov:2012zza} analyses. The left plot in Fig.~\ref{fig:g-2} \cite{Davier:2013wwg} shows the $2\pi$ contribution to $\delta a_\mu^{\mathrm{hvp,LO}}$ obtained from different $\tau$-decay and $e^+e^-$ experiments.

The SM prediction for $a_\mu$ can be decomposed in five types of contributions:
\bea 10^{10} \times a_\mu^{\mathrm{th}}\;  =\;
11\; 658\; 471.895\; 1 \pm 0.008\; 0 && \mathrm{QED}
\nonumber\\ \mbox{}
+\phantom{6}15.4\phantom{95\; 1}\pm 0.1\phantom{08\; 0} && \mathrm{EW}
\nonumber\\ \mbox{}
+ 696.4\phantom{95\; 1} \pm 4.6\phantom{08\; 0}
&& \mathrm{hvp}^{\mathrm{LO}}
\hskip 1cm (701.5\pm 4.7)_\tau \quad (692.3\pm 4.2)_{e^+e^-}
\nonumber\\ \mbox{}
-\phantom{69}9.8\phantom{95\; 1}\pm 0.1\phantom{08\; 0}&& \mathrm{hvp}^{\mathrm{NLO}}
\nonumber\\ \mbox{}
+\phantom{6}10.5\phantom{95\; 1}\pm 2.6\phantom{08\; 0} &&\mathrm{lbl}
\nonumber\\
= \; 11\; 659\; 184.4\phantom{95\; 1} \pm  5.3\phantom{08\; 0} &&
(11\; 659\; 189.5\pm  5.4)_\tau \quad
(11\; 659\; 180.3\pm  4.9)_{e^+e^-}\, .\quad
\eea
The first line gives the very precise QED contribution \cite{Aoyama:2012wj,Aoyama:2012fc,Aoyama:2010yt,Kinoshita:2005sm,Lee:2013sx,Baikov:2012rr,Baikov:2008si,Kataev:2012kn,Aguilar:2008qj,Czarnecki:1998rc,Laporta:1996mq,Laporta:1994md,Li:1992xf,Samuel:1990qf,Broadhurst:1992za}, including the recently computed $O(\alpha^5)$ corrections \cite{Aoyama:2012wj}. The quoted number adopts as input the value of $\alpha$ determined from the ${}^{87}\mathrm{Rb}$ atom; using instead the $a_e$ determination in \eqn{eq:alpha}, one gets the slightly more precise
result $\delta a_\mu^{\mathrm{QED}} = (11\, 658\, 471.884\, 5\pm 0.003\, 7)\times 10^{-10}$
\cite{Aoyama:2012wj}. The pure electroweak correction is shown in the second line \cite{Czarnecki:1995wq,Czarnecki:2001pv,Czarnecki:2002nt,Gribouk:2005ee,Knecht:2002hr,Peris:1995bb,Heinemeyer:2004yq,Kukhto:1992qv,Gnendiger:2013pva}. The leading-order hadronic-vacuum-polarization contribution in the third line is a weighted average of the $\tau$ and $e^+e^-$ determinations, which are shown in parentheses \cite{Davier:2010nc}; the estimated next-to-leading correction is given in the fourth line
\cite{Hagiwara:2006jt,Krause:1996rf}.
Additional QCD uncertainties stem from the smaller {\it light-by-light scattering} contribution (Fig.~\ref{fig:AnMagMom}.d), given in the fifth line \cite{Prades:2009tw}, which needs to be theoretically evaluated
\cite{Prades:2009tw,Bijnens:2007pz,Melnikov:2003xd,Knecht:2001qg,Knecht:2001qf,Blokland:2001pb,Bijnens:2001cq,Hayakawa:1997rq,RamseyMusolf:2002cy,Engel:2013kda,Goecke:2010if,Blum:2013qu,Hoferichter:2013ama}.

The final SM prediction differs from the experimental value by $3.0\,\sigma$. The $\tau$ estimate of the hadronic vacuum polarization results in a smaller deviation of $2.3\,\sigma$, while using $e^+e^-$ data alone increases the discrepancy to $3.6\,\sigma$.
The right plot in Fig.~\ref{fig:g-2} shows a recent compilation of SM predictions \cite{Zhang:2013oaw}, using slightly different estimates of $\delta a_\mu^{\mathrm{hvp,LO}}$ \cite{Davier:2010nc,Jegerlehner:2009ry,Hagiwara:2011af,Hagiwara:2006jt,Jegerlehner:2011ti}.
New precise $e^+e^-$ and $\tau$ data sets are needed to settle the true value of
$a_\mu^{\mathrm{th}}$.
Improved predictions are needed to match the aimed $10^{-10}$ accuracy of the
proposed muon experiments at Fermilab \cite{FNAL989} and J-PARC \cite{J-PARC}.

The $\tau$ anomalous magnetic moment has an enhanced sensitivity to new physics because of the large $\tau$ mass. In the SM, the interesting electroweak correction roughly scales as $(m_\tau/m_\mu)^2=283$. Using the same decomposition as before, the different contributions to $a_\tau$ take the following values \cite{Eidelman:2007sb}:
\bea 10^{8} \times a_\tau^{\mathrm{th}}\;  =\;
117\; 324\phantom{.4} \pm 2\phantom{.7} && \mathrm{QED}\quad
\nonumber\\ \mbox{}
+ \phantom{3}47.4\pm 0.5 && \mathrm{EW}
\nonumber\\ \mbox{}
+ 337.5 \pm 3.7 && \mathrm{hvp}^{\mathrm{LO}}
\nonumber\\ \mbox{}
+\phantom{33}7.6\pm 0.2 && \mathrm{hvp}^{\mathrm{NLO}}
\nonumber\\ \mbox{}
+\phantom{33}5\phantom{.4}\pm 3\phantom{.7} &&\mathrm{lbl}
\nonumber\\
= \; 117\; 721\phantom{.4} \pm  5\phantom{.7} &&\hskip -0.95cm .
\eea
The big enhancement of the electroweak contribution has another welcome feature: it is only a factor of seven smaller than the hadronic one, compared to a factor of 45 for $a_\mu$.
The electroweak contribution to $a_\tau$ is ten times larger than the present hadronic uncertainty, while this factor is only three for the muon case. This makes $a_\tau$ a much cleaner test of new physics contributions.

Unfortunately, the very short $\tau$ lifetime makes very difficult to measure $a_\tau$ with a precision good enough to perform a significative test. The spin-precession technique adopted in the muon $g-2$ experiments is no-longer feasible. Instead, one measures the production of $\tau^+\tau^-$ pairs at different high-energy processes and compares the cross sections with the QED predictions
\cite{Abdallah:2003xd,Achard:2004jj,Acciarri:1998iv,Ackerstaff:1998mt,Escribano:1996wp,Grifols:1990ha,delAguila:1990jg,Domokos:1985rp,Silverman:1982ft}.
The 95\% CL limit quoted by the PDG \cite{Beringer:1900zz}
was derived by the DELPHI collaboration from measurements of $\sigma(e^+e^-\to e^+e^-\tau^+\tau^-)$ at $\sqrt{s}$ between 183 and 208 GeV at LEP2 \cite{Abdallah:2003xd}:
\bel{eq:a_tau}
-0.052\; < \; a_\tau\; < \;  0.013\, .
\ee
The present sensitivity is one order of magnitude worse than the first QED contribution, $\delta a_\tau^{\mathrm{QED,LO}} = \frac{\alpha}{2\pi}$ \cite{Schwinger:1948iu}. Thus, $a_\tau$ is still essentially unknown experimentally.

Several methods have been suggested to improve the measurement of $a_\tau$. For instance,
using precise $\tau^-\to\nu_\tau\ell^-\bar\nu_\ell\gamma$ ($\ell = e,\mu$) data from high-luminosity B factories \cite{Fael:2013ij}, or
taking advantage of the radiation amplitude zero which occurs in this decay at the high-energy end of the lepton distribution
for $g^\gamma_\tau =2$ \cite{Laursen:1983sm}. This requires very good energy resolution, high statistics and a good control of systematic uncertainties \cite{Eidelman:2007sb}. A similar method to study $a_\tau$ using radiative $W^-\to\tau^-\bar\nu_\tau\gamma$ decays and potentially very high data samples at LHC was suggested in Ref.~\cite{Samuel:1992fm}. Some sensitivity to $a_\tau$ could also be achieved through $\gamma\gamma\to\tau^+\tau^-$ collisions; \ie $e^+e^-\to e^+\gamma^*\gamma^* e^-\to e^+\tau^+\tau^-e^-$ \cite{Cornet:1995pw,Billur:2013rva}.
Another, more fancy, possibility could be to use polarized $\tau$ leptons from $B^+\to\tau^+\nu_\tau$ decays, precess their spin in a bent crystal and measure the final $\tau$ polarization through the angular distribution of the daughter lepton in the $\tau^-\to\nu_\tau\ell^-\bar\nu_\ell$ decays \cite{Samuel:1990su}. The channelling of a short-lived polarized particle through a bent crystal \cite{Kim:1982ry}, has been already tested successfully and used to measure the magnetic moment of the $\Sigma^+$ hyperon \cite{Chen:1992wx}.
The feasibility of these ideas at a Super-B factory or at a hadron collider remains to be investigated \cite{Eidelman:2007sb}.

A determination of the dipole form factor $F_2(s)$ can be performed in $e^+e^-\to\tau^+\tau^-$ by analysing the angular distribution of the $\tau$ decay products,
which contains information on the $\tau^+\tau^-$ polarizations \cite{Bernabeu:2008ii,Bernabeu:2007rr}.
With unpolarized electron beams, the imaginary part of the form factor could be accessed by measuring the normal (to the scattering plane) polarization of a single $\tau$, while the real part would require measuring the spin correlations of both taus. Using polarized electron beams, the real part could also be obtained by measuring the transverse and longitudinal polarizations of a single $\tau$. With $10^{12}$ $\tau^+\tau^-$ pairs at a future Super-B factory, the estimated sensitivity is of the order of $10^{-6}$ \cite{Bernabeu:2008ii,Bernabeu:2007rr}.

Using an effective Lagrangian, invariant under the SM gauge group, and writing the lowest-dimension ($D=6$) operators contributing to $a_\tau$, it is possible to combine experimental information from $\tau$ production at LEP1, LEP2 and SLD with $W^-\to\tau^-\bar\nu_\tau$ data from LEP2 and $p\bar p$ colliders. This allows one to set a stronger model-independent bound on new-physics contributions to $a_\tau$ (95\% CL) \cite{GonzalezSprinberg:2000mk}:
\bel{eq:atau-new}
- 0.007\; <\; a_\tau^{\mathrm{New\, Phys}}\; <\; 0.005\, .
\ee

\subsection{Electric Dipole Moments}
\label{susec:EDMs}

The strong experimental limits on the electron \cite{Baron:2013eja} 
and muon \cite{Bennett:2008dy} electric dipole moments
\be\label{eq:edm-e}
|d_e^\gamma|\; <\; 8.7\times 10^{-29}     
\; e\, \mathrm{cm}\quad (90\%\;\mathrm{CL})\, ,
\ee
\bel{eq:edm-mu}
d_\mu^\gamma\; =\; (-0.1\pm 0.9)\times 10^{-19}\; e\, \mathrm{cm}\, ,
\ee
provide stringent constraints on models of new physics beyond the SM \cite{Ginges:2003qt,Pospelov:2005pr,Raidal:2008jk,Fukuyama:2012np,Engel:2013lsa,Jung:2013mg,Jung:2013hka}.

The status is very different for the $\tau$ lepton, owing to the much more difficult experimental conditions. Several $\tau^+\tau^-$ production processes have been investigated: $e^+e^-\to \tau^+\tau^-$ \cite{Albrecht:2000yg,delAguila:1990jg}, $e^+e^-\to \tau^+\tau^-\gamma$ \cite{Acciarri:1998iv},
$e^+e^-\to e^+e^-\tau^+\tau^-$ \cite{Abdallah:2003xd,Achard:2004jj,Cornet:1995pw}, $Z\to\tau^+\tau^-$ \cite{Escribano:1996wp} and $Z\to\tau^+\tau^-\gamma$ \cite{Ackerstaff:1998mt,Grifols:1990ha}. However, since the CP-violating form factor $F_3(s)$ does not interfere with the dominant CP-even amplitudes,
the cross sections depend quadratically on $d_\tau^\gamma$, suppressing strongly the sensitivity to the CP-odd component.

The Belle collaboration has searched for CP-violating effects in the process $e^+e^-\to\gamma^*\to\tau^+\tau^-$, using triple momentum and spin correlations which are linear in $F_3(s)$ \cite{Bernreuther:1993nd,Bernreuther:1996dr}. They obtain the 95\% CL constraints \cite{Inami:2002ah}:
\bea\label{eq:edm-tau}
-0.22\; <\; \mathrm{Re}\, (d_\tau^\gamma) \; < \; 0.45 &&\quad (10^{-16}\; e\, \mathrm{cm})\, ,
\nonumber\\
-0.25\; <\; \mathrm{Im}\, (d_\tau^\gamma) \; < \; 0.08 &&\quad (10^{-16}\; e\, \mathrm{cm})\, .
\eea
Actually, these numbers refer to the real and imaginary parts of the corresponding form factor at the squared energy of the $\tau^+\tau^-$ system. This $s$ dependence is expected to be small and has been ignored. $\mathrm{Re}\, (d_\tau^\gamma)$ is obtained from a CP-odd and T-odd observable, while the measurement of $\mathrm{Im}\, (d_\tau^\gamma)$ requires a CP-odd and T-even one.

With a longitudinally polarized electron beam, one could build a CP-odd asymmetry, associated with the normal (to the scattering plane) polarization of a single $\tau$, proportional to $\mathrm{Re}\, (d_\tau^\gamma)$ \cite{Bernabeu:2006wf}. The estimated sensitivity at a Super-B factory ranges from $10^{-18}$ to $10^{-19}\; e$ cm, assuming integrated luminosities between 2 and 75 ab${}^{-1}$ \cite{Bernabeu:2006wf}.

\subsection{Weak Dipole Moments}
\label{subsec:WDM}

The anomalous weak dipole moments of the $\tau$ lepton are usually defined through the effective $Z$ couplings \cite{Beringer:1900zz}:
\bel{eq:wdm}
\cL_{\mathrm{wdm}}^Z\; =\; -\frac{1}{2\sin{\theta_W}\cos{\theta_W}}\; Z_\mu\;
\bar\tau\,\left[ i\, \alpha^W_\tau\, \frac{e}{2 m_\tau}\,\sigma^{\mu\nu} q_\nu +
d^W_\tau\,\sigma^{\mu\nu}\gamma_5\, q_\nu \right]\,\tau\, .
\ee
Both the weak magnetic term $\alpha^W_\tau$ and the CP-violating weak electric term $d^W_\tau$ have been investigated with LEP data
\cite{Heister:2002ik,Acciarri:1998zc,Ackerstaff:1996gy}.
The SM predicts non-zero contributions to $\alpha^W_\tau$ and $d^W_\tau$, through quantum effects, which are well below the present experimental sensitivity \cite{Bernreuther:1988jr,Bernabeu:1994wh}. This opens the possibility to look for deviations from the SM.

Analysing the complete differential cross section for the production and decay of the $\tau$ leptons in $e^+e^-\to\tau^+\tau^-$, taking into account the correlations between the two $\tau$ spins, one is sensitive to the real and imaginary parts of $\alpha^W_\tau$ and $d^W_\tau$
\cite{Bernreuther:1993nd,Bernreuther:1996dr,Bernreuther:1988jr,Bernreuther:1989kc,Bernabeu:1994wh,Bernabeu:1993er,Bernreuther:1991xe}. Assuming that the vector and axial-vector $Z\tau^+\tau^-$ couplings take their SM values
and using optimal polarimeters to recover the $\tau$ spin information \cite{Bernabeu:1990na,Davier:1992nw,Sanchez:1997kp},
the ALEPH collaboration has obtained the 95\% CL limits \cite{Heister:2002ik}:
\bel{eq:alpha_W_tau}
\left|\mathrm{Re}\, (\alpha^W_\tau)\right|\; <\; 1.1\times 10^{-3}\, ,
\qquad\qquad\qquad
\left|\mathrm{Im}\, (\alpha^W_\tau)\right|\; <\; 2.7\times 10^{-3}\, ,
\ee
\bel{eq:d_W_tau}
\left|\mathrm{Re}\, (d^W_\tau)\right|\; <\; 0.50\times 10^{-17}\; e\,\mathrm{cm}\, ,
\qquad\qquad\qquad
\left|\mathrm{Im}\, (d^W_\tau)\right|\; <\; 1.1\times 10^{-17}\; e\,\mathrm{cm}\, .
\ee

T-odd signals can be also generated through
a relative phase between the vector and axial-vector couplings
of the $Z$ to the $\tau^+\tau^-$ pair \cite{Bernabeu:1990na}.
This effect, which in the  SM appears \cite{Bernabeu:1989ct} at the
one-loop level through absorptive parts in the electroweak amplitudes,
gives rise \cite{Bernabeu:1990na} to a spin-spin correlation associated with the
transverse (within the production plane) and normal (to the production
plane) polarization components of the two taus. Ignoring possible contributions from weak dipole moments, the analysis of the ALEPH data gives the result \cite{Barate:1997mz}
\bel{eq:C-TN}
C_{TN}\; =\; -\frac{2\,\mathrm{Im}(v_\tau a_\tau^*)}{v_\tau^2 + a_\tau^2}\; =\;
0.08\pm 0.13\pm 0.04\, .
\ee
%


A possible weak magnetic $W\tau\nu_\tau$ coupling of the form
\bel{eq:wdmW}
\cL_{\mathrm{wdm}}^W\; =\; -\frac{g}{2\sqrt{2}}\; i\, \frac{\kappa^W_\tau}{2 m_\tau}
\; W_\mu\;
\left[\bar\tau\sigma^{\mu\nu} q_\nu (1-\gamma_5)\nu_\tau\right]\, +\,\mathrm{h.c.}
\ee
was also searched for by the DELPHI collaboration, fitting the distribution of the final decay products in $e^+e^-\to\tau^+\tau^-$ events. The measured value of the anomalous tensor coupling $\kappa_\tau^W$ corresponds to the following 90\% CL allowed interval
\cite{Abreu:2000sg}:
\bel{eq:kappa_W}
-0.096\, <\, \kappa_\tau^W\, <\, 0.037\, .
\ee

Using the electroweak $SU(2)_L\otimes U(1)_Y$ gauge symmetry, the dipole interactions in Eqs.~\eqn{eq:em_ff}, \eqn{eq:wdm} and \eqn{eq:wdmW} can all be written in terms of only two dimension-6 operators:
$\cO_B = g'\, B^{\mu\nu}(\bar L_\tau \phi\,\sigma_{\mu\nu}\tau_R)$ and
$\cO_W = g\, W_i^{\mu\nu}(\bar L_\tau \tau^i\phi\,\sigma_{\mu\nu}\tau_R)$,
where $L_\tau^T = (\nu_\tau,\tau)_L$ and $\phi$ are the tau and Higgs $SU(2)_L$ doublets, respectively,  $B^{\mu\nu}$ and $W_i^{\mu\nu}$ the $U(1)_Y$ and $SU(2)_L$ field strength tensors, $g'$ and $g$ the corresponding gauge couplings and $\tau^i$ the Pauli matrices. Analysing the experimental information in terms of these two operators, one gets the result in Eq.~\eqn{eq:atau-new} and a stronger 95\% CL bound on $\kappa_\tau^W$ \cite{GonzalezSprinberg:2000mk}:
\bel{eq:weakEff}
-0.003\, <\, \kappa_\tau^W\, <\, 0.004\, .
\ee

\section{CP violation}
\label{sec:CPV}

In the three-generation SM, the violation of the CP symmetry originates from the single phase naturally occurring in the quark mixing matrix \cite{Kobayashi:1973fv}. Therefore, CP violation is predicted to be absent in the lepton sector (for massless neutrinos). However, the fundamental origin of the Kobayashi--Maskawa phase is still unknown. Obviously, CP violation could well be a sensitive probe for new physics.

The electroweak dipole moments $d_\tau^\gamma$ and $d^W_\tau$
test CP violation in $\tau$ production, but violations of the CP symmetry could also happen in the $\tau$ decay amplitude. In fact, one would naively expect
CP-violating effects to be larger in $\tau$ decay than in $\tau^+\tau^-$ production. Since the decay of the $\tau$ proceeds through a weak interaction, these effects could be ${\cal O}(1)$ or ${\cal O}(10^{-3})$,
if the leptonic CP violation is {\it weak} or {\it milliweak}.
A variety of CP-violating observables (rate, angular and polarization asymmetries, triple products, Dalitz distributions, etc.) can be exploited to search for violations of the CP symmetry \cite{Bigi:2012kz,Kiers:2012fy}.

With polarized electron (and/or positron) beams, one could use the
longitudinal polarization vectors of the incident leptons to construct
T-odd rotationally invariant products \cite{Tsai:1994rc}. CP could be tested by comparing these T-odd products in $\tau^-$ and $\tau^+$ decays.
In the absence of beam polarization, CP violation can still be tested
through $\tau^+\tau^-$ correlations \cite{Goozovat:1991nu}. Studying the final decays of the $\tau$ decay products, it is possible to separate possible CP-odd effects in the $\tau^+\tau^-$ production and in the $\tau$ decay \cite{Nelson:1993zv}. For instance, one can study the chain process
$e^+e^-\to\tau^+\tau^-\to (\rho^+\bar\nu_\tau)(\rho^-\nu_\tau)\to
(\pi^+\pi^0\bar\nu_\tau)(\pi^-\pi^0\nu_\tau)$. The distribution of the
final pions provides information on the $\rho$ polarization, which allows
to test for possible CP-violating effects in the
$\tau\to\rho\nu_\tau$ decay.

CP violation can also be tested through rate asymmetries, \ie comparing the partial fractions
$\Gamma(\tau^-\to X^-)$ and $\Gamma(\tau^+\to X^+)$. However, this kind
of signal requires the presence of strong final-state interactions in the
decay amplitude.
Another possibility is to study T-odd (CPT-even) asymmetries in the
angular distributions of the final hadrons in semileptonic
$\tau$ decays \cite{Kuhn:1996dv}.
Explicit studies of the decay modes
$\tau^-\to K^-\pi^-\pi^+\nu_\tau$ \cite{Kilian:1994ub,Kiers:2008mv}, $\tau^-\to \pi^- K^- K^+\nu_\tau$ \cite{Kilian:1994ub},
$\tau^-\to (3\pi)^-\nu_\tau$  \cite{Datta:2006kd,Choi:1994ch} and
$\tau^-\to (4\pi)^-\nu_\tau$ \cite{Datta:2006kd} show that
sizeable CP-violating effects could be generated in some models of CP violation
involving several Higgs doublets or left-right symmetry.

The $\tau^+\to\pi^+K_S\bar\nu_\tau\, (\ge 0 \pi^0)$ rate asymmetry recently measured by BaBar \cite{BABAR:2011aa},
\be
\mathcal{A}_\tau\;\equiv\;
\frac{\Gamma(\tau^+\to\pi^+K_S\bar\nu_\tau)-\Gamma(\tau^-\to\pi^-K_S\nu_\tau)}{
\Gamma(\tau^+\to\pi^+K_S\bar\nu_\tau)+\Gamma(\tau^-\to\pi^-K_S\nu_\tau)}\; =\;
(-0.36\pm 0.23\pm 0.11)\%\, ,
\ee
differs by $2.8\,\sigma$ from the expected value due to $K^0$--$\bar K^0$ mixing, $\mathcal{A}_\tau = (0.36\pm 0.01)\%\,$ \cite{Bigi:2005ts,Calderon:2007rg,Grossman:2011zk}.
The Belle collaboration has also searched for a CP signal in this decay mode through a difference in the $\tau^\pm$ angular distributions.
Specifically, they measure the difference between the mean values of $\cos{\beta}\cos{\psi}$ for $\tau^+$ and $\tau^-$ events, evaluated in bins
of the total invariant hadronic mass, where the angles $\beta$ and $\psi$ denote the directions of the $K_S$ and $\tau$, respectively, with respect to the one of the $e^+e^-$ centre-of-mass system, as seen from the hadronic rest frame \cite{Kuhn:1996dv}.
Belle finds a null result at the 0.2--0.3\% level \cite{Bischofberger:2011pw},
improving the previous CLEO limits by one order of magnitude \cite{Bonvicini:2001xz}.
More precise experimental analyses are needed to clarify the compatibility between the BaBar and Belle results.

\section{Tau Production in B and D Decays}
\label{sec:B}

Heavy meson decays into final states containing $\tau$ leptons are a good laboratory to look for new physics related to the fermion mass generation. Decays such as $B^-\to\tau^-\bar\nu_\tau$, $B\to D^{(*)}\tau^-\bar\nu_\tau$, $B_c^-\to\tau^-\bar\nu_\tau$ or $D_s^-\to\tau^-\bar\nu_\tau$ involve the heaviest elementary fermions that can be directly produced at flavour factories, providing important information about the underlying dynamics mediating these processes.

An excess of events in two $b\to c\,\tau^-\bar\nu_\tau$ transitions has been reported by BaBar \cite{Lees:2012xj}.
Including the previous Belle measurements~\cite{Bozek:2010xy},
the experimental ratios  ($\ell=e,\mu$)
\bea\label{eq:Babar}
 R(D)&\!\! \equiv &\!\!
\frac{\Br(\bar B\to D\,\tau^-\bar\nu_\tau)}{\Br(\bar B\to D\,\ell^-\bar\nu_\ell)}
\; = \; 0.438\pm 0.056\, ,
\nonumber\\[5pt]
R(D^*)&\!\! \equiv &\!\!
\frac{\Br(\bar B\to D^*\tau^-\bar\nu_\tau)}{\Br(\bar B\to D^*\ell^-\bar\nu_\ell)}
\; = \; 0.354\pm 0.026\, ,
\eea
are significantly higher than the SM expectations, $R(D)=0.296\pm 0.017$ and $R(D^*)=0.252\pm 0.004$~\cite{Fajfer:2012jt,Kamenik:2008tj,Sakaki:2012ft,Becirevic:2012jf,Bailey:2012jg}. If confirmed, this could signal new-physics contributions violating lepton-flavour universality.

A sizeable deviation from the SM was previously observed in the leptonic decay
$B^-\to\tau^-\bar\nu_\tau$. However, Belle~\cite{Adachi:2012mm} finds now
a much lower value in agreement with the SM; combined with the BaBar result~\cite{Lees:2012ju}, it gives the average
\bel{eq:BtauNu}
\Br(B^-\to\tau^-\bar\nu_\tau)\; =\; (1.15\pm 0.23)\times 10^{-4}\, ,
\ee
to be compared with the SM expectation
$(0.733\,{}^{+\,0.121}_{-\,0.073})\times 10^{-4}$~\cite{Charles:2004jd}.

These results are intriguing enough to trigger the theoretical interest.
The enhancement of  $\tau$ production could be generated by new physics contributions with couplings  proportional to fermion masses. In particular, it could be associated with the exchange of a charged scalar within two-Higgs-doublet models
\cite{Jung:2010ik,Fajfer:2012jt,Sakaki:2012ft,Celis:2012dk,Crivellin:2012ye,Nierste:2008qe,Tanaka:1994ay,Tanaka:2012nw}.
Although the Babar data rules out the usually adopted ``type II'' scenario~\cite{Lees:2012xj,Fajfer:2012jt}, these measurements could be accommodated \cite{Celis:2012dk} within the more general framework of the ``Aligned Two-Higgs-Doublet Model''~(A2HDM)~\cite{Pich:2009sp}. However, the fit to the $B$ decay data suggests
that the charged scalar should couple differently to the different quark families; \ie
one needs a coupling proportional to the quark mass but with a family-dependent proportionality factor. Since charged scalar contributions to $R(D^*)$ are subdominant compared with the vector contributions, large Yukawa couplings are required to explain the observed excess in $R(D^*)$, thus generating a tension with present bounds from $D_{(s)}$ leptonic decays  \cite{Jung:2010ik}. If $R(D^*)$ is not considered in the fit, all other observables can be consistently described within the family-universal A2HDM \cite{Celis:2012dk}.

Using only measurements of branching fractions, it is not possible to disentangle charged scalar contributions in $B\to D^{(*)}\tau^-\bar\nu_\tau$ from other kinds of new physics
\cite{Deshpande:2012rr,He:2012zp}. However, $B\to D^{(*)}\tau^-\bar\nu_\tau$ decays have a rich three-body kinematics and spin structure in the final state that has not been exploited so far. Several observables involving angular distributions, polarization fractions and momentum-transfer dependence would provide, if measured, crucial information to discriminate between different new-physics scenarios and clarify the possible role of charged scalar contributions in these processes
\cite{Fajfer:2012vx,Sakaki:2012ft,Celis:2012dk,Nierste:2008qe,Tanaka:1994ay,Tanaka:2012nw,Datta:2012qk,Duraisamy:2013pia,Hagiwara:1989gza,Korner:1989qb,Chen:2005gr}.
Another possibility would be to measure the leptonic decay $B_c^-\to\tau^-\bar\nu_\tau$. Large enhancements over the SM prediction, $\Br(B_c^-\to\tau^-\bar\nu_\tau)= (1.94\pm 0.18)\%$ \cite{McNeile:2012qf}, should be observed if the current excess in $R(D^*)$ is due to charged scalar contributions \cite{Celis:2012dk}.

\section{Lepton-Flavour Violation}
\label{sec:LFV}

We have clear experimental evidence that neutrinos are massive particles and there is mixing in the lepton sector. The solar, atmospheric, accelerator and reactor neutrino
data lead to a consistent pattern of oscillation parameters \cite{Beringer:1900zz}. The main recent advance is the establishment of a sizeable non-zero value of $\theta_{13}$, both in accelerator (Minos \cite{Adamson:2013ue}, T2K \cite{Abe:2011sj})
and reactor experiments (Daya Bay \cite{An:2012eh}, Double-Chooz \cite{Abe:2011fz}, Reno \cite{Ahn:2012nd}),
with a statistical significance which reaches the $10\,\sigma$ at Daya Bay~\cite{An:2012eh}.
This increases the interest for a next-generation of long-baseline neutrino experiments to measure the CP-violating phase $\delta_{\mathrm{CP}}$ and resolve the neutrino mass hierarchy.


A global analysis, combining the full set of neutrino data, leads to the following preferred ranges for the oscillation parameters \cite{GonzalezGarcia:2012sz}:
$$
 \Delta m^2_{21}\; = \; \left( 7.45\, {}^{+\, 0.19}_{-\, 0.16}\right)
 \cdot 10^{-5}\;\mathrm{eV}^2\, ,
\qquad\qquad\qquad
\delta_{\mathrm{CP}}\; =\; \left(265\, {}^{+\, 56}_{-\, 61}\right)^\circ\, ,
$$
$$
\Delta m^2_{31}\; = \;
\left( 2.417\, {}^{+\, 0.013}_{-\, 0.013}\right) \cdot 10^{-3}\;\mathrm{eV}^2 \, ,
\qquad\qquad
\Delta m^2_{32}\; = \;
\left( -2.410\, {}^{+\, 0.062}_{-\, 0.062}\right) \cdot 10^{-3}\;\mathrm{eV}^2\, ,
$$
$$
\sin^2{\theta_{12}}\; =\;  0.306\, {}^{+\, 0.012}_{-\, 0.012}\, ,
\qquad\qquad\qquad
\sin^2{\theta_{13}}\; =\; 0.0229\, {}^{+\, 0.0020}_{-\, 0.0019}\, ,
$$
\bel{nu_mix}
\sin^2{\theta_{23}}\; =\;  0.446\, {}^{+\, 0.007}_{-\, 0.007}\;\oplus\; 0.587\, {}^{+\, 0.032}_{-\, 0.037}\, ,
\ee
where $\Delta m^2_{ij}\equiv m^2_i - m^2_j$ are the mass-squared
differences between the neutrino mass eigenstates and 
$\theta_{ij}$ the corresponding angles in the standard
three-flavour parametrization of the neutrino mixing matrix \cite{Beringer:1900zz}. 
As atmospheric mass-squared splitting the fit uses $\Delta m^2_{31}$ for {\it normal ordering} and $\Delta m^2_{32}$ for {\it inverted ordering}.
The statistical significance of the $\delta_{\mathrm{CP}}$ determination is low, being compatible with any value between $0^\circ$ and $360^\circ$ at the $3\,\sigma$ level.

Non-zero neutrino masses constitute a clear indication of new physics beyond the SM. Right-handed neutrinos are an obvious possibility to incorporate Dirac neutrino masses; however, the $\nu_{iR}$ fields would be $SU(3)_C\otimes SU(2)_L\otimes U(1)_Y$
singlets, without any SM interaction. If such objects do exist, it would seem natural to expect that they are able to communicate with the rest of the world through some still unknown dynamics. Moreover, the SM gauge symmetry would allow for a right-handed Majorana
neutrino mass matrix of arbitrary size, because it is not related to the ordinary Higgs mechanism. A Majorana mass term mixes neutrinos and anti-neutrinos, violating
lepton number by two units. Clearly, new physics is called for.

Adopting a more general effective field theory language, without any assumption about the existence of right-handed neutrinos or any other new particles, one can write the most general $SU(3)_C\otimes SU(2)_L\otimes U(1)_Y$ invariant Lagrangian, in terms of the known
low-energy fields (left-handed neutrinos only). The SM is the unique answer with dimension four. The first contributions from new physics appear through dimension-5 operators, and have also a unique form which violates lepton number by two units \cite{Weinberg:1979sa}:
\bel{eq:WE} \Delta \cL\; =\; - {c_{ij}\over\Lambda_{\mathrm{NP}}}\; \bar
L_i\,\tilde\phi\, \tilde\phi^t\, L_j^c \; + \; \mathrm{h.c.}\, , \ee
where $L_i$ denotes the $i$-flavoured $SU(2)_L$ lepton doublet,
$\tilde\phi \equiv i\,\tau_2\,\phi^*$ is the $\mathcal{C}$-conjugate of the SM Higgs doublet $\phi$ and $L_i^c \equiv \mathcal{C} \bar L_i^T$.
Similar operators with quark fields are forbidden, due to their different hypercharges, while higher-dimension operators would be suppressed by higher powers of the new-physics scale $\Lambda_{\mathrm{NP}}$.
After spontaneous symmetry breaking, $\langle\phi^{(0)}\rangle = v/\sqrt{2}$, $\Delta \cL$
generates a Majorana mass term for the left-handed neutrinos, with\footnote{
This relation generalizes the well-known see-saw mechanism
($m_{\nu_{L}}\sim m^2/\Lambda_{\mathrm{NP}}$) \cite{GMRS:79,YA:79}.}
$M_{ij} = c_{ij} v^2/\Lambda_{\mathrm{NP}}$. Thus, Majorana neutrino masses should
be expected on general symmetry grounds. Taking $m_\nu\gtrsim 0.05$~eV, as suggested by atmospheric neutrino data, one gets
$\Lambda_{\mathrm{NP}}/c_{ij}\lesssim 10^{15}$~GeV, amazingly close to the
expected scale of Gran Unification.

With non-zero neutrino masses, the leptonic charged-current interactions involve a flavour mixing matrix $V_L$. The data on neutrino oscillations imply the following $3\,\sigma$ CL ranges for the magnitudes of the different $V_L$ entries \cite{GonzalezGarcia:2012sz}:
\bel{eq:V_L}
|V_L|_{3\sigma}\; =\; \left(
\begin{array}{ccc}
0.799\to 0.844\quad & 0.515\to 0.581\quad & 0.130\to 0.170 \\
0.214\to 0.525\quad & 0.427\to 0.706\quad & 0.598\to 0.805 \\
0.234\to 0.536\quad & 0.452\to 0.721\quad & 0.573\to 0.787
\ea\right)\, .
\ee
Therefore the mixing among leptons appears to be very different from the one in the quark sector. The number of relevant phases characterizing the matrix $V_L$ depends on the Dirac or Majorana nature of neutrinos, because if one rotates a Majorana
neutrino by a phase, this phase will appear in its mass term which
will no longer be real. With only three Majorana (Dirac) neutrinos,
the $3\times 3$ matrix $V_L$ involves six (four) independent parameters: three mixing angles and three (one) phases.


The smallness of neutrino masses implies a strong suppression of neutrinoless
lepton-flavour-violating  processes, which can be avoided in models with sources of lepton-flavour violation (LFV) not related to $m_{\nu_i}$.
LFV processes have the potential to probe physics at scales much higher than the TeV
\cite{Raidal:2008jk,Marciano:2008zz,Barbieri:1995tw,Hisano:1997tc,Hisano:1998cx,Hisano:1998fj,Huitu:1997bi,Raidal:1997hq,Ilakovac:1999md,Kuno:1999jp,Sato:2000zh,Sato:2000ff,Black:2002wh,Cvetic:2002jy,Masiero:2002jn,Illana:2003pj,Cirigliano:2004tc,Cirigliano:2005ck,Choudhury:2005jh,Arganda:2005ji,Yaguna:2005qn,Chen:2006hp,Cirigliano:2006su,Cirigliano:2006nu,Agashe:2006iy,Antusch:2006vw,Choudhury:2006sq,Blanke:2007db,Arganda:2007jw,Arganda:2008jj,Herrero:2009tm,Cirigliano:2009bz,Li:2009yr,Benbrik:2008ik,Arhrib:2009xf,Liu:2009su,Li:2010vf,Hisano:2010es,delAguila:2011wk,Kaneko:2011qi}. The LFV scale can be constrained imposing the requirement of a viable leptogenesis.
Recent studies within different new-physics scenarios find interesting correlations between $\mu$ and $\tau$ LFV decays, with $\mu\to e\gamma$ often expected to be close to the present exclusion limit.

\begin{table}[tb]\centering
\caption{Best published limits on lepton-flavour-violating decays \cite{Decamp:1991uy,Abreu:1996mj,Adriani:1993sy,Akers:1995gz,Aaij:2013cby,Adam:2013mnn,Bellgardt:1987du,Bolton:1988af,Lees:2010ez,Miyazaki:2012mx,Bertl:2006up}.}
\label{table:LFV}\vspace{0.2cm}  
\renewcommand{\arraystretch}{1.1} 
\begin{tabular}{ll@{\hspace{.8cm}}ll@{\hspace{.6cm}}ll@{\hspace{.8cm}}ll@{\hspace{.8cm}}ll}
\hline \\[-4mm]
\multicolumn{6}{l}{$\rm{Br}(\mu^-\to X^-)\cdot 10^{12}\qquad\quad (90\%\;\mathrm{CL})$}
\\ \\[-4mm] \hline \\[-4mm]
 $e^-\gamma$ & $0.57$ &
 $e^-2\gamma$ & $72$ &
 $e^-e^-e^+$ & $\phantom{1}1.0$ &&&&
 \\ \\[-4mm] \hline\hline \\[-4mm]
 \multicolumn{6}{l}{$\rm{Br}(\tau^-\to X^-)\cdot 10^{8}\qquad\quad (90\%\;\mathrm{CL})$}
 \\ \\[-4mm] \hline \\[-4mm]
 $e^-\gamma$ & $3.3$ &
 $e^-e^+e^-$ & $\phantom{1}2.7$ &
 $e^-\mu^+\mu^-$ & $2.7$ &
 $e^-e^-\mu^+$ & $1.5$ &
 $e^-\pi^0$ & $\phantom{1}8.0$
 \\
 $\mu^-\gamma$ & $4.4$ &
 $\mu^-e^+e^-$ & $\phantom{1}1.8\quad\;$ &
 $\mu^-\mu^+\mu^-$ & $2.1$ &
 $\mu^-\mu^-e^+$ & $1.7$ &
 $\mu^-\pi^0$ & $11$
 \\
 $e^-\eta$ & $9.2$ &
 $e^-\eta'$ & $16$ &
 $e^-\rho^0$ & $1.8$ &
 $e^-\omega$ & $4.8$ &
 $e^-\phi$ & $\phantom{1}3.1$
 \\
 $\mu^-\eta$ & $6.5$ &
 $\mu^-\eta'$ & $13$ &
 $\mu^-\rho^0$ & $1.2$ &
 $\mu^-\omega$ & $4.7$ &
 $\mu^-\phi$ & $\phantom{1}8.4$
 \\
 $e^-K_S$ & $2.6$ &
 $e^-K^{* 0}$ & $\phantom{1}3.2$ &
 $e^-\bar K^{* 0}$ & $3.4$ &
 $e^-K^+\pi^-$ & $3.1$ &
 $e^-\pi^+K^-$ & $\phantom{1}3.7$
 \\
 $\mu^-K_S$ & $2.3$ &
 $\mu^-K^{*0}$ & $\phantom{1}5.9$ &
 $\mu^-\bar K^{*0}$ & $7.0$ &
 $\mu^-K^+\pi^-$ & $4.5$ &
 $\mu^-\pi^+K^-$ & $\phantom{1}8.6$
 \\
 $e^-K_SK_S$ & $7.1$ &
 $e^-K^+K^-$ & $\phantom{1}3.4$ &
 $e^-\pi^+\pi^-$ & $2.3$ &
 \\
 $\mu^-K_SK_S$ & $8.0$ &
 $\mu^-K^+K^-$ & $\phantom{1}4.4$ &
 $\mu^-\pi^+\pi^-$ & $2.1$ &
\\
\multicolumn{3}{l}{$e^-f_0(980)\to e^-\pi^+\pi^-$} & $\phantom{1}3.2$ &
\multicolumn{3}{l}{$\!\!\!\mu^-f_0(980)\to \mu^-\pi^+\pi^-$} & $3.4$
 \\ \\[-4mm] \hline\hline \\[-4mm]
 \multicolumn{6}{l}{$\rm{Br}(Z\to X^0)\cdot 10^{6}\qquad\quad (95\%\;\mathrm{CL})$}
 \\ \\[-4mm] \hline \\[-4mm]
 $e^\pm\mu^\mp$ & $1.7$ &
 $e^\pm\tau^\mp$ & $9.8$ &
 $\mu^\pm\tau^\mp$ & $12$ &
 \\ \\[-4mm] \hline\hline \\[-4mm]
 \multicolumn{6}{l}{$\rm{Br}(B^0_{(s)}\to X^0)\cdot 10^{8}\qquad\quad (95\%\;\mathrm{CL})$}
 \\ \\[-4mm] \hline \\[-4mm]
 $B^0\to e^\pm\mu^\mp$ & $0.37$ &
 $B^0_s\to e^\pm\mu^\mp$ & $1.4$ &
 \\ \\[-4mm] \hline\hline \\[-4mm]
 \multicolumn{6}{l}{$\rm{Br}(\mu^- + N\to e^- + N)\cdot 10^{12}\qquad\quad (90\%\;\mathrm{CL})$}
 \\ \\[-4mm] \hline \\[-4mm]
 Au & $0.7$ & Ti & $4.3$ & Pb & 46 &
 \\ \\[-4mm] \hline
\end{tabular}\end{table}

Table \ref{table:LFV} shows the best published limits on LFV decays of $Z$ bosons \cite{Decamp:1991uy,Abreu:1996mj,Adriani:1993sy,Akers:1995gz},
B mesons \cite{Aaij:2013cby},
muons \cite{Adam:2013mnn,Bellgardt:1987du,Bolton:1988af} and taus
\cite{Lees:2010ez,Miyazaki:2012mx}, together with the present experimental constraints on
$\mu\to e$ conversions in muonic atoms \cite{Bertl:2006up}.
The B factories are pushing the experimental limits on neutrinoless LFV
$\tau$ decays to the $10^{-8}$ level, increasing in a drastic way the sensitivity to new physics scales.
A rather competitive upper bound on $\tau\to 3\mu$ has been also obtained
at LHCb, $\mathrm{Br}(\tau^-\to \mu^-\mu^+\mu^-) < 8.0\times 10^{-8}$ \cite{Aaij:2013fia}, showing the potential of the high-statistics collider data in some particular decay modes. Future experiments could improve further some limits to the $10^{-9}$
level, allowing to explore interesting and  totally unknown phenomena.

Complementary information is provided by the MEG experiment, which
has recently set a very stringent limit on LFV in muon decay,
$\mathrm{Br}(\mu^+\to e^+\gamma)< 5.7\times 10^{-13}$ (90\% CL) \cite{Adam:2013mnn},
improving their previous best upper limit by a factor of four. With further data still being acquired (the final number of stopped muons is expected to double the sample analysed so far) and the upgrade program currently underway, MEG aims to enhance the sensitivity by one order of magnitude, down to the $6\times 10^{-14}$ level \cite{Baldini:2013ke}.
A possible $10^{4}$ improvement in $\mu\to 3 e$, reaching a sensitivity of $10^{-16}$, is also under study at PSI \cite{Blondel:2013ia},
and ongoing projects at J-PARC \cite{Kuno:2013mha} and FNAL \cite{Abrams:2012er}
aim to study $\mu\to e$ conversions in muonic atoms at the $10^{-16}$ level, a factor $10^4$ better than the current experimental limit from SINDRUM-II at PSI \cite{Bertl:2006up}. New proposals to reach further sensitivities around $10^{-18}$ are also being discussed \cite{Knoepfel:2013ouy,Barlow:2011zza}.


\begin{table}[tb]\centering
\caption{Best published limits on lepton-number-violating
decays \cite{Lees:2010ez,Miyazaki:2012mx,Aaij:2013fia,Miyazaki:2005ng,Abbiendi:1998nj,BABAR:2012aa,Aaij:2011ex,Seon:2011ni}.}
\label{table:LNV}\vspace{0.2cm}  
\renewcommand{\arraystretch}{1.1} 
\begin{tabular}{ll@{\hskip .85cm}ll@{\hskip .85cm}ll@{\hskip .85cm}ll@{\hskip .85cm}ll}
\hline \\[-4mm]
\multicolumn{6}{l}{$\rm{Br}(\tau^-\to X^-)\cdot 10^{8}\qquad\quad (90\%\;\mathrm{CL})$}
\\ \\[-4mm] \hline \\[-4mm]
 $e^+\pi^-\pi^-$ & $2.0$ &
 $e^+K^-K^-$ & $3.3$ &
 $e^+\pi^-K^-$ & $3.2$ &
 $\pi^-\Lambda$ & $\phantom{1}7.2$ &
 $\mu^-\mu^-p$ & $44$
\\
 $\mu^+\pi^-\pi^-$ & $3.9$ &
 $\mu^+K^-K^-$ & $4.7$ &
 $\mu^+\pi^-K^-$ & $4.8$ &
 $\pi^-\overline\Lambda$ & $14\phantom{.0}$ &
 $\mu^+\mu^-\bar p$ & $33$
 \\ \\[-4mm] \hline\hline \\[-4mm]
\multicolumn{6}{l}{$\rm{Br}(B^-\to X^-)\cdot 10^{7}\qquad\quad (95\%\;\mathrm{CL}; \quad
* = 90\%\;\mathrm{CL})$}\\
\\[-4mm] \hline \\[-4mm]
$\pi^+e^-e^-$ & $0.23^*$ & $\pi^+ e^-\mu^-$ & $1.5^*$ &
$\pi^+\mu^-\mu^-$ & $0.13$ & $D^+\mu^-\mu^-$ & $\phantom{1}6.9$ &
$D^+e^-e^-$ & $26^*$
\\
$K^+e^-e^-$ & $0.30^*$ & $K^+ e^-\mu^-$ & $1.6^*$ &
$K^+\mu^-\mu^-$ & $0.54$ & $D^+_s\mu^-\mu^-$ & $\phantom{1}5.8$ &
$D^+e^-\mu^-$ & $18^*$
\\
$\rho^+e^-e^-$ & $1.7^*$ & $\rho^+ e^-\mu^-$ & $4.7^*$ &
$\rho^+\mu^-\mu^-$ & $4.2^*$ & $D^{*+}\mu^-\mu^-$ & $24$ &
\\
$K^{*+}e^-e^-$ & $4.0^*$ & $K^{*+} e^-\mu^-$ & $3.0^*$ &
$K^{*+}\mu^-\mu^-$ & $5.9^*$ & $D^0\pi^+\mu^-\mu^-$ & $15$ &
\\
$e^-\Lambda$ & $0.32^*$ & $\mu^-\Lambda$ & $0.6^*$ &
$e^-\overline\Lambda$ & $0.8^*$ & $\mu^-\overline\Lambda$ & $\phantom{1}0.6^*$
\\ \\[-4mm] \hline\hline \\[-4mm]
\multicolumn{6}{l}{$\rm{Br}(Z\to X^0)\cdot 10^{6}\qquad\quad (95\%\;\mathrm{CL})$} \\ \\[-4mm] \hline \\[-4mm]
$e^-p$ & $1.8$ &
$\mu^-p$ & $1.8$ &
\\ \\[-4mm] \hline
\end{tabular}\end{table}

The upper limits on LFV decays of the $\mu$ and the $\tau$ provide also interesting constraints on possible LFV couplings of the Higgs-like boson discovered at the LHC. The present bounds on $\tau^-\to\ell^-\gamma$ and $\tau^-\to\ell^-\ell^+\ell'^-$ ($\ell^{(\prime)}= e,\mu$)
still allow for sizeable $H\to\tau^\pm\ell^\mp$ decay rates of $\cO(10\%)$ \cite{Blankenburg:2012ex,Harnik:2012pb,Davidson:2012ds}.
Semileptonic transitions like $\tau\to\ell\pi\pi$ are sensitive to LFV scalar couplings while decays such as $\tau\to\ell\eta^{(\prime)}$ probe pseudoscalar couplings, thus providing a useful low-energy handle to disentangle possible Higgs  LFV signals at the LHC \cite{Celis:2013xja,Petrov:2013vka,Daub:2012mu}.

Lepton-number violation has also been tested in $\tau$ \cite{Lees:2010ez,Miyazaki:2012mx,Aaij:2013fia,Miyazaki:2005ng}, $Z$ \cite{Abbiendi:1998nj}
and meson \cite{Beringer:1900zz,BABAR:2012aa,Aaij:2011ex,Seon:2011ni,Edwards:2002kq} decays with sensitivities approaching in some cases the $10^{-8}$ level. The strongest published limits are shown in table~\ref{table:LNV}. These bounds provide useful constraints on models of new physics involving Majorana neutrinos with masses in the GeV range
\cite{Ali:2001gsa,Atre:2005eb,Ivanov:2004ch,Atre:2009rg,Helo:2010cw,Zhang:2010um,Bao:2012vq,Cvetic:2010rw,Cvetic:2012hd,Quintero:2011yh,Quintero:2012jy,Castro:2013jsn}.

\section{Tau Physics at the LHC}
\label{sec:LHC}

The study of processes with $\tau$ leptons in the final state is an important part of the LHC program. Owing to their high momenta, tightly collimated decay products and low multiplicity, $\tau$ leptons provide excellent signatures to probe new physics at high-energy colliders.  Moreover, since $\tau$ decays are fully contained within the detector, the distribution of the $\tau$ decay products has precious polarization information.

The $\tau$ signal has been already exploited successfully at the LHC to measure
$W$ \cite{Aad:2011fu,CMS:2011asa}, $Z$ \cite{Aad:2011kt,Chatrchyan:2011nv,Aaij:2012bi} and top \cite{Aad:2012vip,Chatrchyan:2013kff} production cross sections, through their subsequent decays to $\tau$ leptons
($W^-\to\tau^-\bar\nu_\tau$, $Z\to\tau^+\tau^-$, $t\to b\,\tau^+\nu_\tau$),
in good agreement with the theoretical predictions \cite{Melnikov:2006kv,Anastasiou:2003ds,Catani:2012qa,Martin:2009iq,Moch:2012mk,Moch:2008qy,Kidonakis:2010dk}. These SM processes have relatively large cross sections and are the largest sources of $\tau$ leptons at the LHC. They constitute important backgrounds on new-physics searches, which need to be carefully studied.
Similar analyses were done previously at UA1 \cite{Albajar:1986fn} and the Tevatron \cite{Abe:1991fb,Abbott:1999pk}.

ATLAS has reported the first $\tau$ polarization measurement ever made at hadron colliders, using the $\tau^-\to \rho^-\nu_\tau\to\pi^-\pi^0\nu_\tau$ decay in $W^-\to\tau^-\bar\nu_\tau$ \cite{Aad:2012cia}.
The angle between the $\tau$ flight direction and the hadronic decay products in the $\tau$ rest frame is sensitive to the $\tau$ polarization \cite{Rouge:1990kv,Hagiwara:1989fn}, but its measurement requires the $\tau$ energy which is difficult to reconstruct in $W^-\to\tau^-\bar\nu_\tau$. ATLAS uses instead the angle between the flight directions of the $\rho^-$ and the $\pi^-$, which is related to the kinematics of the two final pions. Using the correlation between the $\tau$ and $\rho$ polarizations, this allows to determine
$\cP_\tau = -1.06 \pm 0.04\, {}^{+\, 0.05}_{-\, 0.07}$  \cite{Aad:2012cia}, in agreement with the SM. This method is independent of the mode of $\tau$ production and could be applied to the characterization of new phenomena at the LHC. In particular, $\cP_\tau$ may be used as a discriminating variable in searches for new particles.

The $\tau$ is the heaviest lepton coupling to the Higgs; with $M_H = 126$~GeV, the decay $H\to\tau^+\tau^-$ has the fourth largest Higgs branching ratio. Thus, this decay mode plays a very important role to establish the nature of the recently discovered Higgs-like particle \cite{Aad:2012tfa,Chatrchyan:2012ufa}.
The experimental analyses are quantified in terms of the signal-strength parameter, measuring the product of Higgs production cross section and branching ratio, normalized to the SM prediction. In the $H\to\tau^+\tau^-$ mode ATLAS quotes $\mu_{\tau\tau} = 0.7\,{}^{+\, 0.7}_{-\, 0.6}$ \cite{Aad:2012tfa}, while CMS finds $\mu_{\tau\tau} = 1.10\pm 0.41$ \cite{Chatrchyan:2012ufa}. The ATLAS result is consistent with either the SM or the absence of a $H\tau^+\tau^-$ coupling, while the significance of the CMS value is still below $3\,\sigma$. Adding all detected Higgs decay channels ($b\bar b$, $W^*W$, $Z^*Z$, $\tau^+\tau^-$ and $\gamma\gamma$), ATLAS finds $\mu = 1.23\pm 0.18$ and CMS $\mu = 0.80\pm 0.14$, which lead to an average signal strength $\langle\mu\rangle = 0.96\pm 0.11$, in perfect agreement with the SM.

The new boson appears to couple to the known gauge bosons with the strength expected for the SM Higgs \cite{Pich:2013vta,Ellis:2013lra,Giardino:2013bma,Cheung:2013kla,Falkowski:2013dza}, although ATLAS observes an excess of events in the $2\gamma$ decay channel, compared with the SM expectation. Moreover, its fermionic couplings seem compatible with a linear dependence with the fermion mass, scaled by the electroweak scale $v\approx 246$~GeV. Thus, it has the properties expected for a Higgs-like particle, related with the spontaneous breaking of the electroweak symmetry. An obvious question to address is whether it corresponds to the unique Higgs boson incorporated in the SM, or it is just the first signal of a much richer scalar sector \cite{Celis:2013rcs}.

Present searches for new phenomena, taking advantage of the $\tau$ signal,
include bounds on narrow resonances \cite{Aad:2012ypy,Chatrchyan:2012hd}, third generation leptoquarks \cite{ATLAS:2013oea,Chatrchyan:2012sv},
supersymmetric neutral ($H\to\tau^+\tau^-$) \cite{ATLAS:2011mha,Chatrchyan:2012vp} and charged ($H^+\to \tau^+\nu_\tau$)
\cite{Aad:2012tj,Chatrchyan:2012vca} Higgses, other supersymmetric particles \cite{ATLAS:2012ht,Chatrchyan:2013dsa}
and the BaBar constraints on a light CP-odd neutral scalar ($\Upsilon\to\gamma A^0\to\gamma\tau^+\tau^-$) \cite{Lees:2012te}. Significant improvements are to be expected with the increasing energy and luminosity of the LHC and the use of more refined tools
to disentangle different new-physics scenarios.

\section{Outlook}
\label{sec:outlook}

The flavour structure of the SM is one of the main pending questions in our understanding of weak interactions. Although we do not know the reason of the observed family replication, we have learned experimentally that the number of SM fermion generations is just three
(and no more). Therefore, we must study as precisely as possible the few existing flavours
to get some hints on the dynamics responsible for their observed structure.

The $\tau$ turns out to be an ideal laboratory to test the SM. It is a lepton, which means clean physics, and moreover it is heavy enough to produce a large variety of decay modes.
Naively, one would expect the $\tau$ to be much more sensitive than the $e$ or the $\mu$ to new physics related to the flavour and mass-generation problems.
QCD studies can also benefit a lot from the existence of this heavy lepton, able to decay into hadrons. Owing to their semileptonic character, the hadronic $\tau$ decays provide a powerful tool to investigate the low-energy effects of the strong interactions in rather simple conditions.

Our knowledge of the $\tau$ properties has been considerably improved with the clean $\tau^+\tau^-$ data samples collected at LEP and the large statistics accumulated at the B factories. Lepton universality has been tested to rather good accuracy, both in the charged and neutral current sectors. The Lorentz structure of the leptonic $\tau$ decays is not yet determined, but useful constraints on new-physics contributions have been established.  The quality of the hadronic $\tau$ decay data has made possible to perform quantitative QCD tests and determine the strong coupling constant very accurately, providing a nice experimental verification of asymptotic freedom. A quite competitive determination of the Cabibbo mixing has also been obtained from $\tau$ decays into final states with strangeness, and significant improvements could be expected with the larger statistics of future flavour factories.

Searches for non-standard phenomena have been pushed to the limits that the existing data samples allow to investigate. The first hints of new physics beyond the SM have emerged recently, with convincing evidence of neutrino oscillations from solar, atmospheric, accelerator and reactor neutrino experiments. The existence of lepton flavour violation opens a very interesting window to unknown phenomena, which we are just starting
to explore. It seems possible to push the present limits on neutrinoless $\tau$ decays beyond the $10^{-8}$ or even $10^{-9}$ level. At the same time, new neutrino oscillation experiments will investigate whether CP violating phases are also present in the lepton mixing matrix.

At present, all experimental results on the $\tau$ lepton seem consistent with the SM, with the exception of the anomalous excess of $\tau$ production observed in some $B$ decays,
the CP rate asymmetry in $\tau\to\pi K_S\nu_\tau$
and the slight violation of universality in $W^-\to\tau^-\bar\nu_\tau$. These effects are, however, not very significant statistically. There is large room for improvements, which require much larger samples of $\tau$ events and a better control of systematic uncertainties. Future $\tau$ experiments will probe the SM to a much deeper level of sensitivity and will explore the frontier of its possible extensions.

While the $\tau$ lepton continues being an increasingly precise laboratory to perform relevant tests of QCD and the electroweak theory, we are witnessing the opening of a new era with this heavy lepton becoming now a superb tool in searches for new phenomena at high-energy colliders. Being one of the fermions most strongly coupled to the scalar sector, the $\tau$ plays a very important role in testing the Higgs properties and the dynamics behind the electroweak symmetry breaking.  The ongoing LHC program will be complemented with refined low-energy measurements at Belle-II, Bes-III and, perhaps, a future Super Tau-Charm Factory, and more precise muon experiments. There is an exciting future ahead of us and unexpected surprises may arise, probably establishing the existence of new physics beyond the SM and offering clues to the problems of mass generation, fermion mixing and family replication.

\section*{Acknowledgments}

I would like to thank Martin Jung, Jorge Portol\'es and Pablo Roig
for useful comments which helped to improve the manuscript.
This work has been supported in part by the Spanish Government and EU funds for regional development [grants FPA2011-23778 and CSD2007-00042 (Consolider Project CPAN)], and the Generalitat Valenciana [PrometeoII/2013/007].



\begin{thebibliography}{99}
\itemsep -2pt

\bibitem{PE:75} M.L. Perl \etal\ (MARK I), \myJournal{\PRL}{35}{1975}{1489}


\bibitem{PE:80} M. Perl, \myJournal{\ARNPS}{30}{1980}{299}

\bibitem{HP:88} K.G. Hayes and M.L. Perl, \myJournal{\PRD}{38}{1988}{3351}

\bibitem{BS:88} B.C. Barish and R. Stroynowski, \myJournal{\PREP}{157}{1988}{1}

\bibitem{GP:88} K.K. Gan and M.L. Perl, \myJournal{\MP}{A 3}{1988}{531}

\bibitem{KI:88} C. Kiesling,
  in {\it High Energy Electron--Positron Physics}, eds. A.~Ali and P.~S\"oding,
  Advanced Series on Directions in High Energy Physics -- Vol.~1
  (World Scientific, Singapore, 1988)  p.~177

\bibitem{PI:90} A. Pich, \myJournal{\MPL}{A 5}{1990}{1995}

\bibitem{PI:92} A. Pich, 
  in {\it Heavy Flavours}, eds. A.J. Buras and M. Lindner,
  Advanced Series on Directions in High Energy Physics -- Vol.~10
  (World Scientific, Singapore, 1992) p.~375

\bibitem{PE:92} M. Perl, \myJournal{\RPP}{55}{1992}{653}

\bibitem{RI:92} K. Riles, \myJournal{\MP}{A 7}{1992}{7647}

\bibitem{WS:93} A.J. Weinstein and R. Stroynowski, \myJournal{\ARNPS}{43}{1993}{457}

\bibitem{GP:96} S. Gentile and M. Pohl, \myJournal{\PREP}{274}{1997}{287}

\bibitem{PI:98} A. Pich, 
  in {\it Heavy Flavours II},
  eds. A.J. Buras and M. Lindner, Advanced Series on Directions in High Energy Physics, Vol. 15 (World Scientific, Singapore, 1998), p.~453

\bibitem{Pich:1999uk}
  A.~Pich,
  {\it Int. J. Mod. Phys.} A 15S1 (2000) 157

\bibitem{ST:00} A. Stahl, {\it Physics with Tau Leptons}, Springer Tracts in Modern Physics 160 (Springer, Berlin, 2000)

\bibitem{PI:06} A. Pich, 
   \MP\ A 21 (2006) 5652

\bibitem{Davier:2005xq}
  M.~Davier, A.~Hocker and Z.~Zhang,
  {\it Rev. Mod. Phys.}  78 (2006) 1043

\bibitem{Pich:2007cu}
  A.~Pich, 
  {\it Nucl. Phys. Proc. Suppl.}  169 (2007) 393

\bibitem{Pich:2009zza}
  A.~Pich, I.~Boyko, D.~Dedovich and I.I.~Bigi,
  {\it Int. J. Mod. Phys.} A  24S1 (2009) 715

\bibitem{Pich:2013kg}
  A.~Pich, 
  arXiv:1301.4474 [hep-ph]



\bibitem{TS:71} Y.S. Tsai, \myJournal{\PRD}{4}{1971}{2821 [{\it Err\/}: D 13 (1976) 771]}

\bibitem{HFAG} Heavy Flavor Averaging Group, 
arXiv:1207.1158 [hep-ex];
http://www.slac.stanford.edu/xorg/hfag/

\bibitem{Beringer:1900zz} Particle Data Group,
  \PRD\ 86 (2012) 010001;
  http://pdg.lbl.gov/


\bibitem{Behrends:1955mb}
  R.E.~Behrends, R.J.~Finkelstein and A.~Sirlin,
  {\it Phys. Rev.} 101 (1956) 866

\bibitem{Berman:1958ti}
  S.M.~Berman,
  {\it Phys. Rev.} 112 (1958) 267

\bibitem{Kinoshita:1958ru}
  T.~Kinoshita and A.~Sirlin,
  {\it Phys. Rev.} 113 (1959) 1652

\bibitem{Roos:1971mj}
  M.~Roos and A.~Sirlin,
  {\it Nucl. Phys.} B 29 (1971) 296

\bibitem{Marciano:1988vm}
  W.J.~Marciano and A.~Sirlin,
  {\it Phys. Rev. Lett.} 61 (1988) 1815

\bibitem{Ferroglia:1999tg}
  A.~Ferroglia, G.~Ossola and A.~Sirlin,
  {\it Nucl. Phys.} B 560 (1999) 23

\bibitem{vRS:99}
T. van Ritbergen and R.G. Stuart, {\it Phys. Rev. Lett.} 82 (1999)
488; {\it Nucl. Phys.} B 564 (2000) 343;
%
  {\it Phys. Lett.} B 437 (1998) 201

\bibitem{Steinhauser:1999bx}
  M.~Steinhauser and T.~Seidensticker,
  {\it Phys. Lett.} B 467 (1999) 271

\bibitem{PC:08}
A. Pak and A. Czarnecki, {\it Phys. Rev. Lett.} 100 (2008) 241807

\bibitem{Ferroglia:2013dga}
  A.~Ferroglia, C.~Greub, A.~Sirlin and Z.~Zhang,
  {\it Phys. Rev.} D 88 (2013) 033012

\bibitem{Fael:2013pja}
  M.~Fael, L.~Mercolli and M.~Passera,
  {\it Phys. Rev.} D 88 (2013) 093011


\bibitem{MuLan:10} MuLan Collaboration, {\it Phys. Rev. Lett} 106 (2011) 041803

\bibitem{Belous:2013dba} Belle Collaboration,
  arXiv:1310.8503 [hep-ex]

\bibitem{Aubert:2009qj} BaBar Collaboration,
  {\it Phys. Rev. Lett.}  105 (2010) 051602


\bibitem{Voloshin:2002mv}
  M.B.~Voloshin,
  {\it Phys. Lett.} B 556 (2003) 153

\bibitem{Smith:1993vp}
  B.H.~Smith and M.B.~Voloshin,
  {\it Phys. Lett.} B 324 (1994) 117
   [Erratum-ibid. B 333 (1994) 564]

\bibitem{RuizFemenia:2001fa}
  P.~Ruiz-Femenia and  A.~Pich,
  {\it Phys. Rev.} D 64 (2001) 053001

\bibitem{Asner:2008nq} D.M.~Asner {\it et al.},
  {\it Int. J. Mod. Phys.} A 24 (2009) S1

\bibitem{Mo} X.H. Mo, 
 talk at the 12th International Workshop on Tau Lepton Physics (TAU2012), Nagoya, Japan, September 2012 [http://tau2012.hepl.phys.nagoya-u.ac.jp/]


\bibitem{Marciano:1993sh}
  W.J.~Marciano and  A.~Sirlin,
  {\it Phys. Rev. Lett.} 71 (1993) 3629

\bibitem{Finkemeier:1995gi}
  M.~Finkemeier,
  {\it Phys. Lett.} B 387 (1996) 391

\bibitem{Cirigliano:2007ga}
  V.~Cirigliano and  I.~Rosell,
  {\it JHEP} 0710 (2007) 005;
%
  {\it Phys. Rev. Lett.} 99 (2007) 231801

\bibitem{Czapek:1993kc} G.~Czapek {\it et al.},
  {\it Phys. Rev. Lett.}  70 (1993) 17

\bibitem{Britton:1992pg}  D.I.~Britton {\it et al.},
  {\it Phys. Rev. Lett.}  68 (1992) 3000

\bibitem{Bryman:1985bv} D.A.~Bryman {\it et al.},
  {\it Phys. Rev.} D 33 (1986) 1211;
%
  {\it Phys. Rev. Lett.}  50 (1983) 7

\bibitem{Lazzeroni:2012cx} NA62 Collaboration,
  {\it Phys. Lett.} B 719 (2013) 326,
%
  B 698 (2011) 105

\bibitem{Ambrosino:2009aa} KLOE Collaboration,
  {\it Eur. Phys. J.} C 64 (2009) 627
   [Erratum-ibid. 65 (2010) 703]

\bibitem{Pocanic:2009gd} PEN Collaboration,
  {\it AIP Conf. Proc.} 1182 (2009) 698

%
\bibitem{Malbrunot:2012zz}  PIENU Collaboration,
  {\it AIP Conf. Proc.} 1441 (2012) 564

\bibitem{Cirigliano:2011ny}
  V.~Cirigliano, G.~Ecker, H.~Neufeld, A.~Pich and  J.~Portol\'es,
  {\it Rev. Mod. Phys.} 84 (2012) 399

\bibitem{Antonelli:2010yf} M.~Antonelli {\it et al.},
  {\it Eur. Phys. J.} C 69 (2010) 399

\bibitem{Decker:1994dd}
  R.~Decker and  M.~Finkemeier,
  {\it Phys. Lett.} B 334 (1994) 199;
%
   {\it Nucl. Phys.} B 438 (1995) 17;
   {\it Nucl. Phys. B (Proc. Suppl.)} 40 (1995) 453

\bibitem{Schael:2013ita}
 ALEPH, DELPHI, L3 and OPAL Collaborations and LEP Electroweak Working Group,
  {\it Phys. Rept.} 532 (2013) 119

\bibitem{Filipuzzi:2012mg}
  A.~Filipuzzi, J.~Portol\'es and M.~Gonz\'alez-Alonso,
  {\it Phys. Rev.} D 85 (2012) 116010

\bibitem{Jung:2010ik}
  M.~Jung, A.~Pich and P.~Tuz\'on,
  {\it JHEP} 1011 (2010) 003


\bibitem{MI:50} L. Michel, \myJournal{Proc. Phys. Soc.}{A 63}{1950}{514, 1371}

\bibitem{BM:57} C. Bouchiat and L. Michel, \myJournal{\PREV}{106}{1957}{170}

\bibitem{KS:57} T. Kinoshita and A. Sirlin, \myJournal{\PREV}{107}{1957}{593,
      108 (1957) 844}

\bibitem{SCH:83} F. Scheck, {\it Leptons, Hadrons and Nuclei}
   (North-Holland, Amsterdam, 1983);
   {\it\PREP} 44 (1978) 187
%

\bibitem{FGJ:86} W. Fetscher, H.-J. Gerber and K.F. Johnson, \myJournal{\PL}{B173}
    {1986}{102}

\bibitem{FG:93} W. Fetscher and H.-J. Gerber,
  in {\it Precision Tests of the Standard Electroweak Model}, ed. P.~Langacker,
  Advanced Series on Directions in High Energy Physics -- Vol.~14
  (World Scientific, Singapore, 1995), p.~657

\bibitem{PS:95} A. Pich and J.P. Silva, \myJournal{\PRD}{52}{1995}{4006}

\bibitem{Rouge:2000um}
  A.~Rouge,
  {\it Eur. Phys. J.} C 18 (2001) 491

\bibitem{JA:66} C. Jarlskog, \myJournal{\it Nucl. Phys.}{75}{1966}{659}

\bibitem{RO:82}
   L.Ph. Roesch {\it et al.}, {\it Helv. Phys. Acta} 55 (1982) 74

\bibitem{JO:88}
  A. Jodidio {\it et al.}, \PRD\ 37 (1988) 237, D 34 (1986) 1967

\bibitem{FE:84} W. Fetscher, \PLB\ 140 (1984) 117

\bibitem{Mursula:1982em}
  K.~Mursula, M.~Roos and F.~Scheck,
  {\it Nucl. Phys.} B 219 (1983) 321

\bibitem{TWIST:2011aa}   TWIST Collaboration,
  {\it Phys. Rev.} D 85 (2012) 092013,
%
  D 84 (2011) 032005,
%
  D 78 (2008) 032010;
%
  {\it Phys. Rev. Lett.} 106 (2011) 041804

\bibitem{Gagliardi:2005fg}
  C.A.~Gagliardi, R.E.~Tribble and N.J.~Williams,
  {\it Phys. Rev.} D 72 (2005) 073002

\bibitem{Mishra:1990yf} S.R.~Mishra {\it et al.},
  {\it Phys. Lett.} B 252 (1990) 170

\bibitem{Vilain:1996yf} CHARM-II Collaboration,
  {\it Phys. Lett.} B 364 (1995) 121

\bibitem{FG:12} W. Fetscher and H.-J. Gerber, 
  in \cite{Beringer:1900zz}

\bibitem{KST:73} S. Kawasaki, T. Shirafuji and S.Y. Tsai, {\it Prog. Theor. Phys.} 49
    (1973) 1656

\bibitem{PS:77} S.-Y. Pi and A.I. Sanda, \myJournal{\em Ann. Phys. (N.Y.)}{106}{1977}{171}

\bibitem{KW:84} H.K. K\"uhn and F. Wagner, \myJournal{\NPB}{236}{1984}{16}

\bibitem{NE:91} C.A. Nelson, \myJournal{\PRD}{43}{1991}{1465, D 40 (1989) 123
    [{\it Err\/}: D 41 (1990) 2327]; \PRL\ 62 (1989) 1347}

\bibitem{GN:91} S. Goozovat and C.A. Nelson, \myJournal{\PRD}{44}{1991}{2818}

\bibitem{FE:90} W. Fetscher, \myJournal{\PRD}{42}{1990}{1544}

\bibitem{Bernabeu:1990na}
  J.~Bernab\'eu, N.~Rius and A.~Pich,
  {\it Phys. Lett.} B 257 (1991) 219

\bibitem{Alemany:1991ki}
  R.~Alemany, N.~Rius, J.~Bernab\'eu, J.J.~G\'omez-Cadenas and A.~Pich,
  {\it Nucl. Phys.} B 379 (1992) 3


\bibitem{Davier:1992nw}
  M.~Davier, L.~Duflot, F.~Le Diberder and A.~Rouge,
  {\it Phys. Lett.} B 306 (1993) 411


\bibitem{Abe:1997dy} SLD Collaboration,
  {\it Phys. Rev. Lett.} 78 (1997) 4691

\bibitem{SV:96} A. Stahl and H. Voss, \myJournal{\ZPC}{74}{1997}{73}

\bibitem{Heister:2001me} ALEPH Collaboration,
  {\it Eur. Phys. J.} C 22 (2001) 217

\bibitem{Abreu:2000sg} DELPHI Collaboration,
  {\it Eur. Phys. J.} C 16 (2000) 229

\bibitem{Ackerstaff:1998yk} OPAL Collaboration,
  {\it Eur. Phys. J.} C 8 (1999) 3

\bibitem{Acciarri:1998as} L3 Collaboration,
  {\it Phys. Lett.} B 438 (1998) 405,
%
  B 377 (1996) 313



\bibitem{Albrecht:1997gn} ARGUS Collaboration,
  {\it Phys. Lett.} B 431 (1998) 179,
%
   B 349 (1995) 576,
%
  B 341 (1995) 441,
%
  B 316 (1993) 608,
%
  B 246 (1990) 278

\bibitem{Alexander:1997bv} CLEO Collaboration,
  {\it Phys. Rev.} D 56 (1997) 5320,
%
  D 32 (1985) 2468;
%
  {\it Phys. Rev. Lett.} 78 (1997) 4686

\bibitem{Janssen:1989wg} Crystal Ball Collaboration,
  {\it Phys. Lett.} B 228 (1989) 273

\bibitem{Ford:1987ha} W.T.~Ford {\it et al.},
  {\it Phys. Rev.} D 36 (1987) 1971

\bibitem{Danneberg:2005xv} N.~Danneberg {\it et al.},
  {\it Phys. Rev. Lett.} 94 (2005) 021802

\bibitem{Balke:1988by} B.~Balke {\it et al.},
  {\it Phys. Rev.} D 37 (1988) 587

\bibitem{Beltrami:1987ne} I.~Beltrami {\it et al.},
  {\it Phys. Lett.} B 194 (1987) 326

\bibitem{Burkard:1985wn} H.~Burkard {\it et al.},
  {\it Phys. Lett.} B 160 (1985) 343,
%
  B 150 (1985) 242,
%
  B 129 (1983) 260

\bibitem{Derenzo:1969za}
  S.E.~Derenzo,
  {\it Phys. Rev.} 181 (1969) 1854

\bibitem{Stahl:12} A. Stahl, 
  in \cite{Beringer:1900zz}

\bibitem{Stahl:1993yk}
  A.~Stahl,
  {\it Phys. Lett.} B 324 (1994) 121

\bibitem{Mursula:1984zb}
  K.~Mursula and F.~Scheck,
  {\it Nucl. Phys.} B 253 (1985) 189

\bibitem{Thurn:1993qw}
  H.~Thurn and H.~Kolanoski,
  {\it Z. Phys.} C 60 (1993) 277

\bibitem{Asner:1999kj} CLEO Collaboration,
  {\it Phys. Rev.} D 61 (2000) 012002,
%
  D 55 (1997) 7291

\bibitem{Ackerstaff:1997dv} OPAL Collaboration,
  {\it Z. Phys.} C 75 (1997) 593
%
  {\it Z. Phys.} C 67 (1995) 45


\bibitem{Albrecht:1992ka} ARGUS Collaboration,
  {\it Z. Phys.} C 58 (1993) 61;
%
  {\it Phys. Lett.} B 250 (1990) 164




\bibitem{Pich:2012sx}
  A.~Pich,
  arXiv:1201.0537 [hep-ph]

\bibitem{ALEPH:2005ab}
  ALEPH, DELPHI, L3, OPAL, SLD, LEP Electroweak Working Group, SLD Electroweak Group and SLD Heavy Flavour Group Collaborations,
  {\it Phys. Rept.} 427 (2006) 257;
http://www.cern.ch/LEPEWWG/

\bibitem{Barate:1997mz} ALEPH Collaboration,
  {\it Phys. Lett.} B 405 (1997) 191

\bibitem{Abreu:1997vp} DELPHI Collaboration,
  {\it Phys. Lett.} B 404 (1997) 194

\bibitem{ALEPH:2010aa}
  ALEPH, CDF, D0, DELPHI, L3, OPAL, SLD, LEP Electroweak Working Group, Tevatron Electroweak Working Group and SLD Electroweak and Heavy Flavour Groups Collaborations,
  arXiv:1012.2367 [hep-ex]

\bibitem{Aad:2012tfa} ATLAS Collaboration,
  {\it Phys. Lett.} B 716 (2012) 1,
%
  B 726 (2013) 88

\bibitem{Chatrchyan:2012ufa} CMS Collaboration,
  {\it Phys. Lett.} B 716 (2012) 30;
%
 {\it JHEP} 1306 (2013) 081;
%
   CMS-PAS-HIG-13-005 (2013)


\bibitem{Buskulic:1994hi} ALEPH Collaboration,
  {\it Phys. Lett.} B 346 (1995) 379
   [Erratum-ibid.\ B 363 (1995) 265],
%
B 321 (1994) 168

\bibitem{Vilain:1993xb} CHARM-II Collaboration,
  {\it Phys. Lett.} B 320 (1994) 203

\bibitem{Kodama:2000mp} DONUT Collaboration,
  {\it Phys. Lett.} B 504 (2001) 218;
%
  {\it Phys. Rev.} D 78 (2008) 052002


\bibitem{PI:89} A. Pich, 
 in Proc. Tau-Charm Factory Workshop (SLAC, 1989), ed. L.V. Beers,
 SLAC-Report-343 (1989) p.~416

\bibitem{ThS:71} H.B. Thacker and J.J. Sakurai, \myJournal{\PLB}{36}{1971}{103}

\bibitem{Braaten:1990ef}
  E.~Braaten and C.-S.~Li,
  {\it Phys. Rev.} D 42 (1990) 3888

\bibitem{Erler:2002mv}
  J.~Erler,
  {\it Rev. Mex. Fis.}  50 (2004) 200

\bibitem{Gilman:1977cc}
  F.J.~Gilman and D.H.~Miller,
  {\it Phys. Rev.} D 17 (1978) 1846

\bibitem{Gilman:1984ry}
  F.J.~Gilman and S.H.~Rhie,
  {\it Phys. Rev.} D 31 (1985) 1066

\bibitem{Gilman:1987my}
  F.J.~Gilman,
  {\it Phys. Rev.} D 35 (1987) 3541

\bibitem{Kuhn:1990ad}
  J.H.~K\"uhn and A.~Santamaria,
  {\it Z. Phys.} C 48 (1990) 445

\bibitem{Eidelman:1990pb}
  S.I.~Eidelman and V.N.~Ivanchenko,
  {\it Phys. Lett.} B 257 (1991) 437

\bibitem{Narison:1993sx}
  S.~Narison and A.~Pich,
  {\it Phys. Lett.} B 304 (1993) 359

\bibitem{Kuhn:1992nz}
  J.H.~K\"uhn and E.~Mirkes,
  {\it Z. Phys.} C 56 (1992) 661
   [Erratum-ibid.\ C 67 (1995) 364];
%
  {\it Phys. Lett.} B 286 (1992) 381


\bibitem{WE:79} S. Weinberg, \myJournal{Physica}{A 96}{1979}{327}

\bibitem{Gasser:1984gg}
  J.~Gasser and H.~Leutwyler,
  {\it Nucl. Phys.} B 250 (1985) 465,
%
  517,
%
  539

\bibitem{Ecker:1994gg}
  G.~Ecker,
  {\it Prog. Part. Nucl. Phys.} 35 (1995) 1

\bibitem{Pich:1995bw}
  A.~Pich,
  {\it Rept. Prog. Phys.} 58 (1995) 563

\bibitem{Bernard:2006gx}
  V.~Bernard and U.-G.~Meissner,
  {\it Ann. Rev. Nucl. Part. Sci.} 57 (2007) 33

\bibitem{WZ:71} J. Wess and B. Zumino, \myJournal{\PLB}{37}{1971}{95}

\bibitem{WI:83} E. Witten, \myJournal{\NPB}{223}{1983}{422}

\bibitem{Colangelo:1996hs}
  G.~Colangelo, M.~Finkemeier and R.~Urech,
  {\it Phys. Rev.} D 54 (1996) 4403

\bibitem{Ecker:2002cw}
  G.~Ecker and R.~Unterdorfer,
  {\it Eur. Phys. J.} C 24 (2002) 535

\bibitem{Unterdorfer:2002zg}
  R.~Unterdorfer,
  {\it JHEP} 0207 (2002) 053

\bibitem{Davoudiasl:1995ed}
  H.~Davoudiasl and M.B.~Wise,
  {\it Phys. Rev.} D 53 (1996) 2523

\bibitem{Fischer:1979fh}
  R.~Fischer, J.~Wess and F.~Wagner,
  {\it Z. Phys.} C 3 (1979) 313

\bibitem{Kramer:1984pm}
  G.~Kramer and W.F.~Palmer,
  {\it Z. Phys.} C  25 (1984) 195,
%
  39 (1988) 423

\bibitem{Pich:1987qq}
  A.~Pich,
  {\it Phys. Lett.} B 196 (1987) 561

\bibitem{Braaten:1987jh}
  E.~Braaten, R.J.~Oakes and S.-M.~Tse,
  {\it Phys. Rev.} D 36 (1987) 2188;
%
  {\it Int. J. Mod. Phys.} A 5 (1990) 2737

\bibitem{GomezCadenas:1990uj}
  J.J.~G\'omez-Cadenas, M.C.~Gonz\'alez-Garc\'{\i}a and A.~Pich,
  {\it Phys. Rev.} D 42 (1990) 3093

\bibitem{Jadach:1990mz} S.~Jadach {\it et. al.},
  {\it Comput. Phys. Commun.} 64 (1991) 275,
%
  76 (1993) 361

\bibitem{Decker:1992kj}
  R.~Decker, E.~Mirkes, R.~Sauer and Z.~Was,
  {\it Z. Phys.} C 58 (1993) 445

\bibitem{Decker:1992rj}
  R.~Decker and E.~Mirkes,
  {\it Phys. Rev.} D 47 (1993) 4012;
%
  {\it Z. Phys.} C 57 (1993) 495

\bibitem{Decker:1993ay}
  R.~Decker, M.~Finkemeier and E.~Mirkes,
  {\it Phys. Rev.} D 50 (1994) 6863

\bibitem{Decker:1994af}
  R.~Decker, M.~Finkemeier, P.~Heiliger and H.H.~Jonsson,
  {\it Z. Phys.} C 70 (1996) 247

\bibitem{Finkemeier:1995sr}
  M.~Finkemeier and E.~Mirkes,
  {\it Z. Phys.} C 69 (1996) 243,
%
  C 72 (1996) 619

\bibitem{Portoles:2007cx}
  J.~Portol\'es,
  {\it Nucl. Phys. Proc. Suppl.}  169 (2007) 3,
%
   144 (2005) 3,
%
   98 (2001) 210

\bibitem{Ecker:1988te}
  G.~Ecker, J.~Gasser, A.~Pich and E.~de Rafael,
  {\it Nucl. Phys.} B 321 (1989) 311

\bibitem{Ecker:1989yg}
  G.~Ecker, J.~Gasser, H.~Leutwyler, A.~Pich and E.~de Rafael,
  {\it Phys. Lett.} B 223 (1989) 425

\bibitem{Pich:2002xy}
  A.~Pich,
  in {\it Phenomenology of Large $N_C$ QCD}, ed. R. Lebed, Proc. Inst for Nuclear Theory -- Vol. 12 (World Scientific, Singapore, 2002), p.239

\bibitem{RuizFemenia:2003hm}
  P.D.~Ruiz-Femen\'{\i}a, A.~Pich and J.~Portol\'es,
  {\it JHEP} 0307 (2003) 003

\bibitem{Cirigliano:2006hb}
  V.~Cirigliano {\it et al.}, 
  {\it Nucl. Phys.} B 753 (2006) 139;
%
  {\it JHEP} 0504 (2005) 006;
%
  {\it Phys. Lett.} B 596 (2004) 96

\bibitem{Pich:2010sm}
  A.~Pich, I.~Rosell and J.J.~Sanz-Cillero,
  {\it JHEP} 1102 (2011) 109,
%
  0408 (2004) 042

\bibitem{Pich:2008jm}
  A.~Pich, I.~Rosell and J.J.~Sanz-Cillero,
  {\it JHEP} 0807 (2008) 014,
%
   0701 (2007) 039

\bibitem{Rosell:2005ai}
  I.~Rosell, P.~Ruiz-Femen\'{\i}a and J.~Portol\'es,
  {\it JHEP} 0512 (2005) 020

\bibitem{Kampf:2011ty}
  K.~Kampf and J.~Novotny,
  {\it Phys. Rev.} D 84 (2011) 014036


\bibitem{Cirigliano:2011tm}
  V.~Cirigliano and H.~Neufeld,
  {\it Phys. Lett.} B 700 (2011) 7

\bibitem{Marciano:2004uf}
  W.~J.~Marciano,
  {\it Phys. Rev.\ Lett.} 93 (2004) 231803

\bibitem{Colangelo:2010et} G.~Colangelo {\it et al.},
  {\it Eur. Phys. J.} C 71 (2011) 1695

\bibitem{Towner:2010zz}
  I.S.~Towner and J.C.~Hardy,
  {\it Rept. Prog. Phys.} 73 (2010) 046301


\bibitem{Cirigliano:2002pv}
  V.~Cirigliano, G.~Ecker and H.~Neufeld,
  {\it JHEP} 0208 (2002) 002;
%
  {\it Phys. Lett.} B 513 (2001) 361

\bibitem{Davier:2009ag} M.~Davier {\it et al.},
  {\it Eur. Phys. J.} C 66 (2010) 127

\bibitem{Antonelli:2013usa}
  M.~Antonelli, V.~Cirigliano, A.~Lusiani and E.~Passemar,
  {\it JHEP} 1310 (2013) 070

\bibitem{Flores-Baez:2013eba}
  F.V.~Flores-B\'aez, J.R.~Morones-Ibarra, G.C.~Quiroz and D.R.~Balbuena,
  {\it Phys. Rev.} D 88 (2013) 073009

\bibitem{Guerrero:1997ku}
  F.~Guerrero and A.~Pich,
  {\it Phys. Lett.} B 412 (1997) 382

\bibitem{GomezDumm:2000fz}
  D. G\'omez Dumm, A.~Pich and J.~Portol\'es,
  {\it Phys. Rev.} D 62 (2000) 054014

\bibitem{Pich:2001pj}
  A.~Pich and J.~Portol\'es,
  {\it Phys. Rev.} D 63 (2001) 093005;
%
  {\it Nucl. Phys. Proc. Suppl.} 121 (2003) 179

\bibitem{SanzCillero:2002bs}
  J.J.~Sanz-Cillero and A.~Pich,
  {\it Eur. Phys. J.} C 27 (2003) 587

\bibitem{Hooft:74} G. 't Hooft, {\it Nucl. Phys.} B 72 (1974) 461, 75 (1974) 461

\bibitem{Witten:79} E. Witten, {\it Nucl. Phys.} B 160 (1979) 57

\bibitem{Dominguez:2001zu}
  C.A.~Dom\'{\i}nguez,
  {\it Phys. Lett.} B 512 (2001) 331

\bibitem{Bruch:2004py}
  C.~Bruch, A.~Khodjamirian and J.H.~K\"uhn,
  {\it Eur. Phys. J.} C 39 (2005) 41

\bibitem{Omnes:1958hv}
  R.~Omn\`es,
  {\it Nuovo Cim.} 8 (1958) 316

\bibitem{Barate:1997hv} ALEPH Collaboration,
  {\it Z. Phys.} C 76 (1997) 15

\bibitem{Anderson:1999ui} CLEO Collaboration,
  {\it Phys. Rev.} D 61 (2000) 112002

\bibitem{Barkov:1985ac} L.M.~Barkov {\it et al.},
  {\it Nucl. Phys.} B 256 (1985) 365

\bibitem{Amendolia:1986wj}
  S.R.~Amendolia {\it et al.}, 
  {\it Nucl. Phys.} B 277 (1986) 168

\bibitem{Fujikawa:2008ma} Belle Collaboration,
  {\it Phys. Rev.} D 78 (2008) 072006

\bibitem{Dumm:2013zh}
  D. G\'omez Dumm and P.~Roig,
 {\it Eur. Phys. J.} C 73 (2013) 2528

\bibitem{Lees:2012cj} BaBar Collaboration,
  {\it Phys. Rev.} D 86 (2012) 032013;
%
  {\it Phys. Rev. Lett.} 103 (2009) 231801

\bibitem{Akhmetshin:2006bx} CMD-2 Collaboration,
  {\it Phys. Lett.} B 648 (2007) 28;
%
  {\it JETP Lett.} 84 (2006) 413,
%
  82 (2005) 743

\bibitem{Bisello:1988hq} DM2 Collaboration,
  {\it Phys. Lett.} B 220 (1989) 321

\bibitem{Babusci:2012rp} KLOE Collaboration,
  {\it Phys.\ Lett.} B 720 (2013) 336,
%
  B 700 (2011) 102,
%
  B 670 (2009) 285,
%
  B 606 (2005) 12

\bibitem{Achasov:2006vp} SND Collaboration,
  {\it JETP} 103 (2006) 380,
%
  101 (2005) 1053

\bibitem{Schael:2005am} ALEPH Collaboration,
  {\it Phys. Rept.} 421 (2005) 191


\bibitem{Binner:1999bt}
  S.~Binner, J.H.~K\"uhn and K.~Melnikov,
  {\it Phys. Lett.} B 459 (1999) 279

\bibitem{Rodrigo:2001kf} G.~Rodrigo {\it et al.},
  {\it Eur. Phys. J.} C 24 (2002) 71,
%
  C 27 (2003) 563,
%
  C 33 (2004) 333,
%
  C 39 (2005) 411,
%
  C 47 (2006) 617

\bibitem{Actis:2010gg}
  S.~Actis {\it et al.},
  {\it Eur. Phys. J.} C 66 (2010) 585

\bibitem{Druzhinin:2011qd}
  V.P.~Druzhinin, S.I.~Eidelman, S.I.~Serednyakov and E.P.~Solodov,
  {\it Rev. Mod. Phys.} 83 (2011) 1545

\bibitem{Davier:2010nc}
  M.~Davier, A.~Hoecker, B.~Malaescu and Z.~Zhang,
  {\it Eur. Phys. J.} C 71 (2011) 1515
   [Erratum-ibid. C 72 (2012) 1874]


\bibitem{Jamin:2008qg}
  M.~Jamin, A.~Pich and J.~Portol\'es,
  {\it Phys. Lett.} B 664 (2008) 78,
%
  B 640 (2006) 176

\bibitem{Boito:2010me}
  D.R.~Boito, R.~Escribano and M.~Jamin,
  {\it JHEP} 1009 (2010) 031;
%
  {\it Eur. Phys. J.} C 59 (2009) 821

\bibitem{Moussallam:2007qc}
  B.~Moussallam,
  {\it Eur. Phys. J.} C 53 (2008) 401


\bibitem{Jamin:2001zq}
  M.~Jamin, J.A.~Oller and A.~Pich,
  {\it Nucl.\ Phys.} B 622 (2002) 279,
%
  B 587 (2000) 331

\bibitem{Jamin:2006tj}
  M.~Jamin, J.A.~Oller and A.~Pich,
  {\it Phys. Rev.} D 74 (2006) 074009;
%
  {\it JHEP} 0402 (2004) 047;
%
  {\it Eur. Phys. J} C 24 (2002) 237

\bibitem{Buettiker:2003pp}
  P.~Buettiker, S.~Descotes-Genon and B.~Moussallam,
  {\it Eur. Phys. J.} C 33 (2004) 409

\bibitem{Epifanov:2007rf} Belle Collaboration,
  {\it Phys. Lett.} B 654 (2007) 65

\bibitem{Adametz:2011zz}
  A.~Adametz [BaBar Collaboration],
  {\it Nucl. Phys. Proc. Suppl.} 218 (2011) 134

\bibitem{Escribano:2013bca}
  R.~Escribano, S.~Gonz\'alez-Sol\'{\i}s and P.~Roig,
  {\it JHEP} 1310 (2013) 039

\bibitem{delAmoSanchez:2010pc} BaBar Collaboration,
  {\it Phys. Rev.} D 83 (2011) 032002

\bibitem{Inami:2008ar} Belle Collaboration,
  {\it Phys. Lett.} B 672 (2009) 209

\bibitem{Lees:2012ks} BaBar Collaboration,
  {\it Phys. Rev.} D 86 (2012) 092010,
%
  D 77 (2008) 112002



\bibitem{Bramon:1987zb}
  A.~Bramon, S.~Narison and A.~Pich,
  {\it Phys. Lett.} B 196 (1987) 543

\bibitem{Tisserant:1982fc}
  S.~Tisserant and T.N.~Truong,
  {\it Phys. Lett.} B 115 (1982) 264

\bibitem{Neufeld:1994eg}
  H.~Neufeld and H.~Rupertsberger,
  {\it Z. Phys.} C 68 (1995) 91

\bibitem{Descotes-Genon:2013uya}
  S.~Descotes-Genon, E.~Kou and B.~Moussallam,
  arXiv:1303.2879 [hep-ph]

\bibitem{Nussinov:2008gx}
  S.~Nussinov and A.~Soffer,
  {\it Phys. Rev.} D 78 (2008) 033006

\bibitem{Paver:2010mz}
  N.~Paver and Riazuddin,
  {\it Phys. Rev.} D 82 (2010) 057301


\bibitem{Shekhovtsova:2012ra} O.~Shekhovtsova {\it et al.},
  {\it Phys. Rev.} D 86 (2012) 113008;
%
  arXiv:1301.1964 [hep-ph]

\bibitem{Dumm:2009va}
  D. G\'omez Dumm, P.~Roig, A.~Pich and J.~Portol\'es,
  {\it Phys. Lett.} B 685 (2010) 158

\bibitem{GomezDumm:2003ku}
  D. G\'omez Dumm, A.~Pich and J.~Portol\'es,
  {\it Phys. Rev.} D 69 (2004) 073002

\bibitem{Barate:1998uf} ALEPH Collaboration,
  {\it Eur. Phys. J.} C 4 (1998) 409

\bibitem{Buskulic:1993sv} ALEPH Collaboration,
  {\it Phys. Lett.} B 307 (1993) 209

\bibitem{Browder:1999fr} CLEO Collaboration,
  {\it Phys. Rev.} D 61 (2000) 052004,
%
  012002



\bibitem{Kuhn:1982di}
  H.-K.~K\"uhn and F.~Wagner,
  {\it Nucl. Phys.} B 236 (1984) 16

\bibitem{Feindt:1990ev}
  M.~Feindt,
  {\it Z. Phys.} C 48 (1990) 681

\bibitem{Dumm:2009kj}
  D. G\'omez Dumm, P.~Roig, A.~Pich and J.~Portol\'es,
  {\it Phys. Rev.} D 81 (2010) 034031

\bibitem{Roig:2013ts}
  P.~Roig, 
  arXiv:1301.7626 [hep-ph]


\bibitem{Dumm:2012vb}
  D. G\'omez Dumm and P.~Roig,
  {\it Phys. Rev.} D 86 (2012) 076009

\bibitem{Finkemeier:1996hh}
  M.~Finkemeier, J.H.~K\"uhn and E.~Mirkes,
  {\it Nucl. Phys. Proc. Suppl.} 55C (1997) 169

\bibitem{Davier:2008sk}
  M.~Davier {\it et al.}, 
  {\it Eur. Phys. J.} C 56 (2008) 305

\bibitem{Dai:2013joa}
  L.Y.~Dai, J.~Portol\'es and O.~Shekhovtsova,
 {\it Phys. Rev.} D 88 (2013) 056001

\bibitem{Guo:2010dv}
  Z.-H.~Guo and P.~Roig,
  {\it Phys. Rev.} D 82 (2010) 113016

\bibitem{Guevara:2013wwa}
  A.~Guevara, G.~L\'opez~Castro and P.~Roig,
  {\it Phys. Rev.} D 88 (2013) 033007



\bibitem{Lees:2012de} BaBar Collaboration,
  {\it Phys. Rev.} D 86 (2012) 092013

\bibitem{Ryu:2013lca}
  S.~Ryu [Belle Collaboration],
  arXiv:1302.4565 [hep-ex]

\bibitem{Nugent:2013ij}
  I.~M.~Nugent [BaBar Collaboration],
  arXiv:1301.7105 [hep-ex]



\bibitem{Narison:1988ni}
  S.~Narison and A.~Pich,
  {\it Phys. Lett.} B 211 (1988) 183

\bibitem{Braaten:1988hc}
  E.~Braaten,
  {\it Phys. Rev. Lett.} 60 (1988) 1606;
%
  {\it Phys. Rev.} D 39 (1989) 1458

\bibitem{Braaten:1991qm}
  E.~Braaten, S.~Narison and A.~Pich,
  {\it Nucl. Phys.} B 373 (1992) 581

\bibitem{Shifman:1978bx}
  M.A.~Shifman, A.I.~Vainshtein and V.I.~Zakharov,
  {\it Nucl. Phys.} B 147 (1979) 385,
%
  448,
%
   519

\bibitem{Baikov:2008jh}
  P.A.~Baikov, K.G.~Chetyrkin and J.H.~K\"uhn,
  {\it Phys. Rev. Lett.} 101 (2008) 012002


\bibitem{Beneke:2008ad}
  M.~Beneke and M.~Jamin,
  {\it JHEP} 0809 (2008) 044

\bibitem{Beneke:2012vb}
  M.~Beneke, D.~Boito and M.~Jamin,
  {\it JHEP} 1301 (2013) 125

\bibitem{Caprini:2011ya}
  I.~Caprini and J.~Fischer,
  {\it Phys. Rev.} D 84 (2011) 054019;
%
  {\it Eur. Phys. J.} C 64 (2009) 35

\bibitem{Abbas:2012fi}
  G.~Abbas, B.~Ananthanarayan, I.~Caprini and J.~Fischer,
  {\it Phys. Rev.} D 87 (2013) 014008,
  D 88 (2013) 034026

\bibitem{Abbas:2012py}
  G.~Abbas, B.~Ananthanarayan and I.~Caprini,
  {\it Phys. Rev.} D 85 (2012) 094018

\bibitem{Groote:2012jq}
  S.~Groote, J.G.~Korner and A.A.~Pivovarov,
  {\it Phys. Part. Nucl.} 44 (2013) 285

\bibitem{Maltman:2008nf}
  K.~Maltman and T.~Yavin,
  {\it Phys. Rev.} D 78 (2008) 094020

\bibitem{Boito:2012cr}
  D.~Boito {\it et al.},
  {\it Phys. Rev.} D 85 (2012) 093015,
%
  D 84 (2011) 113006

\bibitem{Menke:2009vg}
  S.~Menke,
  arXiv:0904.1796 [hep-ph]

\bibitem{Narison:2009vy}
  S.~Narison,
  {\it Phys. Lett.} B 673 (2009) 30

\bibitem{Cvetic:2010ut}
  G.~Cvetic, M.~Loewe, C.~Mart\'{\i}nez and C.~Valenzuela,
  {\it Phys. Rev.} D 82 (2010) 093007

\bibitem{Pich:2011bb}
  A.~Pich,
  arXiv:1107.1123 [hep-ph];
  {\it Nucl. Phys. Proc. Suppl.} 218 (2011) 89,
%
   39BC (1995) 326;
%
  {\it Acta Phys. Polon. Supp.} 3 (2010) 165

\bibitem{Pich:2013sqa}
  A.~Pich,
  {\it PoS ConfinementX} (2012) 022

\bibitem{Adler:1974gd}
  S.L.~Adler,
  {\it Phys. Rev.} D 10 (1974) 3714

\bibitem{Gorishnii:1990vf}
  S.G.~Gorishnii, A.L.~Kataev and S.A.~Larin,
  {\it Phys. Lett.} B 259 (1991) 144

\bibitem{Surguladze:1990tg}
  L.R.~Surguladze and M.A.~Samuel,
  {\it Phys. Rev. Lett.} 66 (1991) 560
   [Erratum-ibid.  66 (1991) 2416]

\bibitem{LeDiberder:1992te}
  F.~Le Diberder and A.~Pich,
  {\it Phys. Lett.} B 286 (1992) 147

\bibitem{Pivovarov:1991rh}
  A.A.~Pivovarov,
  {\it Z. Phys.} C 53 (1992) 461

\bibitem{Kataev:1995vh}
  A.L.~Kataev and V.V.~Starshenko,
  {\it Mod. Phys. Lett.} A 10 (1995) 235

\bibitem{Cvetic:2001ws}
  G.~Cvetic, C.~Dib, T.~Lee and I.~Schmidt,
  {\it Phys. Rev.} D 64 (2001) 093016

\bibitem{Ball:1995ni}
  P.~Ball, M.~Beneke and V.M.~Braun,
  {\it Nucl. Phys.} B 452 (1995) 563

\bibitem{Neubert:1995gd}
  M.~Neubert,
  {\it Nucl. Phys.} B 463 (1996) 511

\bibitem{Altarelli:1994vz}
  G.~Altarelli, P.~Nason and G.~Ridolfi,
  {\it Z. Phys.} C 68 (1995) 257

\bibitem{Jamin:2005ip}
  M.~Jamin,
  {\it JHEP} 0509 (2005) 058

\bibitem{DescotesGenon:2010cr}
  S.~Descotes-Genon and B.~Malaescu,
  arXiv:1002.2968 [hep-ph]

\bibitem{Maxwell:2001he}
  C.J.~Maxwell and A.~Mirjalili,
  {\it Nucl. Phys.} B 611 (2001) 423

\bibitem{Maxwell:1997yw}
  C.J.~Maxwell and D.G.~Tonge,
  {\it Nucl. Phys.} B 535 (1998) 19,
%
   B 481 (1996) 681

\bibitem{LovettTurner:1995ti}
  C.N.~Lovett-Turner and C.J.~Maxwell,
  {\it Nucl. Phys.} B 452 (1995) 188,
%
  B 432 (1994) 147

\bibitem{Maxwell:1989py}
  C.J.~Maxwell and J.A.~Nicholls,
  {\it Phys. Lett.} B 236 (1990) 63

\bibitem{Beneke:1998ui}
  M.~Beneke,
  {\it Phys. Rept.} 317 (1999) 1

\bibitem{Chetyrkin:1993hi}
  K.G.~Chetyrkin and A.~Kwiatkowski,
  {\it Z. Phys.} C 59 (1993) 525

\bibitem{Pich:1999hc}
  A.~Pich and J.~Prades,
  {\it JHEP} 9910 (1999) 004,
%
  9806 (1998) 013

\bibitem{Baikov:2004tk}
  P.A.~Baikov, K.G.~Chetyrkin and J.H.~K\"uhn,
  {\it Phys. Rev. Lett.} 95 (2005) 012003


\bibitem{LeDiberder:1992fr}
  F.~Le Diberder and A.~Pich,
  {\it Phys. Lett.} B 289 (1992) 165

\bibitem{Ackerstaff:1998yj} OPAL Collaboration,
  {\it Eur. Phys. J.} C 7 (1999) 571

\bibitem{Coan:1995nk} CLEO Collaboration,
  {\it Phys. Lett.} B 356 (1995) 580

\bibitem{Girone:1995xb}
  M.~Girone and M.~Neubert,
  {\it Phys. Rev. Lett.} 76 (1996) 3061

\bibitem{Nason:1993ak}
  P.~Nason and M.~Porrati,
  {\it Nucl. Phys.} B 421 (1994) 518

\bibitem{Balitsky:1993ki}
  I.I.~Balitsky, M.~Beneke and V.M.~Braun,
  {\it Phys. Lett.} B 318 (1993) 371

\bibitem{Shifman:2000jv}
  M.A.~Shifman,
  {\it Quark hadron duality},
  in {\it At the frontier of particle physics: Handbook of QCD}, ed. M.A. Shifman, Vol. 3 (World Scientific, Singapore, 2001) p.~1447


\bibitem{Cata:2008ru}
  O.~Cata, M.~Golterman and S.~Peris,
  {\it Phys. Rev.} D 79 (2009) 053002,
  D 77 (2008) 093006;
%
   {\it JHEP} 0508 (2005) 076

\bibitem{Jamin:2011vd}
  M.~Jamin,
  {\it JHEP} 1109 (2011) 141

\bibitem{Rodrigo:1997zd}
  G.~Rodrigo, A.~Pich and A.~Santamaria,
  {\it Phys. Lett.} B 424 (1998) 367

\bibitem{vanRitbergen:1997va}
  T.~van Ritbergen, J.A.~M.~Vermaseren and S.A.~Larin,
  {\it Phys. Lett.} B 400 (1997) 379

\bibitem{Czakon:2004bu}
  M.~Czakon,
  {\it Nucl. Phys.} B 710 (2005) 485

\bibitem{Schroder:2005hy}
  Y.~Schroder and M.~Steinhauser,
  {\it JHEP} 0601 (2006) 051

\bibitem{Chetyrkin:2005ia}
  K.G.~Chetyrkin, J.H.~K\"uhn and C.~Sturm,
  {\it Nucl. Phys.} B 744 (2006) 121


\bibitem{Amoros:1999dp}
  G.~Amoros, J.~Bijnens and P.~Talavera,
  {\it Nucl. Phys.} B 568 (2000) 319

\bibitem{Weinberg:1967kj}
  S.~Weinberg,
  {\it Phys. Rev. Lett.} 18 (1967) 507

\bibitem{Davier:1998dz}
  M.~Davier, L.~Girlanda, A.~Hoecker and J.~Stern,
  {\it Phys. Rev.} D 58 (1998) 096014

\bibitem{GonzalezAlonso:2010xf}
  M.~Gonz\'alez-Alonso, A.~Pich and J.~Prades,
  {\it Phys. Rev.} D 82 (2010) 014019,
%
   D 81 (2010) 074007,
%
   D 78 (2008) 116012

\bibitem{Das:1967it}
  T.~Das, G.S.~Guralnik, V.S.~Mathur, F.E.~Low and J.E.~Young,
  {\it Phys. Rev. Lett.} 18 (1967) 759

\bibitem{Boito:2012nt}
  D.~Boito, M.~Golterman, M.~Jamin, K.~Maltman and S.~Peris,
 {\it Phys. Rev.} D 87 (2013) 094008

\bibitem{Bijnens:2001ps}
  J.~Bijnens, E.~G\'amiz and J.~Prades,
  {\it JHEP} 0110 (2001) 009

\bibitem{Zyablyuk:2004iu}
  K.N.~Zyablyuk,
  {\it Eur. Phys. J.} C 38 (2004) 215

\bibitem{Rojo:2004iq}
  J.~Rojo and J.I.~Latorre,
  {\it JHEP} 0401 (2004) 055

\bibitem{Narison:2004vz}
  S.~Narison,
  {\it Phys. Lett.} B 624 (2005) 223

\bibitem{Peris:2000tw}
  S.~Peris, B.~Phily and E.~de Rafael,
  {\it Phys. Rev. Lett.} 86 (2001) 14

\bibitem{Friot:2004ba}
  S.~Friot, D.~Greynat and E.~de Rafael,
  {\it JHEP} 0410 (2004) 043

\bibitem{Almasy:2008xu}
  A.A.~Almasy, K.~Schilcher and H.~Spiesberger,
  {\it Eur. Phys. J.} C 55 (2008) 237;
%
   {\it Phys. Lett.} B 650 (2007) 179

\bibitem{Masjuan:2007ay}
  P.~Masjuan and S.~Peris,
  {\it JHEP} 0705 (2007) 040

\bibitem{Cirigliano:2002jy}
  V.~Cirigliano, J.F.~Donoghue, E.~Golowich and K.~Maltman,
  {\it Phys. Lett.} B 555 (2003) 71

\bibitem{Cirigliano:2003kc}
  V.~Cirigliano, E.~Golowich and K.~Maltman,
  {\it Phys. Rev.} D 68 (2003) 054013

\bibitem{Dominguez:2003dr}
  C.A.~Dom\'{\i}nguez and K.~Schilcher,
  {\it Phys. Lett.} B 581 (2004) 193

\bibitem{Bordes:2005wv}
  J.~Bordes, C.A.~Dom\'{\i}nguez, J.~Pe\~narrocha and K.~Schilcher,
  {\it JHEP} 0602 (2006) 037

\bibitem{Shintani:2008qe}
  E.~Shintani {\it et al.}, 
  {\it Phys. Rev. Lett.} 101 (2008) 242001

\bibitem{Boyle:2009xi}
  P.A.~Boyle {\it et al.}, 
  {\it Phys. Rev.} D 81 (2010) 014504;
%
  arXiv:1311.0397 [hep-ph]


\bibitem{Gamiz:2004ar} E.~G\'amiz {\it et al.},
  {\it Phys. Rev. Lett.} 94 (2005) 011803;
%
  {\it JHEP} 0301 (2003) 060;
  {\it PoS KAON} (2008) 008;
%
  hep-ph/0610246

\bibitem{Chen:2001qf}
  S.~Chen {\it et al.},  
  {\it Eur. Phys. J.} C 22 (2001) 31;
%
  {\it Nucl. Phys. Proc. Suppl.} 98 (2001) 319

\bibitem{Chetyrkin:1998ej}
  K.G.~Chetyrkin, J.H.~K\"uhn and A.A.~Pivovarov,
  {\it Nucl. Phys.} B 533 (1998) 473

\bibitem{Korner:2000wd}
  J.G.~Korner, F.~Krajewski and A.A.~Pivovarov,
  {\it Eur. Phys. J.} C 20 (2001) 259

\bibitem{Maltman:2006st}
  K.~Maltman and C.E.~Wolfe,
  {\it Phys. Lett.} B 639 (2006) 283

\bibitem{Kambor:2000dj}
  J.~Kambor and K.~Maltman,
  {\it Phys. Rev.} D 62 (2000) 093023,
%
   D 64 (2001) 093014

\bibitem{Maltman:1998qz}
  K.~Maltman,
  {\it Phys. Rev.} D 58 (1998) 093015

\bibitem{Leutwyler:1996qg}
  H.~Leutwyler,
  {\it Phys. Lett.} B 378 (1996) 313

\bibitem{Laiho:2009eu}
  J.~Laiho, E.~Lunghi and R.S.~Van de Water,
  {\it Phys. Rev.} D 81 (2010) 034503;
  http://latticeaverages.org/

\bibitem{Gamiz:2013wn}
  E.~G\'amiz,
  arXiv:1301.2206 [hep-ph]

\bibitem{Bazavov:2012cd} A.~Bazavov {\it et al.},
 {\it Phys. Rev.} D 87 (2013) 073012

\bibitem{Boyle:2013gsa} P.A.~Boyle {\it et al.},
 {\it JHEP} 1308 (2013) 132


\bibitem{Maltman:2008na}
  K.~Maltman,
  {\it Phys. Lett.} B 672 (2009) 257;
%
   {\it Nucl. Phys. Proc. Suppl.} 218 (2011) 146

\bibitem{Maltman:2009bh} K.~Maltman {\it et al.},
  {\it Nucl. Phys. Proc. Suppl.} 189 (2009) 175;
%
  {\it Int. J. Mod. Phys.} A 23 (2008) 3191



\bibitem{Hughes:1999fp}
  V.W.~Hughes and T.~Kinoshita,
  {\it Rev. Mod. Phys.} 71 (1999) S133

\bibitem{Davier:2004gb}
  M.~Davier and W.J.~Marciano,
  {\it Ann. Rev. Nucl. Part. Sci.} 54 (2004) 115

\bibitem{Passera:2004bj}
  M.~Passera,
  {\it J. Phys. G} 31 (2005) R75

\bibitem{Melnikov:2006sr}
  K.~Melnikov and A.~Vainshtein,
  {\it Theory of the muon anomalous magnetic moment},
  Springer Tracts Mod. Phys. -- Vol.~216 (2006) (Springer, Berlin - Heidelberg, 2006)

\bibitem{Miller:2007kk}
  J.P.~Miller, E.~de Rafael and B.L.~Roberts,
  {\it Rept. Prog. Phys.}  70 (2007) 795

\bibitem{Jegerlehner:2009ry}
  F.~Jegerlehner and A.~Nyffeler,
  {\it Phys. Rept.} 477 (2009) 1

\bibitem{Roberts:2010zz}
  B.L.~Roberts and W.J.~Marciano (eds.),
  {\it Lepton dipole moments},
  Advanced series on directions in high energy physics -- Vol. 20 (World Scientific, Singapore, 2009)

\bibitem{Miller:2012opa}
  J.P.~Miller, E.~de~Rafael, B.L.~Roberts and D.~St\"ockinger,
  {\it Ann. Rev. Nucl. Part. Sci.} 62 (2012) 237

\bibitem{Hanneke:2008tm}
  D.~Hanneke, S.~Fogwell and G.~Gabrielse,
  {\it Phys. Rev. Lett.} 100 (2008) 120801

\bibitem{Odom:2006zz}
  B.C.~Odom, D.~Hanneke, B. D'Urso and G.~Gabrielse,
  {\it Phys. Rev. Lett.} 97 (2006) 030801
   [Erratum-ibid.  99 (2007) 039902]

\bibitem{Bennett:2006fi} Muon $g-2$ Collaboration
  {\it Phys. Rev.} D 73 (2006) 072003;
%
   {\it Phys. Rev. Lett.} 92 (2004) 161802,
%
   89 (2002) 101804,
%
  86 (2001) 2227

\bibitem{Aoyama:2012wj} 
  T.~Aoyama, M.~Hayakawa, T.~Kinoshita and M.~Nio,
  {\it Phys. Rev. Lett.} 109 (2012) 111807,
%
111808;
%
  {\it PTEP} 2012 (2012) 01A107

\bibitem{Bouchendira:2010es} R.~Bouchendira {\it et. al.},
  {\it Phys. Rev. Lett.} 106 (2011) 080801

\bibitem{Czarnecki:1995wq}
  A.~Czarnecki, B.~Krause and W.J.~Marciano,
  {\it Phys. Rev.} D 52 (1995) 2619;
%
   {\it Phys. Rev. Lett.} 76 (1996) 3267

\bibitem{Czarnecki:2001pv}
  A.~Czarnecki and W.J.~Marciano,
  {\it Phys. Rev.} D 64 (2001) 013014

\bibitem{Czarnecki:2002nt}
  A.~Czarnecki, W.J.~Marciano and A.~Vainshtein,
  {\it Phys. Rev.} D 67 (2003) 073006
   [Erratum-ibid. D 73 (2006) 119901]

\bibitem{Gribouk:2005ee}
  T.~Gribouk and A.~Czarnecki,
  {\it Phys. Rev.} D 72 (2005) 053016

\bibitem{Knecht:2002hr}
  M.~Knecht, S.~Peris, M.~Perrottet and E.~De Rafael,
  {\it JHEP} 0211 (2002) 003

\bibitem{Peris:1995bb}
  S.~Peris, M.~Perrottet and E.~de Rafael,
  {\it Phys. Lett.} B 355 (1995) 523

\bibitem{Heinemeyer:2004yq}
  S.~Heinemeyer, D.~Stockinger and G.~Weiglein,
  {\it Nucl. Phys.} B 699 (2004) 103,
%
   B 690 (2004) 62

\bibitem{Kukhto:1992qv}
  T.V.~Kukhto, E.A.~Kuraev, Z.K.~Silagadze and A.~Schiller,
  {\it Nucl. Phys.} B 371 (1992) 567

\bibitem{Gnendiger:2013pva}
  C.~Gnendiger, D.~St\"ockinger and H.~St\"ockinger-Kim,
  {\it Phys. Rev.} D 88 (2013) 053005

\bibitem{BM:61}
C. Bouchiat and L. Michel, {\it J. Phys. Radium} 22 (1961) 121

\bibitem{Brodsky:1967sr}
  S.J.~Brodsky and E.~De Rafael,
  {\it Phys. Rev.} 168 (1968) 1620

\bibitem{Davier:2009zi}
  M.~Davier, A.~Hoecker, B.~Malaescu, C.~Z.~Yuan and Z.~Zhang,
  {\it Eur. Phys. J.} C 66 (2010) 1

\bibitem{FloresBaez:2006gf}
  F.V.~Flores-B\'aez, A.~Flores-Tlalpa, G.~L\'opez Castro and G.~Toledo S\'anchez,
  {\it Phys. Rev.} D 74 (2006) 071301;
%
  {\it Nucl. Phys. Proc. Suppl.} 169 (2007) 250

\bibitem{Davier:2003pw}
  M.~Davier, S.~Eidelman, A.~Hoecker and Z.~Zhang,
  {\it Eur. Phys. J.} C 31 (2003) 503,
%
  C 27 (2003) 497

\bibitem{Alemany:1997tn}
  R.~Alemany, M.~Davier and A.~Hoecker,
  {\it Eur. Phys. J.} C 2 (1998) 123

\bibitem{Davier:2013-ARNPS}
 M. Davier, {\it Ann. Rev. Nucl. Part. Sci.} 63 (2013) 407

\bibitem{Lees:2011zi} BaBar Collaboration,
  {\it Phys. Rev.} D 86 (2012) 012008,
%
  D 85 (2012) 112009,
%
   D 77 (2008) 092002,
%
   D 76 (2007) 012008,
%
   092006,
%
   092005,
%
   D 74 (2006) 091103,
%
   111103,
%
   D 73 (2006) 052003,
%
   012005,
%
   D 71 (2005) 052001,
%
   D 70 (2004) 072004

\bibitem{Akhmetshin:2008gz} CMD-2 Collaboration,
  {\it Phys. Lett.} B 669 (2008) 217,
%
   B 595 (2004) 101,
%
   B 578 (2004) 285,
%
   B 562 (2003) 173,
%
   B 527 (2002) 161,
%
   B 509 (2001) 217,
%
   B 489 (2000) 125,
%
   B 475 (2000) 190

\bibitem{Akhmetshin:2013xc}  CMD-3 Collaboration,
  {\it Phys. Lett.} B 723 (2013) 82

\bibitem{Ambrosino:2008gb} KLOE Collaboration,
  {\it Phys. Lett.} B 669 (2008) 223



\bibitem{Achasov:2012zza} SND Collaboration,
  {\it JETP Lett.} 94 (2012) 2,
%
   92 (2010) 80;
%
%
   {\it J. Exp. Theor. Phys.} 109 (2009) 379,
%
   96 (2003) 789;
%
  {\it Phys. Rev.} D 88 (2013) 054013,
%
  D 76 (2007) 072012,
%
   D 74 (2006) 014016,
%
   D 68 (2003) 052006,
%
   D 66 (2002) 032001,
%
   D 65 (2002) 032002,
%
   D 63 (2001) 072002;
%
  {\it Phys. Lett.} B 559 (2003) 171,
%
   B 486 (2000) 29,
%
   B 462 (1999) 365
%

\bibitem{Davier:2013wwg}
  M.~Davier,
  arXiv:1302.1907 [hep-ex]

\bibitem{Aoyama:2012fc}
  T.~Aoyama, M.~Hayakawa, T.~Kinoshita and M.~Nio,
  {\it Phys. Rev.} D 85 (2012) 093013,
%
   033007,
%
  D 84 (2011) 053003,
%
   D 83 (2011) 053002,
%
   053003,
%
   D 82 (2010) 113004,
%
   D 78 (2008) 113006,
%
   D 77 (2008) 053012;
%
   {\it Nucl. Phys.} B 796 (2008) 184,
%
   B 740 (2006) 138;
%
   {\it Phys. Rev. Lett.} 99 (2007) 110406

\bibitem{Aoyama:2010yt}
  T.~Aoyama, K.~Asano, M.~Hayakawa, T.~Kinoshita, M.~Nio and N.~Watanabe,
  {\it Phys. Rev.} D 81 (2010) 053009,
%
   D 78 (2008) 053005

\bibitem{Kinoshita:2005sm}
  T.~Kinoshita and M.~Nio,
  {\it Phys. Rev.} D 73 (2006) 053007,
%
   013003,
%
   D 70 (2004) 113001;
%
   {\it Phys. Rev. Lett.} 90 (2003) 021803

\bibitem{Lee:2013sx}
  R.~Lee, P.~Marquard, A.V.~Smirnov, V.A.~Smirnov and M.~Steinhauser,
  {\it JHEP} 1303 (2013) 162

\bibitem{Baikov:2012rr}
  P.A.~Baikov, K.G.~Chetyrkin, J.H.~K\"uhn and C.~Sturm,
  {\it Nucl. Phys.} B 867 (2013) 182

\bibitem{Baikov:2008si}
  P.A.~Baikov, K.G.~Chetyrkin and C.~Sturm,
  {\it Nucl. Phys. Proc. Suppl.} 183 (2008) 8

\bibitem{Kataev:2012kn}
  A.L.~Kataev,
  {\it Phys. Rev.} D 86 (2012) 013010,
%
   D 74 (2006) 073011;
%
   {\it Phys. Lett.} B 284 (1992) 401
   [Erratum-ibid. B 710 (2012) 710]

\bibitem{Aguilar:2008qj}
  J.-P.~Aguilar, D.~Greynat and E.~De Rafael,
  {\it Phys. Rev.} D 77 (2008) 093010

\bibitem{Czarnecki:1998rc}
  A.~Czarnecki and M.~Skrzypek,
  {\it Phys. Lett.} B 449 (1999) 354

\bibitem{Laporta:1996mq}
  S.~Laporta and E.~Remiddi,
  {\it Phys. Lett.} B 379 (1996) 283,
%
  B 301 (1993) 440

\bibitem{Laporta:1994md}
  S.~Laporta,
  {\it Phys. Lett.} B 328 (1994) 522,
%
   B 312 (1993) 495;
%
  {\it Nuovo Cim.} A 106 (1993) 675

\bibitem{Li:1992xf}
  G.~Li, R.~Mendel and M.A.~Samuel,
  {\it Phys. Rev.} D 47 (1993) 1723

\bibitem{Samuel:1990qf}
  M.A.~Samuel and G.-W.~Li,
  {\it Phys. Rev.} D 44 (1991) 3935
   [Erratum-ibid. D 48 (1993) 1879]

\bibitem{Broadhurst:1992za}
  D.J.~Broadhurst, A.L.~Kataev and O.V.~Tarasov,
  {\it Phys. Lett.} B 298 (1993) 445

\bibitem{Hagiwara:2006jt}
  K.~Hagiwara, A.~D.~Martin, D.~Nomura and T.~Teubner,
  {\it Phys. Lett.} B 649 (2007) 173,
%
   B 557 (2003) 69;
%
  {\it Phys. Rev.} D 69 (2004) 093003

\bibitem{Krause:1996rf}
  B.~Krause,
  {\it Phys. Lett.} B 390 (1997) 392

\bibitem{Prades:2009tw}
  J.~Prades, E.~de Rafael and A.~Vainshtein,
  in Ref.~\cite{Roberts:2010zz}


\bibitem{Bijnens:2007pz}
  J.~Bijnens and J.~Prades,
  {\it Mod. Phys. Lett.} A 22 (2007) 767

\bibitem{Melnikov:2003xd}
  K.~Melnikov and A.~Vainshtein,
  {\it Phys. Rev.} D 70 (2004) 113006

\bibitem{Knecht:2001qg}
  M.~Knecht, A.~Nyffeler, M.~Perrottet and E.~de Rafael,
  {\it Phys. Rev. Lett.} 88 (2002) 071802

\bibitem{Knecht:2001qf}
  M.~Knecht and A.~Nyffeler,
  {\it Phys. Rev.} D 65 (2002) 073034

\bibitem{Blokland:2001pb}
  I.R.~Blokland, A.~Czarnecki and K.~Melnikov,
  {\it Phys. Rev. Lett.} 88 (2002) 071803

\bibitem{Bijnens:2001cq}
  J.~Bijnens, E.~Pallante and J.~Prades,
  {\it Nucl. Phys.} B 626 (2002) 410,
%
  B 474 (1996) 379;
%
   {\it Phys. Rev. Lett.} 75 (1995) 1447
   [Erratum-ibid. 75 (1995) 3781]

\bibitem{Hayakawa:1997rq}
  M.~Hayakawa and T.~Kinoshita,
  {\it Phys. Rev.} D 57 (1998) 465
   [Erratum-ibid. D 66 (2002) 019902];
%
  hep-ph/0112102

\bibitem{RamseyMusolf:2002cy}
  M.J.~Ramsey-Musolf and M.B.~Wise,
  {\it Phys. Rev. Lett.} 89 (2002) 041601

\bibitem{Engel:2013kda}
  K.T.~Engel and M.J.~Ramsey-Musolf,
  arXiv:1309.2225 [hep-ph]

\bibitem{Goecke:2010if}
  T.~Goecke, C.S.~Fischer and R.~Williams,
  {\it Phys. Rev.} D 83 (2011) 094006
   [Erratum-ibid. D 86 (2012) 099901]

\bibitem{Blum:2013qu}
  T.~Blum, M.~Hayakawa and T.~Izubuchi,
  {\it PoS LATTICE 2012} (2012) 022

\bibitem{Hoferichter:2013ama}
  M.~Hoferichter, G.~Colangelo, M.~Procura and P.~Stoffer,
  arXiv:1309.6877 [hep-ph]

\bibitem{Zhang:2013oaw}
  Z.~Zhang,
  arXiv:1302.1896 [hep-ph]

\bibitem{Jegerlehner:2011ti}
  F.~Jegerlehner and R.~Szafron,
  {\it Eur. Phys. J.} C 71 (2011) 1632

\bibitem{Hagiwara:2011af}
  K.~Hagiwara, R.~Liao, A.D.~Martin, D.~Nomura and T.~Teubner,
  {\it J. Phys.} G 38 (2011) 085003

\bibitem{FNAL989} New Muon $(g-2)$ Collaboration, Fermilab Proposal 0989 (2009)\\
http://lss.fnal.gov/archive/test-proposal/0000/fermilab-proposal-0989.pdf

\bibitem{J-PARC} J-PARC $g-2$ Collaboration, J-PARC-PAC2009-12 (2009)\\
http://j-parc.jp/researcher/Hadron/en/pac\_1001/pdf/KEK\_J-PARC-PAC2009-12.pdf


\bibitem{Eidelman:2007sb}
  S.~Eidelman and M.~Passera,
  {\it Mod. Phys. Lett.} A 22 (2007) 159

\bibitem{Abdallah:2003xd} DELPHI Collaboration,
  {\it Eur. Phys. J.} C 35 (2004) 159

\bibitem{Achard:2004jj} L3 Collaboration,
  {\it Phys. Lett.} B 585 (2004) 53

\bibitem{Acciarri:1998iv} L3 Collaboration,
  {\it Phys. Lett.} B 434 (1998) 169

\bibitem{Ackerstaff:1998mt} OPAL Collaboration,
  {\it Phys. Lett.} B 431 (1998) 188

\bibitem{Escribano:1996wp}
  R.~Escribano and E.~Masso,
  {\it Phys. Lett.} B 395 (1997) 369,
%
   B 301 (1993) 419

\bibitem{Grifols:1990ha}
  J.A.~Grifols and A.~Mendez,
  {\it Phys. Lett.} B 255 (1991) 611
   [Erratum-ibid. B 259 (1991) 512]

\bibitem{delAguila:1990jg}
  F.~del Aguila and M.~Sher,
  {\it Phys. Lett.} B 252 (1990) 116

\bibitem{Domokos:1985rp}
  G.~Domokos, S.~Kovesi-Domokos, C.~Vaz and D.~Wurmser,
  {\it Phys. Rev.} D 32 (1985) 247

\bibitem{Silverman:1982ft}
  D.J.~Silverman and G.L.~Shaw,
  {\it Phys. Rev.} D 27 (1983) 1196

\bibitem{Schwinger:1948iu}
  J.S.~Schwinger,
  {\it Phys. Rev.} 73 (1948) 416,
%
  76 (1949) 790

\bibitem{Fael:2013ij}
  M.~Fael, L.~Mercolli and M.~Passera,
  arXiv:1301.5302 [hep-ph]

\bibitem{Laursen:1983sm}
  M.L.~Laursen, M.A.~Samuel and A.~Sen,
  {\it Phys. Rev.} D 29 (1984) 2652
   [Erratum-ibid. D 56 (1997) 3155]

\bibitem{Samuel:1992fm}
  M.A.~Samuel and G.~Li,
  {\it Int. J. Theor. Phys.} 33 (1994) 1471

\bibitem{Cornet:1995pw}
  F.~Cornet and J.I.~Illana,
  {\it Phys. Rev.} D 53 (1996) 1181

\bibitem{Billur:2013rva}
  A.A.~Billur and M.~Koksal,
  arXiv:1306.5620 [hep-ph]

\bibitem{Samuel:1990su}
  M.A.~Samuel, G.-W.~Li and R.~Mendel,
  {\it Phys. Rev. Lett.} 67 (1991) 668
   [Erratum-ibid. 69 (1992) 995]

\bibitem{Kim:1982ry}
  I.J.~Kim,
  {\it Nucl. Phys.} B 229 (1983) 251

\bibitem{Chen:1992wx} E761 Collaboration,
  {\it Phys. Rev. Lett.} 69 (1992) 3286

\bibitem{Bernabeu:2008ii}
  J.~Bernab\'eu, G.A.~Gonz\'alez-Sprinberg and J.~Vidal,
  {\it JHEP} 0901 (2009) 062

\bibitem{Bernabeu:2007rr}
  J.~Bernab\'eu, G.A.~Gonz\'alez-Sprinberg, J.~Papavassiliou and J.~Vidal,
  {\it Nucl. Phys.} B 790 (2008) 160


\bibitem{GonzalezSprinberg:2000mk}
  G.A.~Gonz\'alez-Sprinberg, A.~Santamaria and J.~Vidal,
  {\it Nucl. Phys.} B 582 (2000) 3



\bibitem{Baron:2013eja} ACME Collaboration,
  arXiv:1310.7534 [physics.atom-ph]

\bibitem{Bennett:2008dy} Muon ($g-2$) Collaboration,
  {\it Phys. Rev.} D 80 (2009) 052008


\bibitem{Ginges:2003qt}
  J.S.M.~Ginges and V.V.~Flambaum,
  {\it Phys. Rept.} 397 (2004) 63

\bibitem{Pospelov:2005pr}
  M.~Pospelov and A.~Ritz,
  {\it Annals Phys.} 318 (2005) 119

\bibitem{Raidal:2008jk} M.~Raidal {\it et al.},
  {\it Eur. Phys. J.} C 57 (2008) 13

\bibitem{Fukuyama:2012np}
  T.~Fukuyama,
  {\it Int. J. Mod. Phys.} A 27 (2012) 1230015

\bibitem{Engel:2013lsa}
  J.~Engel, M.J.~Ramsey-Musolf and U.~van Kolck,
  {\it Prog. Part. Nucl. Phys.} 71 (2013) 21

\bibitem{Jung:2013mg}
  M.~Jung,
  {\it JHEP} 1305 (2013) 168

\bibitem{Jung:2013hka}
  M.~Jung and A.~Pich,
  arXiv:1308.6283 [hep-ph]


\bibitem{Albrecht:2000yg} ARGUS Collaboration,
  {\it Phys. Lett.} B 485 (2000) 37

\bibitem{Bernreuther:1993nd}
  W.~Bernreuther, O.~Nachtmann and P.~Overmann,
  {\it Phys. Rev.} D 48 (1993) 78

\bibitem{Bernreuther:1996dr}
  W.~Bernreuther, A.~Brandenburg and P.~Overmann,
  {\it Phys. Lett.} B 391 (1997) 413


\bibitem{Inami:2002ah} Belle Collaboration,
  {\it Phys. Lett.} B 551 (2003) 16


\bibitem{Bernabeu:2006wf}
  J.~Bernab\'eu, G.A.~Gonz\'alez-Sprinberg and J.~Vidal,
  {\it Nucl. Phys.} B 763 (2007) 283,
%
   B 701 (2004) 87



\bibitem{Heister:2002ik} ALEPH Collaboration,
  {\it Eur. Phys. J.} C 30 (2003) 291;
%
  {\it Phys. Lett.} B 346 (1995) 371,
%
   B 297 (1992) 459

\bibitem{Acciarri:1998zc} L3 Collaboration,
  {\it Phys. Lett.} B 426 (1998) 207

\bibitem{Ackerstaff:1996gy} OPAL Collaboration,
  {\it Z. Phys.} C 74 (1997) 403,
%
  C 66 (1995) 31;
%
  {\it Phys. Lett.} B 281 (1992) 405


\bibitem{Bernreuther:1988jr}
  W.~Bernreuther, U.~Low, J.P.~Ma and O.~Nachtmann,
  {\it Z. Phys.} C 43 (1989) 117

\bibitem{Bernabeu:1994wh}
  J.~Bernab\'eu, G.A.~Gonz\'alez-Sprinberg, M.~Tung and J.~Vidal,
  {\it Nucl. Phys.} B 436 (1995) 474

\bibitem{Bernabeu:1993er}
  J.~Bernab\'eu, G.A.~Gonz\'alez-Sprinberg and J.~Vidal,
 {\it Phys. Lett.} B 326 (1994) 168

\bibitem{Bernreuther:1989kc}
  W.~Bernreuther and O.~Nachtmann,
  {\it Phys. Rev. Lett.} 63 (1989) 2787
   [Erratum-ibid. 64 (1990) 1072]

\bibitem{Bernreuther:1991xe}
  W.~Bernreuther, G.W.~Botz, O.~Nachtmann and P.~Overmann,
  {\it Z. Phys.} C 52 (1991) 567

\bibitem{Sanchez:1997kp}
  F.~S\'anchez,
  {\it Phys. Lett.} B 412 (1997) 137


\bibitem{Bernabeu:1989ct}
  J.~Bernab\'eu and N.~Rius,
  {\it Phys. Lett.} B 232 (1989) 127



\bibitem{Kobayashi:1973fv}
  M.~Kobayashi and T.~Maskawa,
  {\it Prog. Theor. Phys.} 49 (1973) 652

\bibitem{Bigi:2012kz}
  I.I.~Bigi,
  arXiv:1210.2968 [hep-ph]

\bibitem{Kiers:2012fy}
  K.~Kiers,
  arXiv:1212.6921 [hep-ph]

\bibitem{Tsai:1994rc}
  Y.S.~Tsai,
  {\it Phys. Rev.} D 51 (1995) 3172

\bibitem{Goozovat:1991nu}
  S.~Goozovat and C.~A.~Nelson,
  {\it Phys. Lett.} B 267 (1991) 128
   [Erratum-ibid. B 271 (1991) 468]

\bibitem{Nelson:1993zv} C.A.~Nelson {\it et al.},
  {\it Phys. Rev.} D 50 (1994) 4544

\bibitem{Kuhn:1996dv}
  J.~H.~K\"uhn and E.~Mirkes,
  Phys.\ Lett.\ B {\bf 398} (1997) 407

\bibitem{Kilian:1994ub}
  U.~Kilian, J.G.~Korner, K.~Schilcher and Y.L.~Wu,
  {\it Z. Phys.} C 62 (1994) 413


\bibitem{Kiers:2008mv}  K.~Kiers {\it et al.},
  {\it Phys. Rev.} D 78 (2008) 113008

\bibitem{Datta:2006kd}
  A.~Datta, K.~Kiers, D.~London, P.J.~O'Donnell and A.~Szynkman,
  {\it Phys. Rev.} D 75 (2007) 074007
   [Erratum-ibid. D 76 (2007) 079902]

\bibitem{Choi:1994ch}
  S.Y.~Choi, K.~Hagiwara and M.~Tanabashi,
  {\it Phys. Rev.} D 52 (1995) 1614


\bibitem{BABAR:2011aa} BaBar Collaboration,
  {\it Phys. Rev.} D 85 (2012) 031102
   [Erratum-ibid. D 85 (2012) 099904]

\bibitem{Bigi:2005ts}
  I.I.~Bigi and A.I.~Sanda,
  {\it Phys. Lett.} B 625 (2005) 47

\bibitem{Calderon:2007rg}
  G.~Calder\'on, D.~Delepine and G. L\'opez Castro,
  {\it Phys. Rev.} D 75 (2007) 076001

\bibitem{Grossman:2011zk}
  Y.~Grossman and Y.~Nir,
  {\it JHEP} 1204 (2012) 002

\bibitem{Bischofberger:2011pw} Belle Collaboration,
  {\it Phys. Rev. Lett.} 107 (2011) 131801

\bibitem{Bonvicini:2001xz} CLEO Collaboration,
  {\it Phys. Rev. Lett.}  88 (2002) 111803


\bibitem{Lees:2012xj} BaBar Collaboration,
  {\it Phys. Rev. Lett.} 109 (2012) 101802

\bibitem{Bozek:2010xy} Belle Collaboration,
  {\it Phys. Rev.} D 82 (2010) 072005;
%
  arXiv:0910.4301 [hep-ex]

\bibitem{Fajfer:2012jt}
  S.~Fajfer, J.F.~Kamenik, I.~Nisandzic and J.~Zupan,
  {\it Phys. Rev. Lett.} 109 (2012) 161801

\bibitem{Kamenik:2008tj}
  J.F.~Kamenik and F.~Mescia,
  {\it Phys. Rev.} D 78 (2008) 014003

\bibitem{Sakaki:2012ft}
  Y.~Sakaki and H.~Tanaka,
  {\it Phys. Rev.} D 87 (2013) 054002

\bibitem{Becirevic:2012jf}
  D.~Becirevic, N.~Kosnik and A.~Tayduganov,
  {\it Phys. Lett.} B 716 (2012) 208

\bibitem{Bailey:2012jg} J.A.~Bailey {\it et al.},
  {\it Phys. Rev. Lett.} 109 (2012) 071802


\bibitem{Adachi:2012mm} Belle Collaboration,
  {\it Phys.  Rev.  Lett.} 110 (2013) 131801,
%
  97 (2006) 251802;
%
  {\it Phys. Rev.} D 82 (2010) 071101

\bibitem{Lees:2012ju} BaBar Collaboration,
  {\it Phys. Rev.} D 88 (2013) 031102,
%
  81 (2010) 051101


\bibitem{Charles:2004jd} CKMfitter Group Collaboration,
  {\it Eur. Phys. J.} C 41 (2005) 1;
  http://ckmfitter.in2p3.fr

\bibitem{Celis:2012dk}
  A.~Celis, M.~Jung, X.-Q.~Li and A.~Pich,
  {\it JHEP} 1301 (2013) 054

\bibitem{Crivellin:2012ye}
  A.~Crivellin, C.~Greub and A.~Kokulu,
  {\it Phys. Rev.} D 86 (2012) 054014

\bibitem{Nierste:2008qe}
  U.~Nierste, S.~Trine and S.~Westhoff,
  {\it Phys. Rev.} D 78 (2008) 015006

\bibitem{Tanaka:1994ay}
  M.~Tanaka,
  {\it Z. Phys.} C 67 (1995) 321

\bibitem{Tanaka:2012nw}
  M.~Tanaka and R.~Watanabe,
  {\it Phys. Rev.} D 87 (2013) 034028,
%
   D 82 (2010) 034027

\bibitem{Pich:2009sp}
  A.~Pich and P.~Tuz\'on,
  {\it Phys. Rev.} D 80 (2009) 091702

\bibitem{Deshpande:2012rr}
  N.G.~Deshpande and A.~Menon,
  {\it JHEP} 1301 (2013) 025

\bibitem{He:2012zp}
  X.-G.~He and G.~Valencia,
  {\it Phys. Rev.} D 87 (2013) 014014


\bibitem{Fajfer:2012vx}
  S.~Fajfer, J.F.~Kamenik and I.~Nisandzic,
  {\it Phys. Rev.} D 85 (2012) 094025

\bibitem{Datta:2012qk}
  A.~Datta, M.~Duraisamy and D.~Ghosh,
  {\it Phys. Rev.} D 86 (2012) 034027

\bibitem{Duraisamy:2013pia}
  M.~Duraisamy and A.~Datta,
  {\it JHEP} 1309 (2013) 059


\bibitem{Hagiwara:1989gza}
  K.~Hagiwara, A.D.~Martin and M.F.~Wade,
  {\it Phys. Lett.} B 228 (1989) 144;
%
   {\it Nucl. Phys.} B 327 (1989) 569

\bibitem{Korner:1989qb}
  J.G.~Korner and G.A.~Schuler,
  {\it Z. Phys.} C 46 (1990) 93,
%
   C 38 (1988) 511
   [Erratum-ibid. C 41 (1989) 690]

\bibitem{Chen:2005gr}
  C.-H.~Chen and C.-Q.~Geng,
  {\it Phys. Rev.} D 71 (2005) 077501

\bibitem{McNeile:2012qf}
  C.~McNeile, C.T.H.~Davies, E.~Follana, K.~Hornbostel and G.P.~Lepage,
  {\it Phys. Rev.} D 86 (2012) 074503



\bibitem{Adamson:2013ue} Minos Collaboration,
  {\it Phys. Rev. Lett.} 110 (2013) 171801,
%
   107 (2011) 181802

\bibitem{Abe:2011sj} T2K Collaboration,
  {\it Phys. Rev. Lett.} 107 (2011) 041801;
%
  {\it Phys. Rev.} D 88 (2013) 032002


\bibitem{An:2012eh} Daya Bay Collaboration,
  {\it Phys. Rev. Lett.} 108 (2012) 171803;
%
  {\it Chin.  Phys.} C 37 (2013) 011001;
%
  arXiv:1310.6732 [hep-ex]

\bibitem{Abe:2011fz} Double-Chooz Collaboration,
  {\it Phys. Rev. Lett.} 108 (2012) 131801

\bibitem{Ahn:2012nd} Reno Collaboration,
  {\it Phys. Rev. Lett.} 108 (2012) 191802



\bibitem{GonzalezGarcia:2012sz}
  M.C.~Gonz\'alez-Garc\'{\i}a, M.~Maltoni, J.~Salvado and T.~Schwetz,
  {\it JHEP} 1212 (2012) 123;\\
  http://www.nu-fit.org

\bibitem{Weinberg:1979sa}
  S.~Weinberg,
  {\it Phys. Rev. Lett.} 43 (1979) 1566

\bibitem{GMRS:79}
  M.~Gell-Mann, P.~Ramond and R.~Slansky,
  {\it Complex Spinors and Unified Theories},
  in {\it Supergravity}, eds. P. van Nieuwenhuizen and D.Z. Freedman
  (North Holland, Stony Brook 1979) p.~315

\bibitem{YA:79}
  T.~Yanagida, in Proc. {\it Workshop on Unified Theory and Baryon Number in
    the Universe}, eds. O.~Sawada and A.~Sugamoto (KEK, 1979) p.~95


\bibitem{Marciano:2008zz}
  W.J.~Marciano, T.~Mori and J.M.~Roney,
  {\it Ann. Rev. Nucl. Part. Sci.} 58 (2008) 315

\bibitem{Barbieri:1995tw}
  R.~Barbieri, L.J.~Hall and A.~Strumia,
  {\it Nucl. Phys.} B 445 (1995) 219

\bibitem{Hisano:1997tc}
  J.~Hisano, D.~Nomura and T.~Yanagida,
  {\it Phys. Lett.} B 437 (1998) 351

\bibitem{Hisano:1998cx}
  J.~Hisano, D.~Nomura, Y.~Okada, Y.~Shimizu and M.~Tanaka,
  {\it Phys. Rev.} D 58 (1998) 116010

\bibitem{Hisano:1998fj}
  J.~Hisano and D.~Nomura,
  {\it Phys. Rev.} D 59 (1999) 116005

\bibitem{Huitu:1997bi}
  K.~Huitu, J.~Maalampi, M.~Raidal and A.~Santamaria,
  {\it Phys. Lett.} B 430 (1998) 355

\bibitem{Raidal:1997hq}
  M.~Raidal and A.~Santamaria,
  {\it Phys. Lett.} B 421 (1998) 250

\bibitem{Ilakovac:1999md}
  A.~Ilakovac,
  {\it Phys. Rev.} D 62 (2000) 036010

\bibitem{Kuno:1999jp}
  Y.~Kuno and Y.~Okada,
  {\it Rev. Mod. Phys.} 73 (2001) 151

\bibitem{Sato:2000zh}
  J.~Sato, K.~Tobe and T.~Yanagida,
  {\it Phys. Lett.} B 498 (2001) 189

\bibitem{Sato:2000ff}
  J.~Sato and K.~Tobe,
  {\it Phys. Rev.} D 63 (2001) 116010

\bibitem{Black:2002wh}
  D.~Black, T.~Han, H.-J.~He and M.~Sher,
  {\it Phys. Rev.} D 66 (2002) 053002

\bibitem{Cvetic:2002jy}
  G.~Cvetic, C.~Dib, C.S.~Kim and J.D.~Kim,
  {\it Phys. Rev.} D 66 (2002) 034008
   [Erratum-ibid. D 68 (2003) 059901]

\bibitem{Masiero:2002jn}
  A.~Masiero, S.K.~Vempati and O.~Vives,
  {\it Nucl. Phys.} B 649 (2003) 189

\bibitem{Illana:2003pj}
  J.I.~Illana and M.~Masip,
  {\it Eur. Phys. J.} C 35 (2004) 365;
%
   {\it Phys. Rev.} D 67 (2003) 035004

\bibitem{Cirigliano:2004tc}
  V.~Cirigliano, A.~Kurylov, M.J.~Ramsey-Musolf and P.~Vogel,
  {\it Phys. Rev. Lett.} 93 (2004) 231802;
%
  {\it Phys. Rev.} D 70 (2004) 075007

\bibitem{Cirigliano:2005ck}
  V.~Cirigliano, B.~Grinstein, G.~Isidori and M.B.~Wise,
  {\it Nucl. Phys.} B 728 (2005) 121

\bibitem{Choudhury:2005jh}
  S.R.~Choudhury, N.~Gaur and A.~Goyal,
  {\it Phys. Rev.} D 72 (2005) 097702

\bibitem{Arganda:2005ji}
  E.~Arganda and M.J.~Herrero,
  {\it Phys. Rev.} D 73 (2006) 055003

\bibitem{Yaguna:2005qn}
  C.E.~Yaguna,
  {\it Int. J. Mod. Phys.} A 21 (2006) 1283

\bibitem{Chen:2006hp}
  C.-H.~Chen and C.-Q.~Geng,
  {\it Phys. Rev.} D 74 (2006) 035010

\bibitem{Cirigliano:2006su}
  V.~Cirigliano and B.~Grinstein,
  {\it Nucl. Phys.} B 752 (2006) 18

\bibitem{Cirigliano:2006nu}
  V.~Cirigliano, G.~Isidori and V.~Porretti,
  {\it Nucl. Phys.} B 763 (2007) 228

\bibitem{Agashe:2006iy}
  K.~Agashe, A.E.~Blechman and F.~Petriello,
  {\it Phys. Rev.} D 74 (2006) 053011

\bibitem{Antusch:2006vw}
  S.~Antusch, E.~Arganda, M.J.~Herrero and A.M.~Teixeira,
  {\it JHEP} 0611 (2006) 090

\bibitem{Choudhury:2006sq}
  S.R.~Choudhury, A.S.~Cornell, A.~Deandrea, N.~Gaur and A.~Goyal,
  {\it Phys. Rev.} D 75 (2007) 055011

\bibitem{Blanke:2007db}
  M.~Blanke, A.J.~Buras, B.~Duling, A.~Poschenrieder and C.~Tarantino,
  {\it JHEP} 0705 (2007) 013

\bibitem{Arganda:2007jw}
  E.~Arganda, M.J.~Herrero and A.M.~Teixeira,
  {\it JHEP} 0710 (2007) 104

\bibitem{Arganda:2008jj}
  E.~Arganda, M.J.~Herrero and J.~Portol\'es,
  {\it JHEP} 0806 (2008) 079

\bibitem{Herrero:2009tm}
  M.J.~Herrero, J.~Portol\'es and A.M.~Rodr\'{\i}guez-S\'anchez,
  {\it Phys. Rev.} D 80 (2009) 015023

\bibitem{Cirigliano:2009bz}
  V.~Cirigliano, R.~Kitano, Y.~Okada and P.~Tuz\'on,
  {\it Phys. Rev.} D 80 (2009) 013002

\bibitem{Li:2009yr}
  Z.-H.~Li, Y.~Li and H.-X.~Xu,
  {\it Phys. Lett.} B 677 (2009) 150

\bibitem{Benbrik:2008ik}
  R.~Benbrik and C.-H.~Chen,
  {\it Phys. Lett.} B 672 (2009) 172

\bibitem{Arhrib:2009xf}
  A.~Arhrib, R.~Benbrik and C.-H.~Chen,
  {\it Phys. Rev.} D 81 (2010) 113003

\bibitem{Liu:2009su}
  W.~Liu, C.-X.~Yue and J.~Zhang,
  {\it Eur. Phys. J.} C 68 (2010) 197

\bibitem{Li:2010vf}
  W.-J.~Li, Y.-Y.~Fan, G.-W.~Liu and L.-X.~Lu,
  {\it Int. J. Mod. Phys.} A 25 (2010) 4827

\bibitem{Hisano:2010es}
  J.~Hisano, S.~Sugiyama, M.~Yamanaka and M.J.S.~Yang,
  {\it Phys. Lett.} B 694 (2011) 380

\bibitem{delAguila:2011wk}
  F.~del Aguila, J.I.~Illana and M.D.~Jenkins,
  {\it JHEP} 1103 (2011) 080,
%
   1009 (2010) 040,
%
   0901 (2009) 080

\bibitem{Kaneko:2011qi} S.~Kaneko {\it et al.},
  {\it Phys. Rev.} D 83 (2011) 115005

\bibitem{Decamp:1991uy} ALEPH Collaboration,
  {\it Phys. Rept.} 216 (1992) 253

\bibitem{Abreu:1996mj} DELPHI Collaboration,
  {\it Z. Phys.} C 73 (1997) 243

\bibitem{Adriani:1993sy} L3 Collaboration,
  {\it Phys. Lett.} B 316 (1993) 427

\bibitem{Akers:1995gz} OPAL Collaboration,
  {\it Z. Phys.} C 67 (1995) 555

\bibitem{Aaij:2013cby} LHCb Collaboration,
  {\it Phys. Rev. Lett.} 111 (2013) 141801

\bibitem{Adam:2013mnn} MEG Collaboration,
  {\it Phys. Rev. Lett.} 110 (2013) 201801,
%
  107 (2011) 171801

\bibitem{Bellgardt:1987du} SINDRUM Collaboration,
  {\it Nucl. Phys.} B 299 (1988) 1

\bibitem{Bolton:1988af} R.D.~Bolton {\it et al.},
  {\it Phys. Rev.} D 38 (1988) 2077

\bibitem{Lees:2010ez}  BaBar Collaboration,
  {\it Phys. Rev.} D 81 (2010) 111101,
%
   79 (2009) 012004;
%
   {\it Phys. Rev. Lett.} 104 (2010) 021802,
%
   103 (2009) 021801,
%
   100 (2008) 071802,
%
   99 (2007) 251803,
%
   98 (2007) 061803,
%
   96 (2006) 041801,
%
   95 (2005) 191801,
%
   041802,
%
   92 (2004) 121801

\bibitem{Miyazaki:2012mx} Belle Collaboration,
  {\it Phys. Lett.} B 719 (2013) 346,
%
   B 699 (2011) 251,
%
   B 692 (2010) 4,
%
   B 682 (2010) 355,
%
   B 672 (2009) 317,
%
   B 666 (2008) 16,
%
   B 664 (2008) 35,
%
   B 660 (2008) 154,
%
   B 648 (2007) 341,
%
   B 640 (2006) 138,
%
   B 639 (2006) 159,
%
   B 622 (2005) 218,
%
   B 613 (2005) 20,
%
   B 589 (2004) 103;
%
   {\it Phys. Rev. Lett.} 93 (2004) 081803,
%
   92 (2004) 171802

\bibitem{Bertl:2006up} SINDRUM II Collaboration,
  {\it Eur. Phys. J.} C 47 (2006) 337;
%
  {\it Phys. Rev. Lett.}  76 (1996) 200;
%
  {\it Phys. Lett.} B 317 (1993) 631


\bibitem{Aaij:2013fia} LHCb Collaboration,
   {\it Phys. Lett.} B 724 (2013) 36

\bibitem{Baldini:2013ke} MEG Collaboration,
  arXiv:1301.7225 [physics.ins-det]

\bibitem{Blondel:2013ia} A. Blondel {\it et al.},
  arXiv:1301.6113 [physics.ins-det]

\bibitem{Kuno:2013mha}
  Y.~Kuno [COMET Collaboration],
  {\it PTEP} 2013 (2013) 022C01

\bibitem{Abrams:2012er} Mu2e Collaboration,
  arXiv:1211.7019 [physics.ins-det]

\bibitem{Knoepfel:2013ouy} Mu2e Collaboration,
  arXiv:1307.1168 [physics.ins-det]

\bibitem{Barlow:2011zza}
  R.J.~Barlow,
  {\it Nucl. Phys. Proc. Suppl.} 218 (2011) 44

\bibitem{Blankenburg:2012ex}
  G.~Blankenburg, J.~Ellis and G.~Isidori,
  {\it Phys. Lett.} B 712 (2012) 386

\bibitem{Harnik:2012pb}
  R.~Harnik, J.~Kopp and J.~Zupan,
  {\it JHEP} 1303 (2013) 026

\bibitem{Davidson:2012ds}
  S.~Davidson and P.~Verdier,
  {\it Phys. Rev.} D 86 (2012) 111701

\bibitem{Celis:2013xja}
  A.~Celis, V.~Cirigliano and E.~Passemar,
  arXiv:1309.3564 [hep-ph]

\bibitem{Petrov:2013vka}
  A.A.~Petrov and D.V.~Zhuridov,
  arXiv:1308.6561 [hep-ph]

\bibitem{Daub:2012mu}
  J.T.~Daub, H.K.~Dreiner, C.~Hanhart, B.~Kubis and U.G.~Meissner,
  {\it JHEP} 1301 (2013) 179

\bibitem{Miyazaki:2005ng} Belle Collaboration,
  {\it Phys. Lett.} B 632 (2006) 51
%

\bibitem{Abbiendi:1998nj} OPAL Collaboration,
  {\it Phys. Lett.} B 447 (1999) 157

\bibitem{BABAR:2012aa} BaBar Collaboration,
  {\it Phys. Rev.} D 85 (2012) 071103,
%
   D 83 (2011) 091101;
%
  arXiv:1310.8238 [hep-ex]

\bibitem{Aaij:2011ex} LHCb Collaboration,
  {\it Phys. Rev. Lett.} 108 (2012) 101601;
%
   {\it Phys. Rev.} D 85 (2012) 112004

\bibitem{Seon:2011ni} Belle Collaboration,
  {\it Phys. Rev.} D 84 (2011) 071106

\bibitem{Edwards:2002kq} CLEO Collaboration,
  {\it Phys. Rev.} D 65 (2002) 111102

\bibitem{Ali:2001gsa}
  A.~Ali, A.V.~Borisov and N.B.~Zamorin,
  {\it Eur. Phys. J} C 21 (2001) 123

\bibitem{Atre:2005eb}
  A.~Atre, V.~Barger and T.~Han,
  {\it Phys. Rev.} D 71 (2005) 113014

\bibitem{Ivanov:2004ch}
  M.A.~Ivanov and S.G.~Kovalenko,
  {\it Phys. Rev.} D 71 (2005) 053004

\bibitem{Atre:2009rg}
  A.~Atre, T.~Han, S.~Pascoli and B.~Zhang,
  {\it JHEP} 0905 (2009) 030

\bibitem{Helo:2010cw}
  J.C.~Helo, S.~Kovalenko and I.~Schmidt,
  {\it Nucl. Phys.} B 853 (2011) 80

\bibitem{Zhang:2010um}
  J.-M.~Zhang and G.-L.~Wang,
  {\it Eur. Phys. J.} C 71 (2011) 1715

\bibitem{Bao:2012vq}
  S.-S.~Bao, H.-L.~Li, Z.-G.~Si and Y.-B.~Yang,
  {\it Commun. Theor. Phys.} 59 (2013) 472

\bibitem{Cvetic:2010rw}
  G.~Cvetic, C.~Dib, S.K.~Kang and C.S.~Kim,
  {\it Phys. Rev.} D 82 (2010) 053010

\bibitem{Cvetic:2012hd}
  G.~Cvetic, C.~Dib and C.S.~Kim,
  {\it JHEP} 1206 (2012) 149

\bibitem{Quintero:2011yh}
  N.~Quintero, G.~L\'opez Castro and D.~Delepine,
  {\it Phys. Rev.} D 84 (2011) 096011
   [Erratum-ibid. D 86 (2012) 079905]

\bibitem{Quintero:2012jy}
  N.~Quintero,
  {\it Phys. Rev.} D 87 (2013) 056005

\bibitem{Castro:2013jsn}
  G.~L\'opez~Castro and N.~Quintero,
  {\it Phys. Rev.} D 87 (2013) 077901,
%
   D 85 (2012) 076006
   [Erratum-ibid. D 86 (2012) 079904]



\bibitem{Aad:2011fu} ATLAS Collaboration,
  {\it Phys. Lett.} B 706 (2012) 276

\bibitem{CMS:2011asa} CMS Collaboration,
  CMS-PAS-EWK-11-002 (2011)

\bibitem{Aad:2011kt} ATLAS Collaboration,
  {\it Phys. Rev.} D 84 (2011) 112006;
%
   ATLAS-CONF-2012-006 (2012)

\bibitem{Chatrchyan:2011nv} CMS Collaboration,
  {\it JHEP} 1108 (2011) 117

\bibitem{Aaij:2012bi} LHCb Collaboration,
  {\it JHEP} 1301 (2013) 111

\bibitem{Aad:2012vip} ATLAS Collaboration,
  {\it Eur. Phys. J.} C 73 (2013) 2328;
%
  {\it Phys. Lett.} B 717 (2012) 89


\bibitem{Chatrchyan:2013kff} CMS Collaboration,
  {\it Eur. Phys. J.} C 73 (2013) 2386;
%
   {\it Phys. Rev.} D 85 (2012) 112007



\bibitem{Melnikov:2006kv}
  K.~Melnikov and F.~Petriello,
  {\it Phys. Rev.} D 74 (2006) 114017;
%
   {\it Phys. Rev. Lett.} 96 (2006) 231803

\bibitem{Anastasiou:2003ds}
  C.~Anastasiou, L.J.~Dixon, K.~Melnikov and F.~Petriello,
  {\it Phys. Rev.} D 69 (2004) 094008

\bibitem{Catani:2012qa}
  S.~Catani, L.~Cieri, D.~de Florian, G.~Ferrera and M.~Grazzini,
  {\it Eur. Phys. J} C 72 (2012) 2195;
%
   {\it Phys. Rev. Lett.} 103 (2009) 082001

\bibitem{Martin:2009iq}
  A.D.~Martin, W.J.~Stirling, R.S.~Thorne and G.~Watt,
  {\it Eur. Phys. J.} C 63 (2009) 189

\bibitem{Moch:2012mk}
  S.~Moch, P.~Uwer and A.~Vogt,
  {\it Phys. Lett.} B 714 (2012) 48

\bibitem{Moch:2008qy}
  S.~Moch and P.~Uwer,
  {\it Phys. Rev.} D 78 (2008) 034003;
%
   {\it Nucl. Phys. Proc. Suppl.} 183 (2008) 75


\bibitem{Kidonakis:2010dk}
  N.~Kidonakis,
  {\it Phys. Rev.} D 82 (2010) 114030


\bibitem{Albajar:1986fn} UA1 Collaboration,
  {\it Phys. Lett.} B 185 (1987) 233
   [Addendum-ibid. B 191 (1987) 462]

\bibitem{Abe:1991fb} CDF Collaboration,
  {\it Phys. Rev. Lett.} 68 (1992) 3398,
%
    79 (1997) 3585,
%
  109 (2012) 192001;
%
    {\it Phys. Lett.} B 639 (2006) 172;
%
   {\it Phys. Rev.} D 75 (2007) 092004

\bibitem{Abbott:1999pk} D0 Collaboration,
  {\it Phys. Rev. Lett.} 84 (2000) 5710;
%
    {\it Phys. Lett.} B 670 (2009) 292,
%
   B 679 (2009) 177;
%
   {\it Phys. Rev.} D 82 (2010) 071102


\bibitem{Aad:2012cia}  ATLAS Collaboration,
  {\it Eur. Phys. J.} C 72 (2012) 2062

\bibitem{Rouge:1990kv}
  A.~Rouge,
  {\it Z. Phys.} C 48 (1990) 75

\bibitem{Hagiwara:1989fn}
  K.~Hagiwara, A.D.~Martin and D.~Zeppenfeld,
  {\it Phys. Lett.} B 235 (1990) 198

\bibitem{Pich:2013vta}
  A.~Pich,
  arXiv:1307.7700 [hep-ph]

\bibitem{Ellis:2013lra}
  J.~Ellis and T.~You,
  {\it JHEP} 1306 (2013) 103

\bibitem{Giardino:2013bma} P.P.~Giardino {\it et al.},
  arXiv:1303.3570 [hep-ph]

\bibitem{Cheung:2013kla}
  K.~Cheung, J.S.~Lee and P.-Y.~Tseng,
  {\it JHEP} 1305 (2013) 134

\bibitem{Falkowski:2013dza}
  A.~Falkowski, F.~Riva and A.~Urbano,
  {\it JHEP} 1311 (2013) 111


\bibitem{Celis:2013rcs}
  A.~Celis, V.~Ilisie and A.~Pich,
  {\it JHEP} 1307 (2013) 053;
%
  arXiv:1310.7941 [hep-ph]

\bibitem{Aad:2012ypy} ATLAS Collaboration,
  {\it Phys. Lett.} B 723 (2013) 15,
%
  B 719 (2013) 242;
%
  ATLAS-CONF-2013-066 (2013)


\bibitem{Chatrchyan:2012hd} CMS Collaboration,
  {\it Phys. Lett.} B 716 (2012) 82

\bibitem{ATLAS:2013oea} ATLAS Collaboration,
 {\it JHEP} 1306 (2013) 033

\bibitem{Chatrchyan:2012sv} CMS Collaboration,
  {\it Phys. Rev. Lett.} 110 (2013) 081801


\bibitem{ATLAS:2011mha} ATLAS Collaboration,
  ATLAS-CONF-2011-132 (2011)

\bibitem{Chatrchyan:2012vp} CMS Collaboration,
  {\it Phys. Lett.} B 713 (2012) 68;
%
  CMS-PAS-HIG-12-050 (2012);
%
  CMS-PAS-HIG-13-021 (2013)

\bibitem{Aad:2012tj} ATLAS Collaboration,
  {\it JHEP} 1206 (2012) 039

\bibitem{Chatrchyan:2012vca} CMS Collaboration,
  {\it JHEP} 1207 (2012) 143

\bibitem{ATLAS:2012ht} ATLAS Collaboration,
  {\it Eur. Phys. J.} C 72 (2012) 2215;
%
   {\it Phys. Lett.} B 714 (2012) 197,
%
   B 714 (2012) 180;
%
   ATLAS-CONF-2013-026 (2013),
%
  ATLAS-CONF-2013-028 (2013)


\bibitem{Chatrchyan:2013dsa} CMS Collaboration,
  {\it Eur. Phys. J.} C 73 (2013) 2493;
%
   CMS-PAS-SUS-11-029 (2011)


\bibitem{Lees:2012te} BaBar Collaboration,
  {\it Phys. Rev.} D 88 (2013) 071102;
%
  {\it Phys. Rev. Lett.} 103 (2009) 181801

\end{thebibliography}
\end{document}